\documentclass[12pt]{article}
\pdfoutput=1
\usepackage[nosort]{cite}
\usepackage[height=8.85in,width=6.45in]{geometry}
\usepackage{pifont}

\usepackage{times}

\usepackage{comment}
\usepackage{booktabs}
\usepackage[utf8]{inputenc}
\usepackage{amsmath}
\usepackage{amssymb}
\usepackage{mathtools}
\numberwithin{equation}{section}
\usepackage{slashed}
\usepackage{braket}
\usepackage[svgnames]{xcolor}
\usepackage{url}
\usepackage[colorlinks,citecolor=DarkGreen,linkcolor=FireBrick,linktocpage]{hyperref}
\usepackage{cite}
\usepackage{graphicx}
\usepackage{float}
\usepackage{tikz}
\usepackage{tikz-cd}
\usepackage{tikzit}

\tikzstyle{junction}=[fill=black, draw=black, shape=circle, inner sep=.8pt]
\tikzstyle{boundarylines}=[draw=blue, thick]

\tikzstyle{boundaryrectangle}=[-, draw=none, fill={rgb,255: red,191; green,191; blue,191}]
\tikzstyle{boundarylines}=[-, draw=blue, thick]
\tikzstyle{bulklines}=[-, draw=red, thick]
\tikzstyle{dd}=[-, dashed]
\tikzstyle{line}=[-, thick]

\usepackage{courier}
\usepackage{bm}

\usepackage{dashbox}
\usepackage{caption}
\usepackage{subcaption}
\usepackage{enumitem}
 
\usepackage{footmisc}
\usepackage{mdframed}

\newenvironment{claim}{  \begin{mdframed}[linecolor=black!0,backgroundcolor=black!10]\noindent\itshape\ignorespaces}{\end{mdframed}}

\usepackage{blkarray}
\usepackage{arydshln}
\usepackage{dsfont}
\usetikzlibrary{arrows,decorations.markings,cd}
\tikzset{
->-/.style args={#1rotate#2}{decoration={markings, mark=at position #1 with {\arrow[scale=1.5,rotate = #2 ]{stealth}}}, postaction={decorate}}
}

\tikzset{line/.style={line width=0.25mm},
curve/.style={line,smooth,tension=1},
->-/.style={decoration={
  markings,
  mark=at position #1 with {\arrow[>=stealth]{>}}},postaction={decorate}},
-<-/.style={decoration={
  markings,
  mark=at position #1 with {\arrow[>=stealth]{<}}},postaction={decorate}},
}

\usepackage{calc}

\usepackage{accents}

\newcommand{\ie}{\begin{equation}\begin{aligned}}
\newcommand{\fe}{\end{aligned}\end{equation}}

\renewcommand{\title}[1]{\vbox{\center\LARGE{#1}}\vspace{5mm}}
\renewcommand{\author}[1]{\vbox{\center#1}\vspace{5mm}}
\newcommand{\address}[1]{\vbox{\center\em#1}}

\makeatletter
\newsavebox{\@brx}
\newcommand{\llangle}[1][]{\savebox{\@brx}{\(\m@th{#1\langle}\)}%
  \mathopen{\copy\@brx\kern-0.5\wd\@brx\usebox{\@brx}}}
\newcommand{\rrangle}[1][]{\savebox{\@brx}{\(\m@th{#1\rangle}\)}%
  \mathclose{\copy\@brx\kern-0.5\wd\@brx\usebox{\@brx}}}
\makeatother

\def\d{\mathsf{d}}

\def\cH{\mathcal{H}}

\def\cC{\mathcal{C}}

\def\cO{\mathcal{O}}

\def\CB{{\mathcal B}}
\def\CC{{\mathcal C}}

\def\CH{{\mathcal H}}

\def\CO{{\mathcal O}}

\newcommand{\Z}{\mathbb{Z}}

\IfFileExists{ying}{
	\usetikzlibrary{external}
	\tikzexternalize[prefix=./figures/]
}

\newcommand{\TV}{\mathrm{TV}}

\newcommand{\qd}{\mathsf{d}}

\newcommand{\x}[4]{
	\begin{gathered}
		\begin{tikzpicture}[baseline=(X.base)]
			\draw [thick, 
			decoration = {markings, mark=at position 0.8 with {\arrow[scale=1]{stealth}}}, postaction=decorate] (-1,0) -- (1,0) node[right] (X) {$#3$};
			\draw [thick, 
			decoration = {markings, mark=at position 0.9 with {\arrow[scale=1]{stealth}}}, postaction=decorate] (0,-0.5) node[below]{$#1$} -- (0,0.5) node[above]{$#2$};
			\fill [
			] (0,0) circle (0.05) node [above right] {\footnotesize $#4$};
		\end{tikzpicture}
	\end{gathered}
}

\usetikzlibrary{calc,shapes.geometric}
\pgfmathparse{atan2(0,1)}
\ifdim\pgfmathresult pt=0pt
\tikzset{declare function={atanXY(\x,\y)=atan2(\y,\x);atanYX(\y,\x)=atan2(\y,\x);}}
\else                 
\tikzset{declare function={atanXY(\x,\y)=atan2(\x,\y);atanYX(\y,\x)=atan2(\x,\y);}}
\fi

\usetikzlibrary{decorations.pathmorphing}

\newcommand{\doubleslash}{/\hspace{-0.05in}/}

\begin{document}

\begin{titlepage}
    \hfill      YITP-SB-2024-18
    \\

\title{Generalized Tube Algebras, Symmetry-Resolved Partition Functions, and Twisted Boundary States}

\author{Yichul Choi${}^{1,2,3}$, Brandon C.\  Rayhaun${}^{1,2}$, and Yunqin Zheng${}^{2,4}$}

        \address{${}^{1}$School of Natural Sciences, Institute for Advanced Study, Princeton, NJ\\
        ${}^{2}$C.\ N.\ Yang Institute for Theoretical Physics, Stony Brook University, Stony Brook, NY\\
        ${}^{3}$Simons Center for Geometry and Physics, Stony Brook University, Stony Brook, NY\\
        ${}^{4}$Kavli Institute for Theoretical Sciences, University of Chinese Academy of Sciences, Beijing}

\abstract

We introduce a class of generalized tube algebras which describe how finite, non-invertible global symmetries of bosonic 1+1d QFTs act on operators which sit at the intersection point of a collection of boundaries and interfaces. 
We develop a 2+1d symmetry topological field theory (SymTFT) picture of boundaries and interfaces which, among other things, allows us to deduce the representation theory of these algebras.
In particular, we initiate the study of a character theory, echoing that of finite groups, and demonstrate how many representation-theoretic quantities can be expressed as partition functions of the SymTFT on various backgrounds, which in turn can be evaluated explicitly in terms of generalized half-linking numbers.

We use this technology to explain how the torus and annulus partition functions of a 1+1d QFT can be refined with information about its symmetries. 
We are led to a vast generalization of Ishibashi states in CFT: to any multiplet of conformal boundary conditions which transform into each other under the action of a symmetry, we associate a collection of generalized Ishibashi states, in terms of which the twisted sector boundary states of the theory and all of its orbifolds can be obtained as linear combinations.
We derive a generalized Verlinde formula involving the characters of the boundary tube algebra which ensures that our formulas for the twisted sector boundary states respect open-closed duality.
Our approach does not rely on rationality or the existence of an extended chiral algebra; however, in the special case of a diagonal RCFT with chiral algebra $V$ and modular tensor category $\mathcal{C}$, our formalism produces explicit closed-form expressions --- in terms of the $F$-symbols and $R$-matrices of $\mathcal{C}$, and the characters of $V$ --- for the twisted Cardy states, and the torus and annulus partition functions decorated by Verlinde lines.

\end{titlepage}

\eject

\setcounter{tocdepth}{2}
\tableofcontents

\section{Introduction}

Recent years have witnessed an increasing appreciation for the role of extended operators and defects in quantum field theories.
Whereas old-fashioned dogma asserts that a conformal field theory can be understood as a list of local operators along with their scaling dimensions and operator product expansions, the modern perspective is that the spectrum of extended objects also probes subtle and useful information about the global structure of a theory which is often inaccessible to the local operator data. See \cite{Poland:2018epd, Billo:2016cpy} for recent reviews. 

An important class of extended operators are those which behave topologically in correlation functions. Such topological operators are celebrated because they share many properties in common with standard symmetries which are taught in quantum field theory courses, and correspondingly they have many of the same kinds of applications \cite{Chang:2018iay, Thorngren:2019iar, Thorngren:2021yso, Komargodski:2020mxz, Kong:2020cie, Verlinde:1988sn, Moore:1988qv, Bhardwaj:2017xup, Carqueville:2012dk, Brunner:2013xna, Lin:2019hks, Ji:2019ugf, Fuchs:2002cm, Frohlich:2006ch, Frohlich:2009gb}. The structures they lead to are therefore often referred to as \emph{generalized global symmetries} \cite{Gaiotto:2014kfa}. Particularly in low dimensions, such symmetries can be sharply characterized using the language of tensor categories \cite{etingof2015tensor}.

The constraining power of topological operators lies in their interaction with the various non-topological sectors of a quantum field theory.
For example, because topological operators can be continuously deformed, one can often think of them as ``acting'' on non-topological objects by fusion or linking. One may attempt to design partition functions which probe these symmetry actions: for example, one way to achieve this is to couple the theory to a background gauge field for the symmetry, though there are others, as we will see. 
We colloquially refer to partition functions which are ``souped-up'' with additional information about the symmetry structure of a theory as being \emph{symmetry-resolved}.

We offer a systematic treatment of these kinds of topics in the context of 1+1d quantum field theories with boundaries, interfaces, and junction operators, though we expect that much of the picture we paint should suitably generalize to higher dimensions as well. 
A companion paper \cite{Choi:2024wfm} uses the machinery developed here to derive a non-invertible symmetry-resolved Affleck-Ludwig-Cardy formula, and applies it to the problem of symmetry-resolved entanglement entropy in 1+1d CFTs.

In the rest of the introduction, we describe our results in more detail.

\subsection{Generalized tube algebras}

The starting point of our analysis is the introduction of \emph{generalized tube algebras} which encode how the topological line operators of a 1+1d quantum field theory act on local operators which appear at the junction of a collection of boundaries and interfaces. 

To motivate our construction, recall that the standard way that a topological line $c$ acts on a local operator $\mathcal{O}(x)$ is by circling a loop of $c$ around $\mathcal{O}(x)$ and shrinking the loop until it has zero size, 
\begin{align}
    \input{figures/encirclinglocalop.tikz}\, ,
\end{align}
producing a new operator $\mathcal{O}'(x)$ in the process. 
For example, in the case that $c=e^{i\alpha \oint \star J}$ with $J$ the Noether current of a $U(1)$ symmetry, then $\mathcal{O}'(x) = e^{i\alpha q_{\mathcal{O}}}\mathcal{O}(x)$, where $q_{\mathcal{O}}$ is the charge of $\mathcal{O}(x)$ with respect to the $U(1)$. 

More generally, if one expands their universe to include twisted sector local operators --- i.e.\ local operators which live at the endpoint of some topological line --- then one can consider more general \emph{lasso operators} \cite{Chang:2018iay},
\begin{align}\label{eqn:lassoopsintro}
    \input{figures/lassointro.tikz}\, ,
\end{align}
where $a$, $b$, $c$, and $d$ are topological line operators, and $y$ and $z$ are suitable topological junctions connecting them. Such lassos can convert operators living in the $a$-twisted sector for some topological line $a$, to operators living in the $b$-twisted sector for some (generally distinct) line $b$. Given a finite collection $\CC$ of line operators which are closed under parallel fusion (and therefore form a fusion category), one can then define a finite-dimensional $C^\ast$ algebra $\mathrm{Tube}(\CC)$ which is generated by lassos of the kind appearing in Equation \eqref{eqn:lassoopsintro} \cite{Lin:2022dhv}. 

Because lassos can move one between different twisted sectors, the tube algebra $\mathrm{Tube}(\CC)$ most naturally acts on the \emph{extended Hilbert space}, 
\begin{align}
    \mathcal{H}_{\mathcal{C}} = \bigoplus_{a\in\mathrm{Irr}(\mathcal{C})} \mathcal{H}_a
\end{align}
which is defined as the direct sum over all the twisted sectors of a theory with respect to a chosen collection of line operators $\CC$ which are closed under fusion.

We would like to write down a suitable generalization of the tube algebra $\mathrm{Tube}(\CC)$ which describes how the topological lines in $\CC$ act on junction operators living at the intersection point of several boundaries and interfaces. As a step towards the more general setup, one can consider the intermediate case of local operators sitting at the endpoint of just a single line operator $I$ which is not necessarily topological. The situation described in the previous paragraphs corresponds to the special case that $I$ is topological and is among the lines in $\CC$.

It will actually be useful to consider not just  $I$, but the entire multiplet $\mathcal{I}$ which is generated  by acting on $I$ with symmetry lines in $\mathcal{C}$ from the left and from the right. The interaction of the symmetry $\CC$ with the not-necessarily topological lines in this multiplet endows $\mathcal{I}$ with the structure of a $(\CC,\CC)$-bimodule category (see e.g.\ \cite{Komargodski:2020mxz,Huang:2021zvu,Choi:2023xjw,Diatlyk:2023fwf,Cordova:2024vsq,Cordova:2024iti,Copetti:2024dcz,Inamura:2024jke} for recent appearances in the physics literature), as we describe in more detail in Section \ref{subsec:modulebimodule}. Briefly, this entails a pair of fusion coefficients $(\widetilde{N}_L)_{cI}^{J}$ and $(\widetilde{N}_R)_{Ic}^J$ which describe the result of left and right fusion of lines in $\CC$ onto lines in $\mathcal{I}$, as well as more subtle data ($\widetilde{F}$-symbols and the middle associator) which encodes recombination rules governing topological junctions between lines in $\CC$ and lines in $\mathcal{I}$.

At the level of pictures then, we can arrive at the correct algebraic structure simply by replacing the topological lines $a$, $d$, and $b$ in the lassos of Equation \eqref{eqn:lassoopsintro} with line operators $I$, $K$, and $J$, respectively, belonging to the multiplet $\mathcal{I}$,
\begin{align}
    \input{figures/lassointerfaceintro.tikz}\, ,
\end{align}
where again, $y$ and $z$ are suitable topological junctions. We denote  the algebra generated by these lassos as $\mathrm{Tube}(\mathcal{I})$. Again, because the lassos are able to convert a junction operator at the end of a line $I\in\mathcal{I}$ to a junction operator at the end of a generally different line $J\in\mathcal{I}$, this algebra most naturally acts on the extended Hilbert space 
\begin{align}
    \mathcal{H}_{\mathcal{I}}=\bigoplus_{I\in\mathrm{Irr}(\mathcal{I})}\mathcal{H}_I
\end{align}
defined as a direct sum over the spaces of defect operators living at the end of the (not necessarily topological) lines $I$ in the multiplet $\mathcal{I}$.

It turns out that the structure of $\mathrm{Tube}(\mathcal{I})$ as an abstract algebra does not depend on the details of the multiplet $\mathcal{I}$ of line operators of $Q$, but only on how they transform with respect to the symmetry lines in $\CC$. (Mathematically, we would say that $\mathrm{Tube}(\mathcal{I})$ only depends on $\mathcal{I}$ through its structure as a $(\CC,\CC)$-bimodule category.) Of course though, the precise action of $\mathrm{Tube}(\mathcal{I})$ on the extended Hilbert space $\mathcal{H}_{\mathcal{I}}$ \emph{will} depend on the details of $\mathcal{I}$. The notation $\mathrm{Tube}(\mathcal{I})$ is justified because, in the special case that $\mathcal{I}$ is chosen to be the multiplet of symmetry lines in $\mathcal{C}$ (described by the ``regular'' $(\CC,\CC)$-bimodule category), the algebra $\mathrm{Tube}(\mathcal{I})$ reduces to $\mathrm{Tube}(\CC)$.

The most general situation involves studying line interfaces $I_i$ (with $i=1,\cdots,n$) between 1+1d quantum field theories $Q_i$ and $Q_{i+1}$, where $Q_i$ is assumed to have a collection of topological line operators $\CC_i$. We write $\mathcal{I}_i$ for the multiplet obtained by acting on $I_i$ with symmetry lines in $\CC_i$ from one side and lines in $\CC_{i+1}$ from the other. If we suppose that the interfaces $I_1,\cdots, I_n$ meet at a point, we may study the space $\mathcal{H}_{I_1\cdots I_n}$ of junction operators which can sit at that point, and also the larger extended Hilbert space 
\begin{align}
    \mathcal{H}_{\mathcal{I}_1\cdots\mathcal{I}_n} = \bigoplus_{I_1\in\mathcal{I}_1}\cdots\bigoplus_{I_n\in\mathcal{I}_n}\mathcal{H}_{I_1\cdots I_n}
\end{align}
which encodes the junction operators between all the lines in the multiplets $\mathcal{I}_1,\cdots,\mathcal{I}_n$.
We obtain a tube algebra $\mathrm{Tube}(\mathcal{I}_1\vert\cdots\vert\mathcal{I}_n)$ which, by definition, is generated by generalized lassos of the form\footnote{Related discussions in the context of the Levin-Wen string-net model \cite{Levin:2004mi} can be found in \cite{Bridgeman:2019wyu,jia2024weak}.} 
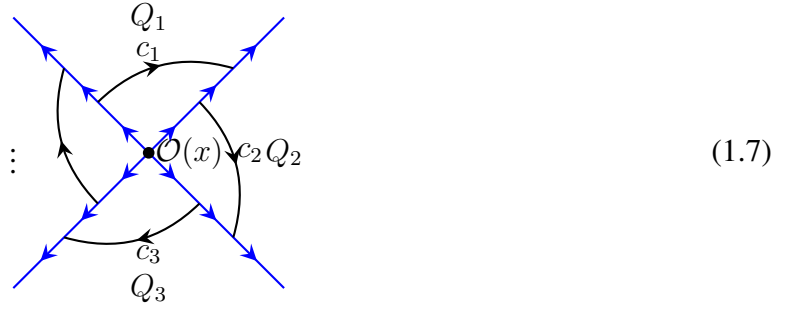
\begin{align}
    \input{figures/junctionlassointro.tikz}
\end{align}
where to avoid clutter, we have suppressed the labels of many of the lines and junctions. (See Equation \eqref{eq:generalizedlassoaction} for the same picture with these labels restored.)

A special case which will be important in our work is the case that $n=2$, and $Q_1$ is taken to be the empty theory.\footnote{More pedantically, we might say that $Q_1$ is the trivially gapped theory with $\CC_1$ taken to be the trivial symmetry generated by the identity line.} Because an interface between a 1+1d QFT $Q_1$ and the empty theory can be thought of as a boundary condition, we may think of $\mathcal{I}_1=:\mathcal{B}_1^\vee$ and $\mathcal{I}_2=:\mathcal{B}_2$ as multiplets of boundary conditions, in which case the extended Hilbert space 
\begin{align}
    \mathcal{H}_{\mathcal{B}_1^\vee\mathcal{B}_2} = \bigoplus_{B_1\in\mathrm{Irr}(\mathcal{B}_1^\vee)}\bigoplus_{B_2\in\mathrm{Irr}(\mathcal{B}_2)}\mathcal{H}_{B_1B_2}
\end{align}
is the space of boundary-changing local operators between boundary conditions in the multiplet $\mathcal{B}_1^\vee$ and boundary conditions in the multiplet $\mathcal{B}_2$.\footnote{The decoration $\vee$ can be ignored by the casual reader. It reflects the fact that the lines in $\CC_1$ act on the boundaries in $\mathcal{B}_1^\vee$ from the right, whereas they act on the boundaries in $\mathcal{B}_2$ from the left. See Section \ref{subsec:modulebimodule} for a more mathematically precise explanation.} This extended Hilbert space is then acted upon naturally by the tube algebra $\mathrm{Tube}(\mathcal{B}_1^\vee\vert\mathcal{B}_2)$ which is generated by boundary lassos of the form 
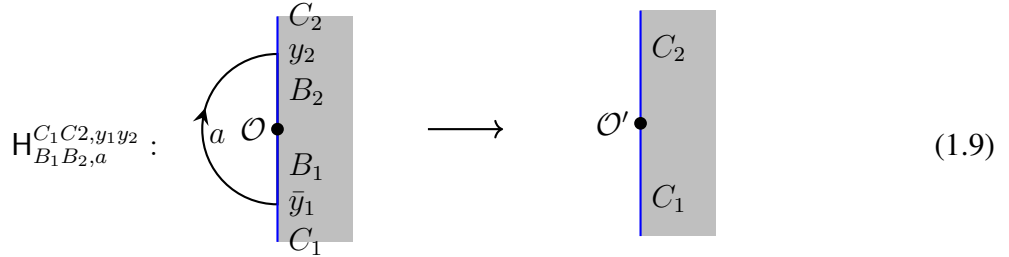
\begin{align}\label{eqn:boundarylassointro}
\mathsf{H}_{B_1B_2,a}^{C_1C2,y_1y_2}:~~~\raisebox{-3.5em}{
    \begin{tikzpicture}
			\fill [gray, opacity=0.5] (0,-1.5) rectangle (1,1.5); 
			\draw[thick, decoration = {markings, mark=at position 0.75 with {\arrow[scale=1.5]{stealth}}}, postaction=decorate]  (0,1) -- (0,-1) arc(270:90:1) --cycle;
			\node[left] at (0, 0) {{$\mathcal{O}$}};
            \draw[->,thick] (2,0) -- (3,0);
			\draw[thick,blue] (0, 1.5) -- (0,-1.5);
			\node[right] at (0, 1.5) {$C_2$};
			\node[right] at (0,-1.5) {$C_1$};
			\node[right] at (0, 0.5) {$B_2$};
			\node[right] at (0,-0.5) {$B_1$};
			\node[right] at (-0, 1) {$y_2$};
			\node[right] at (-0, -1) {$\bar{y}_1$};
			\node[] at (-0.8, -0.1) {$a$};
   \draw[thick, fill=black] (0, 0) circle (2pt);
	\end{tikzpicture}
	\hspace{.3in}
 \raisebox{.6em}{
 \begin{tikzpicture}
			\fill [gray, opacity=0.5] (0,-1.5) rectangle (1,1.5); 
			\node[left] at (0, 0) {{$\mathcal{O}'$}};
			\draw[thick,blue] (0, 1.5) -- (0,-1.5);
			\node[right] at (0, 1) {$C_2$};
			\node[right] at (0,-1) {$C_1$};
                \node[right] at (0,1.5) {};
                \node[right] at (0,-1.5) {};
   \draw[thick, fill=black] (0, 0) circle (2pt);
	\end{tikzpicture}}\, .
 }
\end{align}
These special cases of our generalized tube algebras, which we refer to as \emph{boundary tube algebras}, have been studied in a variety of different contexts \cite{Cordova:2024vsq,Cordova:2024iti,Copetti:2024dcz,2012CMaPh.313..351K,Barter_2022,Copetti:2024onh,Konechny:2024ixa,Bridgeman:2022gdx,Barter:2018hjs}, especially in recent years (though mostly in the case that $\mathcal{B}_1=\mathcal{B}_2$).\footnote{See also \cite{Koide:2023rqd} for a recent discussion on the interplay between non-invertible symmetries and boundary conditions in higher spacetime dimensions, in the context of a four-dimensional Euclidean lattice $\mathbb{Z}_2$ gauge theory \cite{Koide:2021zxj}.}

With this new class of symmetry algebras in hand, a natural question is how the extended junction Hilbert spaces $\mathcal{H}_{\mathcal{I}_1\cdots\mathcal{I}_n}$ of a theory organize into irreducible representations. In order to address this, we appeal to the framework of symmetry topological field theories.

\subsection{Representation theory from the SymTFT}

The basic starting point of the SymTFT approach is the observation that any (say, 1+1d for simplicity of exposition) quantum field theory $Q$ with symmetry $\mathcal{C}$ can be inflated into a 2+1d topological ``Turaev-Viro'' field theory $\TV_{\CC}$  compactified on an interval, with a particular ``Dirichlet'' gapped boundary condition $\mathcal{B}_{\mathrm{reg}}$ imposed on one end of the interval, and a not-necessarily topological ``physical'' boundary condition $\widetilde{Q}$ imposed on the other \cite{Gaiotto:2014kfa,Gaiotto:2020iye,Ji:2019jhk,Apruzzi:2021nmk,Freed:2022qnc, Kong:2015flk, Kong:2017hcw, Kong:2020cie, Kaidi:2022cpf, Antinucci:2022vyk}. One of the virtues of this construction is that it cleanly separates out the kinematic aspects of the physics --- i.e.\ those properties of the physics that hold universally in any theory which has the same symmetry $\mathcal{C}$ --- from the aspects of the physics which are determined dynamically. Indeed heuristically, in the SymTFT picture, the kinematic features are associated with physics occurring in the bulk or near its gapped Dirichlet boundary condition, while the dynamical features are associated with the physical boundary. The SymTFT can give the conceptual clarity necessary to systematically extract the constraints coming from a symmetry in a variety of situations.

For instance, an emerging picture is that the multiplets (representations, charges, etc.) to which $n$-dimensional objects of a quantum field theory belong are labeled by $n+1$-dimensional objects of the bulk SymTFT $\TV_{\CC}$ \cite{Lin:2022dhv,Bhardwaj:2023wzd,Bhardwaj:2023ayw,Bartsch:2023pzl,Bartsch:2023wvv}. The basic intuition behind this claim is that an $n$-dimensional object $R$ of $Q$ is expected to go over in the SymTFT to an $(n+1)$-dimensional object $\mathcal{R}$ of $\TV_{\CC}$ which extends between the two boundaries, and terminates on suitable $n$-dimensional junctions $\widetilde{R}$ and $\underline{R}$ on $\widetilde{Q}$ and $\mathcal{B}_{\mathrm{reg}}$, respectively. On the other hand, the topological operators implementing the symmetry are ``trapped'' on the Dirichlet boundary condition: therefore when they act, they do not modify $\mathcal{R}$, and only affect the junction $\underline{R}$ it forms with the Dirichlet boundary. The different choices for the junction $\underline{R}$ are then interpreted as different members of the multiplet labeled by $\mathcal{R}$. 

\begin{figure}[t]
	\centering
    \raisebox{-73pt}{\begin{tikzpicture}
			\draw [color=black, thick, decoration = {markings, mark=at position 0.5 with {\arrow[scale=1.5]{stealth}}}, postaction=decorate] (2.5,2) -- (2.5,3) node[right]{$a$} -- (2.5,3.75);
			\draw [fill=black] (2.5,2) circle (0.04) node [below] {$\cO$};
			\draw[dashed](2,0) -- (2,3) -- (3,4) -- (3,1) -- cycle;
			\draw (2,0) node[below]{$Q$};
	\end{tikzpicture}
    }
    \quad $\leftrightharpoons$ ~
	\raisebox{-73pt}{
    \begin{tikzpicture}
			\draw (1.35,0.8) node[below]{\small \color{black} $\TV_\cC$};
			\draw[dashed] (0,0) -- (0,3) -- (1,4) -- (1,1) -- cycle;
			\draw (0,0) node[below]{$\mathcal{B}_{\mathrm{reg}}$};
			\draw [color=red, thick, decoration = {markings, mark=at position 0.6 with {\arrow[scale=1.5]{stealth}}}, postaction=decorate] (2.5,2) -- (1.5,2) node[below]{$\mu$} -- (0.5,2);
			\draw [fill=black] (0.5,2) circle (0.04) node [below] {$y$};
			\draw [thick, decoration = {markings, mark=at position 0.5 with {\arrow[scale=1.5]{stealth}}}, postaction=decorate] (0.5,2) -- (0.5,3) node[right]{$a$} -- (0.5,3.75);
			\draw [fill=black] (2.5,2) circle (0.04) node [below] {\color{black} $\widetilde{\cO}$};
			\draw[color=black, preaction={draw=white,line width=3pt}, dashed] (2,0) -- (2,3);
			\draw[color=black, dashed] (2,3) -- (3,4) -- (3,1) -- (2,0);
			\draw (2,0) node[below]{\color{black} $\widetilde{Q}$};
	\end{tikzpicture}}
	\caption{A twisted sector operator $\cO\in\cH_a$ of a 1+1d theory $Q$ decomposes into a triple $(y,\mu,\widetilde{\mathcal{O}})$, where $\mu\in Z(\mathcal{C})$ is a bulk anyon, and $\widetilde{\mathcal{O}}$ and $y$ are suitable junction operators.
	}
	\label{fig:defect.operators}
\end{figure}
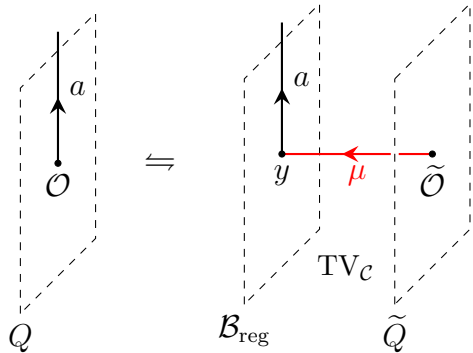

A prototypical example is that a local operator $\mathcal{O}(x)$ inflates into a bulk topological line $\mu$ which terminates on a topological point operator $y$ on the Dirichlet boundary condition, and a not-necessarily topological operator $\widetilde{\mathcal{O}}(x)$ on the physical boundary $\widetilde{Q}$. The generalization of this statement to twisted sector local operators is depicted in Figure \ref{fig:defect.operators}. Thus, irreducible representations of $\mathrm{Tube}(\CC)$ are in one-to-one correspondence with topological line operators of $\TV_{\CC}$, which in turn are in correspondence with simple objects of the modular tensor category $Z(\mathcal{C})$ known as the Drinfeld center of $\CC$. More concisely \cite{evans1995ocneanu,Izumi:2000qa,MUGER2003159}, 
\begin{align}\label{eqn:Rep(Tube(C))}
    \mathrm{Rep}(\mathrm{Tube}(\CC))\cong Z(\CC) \,.
\end{align}

Toggling from $\mathcal{O}(x)$ to a different operator $\mathcal{O}'(x)$ in the $\mathrm{Tube}(\CC)$-multiplet labeled by $\mu$ corresponds to swapping out the topological junction operator $y$ for a different topological junction operator $y'$ on the Dirichlet boundary. Thus, the underlying vector space of the multiplet labeled by $\mu$ is simply $W^\mu=\bigoplus_{a\in\mathrm{Irr}(\CC)}W^\mu_a$, where $W^\mu_a$ is the Hilbert space of topological junction operators that $\mu$ can form with the line $a\in \CC$ on the Dirichlet boundary.

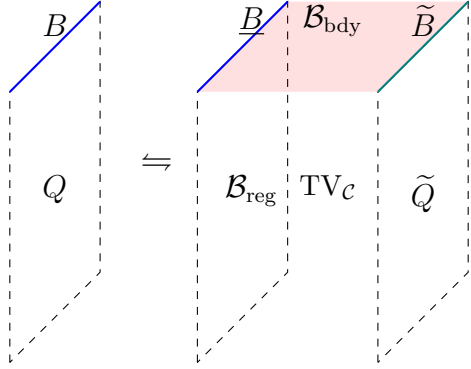
\begin{figure}[t]
	\centering
	\raisebox{-73pt}{\begin{tikzpicture}[scale=1.2]
			\draw[dashed](2,0) -- (2,3) -- (3,4) -- (3,1) -- cycle;
			\draw (2.5,2.2) node[below]{$Q$};
			\draw (2.5,3.5) node[above]{$B$};
			\draw [thick, blue] (2,3) -- (3,4);
	\end{tikzpicture}}
	\quad $\leftrightharpoons$ ~
	\raisebox{-73pt}{\begin{tikzpicture}[scale=1.2]
			\fill[fill=red!40,opacity=.3] (0,3)  -- (2,3)  -- (3,4) -- (1,4) -- cycle;
			\draw (1.45,2.2) node[below]{\small  $\TV_{\mathcal{C}}$};
			\draw[dashed] (0,0) -- (0,3) -- (1,4) -- (1,1) -- cycle;
			
			\draw (.6,2.2) node[below]{$\mathcal{B}_{\mathrm{reg}}$};
			\draw[ dashed] (2,0) -- (2,3);
			\draw[dashed] (2,3) -- (3,4) -- (3,1) -- (2,0);
			\draw (2.5,2.2) node[below]{$\widetilde{Q}$};
			\draw [blue, thick] (0,3) -- (1,4);
			\draw [thick,color=teal] (2,3) -- (3,4);
			\draw (1.5,3.5) node[above] {$\mathcal{B}_{\mathrm{bdy}}$}; 
			\draw  (.6,3.5) node[above] {$\underline{B}$};
			\draw  (2.5,3.5) node[above] {$\widetilde{B}$};
	\end{tikzpicture}}
	\caption{The SymTFT interpretation of a boundary $B$ of a QFT $Q$ with symmetry category  $\CC$.  The boundary condition inflates into a triple $(\underline{B}, \mathcal{B}_{\mathrm{bdy}}, \widetilde{B})$. 
	}
	\label{fig:bdysymtft}
\end{figure}

We would like to develop an analogous  understanding of the representation theory of the generalized tube algebras $\mathrm{Tube}(\mathcal{I}_1\vert\cdots\vert \mathcal{I}_n)$. In order to achieve this, we must first determine how boundaries of a 1+1d QFT, and more generally line interfaces between two 1+1d QFTs, are represented in the SymTFT. 

Let us work with the case of boundaries for ease of exposition. Interfaces can be treated similarly, and we do so in detail in Section \ref{subsec:interfaces}. Consider a boundary condition $B$ of a 1+1d QFT $Q$ with symmetry lines $\CC$, and call $\mathcal{B}_{\mathrm{mul}}$ the multiplet of boundaries to which $B$ belongs. Our basic expectation is that  $B$ should inflate to a topological boundary $\mathcal{B}_{\mathrm{bdy}}$ of the SymTFT $\TV_{\CC}$ which terminates on corners $\widetilde{B}$ and $\underline{B}$ meeting the physical boundary $\widetilde{Q}$ and the Dirichlet boundary $\mathcal{B}_{\mathrm{reg}}$, respectively. See Figure \ref{fig:bdysymtft}. 

How should this topological boundary $\mathcal{B}_{\mathrm{bdy}}$ be characterized? Mathematically it is known that topological boundaries of $\TV_{\CC}$ are in one-to-one correspondence with module categories of $\CC$ \cite{2012CMaPh.313..351K}. On the other hand, any multiplet of boundary conditions in a theory with $\CC$ symmetry also defines a $\CC$-module category. (This is a special case of our earlier claim that multiplets of interfaces form bimodule categories.) Our main proposal is then the following. See \cite{Huang:2023pyk} for an earlier appearance of this idea in the context of diagonal rational conformal field theory.\footnote{Discussions on various boundaries and interfaces in the context of string theory realization of SymTFTs can also be found in \cite{Heckman:2024zdo,Braeger:2024jcj,Baume:2023kkf}.} 

\begin{claim}
    In a 1+1d QFT $Q$ with $\CC$ symmetry and a boundary condition $B$, the topological boundary $\mathcal{B}_{\mathrm{bdy}}$ arising in the SymTFT picture of $B$ agrees as a $\CC$-module category with the multiplet $\mathcal{B}_{\mathrm{mul}}$ to which $B$ belongs. That is, $\mathcal{B}_{\mathrm{bdy}}\cong \mathcal{B}_{\mathrm{mul}}$.
\end{claim}

As a special case, we describe in Section \ref{subsec:symmetricboundariesanomalies} how $\CC$-symmetric boundaries\footnote{By a ``$\CC$-symmetric'' boundary condition, we mean a ``strongly $\CC$-symmetric'' boundary condition in the sense of \cite{Choi:2023xjw}.} of $Q$ inflate to so-called ``magnetic'' topological boundaries \cite{Zhang:2023wlu} of the SymTFT $\TV_{\CC}$. These are by definition boundaries $\mathcal{B}_{\mathrm{bdy}}$ of $\TV_{\CC}$ with the property that the only bulk line which condenses on both the Dirichlet boundary condition and $\mathcal{B}_{\mathrm{bdy}}$ is the trivial line. For example, in the case of a $\mathbb{Z}_2$ symmetry, the SymTFT is the toric code topological order, and a $\mathbb{Z}_2$ symmetric boundary inflates to the usual magnetic (as opposed to electric) boundary of the toric code.

 We henceforth abusively conflate $\mathcal{B}_{\mathrm{bdy}}$ and $ \mathcal{B}_{\mathrm{mul}}$, and write them both as $\mathcal{B}$. This result is in harmony with the general expectation that multiplets of objects in $Q$ are labeled by topological objects in the bulk of one dimension higher. Interfaces are analogous: an interface $I$ between two QFTs inflates to a topological interface $\mathcal{I}$ between the corresponding SymTFTs which is labeled by the multiplet to which $I$ belongs. See Figure \ref{fig:interfacesandwich}.

We are now in a position to describe the irreducible representations of the tube algebras $\mathrm{Tube}(\mathcal{I}_1\vert\cdots\vert\mathcal{I}_n)$.  Taking a hint from the special case of $\mathrm{Tube}(\CC)$, we arrive at the following proposal.

\begin{figure}
\centering
    \input{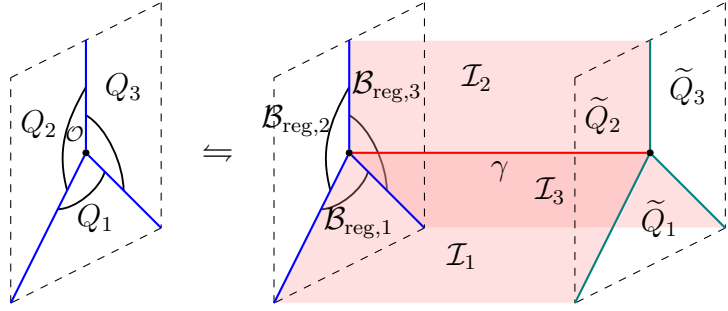}
    \caption{The irreducible representations of $\mathrm{Tube}(\mathcal{I}_1\vert\cdots\vert\mathcal{I}_n)$ are labeled by topological line junctions $\gamma$ between the two-dimensional topological interfaces $\mathcal{I}_1,\cdots,\mathcal{I}_n$. The generalized lassos appear on the topological boundary conditions of the SymTFT. Illustrated in the case of $n=3$. See Figure \ref{fig:junctionsymTFT} for a more detailed version of this figure.}\label{fig:introtubereps}
\end{figure}

\begin{claim}
    The irreducible representations of $\mathrm{Tube}(\mathcal{I}_1\vert\cdots\vert\mathcal{I}_n)$ are in one-to-one correspondence with simple topological line junctions $\gamma$ between the two-dimensional topological interfaces $\mathcal{I}_1$, $\cdots$ , $\mathcal{I}_n$ in the bulk. 
\end{claim}
This is depicted in Figure \ref{fig:introtubereps}.

By swinging the interfaces $\mathcal{I}_1,\cdots,\mathcal{I}_n$ around $\gamma$ and fusing them all together to produce a surface $\mathcal{I}\equiv \mathcal{I}_1\boxtimes_{\CC_2}\mathcal{I}_2\boxtimes_{\CC_3}\cdots\boxtimes_{\CC_n}\mathcal{I}_n$ in $\TV_{\CC_1}$ as in Figure \ref{fig:swingsurfaces}, we see that we can think of $\gamma$ as a topological line which bounds $\mathcal{I}$, or more suggestively, as a topological line interface between $\mathcal{I}$ and the trivial surface $\mathcal{I}_{\mathrm{reg},1}$ of $\TV_{\CC_1}$. This latter description admits a sharp mathematical characterization as the category $\mathrm{Fun}_{(\CC_1,\CC_1)}(\mathcal{I}_{\mathrm{reg},1},\mathcal{I})$ of $(\CC_1,\CC_1)$-bimodule functors from $\mathcal{I}_{\mathrm{reg},1}\to\mathcal{I}$, which is often denoted $Z_{\mathcal{C}_1}(\mathcal{I})$ and called the \emph{center} of $\mathcal{I}$ (see e.g.\ \cite[Definition 2.13]{etingof2010fusion}).\footnote{For the mathematically oriented reader, we note that the cocenter $Z_{\mathcal{C}_1}(\mathcal{I})$ is Morita equivalent to its associated category algebra, which is the generalized tube algebra $\mathrm{Tube}(\mathcal{I}_1\vert\cdots\vert \mathcal{I}_n)$ studied in this paper. We thank the referee for bringing this to our attention.} In other words,
\begin{align}
    \mathrm{Rep}(\mathrm{Tube}(\mathcal{I}_1\vert\cdots\vert\mathcal{I}_n))\cong Z_{\mathcal{C}_1}(\mathcal{I}_1\boxtimes_{\mathcal{C}_2}\mathcal{I}_2\boxtimes_{\mathcal{C}_3}\cdots\boxtimes_{\mathcal{C}_n}\mathcal{I}_n).
\end{align} 
In the special case that $n=1$ and $\mathcal{I}_1=\CC_1=\CC$, this reduces to Equation \eqref{eqn:Rep(Tube(C))}.

To provide an example where this category is more familiar, consider the boundary tube algebras $\mathrm{Tube}(\mathcal{B}^\vee\vert\mathcal{B})$. In this case, the proposal highlighted above tells us that the irreducible representations of $\mathrm{Tube}(\mathcal{B}^\vee\vert\mathcal{B})$ should correspond to simple topological line operators supported on the topological boundary of the SymTFT which is labeled by $\mathcal{B}$. It is known \cite{2012CMaPh.313..351K} that such line operators are objects of $\mathrm{Fun}_{\mathcal{C}}(\mathcal{B},\mathcal{B})^{\mathrm{op}}$, the (opposite of the) category of $\mathcal{C}$-module functors from $\mathcal{B}$ to itself, which leads to the prediction that
\begin{align}\label{eq:introrepboundarytube}
    \mathrm{Rep}(\mathrm{Tube}(\mathcal{B}^\vee\vert\mathcal{B})) \cong \mathrm{Fun}_{\mathcal{C}}(\mathcal{B},\mathcal{B})^{\mathrm{op}}. 
\end{align}
It turns out that Equation \eqref{eq:introrepboundarytube} has been established using more mathematical techniques in earlier literature \cite{Barter_2022}. More physically, recall that the inequivalent ways of orbifolding a symmetry $\CC$ are labeled by module categories of $\CC$ \cite{Bhardwaj:2017xup}: the category $\mathrm{Fun}_{\mathcal{C}}(\mathcal{B},\mathcal{B})^{\mathrm{op}}$ is precisely the ``dual'' symmetry category of the theory $Q/\mathcal{B}$ obtained by orbifolding $\mathcal{B}$.

The SymTFT not only provides us with a convenient conceptual framework, but it further enables us to carry out concrete calculations. We illustrate this in the next subsection, where we apply this understanding of the representation theory of the generalized tube algebras to the analysis of symmetry-resolved partition functions.

\subsection{Symmetry-resolved partition functions}

In a quantum field theory, the Euclidean path integral on a background of the form $M\times S^1$ counts states in the Hilbert space of the theory when it is quantized on $M$, keeping track of their energy. In a conformal field theory, by the state/operator correspondence, such partition functions can often be reinterpreted as counting local operators, keeping track of their quantum numbers with respect to spacetime symmetries. For example, the $T^2$ (torus) partition function of a CFT counts genuine local operators, while the $I\times S^1$ (annulus) partition function of a CFT, with boundary conditions $B_1$ and $B_2$ imposed at either end of the interval, counts boundary changing local operators between $B_1$ and $B_2$. 

We are interested in refining such partition functions, so that they probe the symmetry structure of the theory under consideration. There are at least two ways that one can conceive of doing this.

\begin{enumerate}[label=\arabic*)]
    \item One can define partition functions which keep track of how many operators, at each energy level, transform in some fixed representation of the global symmetry.
    \item One can study partition functions in the presence of a background gauge field for the global symmetry. By Poincaré duality, a background gauge field can be represented by a configuration of codimension-1 topological operators wrapping various cycles of the spacetime manifold. 
\end{enumerate}
The first class of partition functions are often referred to in the literature as partition functions in the ``anyon basis'' (see e.g.\ \cite{Ji:2019jhk,Gaiotto:2020iye,Ji:2021esj}) because of the connection between representations of $\mathrm{Tube}(\CC)$ and anyons of the SymTFT $\TV_{\CC}$. We will instead use the terminology \emph{representation basis} to describe these partition functions, in anticipation of the appearance of the generalized tube algebras, whose irreducible representations are labeled by more general objects in the bulk. On the other hand, we refer to partition functions in the second class as being in the \emph{symmetry basis}. The word ``basis'' here is somewhat an abuse of terminology because, as we describe in Section \ref{subsec:toruslineardependencies} and Section \ref{subsec:lineardependencies}, the symmetry basis partition functions often have linear dependencies.

We restrict our attention to annulus partition functions in the introduction, since their physics is richer than that of torus partition functions. The latter can be treated similarly, and we do so in Section \ref{sec:torus}. Before dealing with the case of a general unitary fusion category symmetry, let us illustrate more explicitly how these two bases are defined  in a 1+1d CFT with a non-anomalous $\mathbb{Z}_2=\langle \eta\rangle$ symmetry and a $\mathbb{Z}_2$ symmetric simple boundary condition $B$. 

Because $B$ is $\mathbb{Z}_2$ symmetric, it is the only simple boundary condition in its multiplet $\mathcal{B}$, and so the extended Hilbert space $\mathcal{H}_{\mathcal{B}^\vee\mathcal{B}}$ coincides with the standard Hilbert space $\mathcal{H}_{BB}$ of boundary local operators on $B$. The boundary tube algebra $\mathrm{Tube}(\mathcal{B}^\vee\vert\mathcal{B})$ in this case is the group algebra $\mathbb{C}[\mathbb{Z}_2]$. Accordingly, the Hilbert space can be decomposed into $\mathbb{Z}_2$ representations
\begin{align}\label{eqn:Z2decomp}
    \mathcal{H}_{BB} = \mathcal{H}_{BB}^{+}\oplus \mathcal{H}_{BB}^-
\end{align}
where here, $\mathcal{H}_{BB}^\pm$ is the space of $\mathbb{Z}_2$ even/odd boundary operators, respectively.

The decomposition in Equation \eqref{eqn:Z2decomp} allows us to define the partition functions in the representation basis as
\begin{align}
    \mathbf{Z}_\pm(\delta) \equiv \mathrm{Tr}_{\mathcal{H}_{BB}^\pm}e^{-H_{\mathrm{op}}\delta},
\end{align}
where $H_{\mathrm{op}}$ is the Hamiltonian on the interval Hilbert space, and $\delta$ is the circumference of the thermal circle. We set the length of the interval to be 1.
On the other hand, we can also produce two partition functions in the symmetry basis, 
\begin{align}
    Z_{BB}(\delta) = \mathrm{Tr}_{\mathcal{H}_{BB}}e^{-H_{\mathrm{op}}\delta} , \ \ \ Z_{BB}^\eta(\delta)=\mathrm{Tr}_{\mathcal{H}_{BB}}\hat\eta e^{-H_{\mathrm{op}}\delta} .
\end{align}
The symmetry basis partition functions can be represented as annulus partition functions in the presence of symmetry line insertions, as in Figure \ref{fig:symbasisZ2}. Naively, one might think that we have missed a partition function, corresponding to wrapping the $\eta$ line along the circle direction of the annulus. However, such a line can be pushed onto $B$ and, since $B$ is symmetric, absorbed by it. Thus, this does not define an independent partition function.

\begin{figure}
    \centering
    \input{figures/Z2symmetrybasis.tikz}
    \caption{The symmetry basis annulus partition functions of a 1+1d QFT $Q$ with a $\mathbb{Z}_2$ symmetry and a symmetric boundary condition $B$. The length of the circle direction is $\delta$. }\label{fig:symbasisZ2}
\end{figure}

Now, in this simple situation, it is clear that these partition functions are related by a change of basis 
\begin{align}
    Z_{BB}(\delta) = \mathbf{Z}_+(\delta)+\mathbf{Z}_-(\delta), \ \ \ Z_{BB}^\eta(\delta) = \mathbf{Z}_+(\delta) - \mathbf{Z}_-(\delta).
\end{align}
More suggestively, we may package these two equalities into the following equation
\begin{align}\label{eqn:Z2characterCOB}
    Z_{BB}^g(\delta) = \chi_+(g) \mathbf{Z}_+(\delta) + \chi_-(g) \mathbf{Z}_-(\delta)
\end{align}
where $\chi_\pm:\mathbb{Z}_2\to\{\pm 1\}$ are the characters of the two irreducible representations of $\mathbb{Z}_2$. 

One of our main results is a generalization of Equation \eqref{eqn:Z2characterCOB} which holds for general unitary fusion categories. 
To explain this result, suppose one has a 1+1d CFT $Q$ with symmetry category $\mathcal{C}$ and two $\CC$-multiplets $\mathcal{B}_1$ and $\mathcal{B}_2$ of conformal boundary conditions. 
We may then define representation basis partition functions $\mathbf{Z}_\alpha(\delta)$ labeled by irreducible representations $\alpha$ of $\mathrm{Tube}(\mathcal{B}_1^\vee\vert\mathcal{B}_2)$. 
Physically, $\alpha$ corresponds to a choice of topological line interface between the gapped boundary conditions $\mathcal{B}_1$ and $\mathcal{B}_2$ of the SymTFT, as in Figure \ref{fig:bdydefect.operators}. 
If one performs a Fourier expansion of $\mathbf{Z}_\alpha(\delta)$ in powers of $q=e^{-\pi \delta}$, then the coefficient of $q^E$ in this expansion is the multiplicity with which the representation $\alpha$ of the boundary tube algebra  appears in the extended Hilbert space $\mathcal{H}_{\mathcal{B}_1^\vee\mathcal{B}_2}$ at energy level $E$. 

On the other hand, we may also study symmetry basis annulus partition functions 
\begin{align}\label{eq:introsymbasisannulus}
    Z_{B_1B_2}^{az_1z_2}(\delta)=\raisebox{-2.5em}
	{\begin{tikzpicture}
			\fill [gray, opacity=0.5] (0,-1) rectangle (-0.6,1); 
			\fill [gray, opacity=0.5] (2,-1) rectangle (2.6,1); 
			\draw[thick, decoration = {markings, mark=at position 0.5 with {\arrow[scale=1.5]{stealth}}}, postaction=decorate] (0,0) -- (2,0);
			\draw[thick, blue] (0,1) -- (0,-1);
			\draw[thick,blue] (2,1) -- (2,-1);
			\node[left] at (0,0.8) {$B_1$};
			\node[right] at (2,0.8) {$B_2$};
			\node[left] at (0,-0.8) {$B_1$};
			\node[right] at (2,-0.8) {$B_2$};
			\node[above] at (1,0) {$a$};
			\node[left] at (0,0) {$\bar{z}_1$};
			\node[right] at (2,0) {$z_2$};
            \draw[dashed] (0,-1) -- (2,-1);
            \draw[dashed] (0,1) -- (2,1);
            \node[] at (1,1) {$\doubleslash$};
            \node[] at (1,-1) {$\doubleslash$};
	\end{tikzpicture}}
\end{align}
where $B_1\in\mathcal{B}_1$ and $B_2\in\mathcal{B}_2$ are boundary conditions, $a$ is a symmetry line, and $\bar{z}_1$ and $z_2$ are suitable topological endpoints of $a$ on the $B_1$ and $B_2$ boundaries. Such partition functions can be represented as traces over the interval Hilbert space $\mathcal{H}_{B_1B_2}$ of suitable elements of the boundary tube algebra $\mathrm{Tube}(\mathcal{B}_1^\vee\vert\mathcal{B}_2)$, as in Equation \eqref{eqn:openstringsymmetrybasis}.

To relate these two classes of partition functions, we introduce ``quantum'' characters $\chi_\alpha$ and $\widetilde{\chi}_\alpha$ of the $\alpha$-representation of $\mathrm{Tube}(\mathcal{B}_1^\vee\vert\mathcal{B}_2)$ via the following partition functions of the SymTFT on a solid torus,
\begin{align}\label{eq:qcharsintro}
\begin{split}
    \left[ \chi_{\alpha} \right]_{\underline{B}_1 \underline{B}_2}^{ay_1 y_2} = \input{figures/characterverlindeintro.tikz} , \ \ \ \ \ [\widetilde{\chi}_\alpha]_{\underline{B}_1\underline{B}_2}^{ay_1y_2}=\input{figures/characterverlindedualintro.tikz}
\end{split}
\end{align}
where we refer to Figure \ref{fig:bdysymtft} for the definition of $\underline{B}_1$ and $\underline{B}_2$. Here, the dashed lines represent a solid torus, and the $T^2$ boundary of the solid torus is divided into 3 regions --- where the topological boundaries $\mathcal{B}_1$, $\mathcal{B}_2$, and $\mathcal{B}_{\mathrm{reg}}$ are imposed --- by the topological line interfaces $\underline{B}_1$, $\underline{B}_2$, and $\alpha$. Using the name ``characters'' for these SymTFT partition functions is justified by the following result, which generalizes the character theory of finite groups.

\begin{claim}
The quantum characters of the boundary tube algebra satisfy an orthogonality relation 
\begin{align}
    \frac{1}{\dim(\CC)^2}\sum_{\underline{B}_1,\underline{B}_2}\sum_{a,y_1,y_2} [\widetilde{\chi}_\alpha]_{\underline{B}_1\underline{B}_2}^{ay_1y_2}~[\chi_\beta]_{\underline{B}_1\underline{B}_2}^{ay_1y_2} = \delta_{\alpha\beta}
\end{align}
where $\dim(\CC)^2 = \sum_{a\in\mathrm{Irr}(\CC)}\d_a^2$. The symmetry basis annulus partition functions of a 1+1d quantum field theory $Q$ with symmetry category $\CC$ can be expressed in terms of the quantum characters and the representation basis partition functions as
\begin{align}
    Z_{B_1B_2}^{ay_1y_2}(\delta) = \sum_{\alpha}[\chi_\alpha]_{\underline{B}_1\underline{B}_2}^{ay_1y_2}\mathbf{Z}_\alpha(\delta),
\end{align}
where the sum runs over irreducible representations of $\mathrm{Tube}(\mathcal{B}_1^\vee\vert\mathcal{B}_2)$. As a result of the orthogonality relation, this equation can be inverted to obtain 
\begin{align}
    \mathbf{Z}_\alpha(\delta) = \frac{1}{\dim(\CC)^2}\sum_{B_1,B_2}\sum_{a,y_1,y_2} [\widetilde{\chi}_\alpha]_{\underline{B}_1\underline{B}_2}^{ay_1y_2} Z_{B_1B_2}^{ay_1y_2} (\delta).
\end{align}
\end{claim}
In Section \ref{sec:matrix}, we explain how these characters $\chi_\alpha$ and $\widetilde{\chi}_\alpha$ can be evaluated in terms of generalized half-linking numbers,\footnote{These numbers were anticipated to exist in \cite{Lin:2022dhv}, though the authors primarily worked with the special case that $a=b=1$ and $\mathcal{B}_1 = \mathcal{B}_2 = \mathcal{B}_{\mathrm{reg}}$.} which are defined as
\begin{align}\label{eqn:halflinkingdefnintro}
	\begin{split}
		{^{\mathcal{B}_1\mathcal{B}_2}}\Psi_{\alpha\beta(\mu xy)}^{(az)(bw)} = \sqrt{\frac{S_{11}}{\qd_a \qd_b}}~\input{figures/psi.tikz} \, , \  {^{\mathcal{B}_1\mathcal{B}_2}}\widetilde{\Psi}_{\alpha\beta(\mu xy)}^{(az)(bw)} = \sqrt{\frac{S_{11}}{\qd_a \qd_b}}~\input{figures/psit.tikz} \,
	\end{split}
\end{align}
where $S_{\mu\nu}$ is the modular S-matrix of the Drinfeld center $Z(\CC)$. 

We recommend that the reader not get lost in the sea of indices, and instead focus on the coarse features of the definition. 
In Equation \eqref{eqn:halflinkingdefnintro}, one should imagine that the plane of the page is a boundary, and everything above the page is filled with the SymTFT $\TV_{\CC}$. The blue lines are topological line interfaces between the two topological boundaries labeled by $\mathcal{B}_1$ and $\mathcal{B}_2$, the red line is a bulk anyon, and the black lines are topological line operators supported on either $\mathcal{B}_1$ or $\mathcal{B}_2$. One extracts numbers from such configurations of lines by shrinking them to a point, in which case one obtains a local boundary operator which must be proportional to the identity: $\Psi$ and $\widetilde{\Psi}$ are defined to be these constants of proportionality as one varies over the lines and junctions appearing in Equation \eqref{eqn:halflinkingdefnintro}.\footnote{We always restrict our attention to \emph{simple} topological boundary conditions of 2+1d TQFTs throughout the paper. Mathematically, these correspond to \emph{indecomposable} $\CC$-module categories, which heuristically correspond to ``irreducible'' multiplets of boundary conditions of 1+1d QFTs.  Any topological point operator on a simple boundary is proportional to the identity operator.}

We study some of the basic properties of these generalized half-linking numbers in Section \ref{subsec:halflinking},  and ultimately deduce the following formula.

\begin{claim}
The characters of the boundary tube algebra, Equation \eqref{eq:qcharsintro}, can be expressed in terms of the generalized half-linking numbers, Equation \eqref{eqn:halflinkingdefnintro}, as
\begin{align}\label{eqn:verlindeintro}
 \left[ \chi_{\alpha} \right]_{\underline{B}_1\underline{B}_2}^{ay_1 y_2}
    = \sqrt{\qd_a} \sum_{\mu x y y'} \frac{
    {^{\mathcal{B}_1\mathcal{B}_{\mathrm{reg}}}}\widetilde{\Psi}_{\underline{B}_1 \underline{B}_1 (\mu xy')}^{1(ay_1)}    {^{\mathcal{B}_1\mathcal{B}_2}}\Psi_{\alpha\alpha (\mu xy)}^{11}
    {^{\mathcal{B}_2\mathcal{B}_{\mathrm{reg}}}}\Psi_{\underline{B}_2 \underline{B}_2 (\mu yy')}^{1(ay_2)}    }{\sqrt{S_{1 \mu}}} \,
\end{align}
where $S_{\mu\nu}$ is the modular S-matrix of the Drinfeld center $Z(\CC)$.
\end{claim}
The astute reader will notice the structural similarity of this formula to the Verlinde formula \cite{Verlinde:1988sn}. And indeed, this is no accident: as we explain in Section \ref{subsec:quantumcharacters}, Equation \eqref{eqn:verlindeintro} precisely reproduces the standard Verlinde formula in the case that $\mathcal{C}$ is a modular tensor category, all boundaries are Dirichlet ($\mathcal{B}_1=\mathcal{B}_2=\mathcal{B}_{\mathrm{reg}}$), and $a=1$ is the identity line.

To illustrate the power of this character theory, let us consider a special class of theories where this formalism can be almost algorithmically carried out to completion. Suppose that $Q$ is a diagonal rational conformal field theory with chiral algebra $V$, and  take $\mathcal{C}$ to be its category of Verlinde lines. Further assume that $\mathcal{B}_1=\mathcal{B}_2=\mathcal{B}_{\mathrm{reg}}$ are the regular $\CC$-module categories corresponding to the multiplet of Cardy boundary conditions (i.e.\ the boundary conditions which preserve the entire extended chiral algebra $V$ of $Q$). 

Under the conditions of the previous paragraph, Equation \eqref{eq:introrepboundarytube} tells us that the representation basis annulus partition functions are labeled by simple lines $\alpha\in\mathrm{Fun}_\CC(\mathcal{B}_{\mathrm{reg}},\mathcal{B}_{\mathrm{reg}})^{\mathrm{op}}\cong \CC$. In fact, they coincide with the conformal $q$-characters $\mathrm{ch}_\alpha(q)$ of $V$,
\begin{align}\label{eqn:qcharacters}
    \mathbf{Z}_\alpha(\delta)=\mathrm{ch}_\alpha(q) = \mathrm{Tr}_{V_\alpha}q^{L_0-c/24}, \ \ \ q=e^{-\pi \delta}
\end{align}
where $V_\alpha$ is the irreducible module of $V$ corresponding to $\alpha\in \CC$. Furthermore, we explain how  the generalized half-linking numbers ${^{\mathcal{B}_{\mathrm{reg}}\mathcal{B}_{\mathrm{reg}}}}\Psi$ and ${^{\mathcal{B}_{\mathrm{reg}}\mathcal{B}_{\mathrm{reg}}}}\widetilde\Psi$, and hence the characters $\chi_\alpha$ and $\widetilde{\chi}_\alpha$ of $\mathrm{Tube}(\mathcal{B}_{\mathrm{reg}}^\vee\vert\mathcal{B}_{\mathrm{reg}})$, admit closed-form expressions when $\CC$ is modular in terms of its $F$-symbols and $R$-matrices (essentially by combining Equations \eqref{eqn:modularhalfbraiding}, \eqref{eq:psi_DD}, \eqref{eq:verlinde}, and \eqref{eq:verlinde_dual}). Since the $F$-symbols and $R$-matrices are known for a large class of modular tensor categories, this allows for the computation of the symmetry basis annulus partition functions defined in Equation \eqref{eq:introsymbasisannulus} for a large swath of diagonal rational conformal field theories. We have not been able to find results of this generality in the literature (although see \cite{Cardy:1989ir,Fuchs:2000cm,Gaberdiel:2002qa,Fuchs:2004dz} for important related work).

For example, the $F$-symbols and $R$-matrices are reported for all unitary modular tensor categories with rank less than or equal to $4$ in \cite{rowell2009classification}. Likewise, most of the diagonal rational conformal field theories with $c\leq 24$ which realize these MTCs as their category of Verlinde lines have been classified, and the conformal $q$-characters $\mathrm{ch}_\alpha(q)$ of their chiral algebras have been computed \cite{Mukhi:2022bte,Rayhaun:2023pgc}. Thus, this technology completely solves the problem of determining also their symmetry basis annulus partition functions $Z_{B_1B_2}^{ay_1y_2}(\delta)$. We demonstrate this explicitly in Section \ref{sec:fib} in the case of rational conformal field theories with two primary operators and modular tensor category given by the Fibonacci category.

Another interesting application of this formalism is to orbifold conformal field theory. It turns out that the SymTFT allows one to easily see that the representation basis annulus partition functions $\mathbf{Z}_\alpha(\delta)$ of a theory $Q$ are the same as those of any of its orbifolds $Q/\mathcal{B}$. Thus, if we define the ``$\mathcal{B}$-orbifolded'' characters $\chi_\alpha^{\mathcal{B}}$ and $\widetilde\chi_\alpha^{\mathcal{B}}$ to coincide with the characters $\chi_\alpha$ and $\widetilde{\chi}_\alpha$ defined in Equation \eqref{eq:qcharsintro}, but replacing the Dirichlet boundary condition $\mathcal{B}_{\mathrm{reg}}$ with the topological boundary condition corresponding to the module category $\mathcal{B}$, then we obtain an expression of the form 
\begin{align}\label{eqn:orbifoldrelationannulusintro}
    {^{Q/\mathcal{B}}}Z^{a'y_1'y_2'}_{B_1'B_2'}(\delta)=\frac{1}{\dim(\CC)^2}\sum_{B_1,B_2}\sum_{a,y_1,y_2}\left(\sum_\alpha [\chi_\alpha^{\mathcal{B}}]_{\underline{B}_1'\underline{B}_2'}^{a'y_1'y_2'}~[\widetilde\chi_\alpha]_{\underline{B}_1\underline{B}_2}^{ay_1y_2} \right) {^Q}Z_{B_1B_2}^{ay_1y_2}(\delta)
\end{align}
which relates the symmetry basis annulus partition functions of the theory $Q/\mathcal{B}$ to those of $Q$. Here, $B_1',B_2'$ run over simple boundary conditions of the multiplets in $Q/\mathcal{B}$ which are ``dual'' to $\mathcal{B}_1$ and $\mathcal{B}_2$ (see Section \ref{subsec:interfaces} for a description of these dual multiplets of boundary conditions). Moreover, $a',y_1',y_2'$ run over topological lines and junctions of the symmetry category of $Q/\mathcal{B}$ which is dual to $\CC$. 

Equation \eqref{eqn:orbifoldrelationannulusintro} is a kind of non-invertible/annulus analog of a more familiar formula which relates the $T^2$ partition function of a theory $Q/G$, with $G$ a finite group, to the $T^2$ partition function of $Q$ (see e.g.\ \cite[Equation (8.17)]{Ginsparg:1988ui}). We also write down a non-invertible generalization of \cite[Equation (8.17)]{Ginsparg:1988ui} which applies to the torus case in Equation \eqref{eq:torus_generalized_orbifolds}.

\subsection{Twisted boundary states}

In the previous subsection, we implicitly worked in the ``open string channel''. That is, we took time to run along the circle direction of the annulus $I\times S^1$, so that the partition functions could be formulated as suitably-defined traces over interval Hilbert spaces. On the other hand, one might expect to obtain a complementary ``closed string'' perspective on these partition functions by taking time to run instead along the interval direction, in which case one works with appropriately defined boundary states and their (regularized) overlaps. 

\subsubsection{Review of boundaries in rational conformal field theory}
Let us recall how this works in the case that one does not incorporate symmetries into the discussion.
In that case, the annulus partition function $Z_{B_1B_2}(\delta)$, without the insertion of any topological lines, can be represented as 
\begin{align}
Z_{B_1B_2}(\delta)=\mathrm{Tr}_{\mathcal{H}_{B_1B_2}}e^{-H_{\mathrm{op}}\delta}
\end{align}
by taking time to run along the thermal circle. This is the open string channel. 

In the closed string approach, a boundary condition $B$ is represented by a (non-normalizable) state $|B\rangle$ in the $S^1$ Hilbert space of the theory, and the annulus partition function admits an expression of the form
\begin{align}
    Z_{B_1B_2}(\delta) = \langle B_1| e^{-H_{\mathrm{cl}}/\delta} |B_2\rangle
\end{align}
where $H_{\mathrm{cl}}$ is the Hamiltonian of the theory on $S^1$. 

The coherence of these two pictures leads to an equation which one might refer to as open-closed duality, also known as the Cardy condition,
\begin{align}\label{eqn:open-closedintro}
    \langle B_1|e^{-H_{\mathrm{cl}}/\delta}|B_2\rangle = \mathrm{Tr}_{\mathcal{H}_{B_1B_2}}e^{-H_{\mathrm{op}}\delta}\,.
\end{align}
Equation \eqref{eqn:open-closedintro} is a kind of annulus analog of modular invariance of the torus partition function.

In the case of diagonal rational conformal field theories, the boundary conditions which preserve the extended chiral algebra $V$ are referred to as \emph{Cardy boundaries} \cite{Ishibashi:1988kg,Cardy:1989ir}. They are in one-to-one correspondence with irreducible representations $V_\alpha$ of $V$, and their corresponding boundary states can be effectively determined. Since our approach to the closed string channel in the presence of non-invertible symmetries is inspired by this classic story, we briefly review it, following \cite{Cardy:2004hm}.

Recall that the full Hilbert space of a diagonal rational conformal field theory decomposes as 
\begin{align}
    \mathcal{H} = \bigoplus_{\alpha} V_\alpha \otimes \overline{V}_{\bar \alpha}
\end{align}
where the $V_\alpha$ are the irreducible representations of $V$.
The condition that a boundary  $B$ preserves conformal invariance imposes that the corresponding boundary state $|B\rangle$ should satisfy 
\begin{align}\label{eqn:conformalboundaryequation}
    (L_n-\bar{L}_{-n})|B\rangle=0
\end{align}
for all $n\in \mathbb{Z}$, where the $L_n$ are the modes of the stress tensor. Preservation of the full chiral algebra $V$ (as opposed to just its Virasoro subalgebra) leads to the stronger condition that\footnote{One can also consider more general conditions involving non-trivial gluings of the left-moving chiral algebra onto the right-moving chiral algebra \cite{Behrend:1999bn}, but we will not need to consider this more general class of boundary states.}
\begin{align}\label{eqn:chiralalgebrapreservation}
    (W_n-(-1)^{h_W}\overline{W}_{-n})|B\rangle = 0
\end{align}
for all $n\in\mathbb{Z}$, where the $W_n$ are the modes of the generators $W(z)$ of the chiral algebra $V$. 

A key insight is that Equation \eqref{eqn:chiralalgebrapreservation} can be solved within the subspaces $V_\alpha\otimes \overline{V}_{\bar \alpha}$ of the full Hilbert space. Indeed, for each $\alpha$, one can define a so-called \emph{Ishibashi state} \cite{Ishibashi:1988kg}
\begin{align}\label{eqn:ishistatedefn}
    |\alpha\rrangle_{\mathrm{Ish}} = \sum_{N=0}^\infty |\alpha;N\rangle \otimes U\overline{|\alpha;N\rangle} \in V_\alpha\otimes \overline{V}_{\bar \alpha}
\end{align}
where $\{|\alpha;N\rangle\}_{N\in \mathbb{Z}^{\geq 0}}$ is an orthonormal basis of $V_\alpha$, and $U$ is an anti-unitary operator satisfying $U\overline{W}_n=(-1)^{h_W}\overline{W}_nU$ for all the generators. It follows from the definition that the regularized overlaps satisfy
\begin{align}\label{eqn:usualishioverlapintro}
    {_{\mathrm{Ish}}}\llangle \alpha|e^{-H_{\mathrm{cl}} /\delta}|\beta\rrangle_{\mathrm{Ish}} = \delta_{\alpha\beta}\mathrm{ch}_\alpha(\tilde{q})=\delta_{\alpha\beta}\sum_\gamma \mathbb{S}_{\alpha\gamma} \mathrm{ch}_\gamma(q)\,, \ \ \ q=e^{-\pi \delta} \,, \tilde{q} = e^{-4\pi/\delta} \,,
\end{align}
where $\mathrm{ch}_\alpha(q)$ are the conformal $q$-characters of $V$, defined in Equation \eqref{eqn:qcharacters}, and $\mathbb{S}_{\alpha\beta}$ is the modular S-matrix associated with the RCFT.

The Ishibashi states are not quite boundary states, but it turns out they are useful building blocks, in the sense that the true Cardy states can be expanded in linear combinations of them, 
\begin{align}
    |a\rangle = \sum_{\beta} C_{a\beta}|\beta\rrangle_{\mathrm{Ish}}\,.
\end{align}
The coefficients $C_{a\beta}$ can be constrained by imposing open-closed duality. Indeed, from the open string picture, compatibility with a Hilbert space interpretation requires that $Z_{ab}(\delta)$ should be expandable in conformal $q$-characters of $V$ with non-negative integral coefficients, i.e.\
\begin{align}
    Z_{ab}(\delta) = \mathrm{Tr}_{\mathcal{H}_{ab}}e^{-H_{\mathrm{op}}\delta} = \sum_\alpha n^\alpha_{ab} \mathrm{ch}_\alpha(q)
\end{align}
with $n^\alpha_{ab}\in\mathbb{Z}^{\geq 0}$. On the other hand, we can directly compute $Z_{ab}(\delta)$ in the closed string picture,
\begin{align}
   Z_{ab}(\delta) =  \langle a|e^{-H_{\mathrm{cl}}/\delta}|b\rangle  = \sum_\beta\left(\sum_{\alpha} C_{a\alpha}^\ast C_{b\alpha} \mathbb{S}_{\alpha\beta}\right) \mathrm{ch}_{\beta}(q)\,.
\end{align}
Compatibility of the two pictures requires that the $C_{a\alpha}$ should satisfy
\begin{align}
    \sum_\alpha C_{a\alpha}^\ast C_{b\alpha}\mathbb{S}_{\alpha\beta} = n_{ab}^\beta \,.
\end{align}
The Verlinde formula,
\begin{align}
    \sum_\alpha \frac{\mathbb{S}_{\alpha a}^\ast \mathbb{S}_{\alpha b}\mathbb{S}_{\alpha\beta}}{\mathbb{S}_{0\alpha}} = N_{b\beta}^a
\end{align}
implies that we may take $C_{a\alpha} = \mathbb{S}_{\alpha a}/\sqrt{\mathbb{S}_{0\alpha}}$ and $n_{ab}^\beta=N_{b\beta}^a$
as a solution to this equation, where the $N_{b\beta}^a$ are the fusion coefficients of the RCFT. In this way, we obtain a boundary state
\begin{align}
    |a\rangle = \sum_{\alpha} \frac{\mathbb{S}_{\alpha a}}{\sqrt{\mathbb{S}_{0\alpha}}}|\alpha\rrangle_{\mathrm{Ish}}
\end{align}
for each irreducible representation of the chiral algebra $V$. Hence, $a,b,\cdots$ take values in the same set as $\alpha,\beta,\cdots$.

\subsubsection{Generalized Ishibashi states and a generalized Verlinde formula}

\begin{figure}
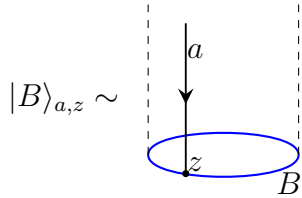

    \centering
    \ctikzfig{figures/twistedboundarystatedefn}
    \caption{A twisted boundary state.}
    \label{fig:twistedboundarystatedefn}
\end{figure}

The Cardy boundary conditions associated with a diagonal RCFT $Q$, which were described in the previous subsection, can be thought of as belonging to a multiplet of boundaries which transform into one another under the action of the Verlinde lines \cite{Verlinde:1988sn} of $Q$. We would like to develop an analogous understanding of states corresponding to boundary conditions which belong to arbitrary multiplets, inside of arbitrary (unitary, compact) conformal field theories with arbitrary fusion category symmetries. We would also like to understand how to characterize the twisted versions of these states, i.e.\ the boundary states which are suitable for describing boundary conditions which have been twisted by a symmetry line, as in Figure \ref{fig:twistedboundarystatedefn}.

The first step is to determine analogs of the Ishibashi states. To explain how we approach this, we recall from \cite{Lin:2022dhv} that, due to Figure \ref{fig:defect.operators}, the $a$-twisted $S^1$ Hilbert space of a conformal field theory admits a decomposition of the form 
\begin{align}
    \mathcal{H}_{a} \cong \bigoplus_{\mu\in\mathrm{Irr}(Z(\CC))} W^\mu_a \otimes \mathcal{V}_\mu
\end{align}
where $W^\mu_a$ is the space of topological junction operators $y$ between the bulk line $\mu$ of the SymTFT and the (Dirichlet) boundary line $a$, and $\mathcal{V}_\mu$ is the space of (not necessarily topological) junction operators $\widetilde{\mathcal{O}}$ on the physical boundary $\widetilde{Q}$ on which the $\mu$ line can terminate. By a version of the state/operator correspondence, $W^\mu_a$ and $\mathcal{V}_\mu$ can be realized as Hilbert spaces of the SymTFT $\TV_{\CC}$ on the disk $D^2$ with suitable decorations, as in Figure \ref{fig:twistedsectorhilbertspaces}. The subspace $W_a^\mu\otimes\mathcal{V}_\mu\subset\mathcal{H}_a$ can be thought of as the subspace of twisted sector states on $S^1$ which transform according to the irreducible representation of $\mathrm{Tube}(\CC)$ labeled by $\mu$ (cf.\ Equation \eqref{eqn:Rep(Tube(C))}).

\begin{figure}
    \centering
    \input{figures/twistedsectorhilbertspaces.tikz}\vspace{-.5cm}
    \caption{On the left, the space $\mathcal{V}_\mu$ is the Hilbert space of $\mathrm{TV}_{\mathcal{C}}$ on a disk $D^2$ with the physical boundary $\widetilde{Q}$ imposed on $\partial D^2$ and $\mu$ puncturing the origin. On the right, the space $W_a^\mu$ is the Hilbert space of $\mathrm{TV}_{\mathcal{C}}$ on a disk $D^2$ with the Dirichlet boundary $\mathcal{B}_{\mathrm{reg}}$ imposed on $\partial D^2$, with $\mu$ puncturing the origin (past-oriented), and with $a$ puncturing the boundary (future-oriented).}\label{fig:twistedsectorhilbertspaces}
\end{figure}

In the spirit of Equation \eqref{eqn:ishistatedefn}, we propose to look for solutions to Equation \eqref{eqn:conformalboundaryequation} within the $\mu$-representation sectors $W^\mu_a\otimes \mathcal{V}_\mu$. In fact, the SymTFT gives us a way to construct them. Indeed, given a multiplet $\mathcal{B}$ of conformal boundary conditions, by passing to the SymTFT, we can obtain a line junction $\widetilde{B}$ between the physical boundary $\widetilde{Q}$ and the topological boundary labeled by $\mathcal{B}$, as in Figure \ref{fig:bdysymtft}. This line junction $\widetilde{B}$ is common to all the boundaries $B$ in the multiplet $\mathcal{B}$. Using $\widetilde{B}$, for each bulk line $\mu$ admitting a topological endpoint $y$ on the boundary $\mathcal{B}$ we can prepare a state in $\mathcal{V}_{\bar\mu}$ via the following Euclidean path integral of the SymTFT on a ``solid cone,''
\begin{align} \label{eq:half_Ishi_intro}
    |y;\mu,\mathcal{B} \rrangle \equiv ~\input{figures/halfishibashistate.tikz}~~~ \in \mathcal{V}_{\bar{\mu}} \,.
\end{align}
Combining this with a choice of topological junction $|\bar x,\bar\mu,\bar a\rangle \in W^{\bar\mu}_{\bar{a}}$, we obtain our desired $a$-twisted generalized Ishibashi states, 
\begin{align}
    |\bar x,\bar \mu, \bar a\rangle \otimes |y;\mu,\mathcal{B}\rrangle \in W_{\bar a}^{\bar \mu}\otimes\mathcal{V}_{\bar \mu}\subset\mathcal{H}_{\bar a}
\end{align}
whose overlaps we show can be expressed as linear combinations of the representation basis partition functions $\mathbf{Z}_\alpha(\delta)$ with coefficients given by  generalized half-linking numbers,
\begin{align} \label{eq:Ishi_overlapintro}
\begin{split}
    &\langle \bar x_1,\bar \mu_1, \bar a_1|\bar x_2,\bar\mu_2,\bar a_2\rangle \llangle y_1;\mu_1, \mathcal{B}_1 | e^{-H_{\mathrm{cl}}/\delta} | y_2;\mu_2, \mathcal{B}_2 \rrangle  \\
    &\hspace{.4in} = \delta_{a_1a_2}\delta_{\mu_1 \mu_2}\delta_{x_1x_2}\sqrt{\frac{\d_\mu \d_a}{S_{11}}} \sum_{\alpha} {^{\mathcal{B}_1\mathcal{B}_2}}\Psi_{\alpha\alpha(\mu y_1y_2)}^{11} \mathbf{Z}_\alpha (\delta) \,.
\end{split}
\end{align}

The reason we refer to these as Ishibashi states is because, if one takes $Q$ to be a diagonal rational conformal field theory, $\mathcal{B}$ to be the multiplet of Cardy boundaries, and $a=1$ to be the identity line, then the spaces $W^\mu_1\otimes\mathcal{V}_\mu$ coincide with the representations of the chiral algebra $V_{\alpha}\otimes \overline{V}_{\bar \alpha}$, and the states $|\bar x,\bar \mu, \bar a\rangle \otimes |y;\mu,\mathcal{B}\rrangle$ go over to the usual Ishibashi states $|\alpha\rrangle_{\mathrm{Ish}}$. Further, Equation \eqref{eq:Ishi_overlapintro} reduces to Equation \eqref{eqn:usualishioverlapintro}.
 They also satisfy the usual condition that $\left( L_n - \bar{L}_{-n} \right) |\bar x,\bar \mu, \bar a\rangle \otimes|y;\mu,\mathcal{B} \rrangle = 0$ for all $n$.

One of the main results of this paper is that these $a$-twisted, generalized Ishibashi states function in much the same way that the usual Ishibashi states do. 

\begin{claim}
    Let $Q$ be a 1+1d conformal field theory with a symmetry category $\CC$ and a multiplet $\mathcal{B}$ of conformal boundary conditions. Every $a$-twisted boundary state $|B\rangle_{a,z}$ in the multiplet can be obtained as a linear combination of the $a$-twisted generalized Ishibashi states as
\begin{equation} \label{eq:twisted_bdy_state_intro}
    \ket{B}_{a,z} = \sum_{\mu x y} \sqrt{\frac{S_{11}}{S_{1\mu}}} {^{\mathcal{B}\mathcal{B}_{\mathrm{reg}}}}\Psi_{\underline{B}\underline{B}(\mu y x)}^{1(az)} |\bar{x}, \bar{\mu}, \bar{a}\rangle \otimes |y;\mu, \mathcal{B} \rrangle \,
\end{equation}
    where the ${^{\mathcal{B}\mathcal{B}_{\mathrm{reg}}}}\Psi_{\underline B \underline B (\mu y x)}^{1(az)}$ are generalized half-linking numbers.
\end{claim}
The proof that these boundary states satisfy open-closed duality invokes our generalized Verlinde formula in the same way that the construction of Cardy states invokes the usual Verlinde formula. Indeed, open-closed duality applied to the symmetry basis annulus partition functions defined in Equation \eqref{eq:introsymbasisannulus} imposes that 
\begin{equation} \label{eq:twisted_Cardy_intro}
    \prescript{}{a,z_1}{\bra{B_1}} \tilde{q}^{\frac12(L_0 + \bar{L}_0 - c/12)} \ket{B_2}_{a,z_2} \stackrel{!}{=} \mathrm{Tr}_{\mathcal{H}_{B_1 B_2}} \left( 
        \mathsf{H}_{B_1 B_2 ,a}^{B_1 B_2, z_1 z_2} q^{L_0 -c/24}
    \right) \equiv Z_{B_1 B_2}^{a z_1 z_2} (\delta) \,,
\end{equation}
where the operator $\mathsf{H}_{B_1B_2,a}^{B_2B_2,z_1z_2}$ was defined in Equation \eqref{eqn:boundarylassointro}. One can expand both sides in the representation basis partition functions $\mathbf{Z}_\alpha$: using Equation \eqref{eq:Ishi_overlapintro} and Equation \eqref{eq:twisted_bdy_state_intro}, one finds that the coefficients of the expansion of the left-hand side are schematically $\Psi^3$, while the coefficients in the expansion of the right-hand side are schematically $\chi$. The equality $\Psi^3\sim \chi$ is precisely our generalized Verlinde formula, Equation \eqref{eqn:verlindeintro}.

\subsection{Applications}

The technology developed in this work has a few consequences. 
\begin{enumerate}
    \item \textbf{Selection rules on bulk/boundary/interface local operators}: 
    We argue that a correlation function $\langle \mathcal{O}_1(t_1)\cdots\mathcal{O}_n(t_n)\rangle$ of boundary local operators $\mathcal{O}_i$ must vanish if the tensor product of their representations $\alpha_i$ under the appropriate boundary tube algebra
    does not contain the trivial representation. In the SymTFT setup, $\alpha_i$ is interpreted as a topological line operator on a topological boundary condition of $\TV_{\CC}$ (see Figure \ref{fig:bdydefect.operators}), and the condition that the tensor product of the corresponding representations contains a copy of the trivial representation translates to the condition that the fusion of the lines $\alpha_i$ contains an identity channel. We further discuss the generalization of this selection rule to correlators involving a mix of bulk, boundary, and interface local operators.
    \item \textbf{Constraining degeneracy on an interval}: Consider a $\CC$-symmetric QFT on an interval with two $\CC$-symmetric boundary conditions belonging to multiplets which are inequivalent as $\CC$-module categories. We show that the dimension of any irreducible representation of the corresponding boundary tube algebra must be greater than one: hence, there is a lower bound on the ground state degeneracy of the interval Hilbert space. In \cite{Cordova:2022lms}, it was argued that similar degeneracies lead to constraints on the central charge of any CFT which mediates a phase transition between two distinct $G$-symmetric symmetry protected topological (SPT) phases. We expect that our results will be useful for extending this idea to non-invertible symmetries.
    \item \textbf{Constraining the fusion rules of conformal line defects/interfaces}: When two conformal line defects $I, I'$ are close and parallel to each other, upon regularizing by a self-energy counterterm, they can be viewed as a direct sum of simple conformal lines with non-negative integer coefficients \cite{Bachas:2007td, Diatlyk:2024zkk}. The SymTFT setup furnishes selection rules on these fusion coefficients. Concretely, in the SymTFT setup, the two defects $I$ and $I'$ are associated with topological interfaces $\mathcal{I}$ and $\mathcal{I}'$ which label the multiplets to which $I$ and $I'$, respectively, belong (see Figure \ref{fig:interfacesandwich}). We argue that any simple conformal defect $I''$ arising in the fusion of $I$ and $I'$ must be associated with a topological interface $\mathcal{I}''$ of the bulk SymTFT that is contained in the decomposition of the fusion $\mathcal{I}\boxtimes \mathcal{I}'$ into simple surfaces. 
    \item \textbf{Constraining line defect RG flows}: Consider a CFT with $\CC$ symmetry, where $\CC$ might be a subcategory of a larger symmetry category of the theory. Given a generically non-topological line defect $I$, an interesting question is what it flows to in the IR, e.g.\ whether the defect is screened \cite{Wu:2023ezm, Aharony:2023amq, Aharony:2022ntz, Cuomo:2022xgw, Cuomo:2021kfm, Cuomo:2021rkm, Nagar:2024mjz, Giombi:2022vnz, Graham:2003nc, Konechny:2012wm, Konechny:2020jym, Konechny:2019wff, Andrei:2018die, Konechny:2015qla, Kormos:2009sk}. We argue that if $I$ is associated with a non-trivial topological interface $\mathcal{I}$ in the bulk SymTFT, then $I$ cannot flow to the trivial line or any decoupled line in the IR.
\end{enumerate}

As was alluded to earlier, in a companion paper \cite{Choi:2024wfm}, we elaborate on another topic: namely, the application of the representation theory of boundary tube algebras to the study of a non-invertible symmetry-resolved version of the Affleck-Ludwig-Cardy formula, as well as its implications for symmetry-resolved entanglement entropy.

\subsection{Outline}

The rest of the paper is organized as follows.
After establishing some category-theoretic tools in Section \ref{sec:review}, and in particular introducing the ``generalized half-linking numbers'' which are ubiquitous in our work, we define in Section \ref{sec:tube} a class of generalized tube algebras which furnish the appropriate algebraic structure for describing the interaction of topological line operators with junction operators sitting at the intersection of a collection of boundaries and interfaces. 
In Section \ref{sec:symTFT}, we then develop a 2+1d symmetry topological field theory (SymTFT) approach to studying the representation theory of these tube algebras.
In Section \ref{sec:matrix}, we show how this SymTFT framework leads to a character theory for tube algebras --- reminiscent of and generalizing the character theory of finite groups --- as well as a generalization of the Verlinde formula.
We leverage this understanding to define symmetry-resolved partition functions of 1+1d QFTs --- focusing on the torus in Section \ref{sec:torus} and the annulus in Sections \ref{sec:annulus} and \ref{sec:boundariesclosedstring} --- which encode information about the interplay of symmetries with twisted sector and boundary-changing local operators. 
Section \ref{sec:annulus} discusses annulus partition functions in the open string channel, whereas Section \ref{sec:boundariesclosedstring} treats the closed string channel.
We illustrate our results explicitly in the context of theories with Fibonacci symmetry in Section \ref{sec:fib}. We then conclude in Section \ref{sec:apps} by sketching a few of the many applications which flow from our work, leaving  more detailed investigations to the future.

\subsection{Notation, conventions, and assumptions}

In much of this work, we assume that $\CC$ acts faithfully for simplicity (for a precise definition, the reader may refer to \cite[Appendix A]{Lin:2022dhv} and \cite[Section 2]{Bhardwaj:2023ayw}). However, this assumption can often be relaxed, particularly in earlier sections.

As a technical comment, we mention that if the fusion category $\mathcal{C}$ contains a self-dual topological line $a \cong \bar{a}$ whose Frobenius-Schur indicator is negative, then the complete isotopy invariance of topological line networks is potentially hindered \cite{Chang:2018iay} unless the conventions are carefully chosen.
To keep our discussions concise, we assume every self-dual line we encounter has a positive Frobenius-Schur indicator throughout this paper.
The general case can be treated, for instance, by using the ``flag'' method of \cite{Kitaev:2005hzj,Simon:2022ohj}.

We often consider partition functions of the Turaev-Viro TQFT $\TV_{\CC}$ on a solid ball with various operator insertions. Our convention is that we normalize the (undecorated) solid ball partition function to be 1 by tuning the boundary Euler counterterm appropriately.

Below, we offer a glossary of various symbols that are used throughout our work.

\begin{footnotesize}
\begin{list}{}{
	\itemsep -1pt
	\labelwidth 23ex
	\leftmargin 13ex	
	}

\item[$\alpha,\beta,\cdots$] Line interfaces between gapped boundary conditions of the SymTFT $\TV_{\CC}$. They also label  representations of the boundary tube algebras $\mathrm{Tube}(\mathcal{B}_1^\vee\vert\mathcal{B}_2)$.
\item[$\mathcal{B}$] A $\mathcal{C}$-module category. Such a category can refer to $\mathcal{C}$-multiplet of boundary conditions of a 1+1d quantum field theory $Q$ (i.e.\ the objects of $\mathcal{B}$ correspond to boundary conditions of $Q$) or represent a topological boundary condition of $\TV_{\CC}$ (i.e.\ the entire category $\mathcal{B}$ labels a boundary condition).
\item[$\mathcal{B}^\vee$] The left (right) $\mathcal{C}$-module category dual to the right (left) $\mathcal{C}$-module category $\mathcal{B}$, see Section \ref{subsec:modulebimodule}.
\item[$\mathcal{B}_{\mathrm{reg}}$] The regular $\mathcal{C}$-module category, whose simple objects are in one-to-one corresponding with simple objects of $\CC$.  When interpreted as a topological boundary condition of the 2+1d TQFT $\TV_{\CC}$, it corresponds to the canonical ``Dirichlet'' boundary condition. 
\item[$B$] A boundary condition of a 1+1d quantum field theory, or an object of a module category $\mathcal{B}$.
\item[$\underline{B}$] Given a boundary condition $B$ of a 1+1d QFT $Q$ with $\CC$ symmetry, $\underline{B}$ is the topological line interface appearing in the SymTFT picture of $B$ in Figure \ref{fig:bdysymtft}. It is a topological line interface between the Dirichlet boundary condition of $\TV_{\CC}$ corresponding to $\mathcal{B}_{\mathrm{reg}}$, and the topological boundary condition of $\TV_{\CC}$ corresponding to the $\CC$-multiplet $\mathcal{B}$ to which $B$ belongs.
\item[$\widetilde{B}$] Given a boundary condition $B$ of a 1+1d QFT $Q$ with $\CC$ symmetry, $\widetilde{B}$ is the not-necessarily-topological line interface appearing in the SymTFT picture of $B$ in Figure \ref{fig:bdysymtft}. It is a line interface between the physical boundary condition $\widetilde{Q}$ of $\TV_{\CC}$, and the topological boundary condition of $\TV_{\CC}$ corresponding to the $\CC$-multiplet $\mathcal{B}$ to which $B$ belongs.
\item[$|B\rangle_{a,z}$] A state in the $S^1$ Hilbert space of a 1+1d QFT twisted by a topological line $a\in \CC$. It is the boundary state describing the boundary condition $B$ with $a$ terminating topologically it with topological junction $z$. See Section \ref{sec:boundariesclosedstring}.
\item[$\mathcal{C}$] A unitary fusion category of topological line operators acting on a 1+1d QFT. All fusion categories are assumed to have positive Frobenius--Schur indicators in this work.
\item[$\CC_A$] The category of right $A$-modules in $\mathcal{C}$, which forms a left $\CC$-module category.
\item[$\mathrm{ch}_\alpha(q)$] The conformal $q$-characters of a rational chiral algebra.
\item[$\mathsf{d}_a$] The quantum dimension of an object $a$ of a fusion category $\mathcal{C}$, see Equation \eqref{eq:F_symbols}.
\item[$\dim(\CC)$] The global dimension of $\CC$, defined as $\dim(\CC)=\sqrt{\sum_{a\in\mathrm{Irr}(\CC)} \d_a^2}$.
\item[$F_{abc}^d$] The $F$-symbols of a fusion category, Equation \eqref{eq:F_symbols}.
\item[$\widetilde{F}_{abB}^{B'}$] The $F$-symbols of a module category, Equation \eqref{eq:FT_symbols}.
\item[$\mathrm{Fun}_{\CC}(\mathcal{B}_1,\mathcal{B}_2)$] The category of $\CC$-module functors between two $\CC$-module categories $\mathcal{B}_1$ and $\mathcal{B}_2$.
\item[$\mathsf{H}_{B_1B_2,c}^{C_1C_2,y_1y_2}$] A boundary lasso operator of $\mathrm{Tube}(\mathcal{B}_1^\vee\vert\mathcal{B}_2)$, defined in Equation \eqref{eq:Hlassoop} and Equation \eqref{eq:Hlassostate}.
\item[$\mathcal{H}_c$] The Hilbert space of twisted sector operators of a 1+1d QFT $Q$, i.e.\ local operators which live at the endpoint of a topological line $c\in\mathcal{C}$. 
\item[$\mathcal{H}_\CC$] The extended Hilbert space, $\mathcal{H}_{\CC} = \bigoplus_{c\in\CC}\mathcal{H}_c$.
\item[$\mathcal{H}_{B_1B_2}$] The Hilbert space of boundary-changing local operators between boundaries $B_1$ and $B_2$ of a 1+1d QFT $Q$.
\item[$\mathcal{H}_{\mathcal{B}_1^\vee\mathcal{B}_2}$] The extended Hilbert space of boundary-changing local operators between boundaries in two $\CC$-multiplets $\mathcal{B}_1$ and $\mathcal{B}_2$, defined as $\mathcal{H}_{\mathcal{B}_1^\vee\mathcal{B}_2}=\bigoplus_{B_1\in\mathcal{B}_1^\vee}\bigoplus_{B_2\in\mathcal{B}_2}\mathcal{H}_{B_1B_2}$.
\item[$\mathcal{I}$] A $(\CC_1,\CC_2)$-bimodule category. Such a category can refer to a $(\CC_1,\CC_2)$-multiplet of interfaces between two 1+1d quantum field theories $Q_1$ and $Q_2$ (i.e.\ the objects of $\mathcal{I}$ correspond to interfaces between $Q_1$ and $Q_2$) or to represent a topological interface between $\TV_{\CC_1}$ and $\TV_{\CC_2}$ (i.e.\ the entire category $\mathcal{I}$ labels a topological interface).
\item[$\mathcal{I}_{\mathrm{reg}}$] The regular $(\CC,\CC)$-bimodule category, whose simple objects are in one-to-one correspondence with simple objects of $\CC$. When thought of as a topological surface operator of $\TV_{\CC}$, it corresponds to the trivial surface.
\item[$I$] A (general non-topological) line operator of a 1+1d QFT $Q$, or an interface between two QFTs. 
\item[$\widetilde{I}$] A topological line junction arising in the SymTFT description of $I$, see Figure \ref{fig:interfacesandwich}.
\item[$\underline{I}$] A (not necessarily topological) line junction arising in the SymTFT description of $I$, see Figure \ref{fig:interfacesandwich}.
\item[$\mathrm{Irr}(\mathcal{X})$] The set of simple objects of a category $\mathcal{X}$.
\item[$\mathsf{L}_{a,c}^{b,dyz}$] A lasso operator of $\mathrm{Tube}(\mathcal{C})$, see Equation \eqref{eq:lasso}.
\item[$N_{ab}^c$] The fusion coefficients of a fusion category.
\item[$\widetilde{N}_{aB}^{B'}$, $\widetilde{N}_{Ba}^{B'}$] The fusion coefficients of a left or right $\CC$-module category, respectively.
\item[$\Omega_{a\mu}^b$] The half-braiding matrix, see Equation \eqref{eq:half-braiding} and Equation \eqref{eq:half-braiding_2}.
\item[$\widetilde{\Psi}$, $\widetilde{\Psi}$] The half-linking numbers, see Equation \eqref{eqn:halflinkingdefn}.
\item[$Q$] A 1+1d unitary compact bosonic quantum field theory, often conformal.
\item[$\widetilde{Q}$] The physical boundary of $\mathrm{TV}_{\mathcal{C}}$ corresponding to a 1+1d QFT $Q$ with $\mathcal{C}$ symmetry. 
\item[$R_{ab}^c$] The R-matrices of a braided fusion category, see Equation \eqref{eq:braidingdef}.
\item[$S_{\mu\nu}$] The modular S-matrix of the Drinfeld center $Z(\mathcal{C})$ of a fusion category $\mathcal{C}$.
\item[$\mathbb{S}_{ab}$] The modular S-matrix of $\mathcal{C}$, in the case that $\mathcal{C}$ is a modular tensor category. (To be distinguished from $S_{\mu\nu}$, which is the modular S-matrix of $Z(\mathcal{C})$.)
\item[$\mathrm{Tube}(\CC)$] The ordinary tube algebra, see Section \ref{subsec:ordinarytube}.
\item[$\mathrm{Tube}(\mathcal{B}_1^\vee\vert\mathcal{B}_2)$] The boundary tube algebra, see Section \ref{subsec:interfacetube}.
\item[$\mathrm{Tube}(\mathcal{I}_1\vert\cdots\vert\mathcal{I}_n)$] The most general class of tube algebras considered in this paper, see Section \ref{subsec:interfacetube}.
\item[$\mathrm{TV}_{\mathcal{C}}$] The Turaev--Viro 2+1d TQFT associated with $\mathcal{C}$, which corresponds as a modular tensor category to $Z(\mathcal{C})$, the Drinfeld center of $\mathcal{C}$. 
\item[$V$] Typically the chiral algebra of a rational conformal field theory.
\item[$\mathcal{V}_\mu$] The Hilbert space of local junction operators living on the physical boundary $\widetilde{Q}$ of $\TV_{\CC}$ at the endpoint of a bulk line $\mu$, see Section \ref{subsec:twistedsectorlocalops}. This (generally infinite-dimensional) space, which is graded by conformal dimensions, can be thought of as encoding the multiplicity with which the irreducible representation of $\mathrm{Tube}(\CC)$ labeled by $\mu$ appears in the extended Hilbert space $\mathcal{H}_\CC$.
\item[$W^\mu$, $W^\mu_a$] The space $W^\mu_a$ is the Hilbert space of topological junction operators between a bulk like $\mu$ in $\TV_{\CC}$ and a boundary line $a$ on the Dirichlet boundary condition. The space $W^\mu=\bigoplus_{a\in \CC}W^\mu_a$ furnishes an irreducible representation of $\mathrm{Tube}(\CC)$, see Section \ref{subsec:twistedsectorlocalops}.
\item[$W^\alpha$, $W^\alpha_{\underline{B}_1\underline{B}_2}$] The space $W^\alpha_{\underline{B}_1\underline{B}_2}$ is the Hilbert space of topological junction operators $x$ appearing in Figure \ref{fig:bdydefect.operators}, and $W^\alpha=\bigoplus_{\underline{B}_1,\underline{B}_2}W^\alpha_{\underline{B}_1\underline{B}_2}$. 
\item[${[\chi_\mu]_{a}^{c,dyz}}$] The character of the $\mu$-representation of $\mathrm{Tube}(\CC)$, see Section \ref{subsec:quantumcharacters}.
\item[${[\chi_\alpha]_{\underline{B}_1\underline{B}_2}^{az_1z_2}}$] A character of the boundary tube algebra $\mathrm{Tube}(\mathcal{B}_1^\vee\vert\mathcal{B}_2)$, see Equation \eqref{eq:bdy_char_solid_torus}.
\item[${[\widetilde\chi_\alpha]_{\underline{B}_1\underline{B}_2}^{az_1z_2}}$] A dual character of the boundary tube algebra $\mathrm{Tube}(\mathcal{B}_1^\vee\vert\mathcal{B}_2)$, see Equation \eqref{eq:dualbdycharacter_solidtorus}.
\item[$Z(\mathcal{C})$] The Drinfeld center of $\mathcal{C}$.
\item[$Z_{\mathcal{C}}(\mathcal{I})$] The center of the $(\mathcal{C},\mathcal{C})$ bimodule category $\mathcal{I}$, see Equation \eqref{eq:centerbimodle}.
\item[$\mathbf{Z}_\mu(\tau)$] The representation basis torus partition functions of a 1+1d QFT, see Equation \eqref{eq:rep_torus}.
\item[$Z_{a}^{b,cij}(\tau)$] The symmetry basis torus partition functions of a 1+1d QFT, see Equation \eqref{eq:torus_symm}.
\item[$\mathbf{Z}_\alpha(\delta)$] The representation basis annulus partition functions of a 1+1d QFT, see Equation \eqref{eq:rep_annulus}.
\item[$Z_{B_1B_2}^{az_1z_2}(\delta)$] The symmetry basis annulus partition functions of a 1+1d QFT, Equation \eqref{eqn:openstringsymmetrybasis}.
\item[$\mu,\nu,\cdots$] A topological line operator (or, anyon) of the SymTFT $\TV_{\CC}$.
\item[$|\mu;y,\mathcal{B}\rrangle$] A generalized (half) Ishibashi state associated with the multiplet $\mathcal{B}$,
\end{list}
\end{footnotesize}

\paragraph{Note added:} As this work was in its final stages of completion, the papers \cite{Copetti:2024onh,Cvetic:2024dzu,Cordova:2024iti,Copetti:2024dcz}, with which our manuscript shares non-negligible overlap, appeared on the arXiv. Moreover, we were recently informed that several further groups have also been studying similar topics: in particular, this paper was submitted in coordination with \cite{Bhardwaj:2024igy,Das:2024qdx,GarciaEtxebarria:2024jfv,Heymann:2024vvf}. We thank the authors of op.\ cit.\ for their cooperation. 
 
\section{Category theory and generalized half-linking}\label{sec:review}

In this section, we lay out the conventions we use for concretely handling various category theoretic objects.
We refer to \cite{Bhardwaj:2017xup,Chang:2018iay} for pedagogical expositions on the relationship between categories and symmetries of 1+1d quantum field theories, and to \cite{Kong:2022cpy} for the categorical approach to 2+1d topological field theories.
We focus on unitary quantum field theories, and the categories that we discuss are assumed to be equipped with appropriate unitary structures.

Below, lines operators, boundary conditions, and interfaces are referred to as being \emph{simple} if the only topological local operator on them is the identity operator.
A simple object (sometimes also called elementary) cannot be written as a sum of other objects.

Many of the facts reviewed below are standard, however there are certain elements of our discussion which, to the best of the authors' knowledge, are new. This includes a more detailed treatment of generalized half-linking numbers, which were anticipated in \cite{Lin:2022dhv}.

\subsection{Fusion categories and braided categories}\label{subsec:fusioncategories}

Fusion categories encode the algebraic properties of a finite set of topological line operators which are closed under fusion. 
They appear both in two-dimensional quantum field theories and on two-dimensional objects in higher-dimensional theories, such as boundary conditions or interfaces of three-dimensional QFTs.
In the former case, the fusion category characterizes a finite (generally non-invertible) global symmetry of the 1+1d quantum field theory \cite{Bhardwaj:2017xup,Chang:2018iay}.

We denote the set of simple objects (i.e.\ simple topological lines) of a fusion category $\mathcal{C}$ as $\mathrm{Irr}(\mathcal{C})$, and write its elements as $a,b,c, \cdots$.
The orientation-reversal of a topological line $a$ is denoted as $\bar{a}$, and is referred to as the dual of $a$.

Given two simple topological lines $a$ and $b$, their fusion is given by 
\begin{align}
    a \otimes b \cong \bigoplus_{c \in \mathrm{Irr}(\mathcal{C})} N_{ab}^c c,
\end{align}
where $N_{ab}^c \in \mathbb{Z}^{\geq 0}$ are referred to as the fusion coefficients.
Given  simple lines $a$, $b$, and $c$, the topological junction operators which can sit at the trivalent junction of the three lines span a complex vector space $\mathrm{Hom}_{\mathcal{C}} ( a \otimes b, c)$ which is $N_{ab}^c$-dimensional.
We fix a set of basis vectors $v_{ab}^{c;i}$ of the fusion spaces $ \mathrm{Hom}_{\mathcal{C}} ( a \otimes b, c)$ , and also a set of dual basis vectors $\bar{v}_{ab}^{c;i}$ of the splitting spaces $\in \mathrm{Hom}_{\mathcal{C}} ( c,  a \otimes b)$, where $i=1, \cdots, N_{ab}^c$. Pictorially, we represent them as
\begin{equation} \label{eq:marked_junctions}
    v_{ab}^{c;i} = \raisebox{-1.8em}{\begin{tikzpicture}
    \draw [thick, decoration = {markings, mark=at position .7 with {\arrow[scale=1.5]{stealth}}}, postaction=decorate] (0,0) to (1,{1/sqrt(3)});
    \draw [thick, decoration = {markings, mark=at position .7 with {\arrow[scale=1.5]{stealth}}}, postaction=decorate] (2,0) to (1,{1/sqrt(3)});
    \draw [thick, decoration = {markings, mark=at position .7 with {\arrow[scale=1.5]{stealth}}}, postaction=decorate] (1,{1/sqrt(3)}) to (1,{sqrt(3)});
    \node at (1+0.2,{1/sqrt(3)+0.1}) {$i$};
    \node at (-0.1,0.15) {$a$};
    \node at (2.1,0.15) {$b$};
    \node at (1.1,{sqrt(3)+0.1}) {$c$};
    \node at (1,{1/sqrt(3)+0.2}) {$\times$};
\end{tikzpicture}} \,, \qquad
\bar{v}_{ab}^{c;i} = \raisebox{-1.8em}{\begin{tikzpicture}
    \draw [thick, decoration = {markings, mark=at position .7 with {\arrow[scale=1.5]{stealth}}}, postaction=decorate]  (1,{2/sqrt(3)}) to (2,{sqrt(3)});
    \draw [thick, decoration = {markings, mark=at position .7 with {\arrow[scale=1.5]{stealth}}}, postaction=decorate] (1,{2/sqrt(3)}) to (0,{sqrt(3)});
    \draw [thick, decoration = {markings, mark=at position .7 with {\arrow[scale=1.5]{stealth}}}, postaction=decorate] (1,0) to (1,{2/sqrt(3)});
    \node at (1+0.2,{2/sqrt(3)-0.2}) {$\bar{\imath}$};
    \node at (2-0.1,{2/sqrt(3)+0.3}) {$b$};
    \node at (0.1,{2/sqrt(3)+0.3}) {$a$};
    \node at (1.1,-0.1) {$c$};
    \node at (1,{2/sqrt(3)-0.2}) {$\times$};
\end{tikzpicture}} \,.
\end{equation}
We mark one of the lines by $\times$ so that it is unambiguous which hom-space the junction belongs to. Otherwise there can be confusion when the picture is rotated (a similar convention was adopted in \cite{Chang:2018iay}).
We choose the basis vectors in such a way that they satisfy the following completeness and orthogonality relations \cite{Kitaev:2005hzj,Barkeshli:2014cna}:
\begin{equation} \label{eq:basis_bdy}
\raisebox{-2em}{\begin{tikzpicture}
\draw [thick, decoration = {markings, mark=at position .7 with {\arrow[scale=1.5]{stealth}}}, postaction=decorate] (0,0) to (0,2);
\node at (-0.2,1.8) {$a$};
\draw [thick, decoration = {markings, mark=at position .7 with {\arrow[scale=1.5]{stealth}}}, postaction=decorate] (1,0) to (1,2);
\node at (1.2,1.8) {$b$};
\end{tikzpicture}}
= \sum_{c \in \mathrm{Irr}(\mathcal{C})} \sum_{i = 1}^{N_{ab}^c} \sqrt{\frac{\qd_c}{\qd_a \qd_b}} \raisebox{-2em}{\begin{tikzpicture}
\draw [thick, decoration = {markings, mark=at position .7 with {\arrow[scale=1.5]{stealth}}}, postaction=decorate] (0,0) to (0.5,0.6);
\node at (-0.2,0.2) {$a$};
\draw [thick, decoration = {markings, mark=at position .7 with {\arrow[scale=1.5]{stealth}}}, postaction=decorate] (1,0) to (0.5,0.6);
\node at (1.2,0.2) {$b$};
\draw [thick, decoration = {markings, mark=at position .7 with {\arrow[scale=1.5]{stealth}}}, postaction=decorate] (0.5,0.6) to (0.5,1.4);
\draw [thick, decoration = {markings, mark=at position .7 with {\arrow[scale=1.5]{stealth}}}, postaction=decorate] (0.5,1.4) to (0.0,2.0);
\draw [thick, decoration = {markings, mark=at position .7 with {\arrow[scale=1.5]{stealth}}}, postaction=decorate] (0.5,1.4) to (1,2);
\node at (-0.2,1.8) {$a$};
\node at (1.2,1.8) {$b$};
\node at (0.8,1) {$c$};
\node at (0.25,1.35) {$\bar{\imath}$};
\node at (0.25,0.65) {$i$};
\node at (0.5,0.7) {$\times$};
\node at (0.5,1.3) {$\times$};
\end{tikzpicture}} \,, \qquad
\raisebox{-2.5em}{\begin{tikzpicture}
\draw [thick, decoration = {markings, mark=at position .6 with {\arrow[scale=1.5]{stealth}}}, postaction=decorate] (0,0.6) arc (-50:50:0.5222);
\draw [thick, decoration = {markings, mark=at position .6 with {\arrow[scale=1.5]{stealth}}}, postaction=decorate] (0,0.6) arc (230:130:0.5222);
\draw  [thick, decoration = {markings, mark=at position .7 with {\arrow[scale=1.5]{stealth}}}, postaction=decorate] (0,0) to (0,0.6);
\draw  [thick, decoration = {markings, mark=at position .7 with {\arrow[scale=1.5]{stealth}}}, postaction=decorate] (0,1.4) to (0,2);
\node at (-0.5,1) {$a$};
\node at (0.5,1) {$b$};
\node at (0.27,0.6) {$\bar{\imath}'$};
\node at (0.2,1.5) {$i$};
\node at (-0.3,0.1) {$c'$};
\node at (-0.2,1.9) {$c$};
\node at (0,0.5) {$\times$};
\node at (0,1.5) {$\times$};
\end{tikzpicture}}
=  \delta_{c c'} \delta_{i i'} \sqrt{\frac{\qd_a \qd_b}{\qd_c}} \raisebox{-2em}{\begin{tikzpicture}
\draw [thick, decoration = {markings, mark=at position .7 with {\arrow[scale=1.5]{stealth}}}, postaction=decorate] (0,0) to (0,2);
\node at (-0.2,1.9) {$c$};
\end{tikzpicture}} \,.
\end{equation}
Here, $\qd_a$, $\qd_b$, and $\qd_c$ are the quantum dimensions of the topological lines $a$, $b$, and $c$, respectively.

The $F$-symbols of a fusion category $\mathcal{C}$ in a chosen basis are defined by
\begin{align}\label{eq:F_symbols}
\begin{split}
 \raisebox{-2em}{\begin{tikzpicture}
    \draw [thick, decoration = {markings, mark=at position .7 with {\arrow[scale=1.5]{stealth}}}, postaction=decorate] (0,0) to (1.2,1.4);
    \draw [thick, decoration = {markings, mark=at position .7 with {\arrow[scale=1.5]{stealth}}}, postaction=decorate] (1.2,1.4) to (1.8,2.1);
    \draw [thick, decoration = {markings, mark=at position .7 with {\arrow[scale=1.5]{stealth}}}, postaction=decorate] (1.2,0) to (1.8,0.7);
    \draw [thick, decoration = {markings, mark=at position .7 with {\arrow[scale=1.5]{stealth}}}, postaction=decorate] (2.4,0) to (1.8,0.7);
    \draw [thick, decoration = {markings, mark=at position .7 with {\arrow[scale=1.5]{stealth}}}, postaction=decorate] (1.8,0.7) to (1.2,1.4);
    \node at (-0.2,0.2) {$a$};
    \node at (1,0.2) {$b$};
    \node at (2.5,0.2) {$c$};
    \node at (1.9,1.9) {$d$};
    \node at (1.3,0.9) {$e$};
    \node at (1.1,1.5) {$i$};
    \node at (2,0.8) {$j$};
    \node[rotate=45] at (1.26,1.47) {$\times$};
    \node[rotate=45] at (1.8-0.06,0.7+0.07) {$\times$};
 \end{tikzpicture}}
 &= \sum_{f \in \mathrm{Irr}(\mathcal{C})} \sum_{k=1}^{N_{ab}^f} \sum_{l=1}^{N_{fc}^d} \left[ 
    F_{abc}^d
 \right]_{(eij)(fkl)}
  \raisebox{-2em}{\begin{tikzpicture}
    \draw [thick, decoration = {markings, mark=at position .7 with {\arrow[scale=1.5]{stealth}}}, postaction=decorate] (0,0) to (0.6,0.7);
    \draw [thick, decoration = {markings, mark=at position .7 with {\arrow[scale=1.5]{stealth}}}, postaction=decorate] (0.6,0.7) to (1.2,1.4);
    \draw [thick, decoration = {markings, mark=at position .7 with {\arrow[scale=1.5]{stealth}}}, postaction=decorate] (1.2,1.4) to (1.8,2.1);
    \draw [thick, decoration = {markings, mark=at position .7 with {\arrow[scale=1.5]{stealth}}}, postaction=decorate] (1.2,0) to (0.6,0.7);
    \draw [thick, decoration = {markings, mark=at position .7 with {\arrow[scale=1.5]{stealth}}}, postaction=decorate] (2.4,0) to (1.2,1.4);
    \node at (-0.2,0.2) {$a$};
    \node at (1.4,0.2) {$b$};
    \node at (2.5,0.2) {$c$};
    \node at (1.9,1.9) {$d$};
    \node at (1.05,0.9) {$f$};
    \node at (1.1,1.5) {$l$};
    \node at (0.45,0.8) {$k$};
    \node[rotate=45] at (1.26,1.47) {$\times$};
    \node[rotate=45] at (0.6+0.06,0.7+0.07) {$\times$};
 \end{tikzpicture}} \,,\\
\raisebox{-2em}{\begin{tikzpicture}
\draw [thick, decoration = {markings, mark=at position .7 with {\arrow[scale=1.5]{stealth}}}, postaction=decorate] (0,0) to (0.6,0.7);
    \draw [thick, decoration = {markings, mark=at position .7 with {\arrow[scale=1.5]{stealth}}}, postaction=decorate] (0.6,0.7) to (1.2,1.4);
    \draw [thick, decoration = {markings, mark=at position .7 with {\arrow[scale=1.5]{stealth}}}, postaction=decorate] (1.2,1.4) to (1.8,2.1);
    \draw [thick, decoration = {markings, mark=at position .7 with {\arrow[scale=1.5]{stealth}}}, postaction=decorate] (1.2,0) to (0.6,0.7);
    \draw [thick, decoration = {markings, mark=at position .7 with {\arrow[scale=1.5]{stealth}}}, postaction=decorate] (2.4,0) to (1.2,1.4);
    \node at (-0.2,0.2) {$a$};
    \node at (1.4,0.2) {$b$};
    \node at (2.5,0.2) {$c$};
    \node at (1.9,1.9) {$d$};
    \node at (1.1,0.9) {$e$};
    \node at (1.1,1.5) {$j$};
    \node at (0.45,0.8) {$i$};
    \node[rotate=45] at (1.26,1.47) {$\times$};
    \node[rotate=45] at (0.6+0.06,0.7+0.07) {$\times$};
 \end{tikzpicture}}
 &= \sum_{f \in \mathrm{Irr}(\mathcal{C})} \sum_{k=1}^{N_{af}^d} \sum_{l=1}^{N_{bc}^f} \left[ 
    F_{abc}^d
 \right]_{(eij)(fkl)}^{-1}
  \raisebox{-2em}{\begin{tikzpicture}
    \draw [thick, decoration = {markings, mark=at position .7 with {\arrow[scale=1.5]{stealth}}}, postaction=decorate] (0,0) to (1.2,1.4);
    \draw [thick, decoration = {markings, mark=at position .7 with {\arrow[scale=1.5]{stealth}}}, postaction=decorate] (1.2,1.4) to (1.8,2.1);
    \draw [thick, decoration = {markings, mark=at position .7 with {\arrow[scale=1.5]{stealth}}}, postaction=decorate] (1.2,0) to (1.8,0.7);
    \draw [thick, decoration = {markings, mark=at position .7 with {\arrow[scale=1.5]{stealth}}}, postaction=decorate] (2.4,0) to (1.8,0.7);
    \draw [thick, decoration = {markings, mark=at position .7 with {\arrow[scale=1.5]{stealth}}}, postaction=decorate] (1.8,0.7) to (1.2,1.4);
    \node at (-0.2,0.2) {$a$};
    \node at (1,0.2) {$b$};
    \node at (2.5,0.2) {$c$};
    \node at (1.9,1.9) {$d$};
    \node at (1.25,0.9) {$f$};
    \node at (1.05,1.5) {$k$};
    \node at (1.95,0.8) {$l$};
    \node[rotate=45] at (1.26,1.47) {${\times}$};
    \node[rotate=45] at (1.8-0.06,0.7+0.07) {$\times$};
 \end{tikzpicture}} \,,
\end{split}
\end{align}
where the inverse $F$-symbols satisfy
\begin{equation}
    \sum_{f \in \mathrm{Irr}(\mathcal{C})} \sum_{k=1}^{N_{ab}^f} \sum_{l=1}^{N_{fc}^d} \left[ 
    F_{abc}^d
    \right]_{(eij)(fkl)} \left[ 
    F_{abc}^d
    \right]_{(fkl)(e'i'j')}^{-1}  = \delta_{ee'} \delta_{ii'} \delta_{jj'} \,.
\end{equation}
The $F$-symbols encode local recombination rules for networks of topological lines on a two-dimensional surface. Consistency imposes a so-called ``pentagon equation'' on the $F$-symbols. These same numbers also govern the recombination rules of topological lines which are connected by the dual junctions, 
\begin{align}
\begin{split}
    \input{figures/Fsplitting1.tikz}&=\sum_{f\in\mathrm{Irr}(\mathcal{C})}\sum_{k=1}^{N_{ab}^f}\sum_{l=1}^{N_{fc}^d} [G_{d}^{abc}]_{(e\bar\imath\bar\jmath)(f\bar k \bar l)}\input{figures/Fsplitting2.tikz}\,,\\
    \input{figures/Fsplitting3.tikz}&=\sum_{f\in\mathrm{Irr}(\mathcal{C})}\sum_{k=1}^{N_{af}^d}\sum_{l=1}^{N_{bc}^f}[G_{d}^{abc}]^{-1}_{(e\bar\imath\bar\jmath)(f\bar k \bar l)}\input{figures/Fsplitting4.tikz}\,.
\end{split}
\end{align}
where $[G_{d}^{abc}]_{(e\bar\imath\bar\jmath)(f\bar k \bar l)} = [F_{abc}^d]^{-1}_{(fkl)(eij)}$. The fusion rules $N_{ab}^c$ of simple topological lines and the $F$-symbols are the defining data of a fusion category.

The fusion and splitting spaces with $\times$ decorating different legs are isomorphic, but it is important to keep track of the precise isomorphism. These isomorphisms are encoded by the following $A$ and $B$-symbols \cite{Kitaev:2005hzj,Bonderson:2007ci},
\begin{align}\label{eq:junctionchangerelations}
\begin{split}
	&\input{figures/A1.tikz}=\sum_{j=1}^{N_{\bar a c}^b}[A_{c}^{ab}]_{\bar\imath j}\input{figures/A2.tikz}\,,~~~~~~\input{figures/B1.tikz}=\sum_{j=1}^{N_{c\bar b}^a} [B_c^{ab}]_{\bar\imath j}\input{figures/B2.tikz}\,,\\
	&\input{figures/A1p.tikz}=\sum_{j=1}^{N_{\bar a c}^b}[A^{c}_{ab}]_{i \bar\jmath }\input{figures/A2p.tikz}\,,~~~~~~\input{figures/B1p.tikz}=\sum_{j=1}^{N_{c\bar b}^a} [B^c_{ab}]_{i \bar\jmath }\input{figures/B2p.tikz}\,.
\end{split}
\end{align}
A straightforward calculation shows that the coefficients are determined by the $F$-symbols and quantum dimensions of $\CC$ as
\begin{align} \label{eq:ABF}
\begin{split}
    &[A_c^{ab}]_{\bar\imath j}=\sqrt{\frac{\d_a\d_b}{\d_c}}[F_{\bar a ab}^b]^{-1}_{1(cj i)}\,, ~~~~~ [B_c^{ab}]_{\bar\imath j}=\sqrt{\frac{\d_a\d_b}{\d_c}}[F^a_{ab\bar b}]_{1(cij)} \,,\\
    &[A^c_{ab}]_{i\bar\jmath }=\sqrt{\frac{\d_a\d_b}{\d_c}}[F_{\bar a ab}^b]_{(cj i)1}\,, ~~~~~ [B^c_{ab}]_{i\bar\jmath }=\sqrt{\frac{\d_a\d_b}{\d_c}}[F^a_{ab\bar b}]^{-1}_{(cij)1} \,.
\end{split}
\end{align}
One useful property that they obey is
\begin{eqnarray}\label{eq:AABB}
    \sum_{j=1}^{N_{\bar{a} c}^{b}} [A^{ab}_c]_{\bar\imath j} [A^c_{ab}]_{i' \bar\jmath} = \sum_{k=1}^{N_{c \bar{b}}^{a}} [B^{ab}_c]_{\bar\imath j} [B^c_{ab}]_{i' \bar\jmath} = \delta_{ii'} \,.
\end{eqnarray}

A fusion category may also admit a braiding if $a\otimes b\cong b\otimes a$ for all $a,b\in \mathrm{Irr}(\mathcal{C})$. 
A braiding is encoded in the specification of $R$-symbols, 
\begin{equation}\label{eq:braidingdef}
\raisebox{-2em}{\begin{tikzpicture}
\draw [thick, decoration = {markings, mark=at position .8 with {\arrow[scale=1.5]{stealth}}}, postaction=decorate] (1,0) to (0.52,0.55);
\draw [thick] (0.52,0.67) arc (-45:45:0.6);
\draw [thick] (0.43,0.63) arc (225:128:0.6);
\draw [thick, preaction={draw=white,line width=6pt}, decoration = {markings, mark=at position .7 with {\arrow[scale=1.5]{stealth}}}, postaction=decorate] (0,0) to (0.52,0.67);
\draw [thick, decoration = {markings, mark=at position .7 with {\arrow[scale=1.5]{stealth}}}, postaction=decorate] (0.5,1.5) to (0.5,2.3);
\node at (1.2,0.2) {$b$};
\node at (-0.2,0.2) {$a$};
\node at (0.7,2.2) {$c$};
\node at (0.65,1.6) {$i$};
\node at (0.5,1.6) {$\times$};
\end{tikzpicture}}
= 
  \sum_{j=1}^{N_{ab}^c} \left[
    R_{ab}^c
 \right]_{ij} \raisebox{-2em}{\begin{tikzpicture}
\draw [thick, decoration = {markings, mark=at position .7 with {\arrow[scale=1.5]{stealth}}}, postaction=decorate] (0,0) to (0.5,1.5);
\draw [thick, decoration = {markings, mark=at position .7 with {\arrow[scale=1.5]{stealth}}}, postaction=decorate] (1,0) to (0.5,1.5);
\draw [thick, decoration = {markings, mark=at position .7 with {\arrow[scale=1.5]{stealth}}}, postaction=decorate] (0.5,1.5) to (0.5,2.3);
\node at (1.2,0.2) {$b$};
\node at (-0.2,0.2) {$a$};
\node at (0.7,2.2) {$c$};
\node at (0.66,1.6) {$j$};
\node at (0.5,1.6) {$\times$};
\end{tikzpicture}} \,, \quad
\raisebox{-2em}{\begin{tikzpicture}
\draw [thick, decoration = {markings, mark=at position .7 with {\arrow[scale=1.5]{stealth}}}, postaction=decorate] (0,0) to (0.52,0.67);
\draw [thick, preaction={draw=white,line width=6pt}, decoration = {markings, mark=at position .8 with {\arrow[scale=1.5]{stealth}}}, postaction=decorate] (1,0) to (0.43,0.63);
\draw [thick] (0.52,0.67) arc (-45:45:0.6);
\draw [thick] (0.43,0.63) arc (225:128:0.6);
\draw [thick, decoration = {markings, mark=at position .7 with {\arrow[scale=1.5]{stealth}}}, postaction=decorate] (0.5,1.5) to (0.5,2.3);
\node at (1.2,0.2) {$b$};
\node at (-0.2,0.2) {$a$};
\node at (0.7,2.2) {$c$};
\node at (0.65,1.6) {$i$};
\node at (0.5,1.6) {$\times$};
\end{tikzpicture}}
= 
  \sum_{j=1}^{N_{ab}^c} \left[
    R_{ba}^c
 \right]_{ij}^{-1} \raisebox{-2em}{\begin{tikzpicture}
\draw [thick, decoration = {markings, mark=at position .7 with {\arrow[scale=1.5]{stealth}}}, postaction=decorate] (0,0) to (0.5,1.5);
\draw [thick, decoration = {markings, mark=at position .7 with {\arrow[scale=1.5]{stealth}}}, postaction=decorate] (1,0) to (0.5,1.5);
\draw [thick, decoration = {markings, mark=at position .7 with {\arrow[scale=1.5]{stealth}}}, postaction=decorate] (0.5,1.5) to (0.5,2.3);
\node at (1.2,0.2) {$b$};
\node at (-0.2,0.2) {$a$};
\node at (0.7,2.2) {$c$};
\node at (0.66,1.6) {$j$};
\node at (0.5,1.6) {$\times$};
\end{tikzpicture}} \,.
\end{equation}
The inverse $R$-symbols satisfy
\begin{equation}
    \sum_{j=1}^{N_{ab}^c}\left[ R_{ab}^c \right]_{ij} \left[ R_{ab}^c \right]_{jk}^{-1} = \delta_{ik} \,.
\end{equation}
The $R$-symbols, together with the $F$-symbols, provide local recombination rules for networks of topological lines in a three-dimensional volume, and consistency requires that they satisfy two hexagon equations \cite{Moore:1988qv}.

\subsection{Module categories and bimodule categories}\label{subsec:modulebimodule}

To study boundaries and interfaces of 1+1d quantum field theories in the presence of (non-invertible) global symmetries given by a fusion category, a central mathematical tool is that of a (bi)module category \cite{ostrik2003module}.
Recent applications in the physics literature can be found, for instance, in \cite{Komargodski:2020mxz,Huang:2021zvu,Choi:2023xjw,Diatlyk:2023fwf,Cordova:2024vsq,Cordova:2024iti,Copetti:2024dcz,Inamura:2024jke}.
\subsubsection{Module categories and boundary conditions}

A $\CC$-multiplet of boundary conditions of a 1+1d theory is a collection of boundaries which transform into each other under the action of the topological line operators in $\CC$. A $\CC$-multiplet $\mathcal{B}$ enjoys the structure of a $\CC$-module category, as we describe below.

A module category is called indecomposable if it cannot be written as a direct sum of two module categories, and we always focus on this case.
By convention, $\mathcal{B}$ is assumed to be a \emph{left} module category over $\mathcal{C}$ unless otherwise stated, meaning that the topological lines of $\mathcal{C}$ act on the boundary conditions from the left.

We denote the set of simple objects (i.e.\ simple boundary conditions) of a module category $\mathcal{B}$ as $\mathrm{Irr}(\mathcal{B})$, and write its elements as $B,B', \cdots$ (or $B_1, B_2 , \cdots$). The simple objects of $\mathcal{B}$ correspond to generically non-topological simple boundary conditions of a 1+1d QFT.
If we bring a topological line operator $a$ close to a boundary condition $B$, one obtains a new boundary condition, which we denote as $a \otimes B$. 
This defines the action of $\mathcal{C}$ on the module category $\mathcal{B}$.
Decomposing $a \otimes B$ into simple boundary conditions, we get
\begin{equation}
    a \otimes B \cong \bigoplus_{B' \in \mathrm{Irr}(\mathcal{B})} \widetilde{N}_{a B}^{B'} B' \,,
\end{equation}
where $\widetilde{N}_{a B}^{B'} \in \mathbb{Z}^{\geq 0}$, and the multiplet of boundary conditions forms a non-negative integer matrix representation (NIM-rep) of the fusion algebra of topological lines \cite{Cardy:1989ir,Behrend:1999bn,Gannon:2001ki,Gaberdiel:2002qa}.

We define the Frobenius-Perron dimensions of simple objects of $\mathcal{B}$ as the unique set of positive real numbers $\mathsf{d}_{B}$ satisfying \cite{etingof2016tensor,Barter_2022}
\begin{equation}
    \mathsf{d}_a \mathsf{d}_{B} = \sum_{B' \in \mathrm{Irr}(\mathcal{B})} \widetilde{N}_{a B}^{B'} \mathsf{d}_{B'} \,, \quad \sum_{B \in \mathrm{Irr}(\mathcal{B})} \mathsf{d}_{B}^2 = \sum_{a \in \mathrm{Irr}(\mathcal{C})} \mathsf{d}_a^2 \,.
\end{equation}
These numbers are analogous to the quantum dimensions of topological lines in a fusion category.
For a conformal boundary condition, the Frobenius-Perron dimension of the boundary is proportional to its $g$-function \cite{PhysRevLett.67.161}, with the proportionality constant being the same for all the boundary conditions belonging to the same NIM-rep.

When a topological line $a$ ends on a junction between two boundary conditions $B$ and $B'$, the topological point operators which can sit at the junction form a complex vector space $\mathrm{Hom}_{\mathcal{B}}(a \otimes B, B')$ of dimension $\widetilde{N}_{aB}^{B'}$.
Similarly to the case of topological lines, we fix a set of basis junction vectors $v_{aB}^{B';i} \in \mathrm{Hom}_{\mathcal{B}}(a \otimes B, B')$ as well as a set of dual basis vectors $\bar{v}_{aB}^{B';i}\in \mathrm{Hom}_{\mathcal{B}}(B', a \otimes B)$, where $i= 1, \cdots, \widetilde{N}_{aB}^{B'}$. In pictures, these are represented as\footnote{We do not draw arrows on boundary conditions (or objects of a module category), and their orientations are understood to be induced from the 1+1d bulk.}
\begin{equation}
    v_{aB}^{B';i} = \raisebox{-2em}{\begin{tikzpicture}
\fill [gray, opacity=0.5] (1,0) rectangle (1.5,2); 
\draw [blue, thick] (1,0) to (1,2);
\node at (1.2,0.2) {$B$};
\draw [thick, decoration = {markings, mark=at position .7 with {\arrow[scale=1.5]{stealth}}}, postaction=decorate] (0,0) to (1,1);
\node at (0.4,0.1) {$a$};
\node at (1.2,1.8) {$B'$};
\node at (0.85,1.15) {$i$};
\end{tikzpicture}} \,, \qquad
\bar{v}_{aB}^{B';i} = \raisebox{-2em}{\begin{tikzpicture}
\fill [gray, opacity=0.5] (1,0) rectangle (1.5,2); 
\draw [blue, thick] (1,0) to (1,2);
\node at (1.2,0.2) {$B'$};
\draw [thick, decoration = {markings, mark=at position .7 with {\arrow[scale=1.5]{stealth}}}, postaction=decorate] (1,1) to (0,2);
\node at (0.4,1.9) {$a$};
\node at (1.2,1.8) {$B$};
\node at (0.85,0.85) {$\bar{i}$};
\end{tikzpicture}}  \,.
\end{equation}
We choose the basis vectors so that they satisfy the followng completeness and orthogonality relations:
\begin{equation} \label{eq:orthogonality_module}
\raisebox{-2em}{\begin{tikzpicture}
\draw [thick, decoration = {markings, mark=at position .7 with {\arrow[scale=1.5]{stealth}}}, postaction=decorate] (0.5,0) to (0.5,2);
\node at (0.3,1.8) {$a$};
\fill [gray, opacity=0.5] (1,0) rectangle (1.5,2); 
\draw [blue, thick] (1,0) to (1,2);
\node at (1.2,1.8) {$B$};
\end{tikzpicture}}
= \sum_{B' \in \mathrm{Irr}(\mathcal{B})} \sum_{i=1}^{\widetilde{N}_{a B}^{B'}} \sqrt{\frac{\qd_{B'}}{\qd_a \qd_B}} \raisebox{-2em}{\begin{tikzpicture}
\fill [gray, opacity=0.5] (1,0) rectangle (1.5,2); 
\draw [blue, thick] (1,0) to (1,2);
\node at (1.2,1.8) {$B$};
\node at (1.2,0.2) {$B$};
\draw [thick, decoration = {markings, mark=at position .7 with {\arrow[scale=1.5]{stealth}}}, postaction=decorate] (0.5,0) to (1,0.6);
\node at (0.4,0.2) {$a$};
\draw [thick, decoration = {markings, mark=at position .7 with {\arrow[scale=1.5]{stealth}}}, postaction=decorate] (1,1.4) to (0.5,2.0);
\node at (0.4,1.8) {$a$};
\node at (1.2,1) {$B'$};
\node at (0.85,1.3) {$\bar{i}$};
\node at (0.85,0.75) {$i$};
\end{tikzpicture}} \,, \quad
\raisebox{-2em}{\begin{tikzpicture}
\fill [gray, opacity=0.5] (1,0) rectangle (1.5,2); 
\draw [blue, thick] (1,0) to (1,2);
\node at (1.2,1.8) {$B''$};
\node at (1.2,0.2) {$B$};
\node at (1.2,1) {$B'$};
\node at (1.15,1.4) {$i'$};
\node at (1.15,0.55) {$\bar{i}$};
\draw [thick, decoration = {markings, mark=at position .6 with {\arrow[scale=1.5]{stealth}}}, postaction=decorate] (1,0.4) arc (230:130:0.8);
\node at (0.5,1) {$a$};
\end{tikzpicture}}
= ~ \delta_{B B''} \delta_{i i'} \sqrt{\frac{\qd_a \qd_{B'}}{\qd_B}} ~~\raisebox{-2em}{\begin{tikzpicture}
\fill [gray, opacity=0.5] (1,0) rectangle (1.5,2); 
\draw [blue, thick] (1,0) to (1,2);
\node at (1.2,1.8) {$B$};
\end{tikzpicture}} \,.
\end{equation}

Module categories possess numbers called $\widetilde{F}$-symbols which are analogous to the $F$-symbols of a fusion category.
They are defined by the equality
\begin{equation}\label{eq:FT_symbols}
\raisebox{-2em}{\begin{tikzpicture}
    \fill [gray, opacity=0.5] (1,0) rectangle (1.5,2); 
    \draw [blue, thick] (1,0) to (1,2);
    \node at (1.2,1.8) {$B''$};
    \node at (1.2,0.2) {$B$};
    \node at (1.2,1) {$B'$};
    \node at (1.15,1.35) {$i$};
    \node at (1.15,0.65) {$j$};
    \draw [thick, decoration = {markings, mark=at position .7 with {\arrow[scale=1.5]{stealth}}}, postaction=decorate] (0.5,0) to (1,0.6);
    \draw [thick, decoration = {markings, mark=at position .7 with {\arrow[scale=1.5]{stealth}}}, postaction=decorate] (0,0) to (1,1.4);
    \node at (-0.1,0.2) {$a$};
    \node at (0.4,0.2) {$b$};
 \end{tikzpicture}}
 = \sum_{c \in \mathrm{Irr}(\mathcal{C})} \sum_{x=1}^{N_{ab}^c} \sum_{k=1}^{\widetilde{N}_{cB}^{B''}} \left[ 
    \widetilde{F}_{abB}^{B''}
 \right]_{(B'ij)(cxk)}
  \raisebox{-2em}{\begin{tikzpicture}
    \fill [gray, opacity=0.5] (1,0) rectangle (1.5,2); 
    \draw [blue, thick] (1,0) to (1,2);
    \node at (1.2,1.8) {$B''$};
    \node at (1.2,0.2) {$B$};
    \node at (1.15,1) {$k$};
    \draw [thick, decoration = {markings, mark=at position .7 with {\arrow[scale=1.5]{stealth}}}, postaction=decorate] (0.5,0) to (0.5,0.7);
    \draw [thick, decoration = {markings, mark=at position .7 with {\arrow[scale=1.5]{stealth}}}, postaction=decorate] (0,0) to (0.5,0.7);
    \draw [thick, decoration = {markings, mark=at position .7 with {\arrow[scale=1.5]{stealth}}}, postaction=decorate] (0.5,0.7) to (1,1);
    \node at (-0.1,0.2) {$a$};
    \node at (0.7,0.2) {$b$};
    \node at (0.7,1.1) {$c$};
    \node at (0.3,0.7) {$x$};
 \end{tikzpicture}} \,.
\end{equation}
The boundary $\widetilde{F}$-symbols allow us to perform local modifications of networks of topological lines in the presence of a boundary. Consistency requires that $\widetilde{F}$-symbols satisfy a module pentagon equation (see, for instance, \cite{Choi:2023xjw}). For completeness, we record here also the following recombination rules governing splitting spaces,
\begin{align}
    \input{figures/FTsplitting1.tikz}=\sum_{c\in\mathrm{Irr}(\mathcal{C})}\sum_{x=1}^{N_{ab}^c}\sum_{k=1}^{\widetilde N_{cB}^{B''}} [\widetilde G_{B''}^{abB}]_{(B'\bar\imath \bar\jmath)(c\bar x\bar k)} \input{figures/FTsplitting2.tikz} \,,
\end{align}
which are determined in terms of the $\widetilde{F}$-symbols as $[\widetilde{G}_{B''}^{abB}]_{(B'\bar\imath\bar\jmath)(c\bar x\bar k)}=[\widetilde{F}_{abB}^{B''}]^{-1}_{(cxk)(B'ij)}$. The NIM-rep coefficients $\widetilde{N}_{aB}^{B'}$ and the $\widetilde{F}$-matrices are the definining data of a module category.

The simplest example of a module category is a fusion category $\mathcal{C}$ viewed as a module category over itself. In this situation, the NIM-rep coefficients $\widetilde{N}$ are identified with the fusion coefficients $N$ of $\CC$, and the $\widetilde{F}$-symbols coincide with the $F$-symbols of $\CC$.
This is called a \emph{regular} module category, and we denote it throughout as $\mathcal{B}_{\mathrm{reg}}$. 

One also has the notion of a right $\mathcal{C}$-module category, which describes the situation that bulk topological lines act on a collection of boundary conditions from the right, as opposed to from the left. Most of the conventions/notations from the case of left module categories admit clear extensions to the case of right module categories. In order to be completely explicit, we write down our conventions for the $\widetilde{F}$-symbols: 
\begin{align}
\input{figures/FTright1.tikz}&=\sum_{c\in\mathrm{Irr}(\mathcal{C})}\sum_{k=1}^{\widetilde N_{Bc}^{B''}}\sum_{x=1}^{N_{ab}^c}[\widetilde{F}_{Bab}^{B''}]_{(B'ij)(ckx)}~\input{figures/FTright2.tikz} \, ,
\end{align}
and similarly
\begin{align}
    \input{figures/FTsplittingright1.tikz} &= \sum_{c\in\mathrm{Irr}(\mathcal{C})}\sum_{k=1}^{\widetilde N_{Bc}^{B''}}\sum_{x=1}^{N_{ab}^c}[\widetilde{G}_{B''}^{Bab}]_{(B'\bar\imath\bar\jmath)(c\bar k\bar x)} ~ \input{figures/FTsplittingright2.tikz} \,,
\end{align}
where $[\widetilde{G}_{B''}^{Bab}]_{(B'\bar\imath\bar\jmath)(c\bar k\bar x)} = [\widetilde{F}_{Bab}^{B''}]^{-1}_{(ckx)(B'ij)}$.

In later sections we often discuss 1+1d quantum field theories defined on an interval.
As briefly mentioned, our convention is that we consider a $\CC$-multiplet of boundary conditions $\mathcal{B}$ to form a left $\CC$-module category.
This is appropriate if we would like to impose the boundary conditions in $\mathcal{B}$ on the right end of an interval, so that the lines in $\CC$ act from the left. However, if we would like to impose the boundary conditions in $\mathcal{B}$ on the \emph{left} side of an interval, we must first convert $\mathcal{B}$ into a right $\mathcal{C}$-module category. This is accomplished by passing to the ``dual category'' $\mathcal{B}^\vee$, which is a \emph{right} module category over $\mathcal{C}$ with the same set of objects (i.e.\ boundary conditions) as $\mathcal{B}$, but with the directions of morphisms  reversed.\footnote{We follow conventions and definitions of \cite{etingof2016tensor}, see their Remark 7.1.5.}
When a topological line $a \in \mathrm{Irr}(\mathcal{C})$ acts from the right on a boundary condition $B \in \mathrm{Irr}(\mathcal{B}^\vee)$, the corresponding NIM-rep coefficients,
\begin{equation}
    B \otimes a \cong \bigoplus_{B' \in \mathrm{Irr}(\mathcal{B}^\vee)} \,{^{\mathcal{B}^\vee}}\hspace{-.025in}\widetilde{N}_{Ba}^{B'} B' \,,
\end{equation}
are given by ${^{\mathcal{B}^\vee}}\hspace{-.025in}\widetilde{N}_{Ba}^{B'} = {^{\mathcal{B}}}\hspace{-.025in}\widetilde{N}_{\bar{a}B}^{B'}$. Similarly, the $\widetilde{F}$-symbols of $\mathcal{B}^\vee$ are related to those of $\mathcal{B}$ as
\begin{align}
   {^{\mathcal{B}^\vee}}\hspace{-.025in}[\widetilde{F}_{B\bar a\bar b}^{B''}]_{(B'ij)(\bar c kx)} = {^{\mathcal{B}}}\hspace{-.02in}[\widetilde{F}_{abB}^{B''}]^{-1}_{(cxk)(B'ji)}\,.
\end{align}

We frequently utilize the correspondence between  module categories over $\mathcal{C}$ and (Morita classes of) algebra objects in $\mathcal{C}$ \cite{ostrik2003module}.
That is, given a (left) module category $\mathcal{B}$ over $\mathcal{C}$, there is an algebra object $A$ in $\mathcal{C}$ such that the category of (right) $A$-modules in $\mathcal{C}$, denoted as $\mathcal{C}_A$, is equivalent to $\mathcal{B}$ as a $\mathcal{C}$-module category, i.e. $\mathcal{B} \cong \mathcal{C}_A$.
Furthermore, given such a module category $\mathcal{B}$ over $\mathcal{C}$, the category of $\mathcal{C}$-module functors from $\mathcal{B}$ to itself, denoted as $\mathcal{C}^*_\mathcal{B} \equiv \mathrm{Fun}_{\mathcal{C}}(\mathcal{B},\mathcal{B})$, is a fusion category, which is isomorphic to the opposite of the fusion category $_A \mathcal{C}_A$ of $A$-$A$-bimodules in $\mathcal{C}$ \cite{etingof2016tensor,Diatlyk:2023fwf}, namely $_A \mathcal{C}_A \cong (\mathcal{C}^*_\mathcal{B})^{\text{op}}$.\footnote{See \cite[Definition 2.1.5.]{etingof2016tensor} for the definition of the opposite of a general monoidal category.
Physically, if a 1+1d quantum field theory $Q$ has a fusion category symmetry $\mathcal{C}$, then its orientation-reversal $\overline{Q}$ has the opposite fusion category $\mathcal{C}^{\text{op}}$ as a symmetry.}
The fusion category $\mathcal{C}^*_\mathcal{B}$ is referred to as the dual category to $\mathcal{C}$ with respect to $\mathcal{B}$.

Module categories $\mathcal{B}$ over $\mathcal{C}$ are in 1-to-1 correspondence with the physically distinct ways of gauging (a subpart of) the fusion category symmetry $\mathcal{C}$, and  $_A \mathcal{C}_A \cong (\mathcal{C}^*_\mathcal{B})^{\text{op}}$ describes the dual symmetry after gauging \cite{Bhardwaj:2017xup}.

Finally, suppose $\mathcal{B}_1$ and $\mathcal{B}_2$ are two potentially distinct module categories over $\mathcal{C}$, whose corresponding algebra objects are $A_1$ and $A_2$ so that $\mathcal{B}_1 \cong \mathcal{C}_{A_1}$ and $\mathcal{B}_2 \cong \mathcal{C}_{A_2}$, respectively.
Then the category $\mathrm{Fun}_\mathcal{C} (\mathcal{B}_1 , \mathcal{B}_2)$ of $\mathcal{C}$-module functors from $\mathcal{B}_1$ to $\mathcal{B}_2$ is isomorphic to the category of $A_1$-$A_2$-bimodules in $\mathcal{C}$, i.e. $\mathrm{Fun}_\mathcal{C} (\mathcal{B}_1 , \mathcal{B}_2) \cong {_{A_1}}\mathcal{C}_{A_2}$ (see \cite[Proposition 7.11.1.]{etingof2016tensor}).
The physical relevance of these concepts will be explained in later sections.

\subsubsection{Bimodule categories and non-topological interfaces}
\label{sec:bimodule_Nontopo}

Consider a generically non-topological interface $I$ between two 1+1d quantum field theories (or between two 1+1d defects/boundaries of a higher-dimensional quantum field theory). We assume that the theory to the left of the interface $I$ has a fusion category symmetry $\CC_L$, and the theory to the right of the interface $I$ has a (potentially distinct) fusion category symmetry  $\mathcal{C}_R$.
Topological line operators from both $\CC_L$ and $\CC_R$ can be fused onto the interface to produce a new interface.
Mathematically, a multiplet of interfaces transforming into themselves under the action of topological lines in $\mathcal{C}_L$ and $\mathcal{C}_R$ forms a $(\mathcal{C}_L, \mathcal{C}_R)$-bimodule category.\footnote{We again focus only on indecomposable bimodule categories which cannot be written as a direct sum of two bimodule categories.}

We denote the set of simple objects (i.e.\ simple interfaces) of a $(\mathcal{C}_L, \mathcal{C}_R)$-bimodule category $\mathcal{I}$ as $\mathrm{Irr}(\mathcal{I})$, and its elements as $I, I', I'', \cdots$ (or $I_1, I_2 , \cdots$).
The bimodule category $\mathcal{I}$ is a left module category over $\mathcal{C}_L$, and it satisfies the corresponding axioms. 
That is, it defines a NIM-rep of the fusion algebra of $\mathcal{C}_L$, and there are boundary $\widetilde{F}$-matrices satisfying the module pentagon equation.
Similarly, $\mathcal{I}$ is at the same time a right module category over $\mathcal{C}_R$.
Concretely, we define left and right $\widetilde{F}$-matrices by
\begin{align}
\begin{split}
\raisebox{-2em}{\begin{tikzpicture}
    \draw [thick, blue, decoration = {markings, mark=at position .7 with {\arrow[scale=1.5]{stealth}}}, postaction=decorate] (1,0) to (1,0.6);
    \draw [thick, blue, decoration = {markings, mark=at position .7 with {\arrow[scale=1.5]{stealth}}}, postaction=decorate] (1,0.6) to (1,1.4);
    \draw [thick, blue, decoration = {markings, mark=at position .7 with {\arrow[scale=1.5]{stealth}}}, postaction=decorate] (1,1.4) to (1,2);    
    \node at (1.3,1.8) {$I''$};
    \node at (1.3,0.2) {$I$};
    \node at (1.3,1) {$I'$};
    \node at (1.15,1.35) {$i$};
    \node at (1.15,0.65) {$j$};
    \draw [thick, decoration = {markings, mark=at position .7 with {\arrow[scale=1.5]{stealth}}}, postaction=decorate] (0.5,0) to (1,0.6);
    \draw [thick, decoration = {markings, mark=at position .7 with {\arrow[scale=1.5]{stealth}}}, postaction=decorate] (0,0) to (1,1.4);
    \node at (-0.1,0.2) {$a$};
    \node at (0.4,0.2) {$b$};
 \end{tikzpicture}}
 &= \sum_{c \in \mathrm{Irr}(\mathcal{C}_L)} \sum_{x=1}^{(N_L)_{ab}^c} \sum_{k=1}^{(\widetilde{N}_L)_{cI}^{I''}} \left[ 
    (\widetilde{F}_L)_{abI}^{I''}
 \right]_{(I'ij)(cxk)}
  \raisebox{-2em}{\begin{tikzpicture}
    \draw [thick, blue, decoration = {markings, mark=at position .7 with {\arrow[scale=1.5]{stealth}}}, postaction=decorate] (1,0) to (1,1);
    \draw [thick, blue, decoration = {markings, mark=at position .7 with {\arrow[scale=1.5]{stealth}}}, postaction=decorate] (1,1) to (1,2);  
    \node at (1.3,1.8) {$I''$};
    \node at (1.3,0.2) {$I$};
    \node at (1.15,1) {$k$};
    \draw [thick, decoration = {markings, mark=at position .7 with {\arrow[scale=1.5]{stealth}}}, postaction=decorate] (0.5,0) to (0.5,0.7);
    \draw [thick, decoration = {markings, mark=at position .7 with {\arrow[scale=1.5]{stealth}}}, postaction=decorate] (0,0) to (0.5,0.7);
    \draw [thick, decoration = {markings, mark=at position .7 with {\arrow[scale=1.5]{stealth}}}, postaction=decorate] (0.5,0.7) to (1,1);
    \node at (-0.1,0.2) {$a$};
    \node at (0.7,0.2) {$b$};
    \node at (0.7,1.1) {$c$};
    \node at (0.3,0.7) {$x$};
 \end{tikzpicture}} \,, \\
 \raisebox{-2em}{\begin{tikzpicture}
    \draw [thick, blue, decoration = {markings, mark=at position .7 with {\arrow[scale=1.5]{stealth}}}, postaction=decorate] (1,0) to (1,0.6);
    \draw [thick, blue, decoration = {markings, mark=at position .7 with {\arrow[scale=1.5]{stealth}}}, postaction=decorate] (1,0.6) to (1,1.4);
    \draw [thick, blue, decoration = {markings, mark=at position .7 with {\arrow[scale=1.5]{stealth}}}, postaction=decorate] (1,1.4) to (1,2);    
    \node at (1-.3,1.8) {$I''$};
    \node at (1-.3,0.2) {$I$};
    \node at (1-.3,1) {$I'$};
    \node at (1-.15,1.35) {$j'$};
    \node at (1-.15,0.65) {$i'$};
    \draw [thick, decoration = {markings, mark=at position .7 with {\arrow[scale=1.5]{stealth}}}, postaction=decorate] (1.5,0) to (1,0.6);
    \draw [thick, decoration = {markings, mark=at position .7 with {\arrow[scale=1.5]{stealth}}}, postaction=decorate] (2,0) to (1,1.4);
    \node at (2.2,0.2) {$b'$};
    \node at (1.7,0.2) {$a'$};
 \end{tikzpicture}}
 &= \sum_{c' \in \mathrm{Irr}(\mathcal{C}_R)} \sum_{x'=1}^{ (N_R)_{b'a'}^{c'}} \sum_{k'=1}^{(\widetilde{N}_R)_{Ic'}^{I''}} \left[ 
    (\widetilde{F}_R)_{Ia'b'}^{I''}
 \right]_{(I'i'j')(c'k'x')}
  \raisebox{-2em}{\begin{tikzpicture}
    \draw [thick, blue, decoration = {markings, mark=at position .7 with {\arrow[scale=1.5]{stealth}}}, postaction=decorate] (1,0) to (1,1);
    \draw [thick, blue, decoration = {markings, mark=at position .7 with {\arrow[scale=1.5]{stealth}}}, postaction=decorate] (1,1) to (1,2); 
    \node at (1-.3,1.8) {$I''$};
    \node at (1-.3,0.2) {$I$};
    \node at (1-.15,1) {$k'$};
    \draw [thick, decoration = {markings, mark=at position .7 with {\arrow[scale=1.5]{stealth}}}, postaction=decorate] (1.5,0) to (1.5,0.7);
    \draw [thick, decoration = {markings, mark=at position .7 with {\arrow[scale=1.5]{stealth}}}, postaction=decorate] (2,0) to (1.5,0.7);
    \draw [thick, decoration = {markings, mark=at position .7 with {\arrow[scale=1.5]{stealth}}}, postaction=decorate] (1.5,0.7) to (1,1);
    \node at (2.2,0.2) {$a'$};
    \node at (1.3,0.2) {$b'$};
    \node at (1.4,1.1) {$c'$};
    \node at (1.7,0.8) {$x'$};
 \end{tikzpicture}} \,,
\end{split}
\end{align}
where the interfaces are drawn in blue.
Here, the lines $a,b$ belong to $\mathrm{Irr}(\mathcal{C}_L)$, the lines $a',b'$ belong to $\mathrm{Irr}(\mathcal{C}_R)$, the numbers $N_L$, $N_R$ are the fusion coefficients of $\mathcal{C}_L$, $\mathcal{C}_R$, and the numbers $\widetilde{N}_L$, $\widetilde{N}_R$ are the NIM-rep coefficients for the left and right module category structures.

\begin{figure}[t]
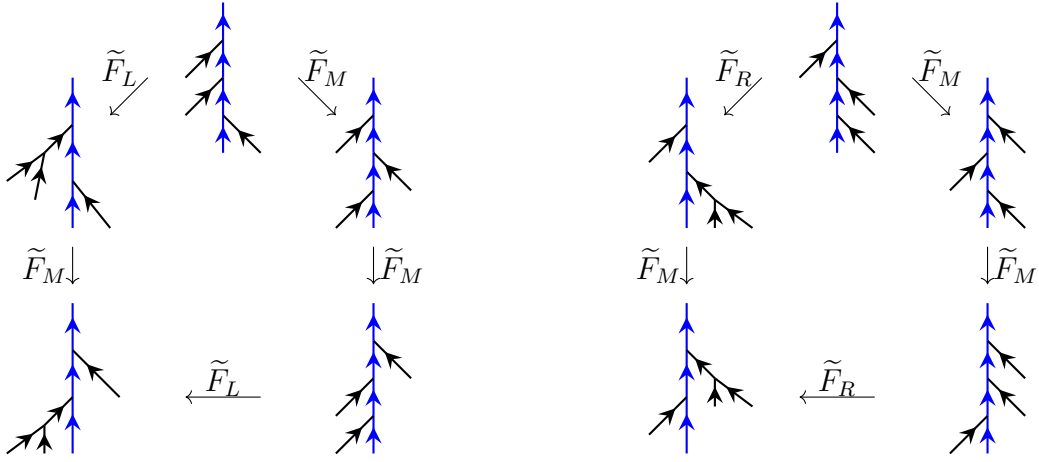

\begin{subfigure}{0.5\textwidth}
\ctikzfig{figures/pentagonmiddle1}
\end{subfigure}
\begin{subfigure}{0.5\textwidth}
\ctikzfig{figures/pentagonmiddle2}
\end{subfigure}
\caption{Pentagon equations satisfied by the middle associator of a bimodule category.   }
\label{fig:pentagonmiddleassociator}
\end{figure}

A bimodule category $\mathcal{I}$ comes equipped with an additional structure  called the middle associator (see, for instance, \cite[Definition 7.1.7.]{etingof2016tensor}).
The middle associator $\widetilde{F}_M$ is defined by the equation
\begin{equation} \label{eq:middle}
    \raisebox{-2em}{\begin{tikzpicture} 
    \draw [thick, blue, decoration = {markings, mark=at position .7 with {\arrow[scale=1.5]{stealth}}}, postaction=decorate] (1,0) to (1,0.6);
    \draw [thick, blue, decoration = {markings, mark=at position .7 with {\arrow[scale=1.5]{stealth}}}, postaction=decorate] (1,0.6) to (1,1.4);
    \draw [thick, blue, decoration = {markings, mark=at position .7 with {\arrow[scale=1.5]{stealth}}}, postaction=decorate] (1,1.4) to (1,2);
    \node at (1.3,1.8) {$I_3$};
    \node at (1-.3,0.2) {$I_1$};
    \node at (1.3,1) {$I_2$};
    \node at (1.15,1.35) {$i$};
    \node at (1-.15,0.65) {$i'$};
    \draw [thick, decoration = {markings, mark=at position .7 with {\arrow[scale=1.5]{stealth}}}, postaction=decorate] (1.5,0) to (1,0.6);
    \draw [thick, decoration = {markings, mark=at position .7 with {\arrow[scale=1.5]{stealth}}}, postaction=decorate] (0,0) to (1,1.4);
    \node at (-0.1,0.2) {$a$};
    \node at (1.7,0.2) {$b'$};
 \end{tikzpicture}}
 = \sum_{I_4 \in \mathrm{Irr}(\mathcal{I})} \sum_{j=1}^{(\widetilde{N}_L)_{a I_1}^{I_4}} \sum_{j'=1}^{(\widetilde{N}_R)_{I_4 b'}^{I_3}} \left[ 
    (\widetilde{F}_M)_{a I_1 b }^{I_3}
 \right]_{(I_2 ii')(I_4 jj')}
  \raisebox{-2em}{\begin{tikzpicture}
    \draw [thick, blue, decoration = {markings, mark=at position .7 with {\arrow[scale=1.5]{stealth}}}, postaction=decorate] (1,0) to (1,0.6);
    \draw [thick, blue, decoration = {markings, mark=at position .7 with {\arrow[scale=1.5]{stealth}}}, postaction=decorate] (1,0.6) to (1,1.4);
    \draw [thick, blue, decoration = {markings, mark=at position .7 with {\arrow[scale=1.5]{stealth}}}, postaction=decorate] (1,1.4) to (1,2);
    \node at (1.3,1.8) {$I_3$};
    \node at (1.3,0.2) {$I_1$};
    \node at (1-.3,1) {$I_4$};
    \node at (1-.15,1.45) {$j'$};
    \node at (1.15,0.65) {$j$};
    \draw [thick, decoration = {markings, mark=at position .7 with {\arrow[scale=1.5]{stealth}}}, postaction=decorate] (2,0) to (1,1.4);
    \draw [thick, decoration = {markings, mark=at position .7 with {\arrow[scale=1.5]{stealth}}}, postaction=decorate] (0.5,0) to (1,0.6);
    \node at (0.4,0.2) {$a$};
    \node at (2.2,0.2) {$b'$};
 \end{tikzpicture}} \,.
\end{equation}
The $\widetilde{F}_M$-symbols, together with $\widetilde{F}_L$ and $\widetilde{F}_R$, satisfy two additional pentagon equations, which are schematically shown in Figure \ref{fig:pentagonmiddleassociator}.
More explicitly, we have
\begin{equation}
    \begin{split}
    &\sum_{k=1}^{({\widetilde{N}_L})_{c I_2}^{I_4}} [(\widetilde{F}_L)_{abI_2}^{I_4}]_{(I_3 i j)(cxk)} [(\widetilde{F}_M)_{cI_1 a'}^{I_4}]_{(I_2 ki')(I_3'mk')} \\&= 
        \sum_{I_2'\in \mathrm{Irr}(\mathcal{I})}  \sum_{l=1}^{(\widetilde{N}_L)_{b I_1}^{I_2'}} \sum_{j'=1}^{(\widetilde{N}_R)_{ I_2' a'}^{I_3}} \sum_{n=1}^{(\widetilde{N}_L)_{a I_2'}^{I_3'}}   [(\widetilde{F}_M)_{bI_1 a'}^{I_3}]_{(I_2 ji')(I_2'l j')}[(\widetilde{F}_M)_{aI_2' a'}^{I_4}]_{(I_3 i j')(I_3'nk')} [(\widetilde{F}_L)_{abI_1}^{I_3'}]_{(I_2'nl)(cxm)} \,, 
    \end{split} 
\end{equation}
and
\begin{equation}
    \begin{split}
        &\sum_{k'=1}^{(\widetilde{N}_R)_{I_1 c'}^{I_3}} [(\widetilde{F}_R)_{I_1 a'b' }^{I_3}]_{(I_2 i'j')(c' k' x')} [(\widetilde{F}_M)_{a I_1 c'}^{I_4}]_{(I_3 ik')(I_2'j l')} \\&= \sum_{I_3'\in \mathrm{Irr}(\mathcal{I})} \sum_{k=1}^{(\widetilde{N}_L)_{a I_2 }^{I_3'}} \sum_{m'=1}^{(\widetilde{N}_R)_{I_3' b'}^{I_4}} \sum_{n'=1}^{(\widetilde{N}_R)_{I_2'a' }^{I_3'}} [(\widetilde{F}_M)_{a I_2 b'}^{I_4}]_{(I_3 i j')(I_3' k m')} [(\widetilde{F}_M)_{a I_1 a'}^{I_3'}]_{(I_2k i')(I_2' j n')} [(\widetilde{F}_R)_{I_2' a' b'}^{I_4}]_{(I_3'n' m')(c' m' x')}   \,.
    \end{split}
\end{equation}

The left and right module category structures and the $\widetilde{F}_M$-symbols are the defining data of a bimodule category $\mathcal{I}$.
The simplest example of a bimodule category is a fusion category $\mathcal{C}$,  viewed as a $(\mathcal{C},\mathcal{C})$-bimodule category via the fact that $\mathcal{C}$ acts on itself from the left and from the right. We will refer to this as the \emph{regular} $(\mathcal{C},\mathcal{C})$-bimodule category, and denote it as $\mathcal{I}_{\mathrm{reg}}$. In this case,  all the $\widetilde{F}_L$-, $\widetilde{F}_R$-, and $\widetilde{F}_M$-symbols reduce to the standard $F$-symbols of $\mathcal{C}$. (Actually, in our conventions, $\widetilde{F}_R$ unfortunately coincides with $F^{-1}$ for the regular bimodule category.)

\subsubsection{Bimodule categories and topological interfaces}
\label{sec:bimodule_topo}

A special class of interfaces that will play an important role is that of topological interfaces, e.g. between two 1+1d QFTs or between two 1+1d topological boundary conditions of a 2+1d TQFT.\footnote{An example of the former case is the interface defined by gauging a finite symmetry in half space \cite{Thorngren:2021yso,Choi:2021kmx,Choi:2022zal}. The latter case will be discussed in more detail in Section \ref{sec:symTFT}. } All the discussions in Section \ref{sec:bimodule_Nontopo} apply. In addition, there is a well-defined fusion structure analogous to that of topological lines \cite{Diatlyk:2023fwf}.
To distinguish them from generic non-topological interfaces, we denote simple topological interfaces using the greek letters $\alpha, \beta, \gamma, \cdots$.

Consider three fusion categories, $\mathcal{C}_1$, $\mathcal{C}_2$, and $\mathcal{C}_3$.
Given three simple topological interfaces, $\alpha$, $\beta$, and $\gamma$ between 1+1d theories (or between 1+1d objects in a higher-dimensional theory) with these fusion category symmetries, we fix a set of basis and dual basis trivalent junction vectors using the similar conventions as in \eqref{eq:marked_junctions} and \eqref{eq:basis_bdy}.
Namely, we define 
\begin{equation}
    v_{\alpha \beta}^{\gamma;i} = \raisebox{-1.8em}{\begin{tikzpicture}
    \draw [thick, blue, decoration = {markings, mark=at position .7 with {\arrow[scale=1.5]{stealth}}}, postaction=decorate] (0,0) to (1,{1/sqrt(3)});
    \draw [thick, blue, decoration = {markings, mark=at position .7 with {\arrow[scale=1.5]{stealth}}}, postaction=decorate] (2,0) to (1,{1/sqrt(3)});
    \draw [thick, blue, decoration = {markings, mark=at position .7 with {\arrow[scale=1.5]{stealth}}}, postaction=decorate] (1,{1/sqrt(3)}) to (1,{sqrt(3)});
    \node at (1+0.2,{1/sqrt(3)+0.1}) {$i$};
    \node at (-0.1,0.15) {$\alpha$};
    \node at (2.1,0.15) {$\beta$};
    \node at (1.1,{sqrt(3)+0.1}) {$\gamma$};
    \node at (1,{1/sqrt(3)+0.2}) {$\times$};
    \node at (0.3,1) {$\mathcal{C}_1$};
    \node at (1.7,1) {$\mathcal{C}_3$};
    \node at (1,0) {$\mathcal{C}_2$};
\end{tikzpicture}} \,, \qquad
\bar{v}_{\alpha \beta}^{\gamma;i} = \raisebox{-1.8em}{\begin{tikzpicture}
    \draw [thick, blue, decoration = {markings, mark=at position .7 with {\arrow[scale=1.5]{stealth}}}, postaction=decorate]  (1,{2/sqrt(3)}) to (2,{sqrt(3)});
    \draw [thick, blue, decoration = {markings, mark=at position .7 with {\arrow[scale=1.5]{stealth}}}, postaction=decorate] (1,{2/sqrt(3)}) to (0,{sqrt(3)});
    \draw [thick, blue, decoration = {markings, mark=at position .7 with {\arrow[scale=1.5]{stealth}}}, postaction=decorate] (1,0) to (1,{2/sqrt(3)});
    \node at (1+0.2,{2/sqrt(3)-0.2}) {$\bar{i}$};
    \node at (2-0.1,{2/sqrt(3)+0.3}) {$\beta$};
    \node at (0.1,{2/sqrt(3)+0.3}) {$\alpha$};
    \node at (1.1,-0.1) {$\gamma$};
    \node at (1,{2/sqrt(3)-0.2}) {$\times$};
    \node at (0.3,{sqrt(3)-1}) {$\mathcal{C}_1$};
    \node at (1.7,{sqrt(3)-1}) {$\mathcal{C}_3$};
    \node at (1,{sqrt(3)}) {$\mathcal{C}_2$};
\end{tikzpicture}} \,.
\end{equation}
Here, $i = 1, \cdots, {N}_{\alpha \beta}^\gamma$, where ${N}_{\alpha \beta}^\gamma$ are the fusion coefficients governing the fusion rule of topological interfaces.
The basis junction vectors are again chosen in such a way that they satisfy the following completeness and orthogonality relations:
\begin{equation} \label{eq:interfaces_comp_ortho}
\raisebox{-2em}{\begin{tikzpicture}
\draw [thick, blue, decoration = {markings, mark=at position .7 with {\arrow[scale=1.5]{stealth}}}, postaction=decorate] (0,0) to (0,2);
\node at (-0.2,1.8) {$\alpha$};
\draw [thick, blue, decoration = {markings, mark=at position .7 with {\arrow[scale=1.5]{stealth}}}, postaction=decorate] (1,0) to (1,2);
\node at (1.2,1.8) {$\beta$};
\end{tikzpicture}}
= \sum_{\gamma, i} \sqrt{\frac{\qd_\gamma}{\qd_\alpha \qd_\beta}} \raisebox{-2em}{\begin{tikzpicture}
\draw [thick, blue, decoration = {markings, mark=at position .7 with {\arrow[scale=1.5]{stealth}}}, postaction=decorate] (0,0) to (0.5,0.6);
\node at (-0.2,0.2) {$\alpha$};
\draw [thick, blue, decoration = {markings, mark=at position .7 with {\arrow[scale=1.5]{stealth}}}, postaction=decorate] (1,0) to (0.5,0.6);
\node at (1.2,0.2) {$\beta$};
\draw [thick, blue, decoration = {markings, mark=at position .7 with {\arrow[scale=1.5]{stealth}}}, postaction=decorate] (0.5,0.6) to (0.5,1.4);
\draw [thick, blue, decoration = {markings, mark=at position .7 with {\arrow[scale=1.5]{stealth}}}, postaction=decorate] (0.5,1.4) to (0.0,2.0);
\draw [thick, blue, decoration = {markings, mark=at position .7 with {\arrow[scale=1.5]{stealth}}}, postaction=decorate] (0.5,1.4) to (1,2);
\node at (-0.2,1.8) {$\alpha$};
\node at (1.2,1.8) {$\beta$};
\node at (0.8,1) {$\gamma$};
\node at (0.25,1.35) {$\bar{i}$};
\node at (0.25,0.65) {$i$};
\node at (0.5,0.7) {$\times$};
\node at (0.5,1.3) {$\times$};
\end{tikzpicture}} \,, \qquad
\raisebox{-2.5em}{\begin{tikzpicture}
\draw [thick, blue, decoration = {markings, mark=at position .6 with {\arrow[scale=1.5]{stealth}}}, postaction=decorate] (0,0.6) arc (-50:50:0.5222);
\draw [thick, blue, decoration = {markings, mark=at position .6 with {\arrow[scale=1.5]{stealth}}}, postaction=decorate] (0,0.6) arc (230:130:0.5222);
\draw  [thick, blue, decoration = {markings, mark=at position .7 with {\arrow[scale=1.5]{stealth}}}, postaction=decorate] (0,0) to (0,0.6);
\draw  [thick, blue, decoration = {markings, mark=at position .7 with {\arrow[scale=1.5]{stealth}}}, postaction=decorate] (0,1.4) to (0,2);
\node at (-0.5,1) {$\alpha$};
\node at (0.5,1) {$\beta$};
\node at (0.2,0.6) {$\bar{i}'$};
\node at (0.15,1.5) {$i$};
\node at (-0.3,0.1) {$\gamma'$};
\node at (-0.2,1.9) {$\gamma$};
\node at (0,0.5) {$\times$};
\node at (0,1.5) {$\times$};
\end{tikzpicture}}
=  \delta_{\gamma \gamma'} \delta_{i i'} \sqrt{\frac{\qd_\alpha \qd_\beta}{\qd_\gamma}} \raisebox{-2em}{\begin{tikzpicture}
\draw [thick, blue, decoration = {markings, mark=at position .7 with {\arrow[scale=1.5]{stealth}}}, postaction=decorate] (0,0) to (0,2);
\node at (-0.2,1.9) {$c$};
\end{tikzpicture}} \,.
\end{equation}
Shrinking a closed, contractible loop of a topological interface on a plane results in a constant multiplying the identity operator, and we define the quantum dimensions $\qd_\alpha$, $\qd_\beta$, $\qd_\gamma$ of topological interfaces as this constant \cite{Diatlyk:2023fwf}.

Conformal line interfaces also have a well-defined fusion structure \cite{Bachas:2007td,Diatlyk:2024zkk}.
In Section \ref{subsec:interfacefusion}, we discuss some implications of global symmetries on the fusion of conformal interfaces.

\subsection{Half-braiding, $\Omega$-symbols, and the Drinfeld center}\label{subsec:halfbraiding}

Much of this work leverages techniques involving the SymTFT associated to a fusion category $\mathcal{C}$, which we sometimes write as $\mathrm{TV}_{\mathcal{C}}$ for ``Turaev-Viro'' \cite{Turaev:1992hq,Barrett:1993ab}. 
This is a 2+1d topological field theory which admits a canonical ``Dirichlet'' boundary condition described by the module category $\mathcal{B}_{\mathrm{reg}}$ whose topological excitations (i.e.\ topological lines on the boundary) are described by $\mathcal{C}$. 
As a category, $\TV_{\CC}$ is realized by a construction known as the Drinfeld center which involves the notion of a half-braiding between the bulk and boundary topological lines of the SymTFT. The half-braiding will be a key ingredient in our discussions in later sections, so we explain its definition and properties below. We relegate a more systematic discussion of the SymTFT, as well as its interplay with boundary conditions and interfaces, to Section \ref{sec:symTFT}.

\subsubsection{Drinfeld center and half-braiding}

Let $\mathcal{C}$ be a fusion category.
The Drinfeld center $Z(\mathcal{C})$ of $\mathcal{C}$ is a modular tensor category \cite{Moore:1988qv}, which characterizes the 2+1d SymTFT for the 1+1d fusion category symmetry $\mathcal{C}$.
The objects in the Drinfeld center $Z(\mathcal{C})$ (i.e., the anyons in the SymTFT) are given by pairs $(Z,\gamma_Z)$, where $Z$ is a (not necessarily simple) object in $\mathcal{C}$ and $\gamma_Z$ is a family of isomorphisms which is called a \emph{half-braiding} (see, for instance, \cite[Definition 7.13.1]{etingof2016tensor}).
A half-braiding $\gamma_Z$ defines, for every object $X$ in $\mathcal{C}$, 
\begin{equation} \label{eq:hb_morphism}
    \gamma_{Z;X} : X \otimes Z \xrightarrow{\sim} Z \otimes X \,,
\end{equation}
such that the following hexagon diagram commutes for arbitrary $X$ and $Y$ in $\mathcal{C}$,
\begin{equation} \label{eq:hexagon_math}
\raisebox{-2em}{\begin{tikzpicture}
\node at (0,0) {$X \otimes (Y \otimes Z)$};
\draw[->] (0.6,0.3) -- (2.5,0.9);
\node at (1.1,0.9) {\scriptsize $\mathrm{id}_X \otimes \gamma_{Z;Y}$};
\node at (3.3,1.2) {$X \otimes (Z \otimes Y)$};
\draw[->] (4.6,1.2) -- (6.6,1.2);
\node at (5.6,1.5) {\scriptsize $\alpha^{-1}_{XZY}$};
\node at (7.9,1.2) {$(X \otimes Z) \otimes Y$};
\draw[->] (9,0.9) -- (10.9,0.3);
\node at (10.5,0.9) {\scriptsize $\gamma_{Z;X} \otimes \mathrm{id}_Y$};
\node at (11.5,0) {$(Z \otimes X) \otimes Y$};
\draw[->] (0.6,-0.3) -- (2.5,-0.9);
\node at (1.1,-0.8) {\scriptsize $\alpha^{-1}_{XYZ}$};
\node at (3.3,-1.3) {$(X \otimes Y) \otimes Z$};
\draw[->] (4.6,-1.3) -- (6.6,-1.3);
\node at (5.6,-1.6) {\scriptsize $\gamma_{Z;X \otimes Y}$};
\node at (7.9,-1.3) {$Z \otimes (X \otimes Y)$};
\draw[->] (9,-0.9) -- (10.9,-0.3);
\node at (10.5,-0.8) {\scriptsize $\alpha^{-1}_{ZXY}$};
\end{tikzpicture}} \,.
\end{equation}
Here, $\alpha_{XYZ} : (X\otimes Y)\otimes Z \rightarrow X \otimes (Y\otimes Z)$ are the associator isomorphisms of $\mathcal{C}$, and are defined for arbitrary objects $X$, $Y$, $Z$ in $\mathcal{C}$.

The forgetful functor,
\begin{equation} \label{eq:forget}
    F: Z(\mathcal{C}) \rightarrow \mathcal{C} \,, \quad (Z,\gamma_{Z}) \mapsto Z \,,
\end{equation}
defines a map from the set of bulk topological lines in the SymTFT to the set of boundary topological lines on the Dirichlet boundary of the SymTFT. Physically, it determines the image of a bulk line when it is brought close to the Dirichlet boundary.

Given a simple bulk topological line $\mu$ in $Z(\mathcal{C})$, we denote its image under the forgetful functor \eqref{eq:forget} as
\begin{equation} \label{eq:forgetmu}
    F(\mu) = \bigoplus_{a \in \mathrm{Irr}(\mathcal{C})} \langle \mu , a \rangle a \,.
\end{equation}
Here, the $\langle \mu , a \rangle \in \mathbb{Z}^{\geq 0}$ are non-negative integers, and they correspond to the dimensions of the bulk-to-boundary topological junction vector spaces $W_a^\mu \equiv \mathrm{Hom}_{\mathcal{C}}(F(\mu),a)$.
We fix a set of basis vectors $u_\mu^{a;x} \in \mathrm{Hom}_{\mathcal{C}}(F(\mu),a)$ as well as  dual basis vectors $\bar{u}_\mu^{a;x} \in \mathrm{Hom}_{\mathcal{C}}(a,F(\mu))$, where $x= 1, \cdots, \langle \mu , a \rangle$.
We pictorially represent them by
\begin{equation} \label{eq:basis_bulk_to_bdy_2}
    u_\mu^{a;x} = \raisebox{-1.8em}{\begin{tikzpicture}
    \draw [thick, decoration = {markings, mark=at position .7 with {\arrow[scale=1.5]{stealth}}}, postaction=decorate] (0,1) to (0,2);
    \draw [red, thick, decoration = {markings, mark=at position .7 with {\arrow[scale=1.5]{stealth}}}, postaction=decorate] (0,0) to (0,1);
    \node at (0.2,1) {$x$};
    \node at (0.2,1.85) {$a$};
    \node at (0.2,0.15) {$\mu$};
\end{tikzpicture}} \,, \qquad
\bar{u}_\mu^{a;x} = \raisebox{-1.8em}{\begin{tikzpicture}
    \draw [red, thick, decoration = {markings, mark=at position .7 with {\arrow[scale=1.5]{stealth}}}, postaction=decorate] (0,1) to (0,2);
    \draw [thick, decoration = {markings, mark=at position .7 with {\arrow[scale=1.5]{stealth}}}, postaction=decorate] (0,0) to (0,1);
    \node at (0.2,1) {$\bar{x}$};
    \node at (0.2,1.85) {$\mu$};
    \node at (0.2,0.15) {$a$};
\end{tikzpicture}} \,,
\end{equation}
where we draw bulk topological lines in red.
We choose the junction basis vectors so that they satisfy the following completeness and orthogonality relations:
\begin{equation} \label{eq:basis_bulk_to_bdy}
\raisebox{-2em}{\begin{tikzpicture}
\draw [red, thick, decoration = {markings, mark=at position .7 with {\arrow[scale=1.5]{stealth}}}, postaction=decorate] (0,0) to (0,2);
\node [] at (-0.2,1.8) {$\mu$};
\end{tikzpicture}}
= 
 \sum_{a \in \mathrm{Irr}(\mathcal{C})} \sum_{x=1}^{\langle \mu , a \rangle} \sqrt{\frac{\qd_a}{\qd_\mu}} \raisebox{-2em}{\begin{tikzpicture}
\draw [red, thick, decoration = {markings, mark=at position .6 with {\arrow[scale=1.5]{stealth}}}, postaction=decorate] (0.6,0) arc (0:90:0.6);
\draw [red, thick, decoration = {markings, mark=at position .6 with {\arrow[scale=1.5]{stealth}}}, postaction=decorate] (0,1.4) arc (-90:0:0.6);
\draw [thick, decoration = {markings, mark=at position .7 with {\arrow[scale=1.5]{stealth}}}, postaction=decorate] (0,0.6) to (0,1.4);
\node [] at (0.8,1.8) {$\mu$};
\node [] at (0.8,0.2) {$\mu$};
\node at (0.3,1) {$a$};
\node at (-0.2,0.6) {$x$};
\node at (-0.2,1.4) {$\bar{x}$};
\end{tikzpicture}} \,, \qquad
\raisebox{-2.5em}{\begin{tikzpicture}
\draw [thick, decoration = {markings, mark=at position .7 with {\arrow[scale=1.5]{stealth}}}, postaction=decorate] (0,0) to (0,0.6);
\draw [thick, decoration = {markings, mark=at position .7 with {\arrow[scale=1.5]{stealth}}}, postaction=decorate] (0,1.4) to (0,2);
\draw [red, thick, decoration = {markings, mark=at position .7 with {\arrow[scale=1.5]{stealth}}}, postaction=decorate] (0,0.6) arc (-90:90:0.4);
\node at (0.3,1.8) {$a$};
\node at (0.3,0.2) {$a'$};
\node at (-0.2,1.4) {$x$};
\node at (-0.2,0.6) {$\bar{x}'$};
\node [] at (0.7,1) {$\mu$};
\end{tikzpicture}}
=  \delta_{a a'} \delta_{x x'} \sqrt{\frac{\qd_\mu}{\qd_a}} \raisebox{-2em}{\begin{tikzpicture}
\draw [thick, decoration = {markings, mark=at position .7 with {\arrow[scale=1.5]{stealth}}}, postaction=decorate] (0,0) to (0,2);
\node at (-0.2,1.9) {$a$};
\end{tikzpicture}} \,.
\end{equation}
Here, $\qd_\mu$ is the quantum dimension of the bulk topological line $\mu$.

\subsubsection{$\Omega$-symbols as matrix elements of half-braidings}

A simple topological line $\mu$ in $Z(\mathcal{C})$ consists of a pair $(F(\mu),\gamma_{F(\mu)})$, where $F(\mu) \in \mathcal{C}$ is the image of $\mu$ under the forgetful functor as in \eqref{eq:forgetmu}, and $\gamma_{F(\mu)} \equiv \gamma_\mu$ is the half-braiding.
Given an arbitrary simple object $a$ in $\mathcal{C}$, the explicit matrix elements of the half-braiding morphism $\gamma_{\mu;a} : a \otimes F(\mu) \rightarrow F(\mu) \otimes a$ in a fixed basis (as in \eqref{eq:basis_bulk_to_bdy_2} and \eqref{eq:basis_bulk_to_bdy}) can be obtained as follows.

First, we pick an arbitrary simple object $b$ in $\mathcal{C}$, and apply the hom-functor $\mathrm{Hom}_{\mathcal{C}}(-,b)$ to the half-braiding morphism $\gamma_{\mu;a}$, which gives
\begin{equation} \label{eq:hb_hom}
    \mathrm{Hom}_{\mathcal{C}}(\gamma_{\mu;a},b) : \mathrm{Hom}_{\mathcal{C}}(F(\mu) \otimes a,b) \rightarrow \mathrm{Hom}_{\mathcal{C}}(a \otimes F(\mu),b) \,, \quad
    v \mapsto v \circ \gamma_{\mu;a} \,.
\end{equation}
$\mathrm{Hom}_{\mathcal{C}}(\gamma_{\mu;a},b)$ is an ordinary linear map between finite-dimensional complex vector spaces.
Moreover, we have decompositions
\begin{align}
\begin{split}
    \mathrm{Hom}_{\mathcal{C}}(F(\mu) \otimes a,b) &= \bigoplus_{c \in \mathrm{Irr}(\mathcal{C})} \mathrm{Hom}_{\mathcal{C}} (F(\mu),c) \otimes \mathrm{Hom}_{\mathcal{C}} (c\otimes a, b) \,,\\
    \mathrm{Hom}_{\mathcal{C}}(a \otimes F(\mu) ,b) &= \bigoplus_{d \in \mathrm{Irr}(\mathcal{C})} \mathrm{Hom}_{\mathcal{C}} (a\otimes d, b) \otimes \mathrm{Hom}_{\mathcal{C}} (d,F(\mu))  \,.
\end{split}
\end{align}
Since we have fixed the basis vectors for the vector spaces appearing on the right-hand side, we can explicitly write down the matrix elements of the linear map \eqref{eq:hb_hom} in that chosen basis. We call these matrix elements $\Omega$-symbols:
\begin{equation} \label{eq:half-braiding}
\raisebox{-2em}{\begin{tikzpicture}
\draw [red, thick, decoration = {markings, mark=at position .4 with {\arrow[scale=1.5]{stealth}}}, postaction=decorate] (1,0) to (0.1,1);
\draw [thick, decoration = {markings, mark=at position .7 with {\arrow[scale=1.5]{stealth}}}, postaction=decorate] (0.1,1) to (0.5,1.5);
\draw [thick] (0.52,0.67) arc (-45:45:0.6);
\draw [thick, decoration = {markings, mark=at position .7 with {\arrow[scale=1.5]{stealth}}}, postaction=decorate] (0,0) to (0.4,0.5);
\draw [thick, decoration = {markings, mark=at position .7 with {\arrow[scale=1.5]{stealth}}}, postaction=decorate] (0.5,1.5) to (0.5,2.3);
\node [red] at (1.2,0.2) {$\mu$};
\node at (-0.2,0.2) {$a$};
\node at (0.7,2.2) {$b$};
\node at (-0.1,0.98) {$x$};
\node at (0.15,1.4) {$c$};
\node at (0.65,1.6) {$i$};
\node at (0.5,1.6) {$\times$};
\end{tikzpicture}}
= 
 \sum_{d \in \mathrm{Irr}(\mathcal{C})} \sum_{y=1}^{\langle \mu , d \rangle} \sum_{j=1}^{N_{ad}^b} \left[
    \Omega_{a\mu}^b
 \right]_{(cxi)(dyj)} \raisebox{-2em}{\begin{tikzpicture}
\draw [thick, decoration = {markings, mark=at position .7 with {\arrow[scale=1.5]{stealth}}}, postaction=decorate] (0,0) to (0.5,1.5);
\draw [thick, decoration = {markings, mark=at position .7 with {\arrow[scale=1.5]{stealth}}}, postaction=decorate] (0.5,1.5) to (0.5,2.3);
\draw [red, thick, decoration = {markings, mark=at position .7 with {\arrow[scale=1.5]{stealth}}}, postaction=decorate] (1,0) to (0.75,0.75);
\draw [thick, decoration = {markings, mark=at position .7 with {\arrow[scale=1.5]{stealth}}}, postaction=decorate] (0.75,0.75) to (0.5,1.5);
\node [red] at (1.2,0.2) {$\mu$};
\node at (-0.2,0.2) {$a$};
\node at (0.7,2.2) {$b$};
\node at (0.9,0.75) {$y$};
\node at (0.85,1.2) {$d$};
\node at (0.66,1.6) {$j$};
\node at (0.5,1.6) {$\times$};
\end{tikzpicture}} \,.
\end{equation}
Here, $a,b,c,d \in \mathrm{Irr}(\mathcal{C})$ are boundary topological lines, $\mu \in \mathrm{Irr}(Z(\mathcal{C}))$ is a bulk topological line, and the various topological junctions are $u_\mu^{c;x} \in \mathrm{Hom}_{\mathcal{C}} (F(\mu),c)$, $v_{ca}^{b;i} \in \mathrm{Hom}_{\mathcal{C}} (c\otimes a,b)$, $u_\mu^{d;y} \in \mathrm{Hom}_{\mathcal{C}} (F(\mu),d)$, and $v_{ad}^{b;j} \in \mathrm{Hom}_{\mathcal{C}} (a\otimes d,b)$.

The $\Omega$-symbols $\left[\Omega_{a\mu}^b \right]_{(cxi)(dyj)}$ in a given basis determine the half-braiding morphism $\gamma_\mu$ for the bulk topological line $\mu$.\footnote{
Such $\Omega$-symbols have been discussed in the context of string net models \cite{Levin:2004mi,PhysRevB.103.195155}, but our conventions are slightly different. See also \cite{Zhang:2023wlu} for the computation of $\Omega$-symbols for $\mathbb{Z}_N \times \mathbb{Z}_N$ Tambara-Yamagami fusion categories.}
We can equivalently write \eqref{eq:half-braiding} as
\begin{equation}
   v_{ca}^{b;i} \circ (u_\mu^{c;x}\otimes \mathrm{id}_a) \circ \gamma_{\mu ; a} = 
 \sum_{d \in \mathrm{Irr}(\mathcal{C})} \sum_{y=1}^{ \langle \mu , d \rangle} \sum_{j=1}^{N_{ad}^b} \left[
    \Omega_{a\mu}^b
 \right]_{(cxi)(dyj)} v_{ad}^{b;j} \circ (\mathrm{id}_a \otimes u_\mu^{d;y}) \,.
\end{equation}
Note that both sides of the equation are vectors in the junction vector space $\mathrm{Hom}_{\mathcal{C}}(a\otimes F(\mu),b)$, and the $\Omega$-symbols are the matrix elements of the half-braiding morphism $\gamma_{\mu;a}$.
Similarly, we define the inverse $\Omega$-symbols by
\begin{equation} \label{eq:inverse_half-braiding}
\raisebox{-2em}{\begin{tikzpicture}
\draw [red, thick, decoration = {markings, mark=at position .4 with {\arrow[scale=1.5]{stealth}}}, postaction=decorate] (0,0) to (0.9,1);
\draw [thick, decoration = {markings, mark=at position .7 with {\arrow[scale=1.5]{stealth}}}, postaction=decorate] (0.9,1) to (0.5,1.5);
\draw [thick] (0.48,0.67) arc (225:135:0.6);
\draw [thick, decoration = {markings, mark=at position .7 with {\arrow[scale=1.5]{stealth}}}, postaction=decorate] (1,0) to (0.6,0.5);
\draw [thick, decoration = {markings, mark=at position .7 with {\arrow[scale=1.5]{stealth}}}, postaction=decorate] (0.5,1.5) to (0.5,2.3);
\node [red] at (-0.2,0.2) {$\mu$};
\node at (1.2,0.2) {$a$};
\node at (0.7,2.2) {$b$};
\node at (1.1,0.98) {$x$};
\node at (0.85,1.4) {$c$};
\node at (0.32,1.6) {$i$};
\node at (0.5,1.6) {$\times$};
\end{tikzpicture}}
= 
 \sum_{d \in \mathrm{Irr}(\mathcal{C})} \sum_{y=1}^{\langle \mu , d \rangle} \sum_{j=1}^{N_{da}^b} \left[
    \Omega_{a\mu}^b
 \right]^{-1}_{(cxi)(dyj)} \raisebox{-2em}{\begin{tikzpicture}
\draw [thick, decoration = {markings, mark=at position .7 with {\arrow[scale=1.5]{stealth}}}, postaction=decorate] (1,0) to (0.5,1.5);
\draw [thick, decoration = {markings, mark=at position .7 with {\arrow[scale=1.5]{stealth}}}, postaction=decorate] (0.5,1.5) to (0.5,2.3);
\draw [red, thick, decoration = {markings, mark=at position .7 with {\arrow[scale=1.5]{stealth}}}, postaction=decorate] (0,0) to (0.25,0.75);
\draw [thick, decoration = {markings, mark=at position .7 with {\arrow[scale=1.5]{stealth}}}, postaction=decorate] (0.25,0.75) to (0.5,1.5);
\node [red] at (-0.2,0.2) {$\mu$};
\node at (1.2,0.2) {$a$};
\node at (0.3,2.2) {$b$};
\node at (0.1,0.75) {$y$};
\node at (0.15,1.2) {$d$};
\node at (0.31,1.6) {$j$};
\node at (0.5,1.6) {$\times$};
\end{tikzpicture}} \,,
\end{equation}
which satisfy
\begin{equation}
     \sum_{d \in \mathrm{Irr}(\mathcal{C})} \sum_{y=1}^{\langle \mu , d \rangle} \sum_{j=1}^{N_{ad}^b} \left[
    \Omega_{a\mu}^b
 \right]_{(cxi)(dyj)} \left[
    \Omega_{a\mu}^b
 \right]^{-1}_{(dyj)(c'x'i')} = \delta_{c c'} \delta_{x x'} \delta_{i i'} \,.
\end{equation}

From \eqref{eq:half-braiding} and \eqref{eq:inverse_half-braiding}, it follows that
\begin{align} \label{eq:half-braiding_2}
\begin{split}
\raisebox{-2em}{\begin{tikzpicture}
\draw [thick, decoration = {markings, mark=at position .4 with {\arrow[scale=1.5]{stealth}}}, postaction=decorate] (0,0) to (2,2);
\draw [red, thick, preaction={draw=white,line width=6pt}, decoration = {markings, mark=at position .4 with {\arrow[scale=1.5]{stealth}}}, postaction=decorate] (2,0) to (0,2);
\node [red] at (2.2,0.2) {$\mu$};
\node at (-0.2,0.2) {$a$};
\end{tikzpicture}}
&= 
 \sum_{b,c,d \in \mathrm{Irr}(\mathcal{C})} \sum_{x =1}^{ \langle \mu ,c \rangle} \sum_{y=1}^{\langle \mu , d \rangle} \sum_{i=1}^{ N^b_{ca}} \sum_{j=1}^{N_{ad}^b} \sqrt{\frac{\qd_b}{\qd_\mu \qd_a}} \left[
    \Omega_{a\mu}^b
 \right]_{(cxi)(dyj)} \raisebox{-2em}{\begin{tikzpicture}
    \draw [thick, decoration = {markings, mark=at position .7 with {\arrow[scale=1.5]{stealth}}}, postaction=decorate] (0,0) to (1,0.7);
    \draw [thick, decoration = {markings, mark=at position .7 with {\arrow[scale=1.5]{stealth}}}, postaction=decorate] (1,0.7) to (1,1.3);
    \draw [thick, decoration = {markings, mark=at position .7 with {\arrow[scale=1.5]{stealth}}}, postaction=decorate] (1,1.3) to (2,2);
    \draw [thick, decoration = {markings, mark=at position .7 with {\arrow[scale=1.5]{stealth}}}, postaction=decorate] (1.4,0.2) to (1,0.7);
    \draw [red, thick, decoration = {markings, mark=at position .7 with {\arrow[scale=1.5]{stealth}}}, postaction=decorate] (2,0) to (1.4,0.2);
    \draw [thick, decoration = {markings, mark=at position .7 with {\arrow[scale=1.5]{stealth}}}, postaction=decorate] (1,1.3) to (0.6,1.8);
    \draw [red, thick, decoration = {markings, mark=at position .7 with {\arrow[scale=1.5]{stealth}}}, postaction=decorate] (0.6,1.8) to (0,2);
    \node [red] at (2.2,0.2) {$\mu$};
    \node at (-0.2,0.2) {$a$};
    \node [red] at (-0.2,1.8) {$\mu$};
    \node at (2.1,1.8) {$a$};
    \node at (0.6,1.4) {$c$};
    \node at (0.75,1) {$b$};
    \node at (1,0.35) {$d$};
    \node at (1.2,1.2) {$\bar{i}$};
    \node at (1.2,0.8) {$j$};
    \node at (0.7,1.95) {$\bar{x}$};
    \node at (1.5,0.4) {$y$};
    \node at (1,0.8) {$\times$};
    \node at (1,1.2) {$\times$};
\end{tikzpicture}} \,, \\
\raisebox{-2em}{\begin{tikzpicture}
\draw [thick, decoration = {markings, mark=at position .4 with {\arrow[scale=1.5]{stealth}}}, postaction=decorate] (2,0) to (0,2);
\draw [red, thick, preaction={draw=white,line width=6pt}, decoration = {markings, mark=at position .4 with {\arrow[scale=1.5]{stealth}}}, postaction=decorate] (0,0) to (2,2);
\node [red] at (-0.2,0.2) {$\mu$};
\node at (2.2,0.2) {$a$};
\end{tikzpicture}}
&= 
  \sum_{b,c,d \in \mathrm{Irr}(\mathcal{C})} \sum_{x =1}^{ \langle \mu ,c \rangle} \sum_{y=1}^{\langle \mu , d \rangle} \sum_{i=1}^{ N^b_{ac}} \sum_{j=1}^{N_{da}^b} \sqrt{\frac{\qd_b}{\qd_\mu \qd_a}} \left[
    \Omega_{a\mu}^b
 \right]^{-1}_{(cxi)(dyj)} \raisebox{-2em}{\begin{tikzpicture}
    \draw [thick, decoration = {markings, mark=at position .7 with {\arrow[scale=1.5]{stealth}}}, postaction=decorate] (2,0) to (1,0.7);
    \draw [thick, decoration = {markings, mark=at position .7 with {\arrow[scale=1.5]{stealth}}}, postaction=decorate] (1,0.7) to (1,1.3);
    \draw [thick, decoration = {markings, mark=at position .7 with {\arrow[scale=1.5]{stealth}}}, postaction=decorate] (1,1.3) to (0,2);
    \draw [thick, decoration = {markings, mark=at position .7 with {\arrow[scale=1.5]{stealth}}}, postaction=decorate] (0.6,0.2) to (1,0.7);
    \draw [red, thick, decoration = {markings, mark=at position .7 with {\arrow[scale=1.5]{stealth}}}, postaction=decorate] (0,0) to (0.6,0.2);
    \draw [thick, decoration = {markings, mark=at position .7 with {\arrow[scale=1.5]{stealth}}}, postaction=decorate] (1,1.3) to (1.4,1.8);
    \draw [red, thick, decoration = {markings, mark=at position .7 with {\arrow[scale=1.5]{stealth}}}, postaction=decorate] (1.4,1.8) to (2,2);
    \node [red] at (-0.2,0.2) {$\mu$};
    \node at (2.2,0.2) {$a$};
    \node [red] at (2.2,1.8) {$\mu$};
    \node at (-0.1,1.8) {$a$};
    \node at (1.4,1.4) {$c$};
    \node at (1.25,1) {$b$};
    \node at (1,0.35) {$d$};
    \node at (0.8,1.2) {$\bar{i}$};
    \node at (0.8,0.8) {$j$};
    \node at (1.3,1.95) {$\bar{x}$};
    \node at (0.5,0.4) {$y$};
    \node at (1,0.8) {$\times$};
    \node at (1,1.2) {$\times$};
\end{tikzpicture}} \,.
\end{split}
\end{align}
These relations allow us to push the bulk topological lines of $\mathrm{TV}_\mathcal{C}$ onto the Dirichlet boundary in the presence of  boundary topological lines.

From the definition of the $\Omega$-symbols, it also follows that a subset of them are determined by the topological spins of bulk anyons.
In particular, the two equations
\begin{equation} \label{eq:omega_theta}
    \left[ \Omega_{a\mu}^{1} \right]_{(\bar{a}x)(\bar{a}y)} = \theta_\mu \delta_{xy}\,, \quad \left[ \Omega_{a\mu}^{1} \right]^{-1}_{(\bar{a}x)(\bar{a}y)} = \theta^*_\mu \delta_{xy} \,,
\end{equation}
hold independent of $a$, where $\theta_\mu \in U(1)$ is the spin of the bulk line $\mu$ defined by 
\begin{eqnarray}
    \raisebox{-3.3em}{\begin{tikzpicture}
\draw[thick, red, decoration = {markings, mark=at position .3 with {\arrow[scale=1.5]{stealth}}}, postaction=decorate] (0,0) .. controls +(0,1.8) and +(0,0.3) .. (0.5,1) .. controls +(0,-0.5) and +(0,0) .. (0.15,0.9);
\draw[thick, red] (0.05,1) .. controls +(-0.1,0) and +(0,0) .. (0,2);
\node[ anchor = north] at (0,0) {$\mu$}; 
\end{tikzpicture}}
~~~ = ~~ \theta_\mu ~
\raisebox{-3.3em}{\begin{tikzpicture}
\draw[red, thick,  anchor = north, decoration = {markings, mark=at position .7 with {\arrow[scale=1.5]{stealth}}}, postaction=decorate] (0,0) -- (0,2); 
\node[ anchor = north] at (0,0) {$\mu$}; 
\end{tikzpicture}} \,.
\end{eqnarray}
We have suppressed $i$ and $j$ indices in \eqref{eq:omega_theta}, and they are chosen to be the identity operator on the topological line $a$.

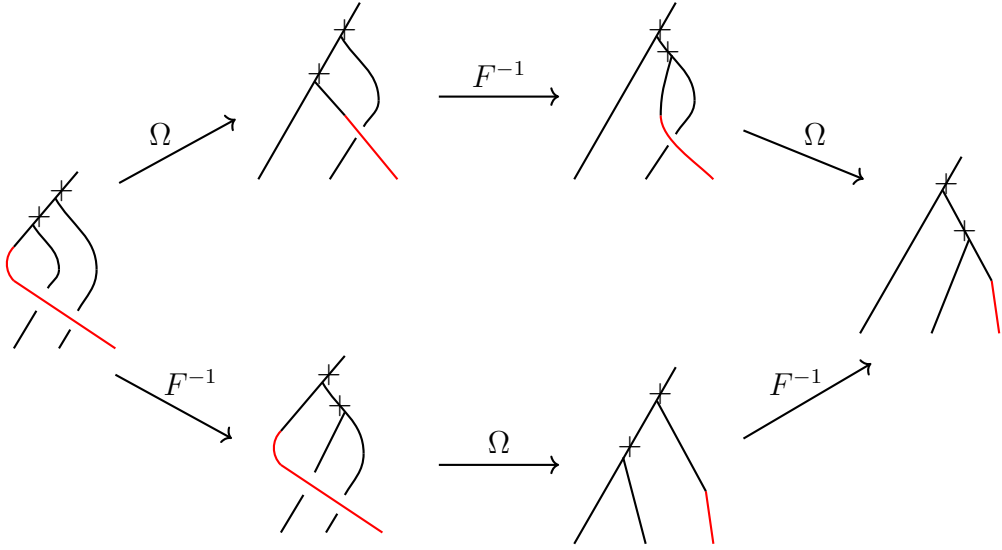
\begin{figure}[t!]
	\centering
	\begin{tikzpicture}[scale=0.2]
		\begin{pgfonlayer}{nodelayer}
			\node [style=none] (0) at (-7.25, 0) {};
			\node [style=none] (1) at (-3, 5) {};
			\node [style=none] (2) at (-7.25, -2.25) {};
			\node [style=none] (3) at (-0.5, -6.75) {};
			\node [style=none] (4) at (-6, 1.5) {};
			\node [style=none] (5) at (-4.5, 3.25) {};
			\node [style=none] (6) at (-4.25, -1.5) {};
			\node [style=none] (7) at (-1.75, -1.5) {};
			\node [style=none] (8) at (-5, -2.75) {};
			\node [style=none] (9) at (-3, -4.25) {};
			\node [style=none] (10) at (-5.75, -4.25) {};
			\node [style=none] (11) at (-3.5, -5.5) {};
			\node [style=none] (12) at (-4.25, -6.75) {};
			\node [style=none] (13) at (-7.25, -6.75) {};
			\node [style=none] (14) at (-0.25, 4.25) {};
			\node [style=none] (15) at (7.5, 8.5) {};
			\node [style=none] (16) at (2.5, 7.5) { $\Omega$};
			\node [style=none] (17) at (9, 4.5) {};
			\node [style=none] (18) at (15.75, 16.25) {};
			\node [style=none] (19) at (14.75, 8.75) {};
			\node [style=none] (20) at (18.25, 4.5) {};
			\node [style=none] (21) at (12.75, 11) {};
			\node [style=none] (22) at (14.5, 14) {};
			\node [style=none] (23) at (14.75, 8.75) {};
			\node [style=none] (24) at (17, 9.75) {};
			\node [style=none] (26) at (16, 8) {};
			\node [style=none] (28) at (13.75, 4.5) {};
			\node [style=none] (29) at (15.5, 7.25) {};
			\node [style=none] (30) at (21, 10) {};
			\node [style=none] (31) at (29, 10) {};
			\node [style=none] (32) at (25, 11.5) {$F^{-1}$};
			\node [style=none] (33) at (30, 4.5) {};
			\node [style=none] (34) at (36.75, 16.25) {};
			\node [style=none] (35) at (35.75, 8.75) {};
			\node [style=none] (36) at (39.25, 4.5) {};
			\node [style=none] (37) at (36.5, 12.75) {};
			\node [style=none] (38) at (35.5, 14) {};
			\node [style=none] (39) at (35.75, 8.75) {};
			\node [style=none] (40) at (38, 9.75) {};
			\node [style=none] (41) at (36.75, 7.5) {};
			\node [style=none] (42) at (34.75, 4.5) {};
			\node [style=none] (43) at (36.25, 6.75) {};
			\node [style=none] (44) at (41.25, 7.75) {};
			\node [style=none] (45) at (49.25, 4.5) {};
			\node [style=none] (46) at (46, 7.5) {$\Omega$};
			\node [style=none] (47) at (49, -5.75) {};
			\node [style=none] (48) at (55.75, 6) {};
			\node [style=none] (49) at (57.75, -2.25) {};
			\node [style=none] (50) at (58.25, -5.75) {};
			\node [style=none] (52) at (54.5, 3.75) {};
			\node [style=none] (54) at (57.75, -2.25) {};
			\node [style=none] (56) at (53.75, -5.75) {};
			\node [style=none] (57) at (56.25, 0.5) {};
			\node [style=none] (58) at (-0.5, -8.5) {};
			\node [style=none] (59) at (7.25, -12.75) {};
			\node [style=none] (60) at (4.5, -9) {$F^{-1}$};
			\node [style=none] (61) at (10.5, -12.25) {};
			\node [style=none] (62) at (14.75, -7.25) {};
			\node [style=none] (63) at (10.5, -14.5) {};
			\node [style=none] (64) at (17.25, -19) {};
			\node [style=none] (66) at (13.25, -9) {};
			\node [style=none] (67) at (14.75, -11) {};
			\node [style=none] (68) at (16, -13.75) {};
			\node [style=none] (69) at (12.75, -15) {};
			\node [style=none] (70) at (14.75, -16.5) {};
			\node [style=none] (71) at (12, -16.5) {};
			\node [style=none] (72) at (14.25, -17.75) {};
			\node [style=none] (73) at (13.5, -19) {};
			\node [style=none] (74) at (10.5, -19) {};
			\node [style=none] (75) at (21, -14.5) {};
			\node [style=none] (76) at (29, -14.5) {};
			\node [style=none] (77) at (25, -13) {$\Omega$};
			\node [style=none] (78) at (30, -19.75) {};
			\node [style=none] (79) at (36.75, -8) {};
			\node [style=none] (80) at (38.75, -16.25) {};
			\node [style=none] (81) at (39.25, -19.75) {};
			\node [style=none] (82) at (35.5, -10.25) {};
			\node [style=none] (83) at (38.75, -16.25) {};
			\node [style=none] (84) at (34.75, -19.75) {};
			\node [style=none] (85) at (33.25, -14) {};
			\node [style=none] (86) at (41.25, -12.75) {};
			\node [style=none] (87) at (49.75, -7.75) {};
			\node [style=none] (88) at (44.75, -9) {$F^{-1}$};
			\node [style=none] (89) at (7, 8.75) {};
			\node [style=none] (90) at (7.5, 8) {};
			\node [style=none] (91) at (8, 8.75) {};
			\node [style=none] (92) at (28.5, 10.5) {};
			\node [style=none] (93) at (28.5, 9.5) {};
			\node [style=none] (94) at (29.5, 10) {};
			\node [style=none] (95) at (48.75, 4.25) {};
			\node [style=none] (96) at (49, 5) {};
			\node [style=none] (97) at (49.75, 4.25) {};
			\node [style=none] (98) at (48.75, -7.75) {};
			\node [style=none] (99) at (49.25, -8.5) {};
			\node [style=none] (100) at (28.5, -14) {};
			\node [style=none] (101) at (28.5, -15) {};
			\node [style=none] (102) at (29.5, -14.5) {};
			\node [style=none] (103) at (6.25, -12.75) {};
			\node [style=none] (104) at (6.75, -12) {};
		\end{pgfonlayer}
		\begin{pgfonlayer}{edgelayer}
			\draw [thick, postaction={
				decorate,
				decoration={
					markings,
					mark=at position 0.4 with {
						\node[scale=1, rotate=45] at (0,0) {$\times$};
					},
					mark=at position 0.75 with {
						\node[scale=1, rotate=45] at (0,0) {$\times$};
					}
			}}] (0.center) to (1.center);
			\draw [red, thick, in=135, out=-135] (0.center) to (2.center);
			\draw [red, thick] (2.center) to (3.center);
			\draw [thick, in=90, out=-60] (4.center) to (6.center);
			\draw [thick, in=90, out=-60] (5.center) to (7.center);
			\draw [thick, in=60, out=-90] (6.center) to (8.center);
			\draw [thick, in=60, out=-90] (7.center) to (9.center);
			\draw [thick] (11.center) to (12.center);
			\draw [thick] (10.center) to (13.center);
			\draw [thick, postaction={
				decorate,
				decoration={
					markings,
					mark=at position 0.6 with {
						\node[scale=1, rotate=45] at (0,0) {$\times$};
					},
					mark=at position 0.85 with {
						\node[scale=1, rotate=45] at (0,0) {$\times$};
					}
			}}] (17.center) to (18.center);
			\draw [red, thick] (19.center) to (20.center);
			\draw [thick] (21.center) to (23.center);
			\draw [thick, in=90, out=-60] (22.center) to (24.center);
			\draw [thick, in=60, out=-90] (24.center) to (26.center);
			\draw [thick] (28.center) to (29.center);
			\draw [thick, postaction={
				decorate,
				decoration={
					markings,
					mark=at position 0.85 with {
						\node[scale=1, rotate=45] at (0,0) {$\times$};
					}
			}}] (33.center) to (34.center);
			\draw [red, thick, in=135, out=-90, looseness=0.75] (35.center) to (36.center);
			\draw [thick, in=90, out=-105] (37.center) to (39.center);
			\draw [thick, in=90, out=-60, postaction={
				decorate,
				decoration={
					markings,
					mark=at position 0.25 with {
						\node[scale=1, rotate=45] at (0,0) {$\times$};
					}
			}}] (38.center) to (40.center);
			\draw [thick, in=60, out=-90] (40.center) to (41.center);
			\draw [thick] (42.center) to (43.center);
			\draw [thick, postaction={
				decorate,
				decoration={
					markings,
					mark=at position 0.85 with {
						\node[scale=1, rotate=45] at (0,0) {$\times$};
					}
			}}] (47.center) to (48.center);
			\draw [red, thick] (49.center) to (50.center);
			\draw [thick, postaction={
				decorate,
				decoration={
					markings,
					mark=at position 0.45 with {
						\node[scale=1, rotate=45] at (0,0) {$\times$};
					}
			}}] (52.center) to (54.center);
			\draw [thick] (56.center) to (57.center);
			\draw [thick, postaction={
				decorate,
				decoration={
					markings,
					mark=at position 0.75 with {
						\node[scale=1, rotate=45] at (0,0) {$\times$};
					}
			}}] (61.center) to (62.center);
			\draw [red, thick, in=135, out=-135] (61.center) to (63.center);
			\draw [red, thick] (63.center) to (64.center);
			\draw [thick, in=90, out=-60, postaction={
				decorate,
				decoration={
					markings,
					mark=at position 0.35 with {
						\node[scale=1, rotate=45] at (0,0) {$\times$};
					}
			}}] (66.center) to (68.center);
			\draw [thick] (67.center) to (69.center);
			\draw [thick, in=60, out=-90] (68.center) to (70.center);
			\draw [thick] (72.center) to (73.center);
			\draw [thick] (71.center) to (74.center);
			\draw [thick, postaction={
				decorate,
				decoration={
					markings,
					mark=at position 0.85 with {
						\node[scale=1, rotate=45] at (0,0) {$\times$};
					}, 
					mark=at position 0.55 with {
						\node[scale=1, rotate=45] at (0,0) {$\times$};
					}
			}}] (78.center) to (79.center);
			\draw [red, thick] (80.center) to (81.center);
			\draw [thick] (82.center) to (83.center);
			\draw [thick] (84.center) to (85.center);
			\draw [thick, ->] (86.center) to (87.center);
			\draw [thick, ->] (58.center) to (59.center);
			\draw [thick, ->] (14.center) to (15.center);
			\draw [thick, ->] (30.center) to (31.center);
			\draw [thick, ->] (44.center) to (45.center);
			\draw [thick, ->] (75.center) to (76.center);
		\end{pgfonlayer}
	\end{tikzpicture}
	\caption{A schematic picture of the hexagon equation obeyed by $\Omega$-symbols. The arrows, which are suppressed in the diagram, are all taken to point upwards. }\label{fig:hexagon}
\end{figure}

The $\Omega$-symbols satisfy a hexagon equation, which is shown schematically in Figure \ref{fig:hexagon}, and unpacks more explicitly as 
\begin{align} \label{eq:hexagon}
\begin{split}
    &~~~~~~\sum_{f \in \mathrm{Irr}(\mathcal{C})} \sum_{y=1}^{\langle \mu , f \rangle} \sum_{k=1}^{N_{af}^e}  \sum_{m=1}^{N_{fb}^g} \left[ \Omega_{a\mu}^e \right]_{(dxi)(fyk)} \left[ F_{afb}^c \right]_{(ekj)(glm)}^{-1} \left[\Omega_{b\mu}^g \right]_{(fym)(hzn)} \\
    &= \sum_{a' \in \mathrm{Irr}(\mathcal{C})} \sum_{o=1}^{N_{da'}^c} \sum_{p=1}^{N_{ab}^{a'}}  \sum_{q=1}^{N_{a' h}^c} \left[ F_{dab}^c \right]_{(eij)(a'op)}^{-1}  \left[ \Omega_{a'\mu}^c \right]_{(dxo)(hzq)} \left[ F_{abh}^c \right]_{(a'pq)(gln)}^{-1} \,.
\end{split}
\end{align}
This furnishes a method for computing $\Omega$-symbols in practice, once the $F$-symbols of the fusion category $\mathcal{C}$ are known.
The hexagon equation \eqref{eq:hexagon} comes from the commutative hexagon diagram \eqref{eq:hexagon_math} that the half-braiding morphisms $\gamma_{\mu ;a}$ satisfy.

\subsubsection{The $\Omega$-symbols of a modular tensor category}

\begin{figure}[t!]
	\centering
	\begin{tikzpicture}[scale=0.35]
		\begin{pgfonlayer}{nodelayer}
			\node [style=none] (0) at (-7.25, 11.75) {};
			\node [style=none] (1) at (-7.25, 9.75) {};
			\node [style=none] (2) at (-8, 8.5) {};
			\node [style=none] (3) at (-5, 5.75) {};
			\node [style=none] (4) at (-6.25, 7.25) {};
			\node [style=none] (5) at (-6.75, 6.75) {};
			\node [style=none] (6) at (-8.25, 5.75) {};
			\node [style=none] (7) at (-4.25, 6.5) { $\color{red}(a,b)$};
			\node [style=none] (8) at (-3.75, 8.75) {};
			\node [style=none] (9) at (1.25, 8.75) {};
			\node [style=none] (10) at (-1.25, 9.75) { $\Omega$};
			\node [style=none] (11) at (4.25, 11.75) {};
			\node [style=none] (12) at (4.25, 9.75) {};
			\node [style=none] (13) at (5.25, 8) {};
			\node [style=none] (14) at (6.5, 5.75) {};
			\node [style=none] (15) at (5.25, 8) {};
			\node [style=none] (16) at (4.25, 9.75) {};
			\node [style=none] (17) at (3.25, 5.75) {};
			\node [style=none] (25) at (-10.75, 2.5) { Unfold};
			\node [style=none] (26) at (-6.25, -0.75) {};
			\node [style=none] (27) at (-6.25, -2.75) {};
			\node [style=none] (28) at (-7, -4) {};
			\node [style=none] (29) at (-4, -6.75) {};
			\node [style=none] (30) at (-5.25, -5.25) {};
			\node [style=none] (31) at (-5.75, -5.75) {};
			\node [style=none] (32) at (-7.25, -6.75) {};
			\node [style=none] (34) at (-6.75, -6) {};
			\node [style=none] (35) at (-6.25, -6.5) {};
			\node [style=none] (36) at (-5.5, -6.75) {};
			\node [style=none] (37) at (-5.25, -7.25) { $a$};
			\node [style=none] (38) at (-3.5, -7) { $b$};
			\node [style=none] (39) at (-9, -6) {};
			\node [style=none] (40) at (-16.75, -8.75) {};
			\node [style=none] (41) at (-19, -6.75) {};
			\node [style=none] (42) at (-19, -8.75) {};
			\node [style=none] (43) at (-19.75, -10) {};
			\node [style=none] (44) at (-16.75, -12.75) {};
			\node [style=none] (45) at (-18, -11.25) {};
			\node [style=none] (46) at (-18.5, -11.75) {};
			\node [style=none] (47) at (-20, -12.75) {};
			\node [style=none] (48) at (-19.5, -12) {};
			\node [style=none] (49) at (-19, -12.5) {};
			\node [style=none] (50) at (-18.25, -12.75) {};
			\node [style=none] (51) at (-18, -13.25) { $a$};
			\node [style=none] (52) at (-16.25, -13) { $b$};
			\node [style=none] (53) at (-13.25, -5.75) { $F^{-1}$};
			\node [style=none] (54) at (-18.5, -9.5) {};
			\node [style=none] (55) at (-18.5, -11.25) {};
			\node [style=none] (56) at (-18, -11.75) {};
			\node [style=none] (57) at (-15.5, -12) {};
			\node [style=none] (58) at (-8.5, -13.25) {};
			\node [style=none] (59) at (-12.5, -14.5) { $R^{-1}$};
			\node [style=none] (60) at (-5, -9.5) {};
			\node [style=none] (61) at (-5, -11.5) {};
			\node [style=none] (62) at (-5.75, -12.75) {};
			\node [style=none] (65) at (-3.75, -12.75) {};
			\node [style=none] (66) at (-6, -15.5) {};
			\node [style=none] (67) at (-5.5, -14.75) {};
			\node [style=none] (68) at (-5, -15.25) {};
			\node [style=none] (69) at (-4.25, -15.5) {};
			\node [style=none] (70) at (-4, -16) { $a$};
			\node [style=none] (71) at (-2.25, -15.75) { $b$};
			\node [style=none] (72) at (-5, -11.5) {};
			\node [style=none] (73) at (-2.75, -15.5) {};
			\node [style=none] (74) at (-2, -13.5) {};
			\node [style=none] (75) at (3, -13.5) {};
			\node [style=none] (76) at (0.5, -15) { $F$};
			\node [style=none] (77) at (5, -9.5) {};
			\node [style=none] (78) at (5, -11.5) {};
			\node [style=none] (79) at (4.25, -12.75) {};
			\node [style=none] (80) at (5.25, -14) {};
			\node [style=none] (81) at (4, -15.5) {};
			\node [style=none] (82) at (4.5, -14.75) {};
			\node [style=none] (83) at (5, -15.25) {};
			\node [style=none] (84) at (5.75, -15.5) {};
			\node [style=none] (85) at (6, -16) { $a$};
			\node [style=none] (86) at (7.75, -15.75) { $b$};
			\node [style=none] (87) at (5, -11.5) {};
			\node [style=none] (88) at (7.25, -15.5) {};
			\node [style=none] (89) at (9, -13.25) {};
			\node [style=none] (90) at (16.25, -11.75) {};
			\node [style=none] (91) at (13, -13.75) { $R$};
			\node [style=none] (92) at (18.75, -7) {};
			\node [style=none] (93) at (18.75, -9) {};
			\node [style=none] (94) at (18, -10.25) {};
			\node [style=none] (96) at (17.75, -13) {};
			\node [style=none] (98) at (18, -10.25) {};
			\node [style=none] (99) at (19.5, -13) {};
			\node [style=none] (100) at (19.75, -13.5) { $a$};
			\node [style=none] (101) at (21.5, -13.25) { $b$};
			\node [style=none] (102) at (18.75, -9) {};
			\node [style=none] (103) at (21, -13) {};
			\node [style=none] (104) at (17, -8) {};
			\node [style=none] (105) at (9, -5.75) {};
			\node [style=none] (106) at (13.75, -5.25) { $F^{-1}$};
			\node [style=none] (107) at (5, -0.25) {};
			\node [style=none] (108) at (5, -2.25) {};
			\node [style=none] (109) at (4.25, -3.5) {};
			\node [style=none] (110) at (4, -6.25) {};
			\node [style=none] (111) at (5.75, -2.75) {};
			\node [style=none] (112) at (5.75, -6.25) {};
			\node [style=none] (113) at (6, -6.75) { $a$};
			\node [style=none] (114) at (7.75, -6.5) { $b$};
			\node [style=none] (115) at (5, -2.25) {};
			\node [style=none] (116) at (7.25, -6.25) {};
			\node [style=none] (117) at (7.5, 6.5) { $\color{red}(a,b)$};
			\node [style=none] (118) at (-7, 4) {};
			\node [style=none] (119) at (-6, 4) {};
			\node [style=none] (120) at (-6, 2) {};
			\node [style=none] (121) at (-7, 2) {};
			\node [style=none] (122) at (-5.5, 2) {};
			\node [style=none] (123) at (-6.5, 0.75) {};
			\node [style=none] (124) at (-7.5, 2) {};
			\node [style=none] (125) at (4.5, 1) {};
			\node [style=none] (126) at (5.5, 1) {};
			\node [style=none] (127) at (4.5, 3) {};
			\node [style=none] (128) at (5.5, 3) {};
			\node [style=none] (129) at (6, 3) {};
			\node [style=none] (130) at (5, 4.25) {};
			\node [style=none] (131) at (4, 3) {};
			\node [style=none] (132) at (8.5, 2.25) { Fold};
		\end{pgfonlayer}
		\begin{pgfonlayer}{edgelayer}
			\draw [thick, in=240, out=90, looseness=0.75] (2.center) to (1.center);
			\draw [thick, postaction={
				decorate,
				decoration={
					markings,
					mark=at position 0.25 with {
						\node[scale=1, rotate=0] at (0,0) {$\times$};
					}
			}}] (1.center) to (0.center);
			\draw [red, thick, in=135, out=-90, looseness=0.25] (2.center) to (3.center);
			\draw [thick, in=45, out=-60] (1.center) to (4.center);
			\draw [thick, in=45, out=-120] (5.center) to (6.center);
			\draw [thick, ->] (8.center) to (9.center);
			\draw [thick, postaction={
				decorate,
				decoration={
					markings,
					mark=at position 0.25 with {
						\node[scale=1, rotate=0] at (0,0) {$\times$};
					}
			}}] (12.center) to (11.center);
			\draw [red, thick, in=135, out=-90, looseness=0.25] (13.center) to (14.center);
			\draw [thick] (12.center) to (15.center);
			\draw [thick] (16.center) to (17.center);
			\draw [thick, in=240, out=90, looseness=0.75, postaction={
				decorate,
				decoration={
					markings,
					mark=at position 0.25 with {
						\node[scale=1, rotate=0] at (0,0) {$\times$};
					}
			}}] (28.center) to (27.center);
			\draw [thick, postaction={
				decorate,
				decoration={
					markings,
					mark=at position 0.25 with {
						\node[scale=1, rotate=0] at (0,0) {$\times$};
					}
			}}] (27.center) to (26.center);
			\draw [thick, in=45, out=-60] (27.center) to (30.center);
			\draw [thick, in=45, out=-120] (31.center) to (32.center);
			\draw [thick, in=150, out=-45] (28.center) to (29.center);
			\draw [thick, in=135, out=-90] (28.center) to (34.center);
			\draw [thick, in=165, out=-45, looseness=1.25] (35.center) to (36.center);
			\draw [thick, ->] (39.center) to (40.center);
			\draw [thick, in=240, out=90, looseness=0.75] (43.center) to (42.center);
			\draw [thick, postaction={
				decorate,
				decoration={
					markings,
					mark=at position 0.25 with {
						\node[scale=1, rotate=0] at (0,0) {$\times$};
					}
			}}] (42.center) to (41.center);
			\draw [thick, in=45, out=-60, postaction={
				decorate,
				decoration={
					markings,
					mark=at position 0.25 with {
						\node[scale=1, rotate=0] at (0,0) {$\times$};
					}
			}}] (42.center) to (45.center);
			\draw [thick, in=45, out=-120] (46.center) to (47.center);
			\draw [thick, in=135, out=-90] (43.center) to (48.center);
			\draw [thick, in=165, out=-45, looseness=1.25] (49.center) to (50.center);
			\draw [thick, in=135, out=-135] (54.center) to (55.center);
			\draw [thick] (55.center) to (56.center);
			\draw [thick, in=135, out=-45, looseness=1.50] (56.center) to (44.center);
			\draw [thick, ->] (57.center) to (58.center);
			\draw [thick, in=240, out=90, looseness=0.75] (62.center) to (61.center);
			\draw [thick, postaction={
				decorate,
				decoration={
					markings,
					mark=at position 0.25 with {
						\node[scale=1, rotate=0] at (0,0) {$\times$};
					}
			}}] (61.center) to (60.center);
			\draw [thick, in=0, out=-135, looseness=0.50] (65.center) to (66.center);
			\draw [thick, in=135, out=-90] (62.center) to (67.center);
			\draw [thick, in=165, out=-45, looseness=1.25] (68.center) to (69.center);
			\draw [thick, in=135, out=-30, looseness=1.50, postaction={
				decorate,
				decoration={
					markings,
					mark=at position 0.3 with {
						\node[scale=1, rotate=0] at (0,0) {$\times$};
					}
			}}] (72.center) to (73.center);
			\draw [thick, ->] (74.center) to (75.center);
			\draw [thick, in=240, out=90, looseness=0.75, postaction={
				decorate,
				decoration={
					markings,
					mark=at position 0.25 with {
						\node[scale=1, rotate=0] at (0,0) {$\times$};
					}
			}}] (79.center) to (78.center);
			\draw [thick, postaction={
				decorate,
				decoration={
					markings,
					mark=at position 0.25 with {
						\node[scale=1, rotate=0] at (0,0) {$\times$};
					}
			}}] (78.center) to (77.center);
			\draw [thick, in=0, out=-90, looseness=0.75] (80.center) to (81.center);
			\draw [thick, in=135, out=-90] (79.center) to (82.center);
			\draw [thick, in=165, out=-45, looseness=1.25] (83.center) to (84.center);
			\draw [thick, in=135, out=-30, looseness=1.50] (87.center) to (88.center);
			\draw [thick, in=-30, out=90] (80.center) to (79.center);
			\draw [thick, ->] (89.center) to (90.center);
			\draw [thick, in=240, out=90, looseness=0.75, postaction={
				decorate,
				decoration={
					markings,
					mark=at position 0.25 with {
						\node[scale=1, rotate=0] at (0,0) {$\times$};
					}
			}}] (94.center) to (93.center);
			\draw [thick, postaction={
				decorate,
				decoration={
					markings,
					mark=at position 0.25 with {
						\node[scale=1, rotate=0] at (0,0) {$\times$};
					}
			}}] (93.center) to (92.center);
			\draw [thick, in=165, out=-45, looseness=1.25] (98.center) to (99.center);
			\draw [thick, in=135, out=-30, looseness=1.50] (102.center) to (103.center);
			\draw [thick, in=-105, out=90] (96.center) to (98.center);
			\draw [thick, ->] (104.center) to (105.center);
			\draw [thick, in=240, out=90, looseness=0.75] (109.center) to (108.center);
			\draw [thick, postaction={
				decorate,
				decoration={
					markings,
					mark=at position 0.25 with {
						\node[scale=1, rotate=0] at (0,0) {$\times$};
					}
			}}] (108.center) to (107.center);
			\draw [thick, in=165, out=-105] (111.center) to (112.center);
			\draw [thick, in=135, out=-30, looseness=1.50, postaction={
				decorate,
				decoration={
					markings,
					mark=at position 0.15 with {
						\node[scale=1, rotate=0] at (0,0) {$\times$};
					}
			}}] (115.center) to (116.center);
			\draw [thick, in=-90, out=90, looseness=0.75] (110.center) to (109.center);
			\draw[thick, ->,decorate,decoration=snake]  (-6.5,4)--(-6.5,1);
			\draw[thick, ->,decorate,decoration=snake]  (5,1)--(5,4);
		\end{pgfonlayer}
	\end{tikzpicture}
	\caption{When $\mathcal{C}$ is  a modular tensor category, the $\Omega$-symbols for the Drinfeld center $Z(\mathcal{C})\cong \mathcal{C} \boxtimes \overline{\mathcal{C}}$ can be obtained in terms of the $F$- and $R$-symbols of $\mathcal{C}$ by using the folding trick. All arrows are taken to point upwards.}\label{fig:CMTC}
\end{figure}
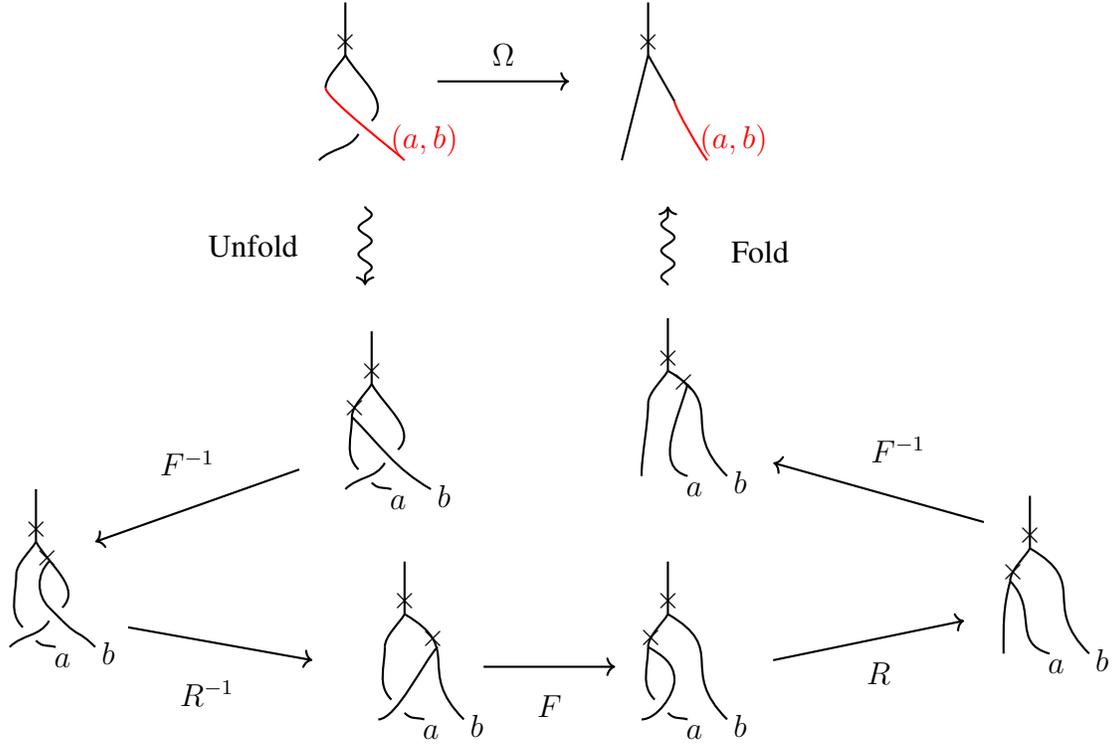

Finally, suppose that the fusion category $\mathcal{C}$ admits the structure of a modular tensor category.
In this case, the $\Omega$-symbols admit a closed-form expression in terms of the $F$- and $R$-symbols of $\mathcal{C}$, as we now describe.

When $\mathcal{C}$ is a modular tensor category, the Drinfeld center is given by $Z(\mathcal{C}) \cong \mathcal{C} \boxtimes \overline{\mathcal{C}}$, where $\overline{\mathcal{C}}$ is equivalent to $\mathcal{C}$ as a fusion category but has its braiding reversed \cite{MUGER2003159}.
Physically, this means that the SymTFT $\mathrm{TV}_\mathcal{C}$ is obtained from the 2+1d (Reshetikhin-Turaev) TQFT based on $\mathcal{C}$ by folding.
The simple topological lines in $Z(\mathcal{C})$ are denoted as $(a, b)$ for $a,b \in \mathrm{Irr}(\mathcal{C})$, and the forgetful functor \eqref{eq:forget} becomes
\begin{equation} \label{eq:MTC_ffunctor}
    F: Z(\mathcal{C}) \cong \mathcal{C} \boxtimes \overline{\mathcal{C}} \rightarrow \mathcal{C} \,, \quad (a,b) \mapsto a \otimes b \,.
\end{equation}
In particular, we have $\mathrm{Hom}_{\mathcal{C}}(F(a, b),c) = \mathrm{Hom}_{\mathcal{C}}(a \otimes b,c)$ and $\langle (a,b), c \rangle = N_{ab}^c$.
The conventions for the basis junction vectors in \eqref{eq:basis_bdy} and \eqref{eq:basis_bulk_to_bdy} are compatible with each other under these identifications.
The $\Omega$-symbols are obtained by applying the folding trick as shown in Figure \ref{fig:CMTC}.
Explicitly, we have
\begin{align}\label{eqn:modularhalfbraiding}
	\begin{split}
		&\left[ \Omega_{c(a,b)}^d \right]_{(exi)(fyj)}  \\
		&= \sum_{g,h \in \mathrm{Irr}(\mathcal{C})} \sum_{k=1}^{N_{ag}^d} \sum_{l=1}^{N_{bc}^g} \sum_{m=1}^{N_{cb}^g} \sum_{n=1}^{N_{ac}^h} \sum_{p=1}^{N_{hb}^d} \sum_{q=1}^{N_{ca}^h} \left[ F_{cab}^d \right]^{-1}_{(hqp)(fjy)}\left[ R_{ca}^h \right]_{nq}  \left[ F_{acb}^d \right]_{(gkm)(hnp)} \left[ R_{bc}^g \right]_{l m}^{-1} \left[ F_{abc}^d \right]^{-1}_{(exi)(gkl)}.
	\end{split}
\end{align}
Therefore, when $\CC$ is an MTC, the half braiding matrix elements, i.e.\ the $\Omega$-symbols, can be explicitly expressed in terms of $F$-symbols and $R$-symbols.
In particular, some of them are given by the $S$-matrix of $\mathcal{C}$, which we denote as $\mathbb{S}_{ab}$,
\begin{equation} \label{eq:Omega_S}
    \left[ \Omega_{a(b,\bar{b})}^a \right]_{11} = \frac{\mathbb{S}_{ab}}{\sqrt{S_{11}} \qd_a \qd_b}  \,, \quad \left[ \Omega_{a(b,\bar{b})}^a \right]_{11}^{-1} = \frac{\mathbb{S}_{ab}^*}{\sqrt{S_{11}} \qd_a \qd_b}  \,.
\end{equation}
Here, $S_{\mu\nu}$ is the $S$-matrix for the bulk topological lines of $Z(\mathcal{C})$, and in particular $
S_{11}^{-1} = \mathbb{S}_{11}^{-2} = \sum_{a \in \mathrm{Irr}(\mathcal{C})} \qd_a^2$.

\subsection{Generalized half-linking}\label{subsec:halflinking}

In Section \ref{subsec:halfbraiding}, we described half-braiding processes involving topological lines on the Dirichlet boundary and topological lines in the bulk $\mathrm{TV}_{\CC}$. 
Here, we discuss a more general braiding process involving topological lines in the bulk $\mathrm{TV}_{\CC}$, and  topological line interfaces between two potentially distinct gapped boundary conditions. 
The constructions we consider here generalize those found in \cite{Lin:2022dhv}, and we follow their methods closely.

\subsubsection{The $\Psi$-symbols}

The (simple) topological boundary conditions of $\mathrm{TV}_{\mathcal{C}}$  are in correspondence with (indecomposable) module categories $\mathcal{B}$ over $\mathcal{C}$ (see, e.g., \cite{Thorngren:2019iar}), or equivalently with Morita equivalence classes of algebra objects $A$ of $\mathcal{C}$. These two descriptions are related by the correspondence $\mathcal{B} \cong \mathcal{C}_{A}$ \cite{ostrik2003module}, as was briefly discussed in Section \ref{subsec:modulebimodule}.\footnote{There is a third description of topological boundary conditions of the SymTFT $\mathrm{TV}_{\mathcal{C}}$ in terms of Lagrangian algebra objects of $Z(\mathcal{C})$ \cite{Kong:2013aya,DavydovMügerNikshychOstrik+2013+135+177,Fuchs:2012dt}, see Section \ref{sec:symTFT}.}  
Throughout this paper, we label  topological boundary conditions of $\mathrm{TV}_\mathcal{C}$ by the corresponding module categories $\mathcal{B}$.

Consider two topological boundaries of the SymTFT described by indecomposable module categories $\mathcal{B}_1$ and $\mathcal{B}_2$ corresponding to algebra objects $A_1$ and $A_2$, respectively. 
Recall that the topological line interfaces between these boundaries are captured by the category $\mathrm{Fun}_{\mathcal{C}}(\mathcal{B}_1,\mathcal{B}_2)$ of $\mathcal{C}$-module functors from $\mathcal{B}_1$ to $\mathcal{B}_2$, or equivalently, by the  category ${_{A_1}}\mathcal{C}_{A_2}$ of $A_1$-$A_2$--bimodules in $\mathcal{C}$. 
We study the \emph{generalized half-linking numbers} defined by the following configurations of bulk, boundary, and boundary changing lines,
\begin{align}\label{eqn:halflinkingdefn}
	\begin{split}
		{^{\mathcal{B}_1\mathcal{B}_2}}\Psi_{\alpha\beta(\mu xy)}^{(az)(bw)} = \sqrt{\frac{S_{11}}{\qd_a \qd_b}}~\input{figures/psi.tikz} \, , \  {^{\mathcal{B}_1\mathcal{B}_2}}\widetilde{\Psi}_{\alpha\beta(\mu xy)}^{(az)(bw)} = \sqrt{\frac{S_{11}}{\qd_a \qd_b}}~\input{figures/psit.tikz} \,.
	\end{split}
\end{align}
Here, $\mu$ is a bulk topological line of $Z(\mathcal{C})$. Further, $a \in {_{A_1}}\mathcal{C}_{A_1} \cong \mathrm{Fun}_{\mathcal{C}}(\mathcal{B}_1,\mathcal{B}_1)^{\mathrm{op}}$ is a topological line on the $\mathcal{B}_1$ boundary, $b\in {_{A_2}}\mathcal{C}_{A_2} \cong \mathrm{Fun}_{\mathcal{C}}(\mathcal{B}_2,\mathcal{B}_2)^{\mathrm{op}}$ is a topological line on the $\mathcal{B}_2$ boundary, and  $\alpha,\beta \in {_{A_2}}\mathcal{C}_{A_1} \cong \mathrm{Fun}_{\mathcal{C}}(\mathcal{B}_2,\mathcal{B}_1)$ are topological line interfaces between the $\mathcal{B}_2$ and $\mathcal{B}_1$ boundaries ($\mathcal{B}_2$ to the left of $\alpha$ and $\mathcal{B}_1$ to the right). Finally,  $x,y,w,z$ are suitable topological point junctions as defined in earlier subsections.

The reason each such diagram defines a number is that the lines can be shrunk to a topological local operator on the $\mathcal{B}_1$ or $\mathcal{B}_2$ boundary, which must be proportional to the identity operator due to the simplicity of $\mathcal{B}_1$ and $\mathcal{B}_2$; the half-linking numbers ${^{\mathcal{B}_1\mathcal{B}_2}}\Psi_{\alpha\beta(\mu xy)}^{(az)(bw)}$ and ${^{\mathcal{B}_1\mathcal{B}_2}}\widetilde{\Psi}_{\alpha\beta(\mu xy)}^{(az)(bw)}$ are defined as these constants of proportionality. 
When $\mathcal{B}_1=\mathcal{B}_2=\mathcal{B}_{\mathrm{reg}}$, we suppress them from the notation.
We also suppress some of the junctions from the notations when there is a unique and canonical choice for them (e.g. an identity operator).

The reason that we name the configurations in \eqref{eqn:halflinkingdefn} generalized half-linking numbers is that they reduce to the half-linking numbers already defined in \cite{Lin:2022dhv}, by restricting to the special case $\mathcal{B}_1 \cong \mathcal{B}_2 \cong \mathcal{B}_{\mathrm{reg}}$ and $a = b = 1$.

\subsubsection{Relationship with $\Omega$-symbols when $\mathcal{B}_1 \cong \mathcal{B}_2 \cong \mathcal{B}_{\mathrm{reg}}$}

When both topological boundaries are taken to be the Dirichlet boundary condition corresponding to the regular module category $\mathcal{B}_1\cong\mathcal{B}_2\cong \mathcal{B}_{\mathrm{reg}}$ (but still allowing $\alpha$,  $\beta$, $a$, and $b$ to be generic), then the generalized half-linking numbers reduce to the half-braiding numbers given by the $\Omega$-symbols, up to some quantum dimensions.
Indeed, by first using \eqref{eq:half-braiding} (or \eqref{eq:inverse_half-braiding}) in the definition of the generalized half-linking numbers \eqref{eqn:halflinkingdefn}, and then shrinking the remaining topological lines to a point using the conventions laid out in \eqref{eq:basis_bdy} and \eqref{eq:basis_bulk_to_bdy} as well as \eqref{eq:AABB},
we obtain 
\begin{align} \label{eq:psi_DD}
\begin{split}
    \Psi_{\alpha\beta(\mu xy)}^{(az)(bw)} &= \sqrt{ \frac{S_{11} \qd_\mu \qd_\alpha \qd_\beta}{\qd_a \qd_b}} \left[ \Omega_{\alpha\mu}^\beta \right]_{(byw)(axz)}\,, \\
    \widetilde{\Psi}_{\alpha\beta(\mu xy)}^{(az)(bw)} &= \sqrt{ \frac{S_{11} \qd_\mu \qd_\alpha \qd_\beta}{\qd_a \qd_b} } \left[ \Omega_{\alpha\mu}^\beta \right]^{-1}_{(axz)(byw)} \,.
\end{split}
\end{align}
Hence, the generalized half-linking numbers \eqref{eqn:halflinkingdefn} may also be thought of as a generalization of the half-braiding ($\Omega$-symbols).\footnote{Alternatively, one can define a generalization of the $\Omega$-symbols, i.e.\ a generalized half-braiding matrix, by allowing $b$ and $a$ in \eqref{eq:half-braiding} to be topological interfaces between two distinct boundaries $\mathcal{B}_{1}$ and $\mathcal{B}_2$. The generalized half-linking numbers \eqref{eqn:halflinkingdefn} contain the same information as such generalized half-braiding numbers, so we do not bother introducing additional symbols.}

In the special case where $\mathcal{C}$ is itself a modular tensor category, combined with \eqref{eq:Omega_S}, we obtain
\begin{equation} \label{eq:psi_S}
    \Psi_{aa(b,\bar{b})}^{11} = \mathbb{S}_{ab} \,, \quad \widetilde{\Psi}_{aa(b,\bar{b})}^{11} = \mathbb{S}_{ab}^* \,.
\end{equation}
This can also be derived from ``unfolding'' the definition \eqref{eqn:halflinkingdefn} using $Z(\mathcal{C}) \cong \mathcal{C} \boxtimes \overline{\mathcal{C}}$.

Because of this relationship between generalized half-linking numbers and half-braiding numbers, it is clear that many of the properties enjoyed by the $\Omega$-symbols will have close analogs for the $\Psi$-symbols. Perhaps the most important property to establish, for the purposes of being able to actually compute the generalized half-linking numbers, is a suitable generalization of the hexagon equation \eqref{eq:hexagon}. In fact, it is clear that one can immediately write one down, once one understands the correct generalization of the $F$-symbols which govern recombination rules of line interfaces between different topological boundaries of $\TV_{\CC}$. We leave a detailed investigation of this question to future work.

\subsubsection{Orthogonality relations}

The generalized half-linking numbers obey orthogonality relations,
\begin{align}\label{eqn:halflinkingorthogonality}
\begin{split}
    \sum_{\alpha, \beta \in \mathrm{Irr}(\mathcal{I})} \sum_{z=1}^{(\widetilde{N}_{R})_{\alpha a}^{\beta}}  \sum_{w=1}^{(\widetilde{N}_L)_{b\alpha}^{\beta}}{^{\mathcal{B}_1\mathcal{B}_2}}\Psi_{\alpha\beta(\mu x y)}^{(az)(bw)} {^{\mathcal{B}_1\mathcal{B}_2}}\widetilde{\Psi}_{\alpha\beta(\mu' x' y')}^{(az)(bw)} &=  \delta_{\mu\mu'} \delta_{xx'} \delta_{yy'} \,, \\ 
   \sum_{\mu \in \mathrm{Irr}(Z(\CC))} \sum_{x=1}^{\braket{\mu, a}} \sum_{y=1}^{\braket{\mu, b}}{^{\mathcal{B}_1\mathcal{B}_2}}\Psi_{\alpha\beta(\mu x y)}^{(az)(bw)} {^{\mathcal{B}_1\mathcal{B}_2}}\widetilde{\Psi}_{\alpha' \beta' (\mu x y)}^{(az')(bw')} &= \delta_{\alpha\alpha'}\delta_{\beta\beta'}\delta_{zz'}\delta_{ww'} \,,
\end{split}
\end{align}
where $\mathcal{I}$ is the multiplet of topological interfaces between $\mathcal{B}_2$ and $\mathcal{B}_1$ on the left and right respectively.  That is, $\mathcal{I}$ is a ${_{A_2}}\mathcal{C}_{A_2}$-${_{A_1}}\mathcal{C}_{A_1}$-bimodule category.

We relegate the proof of the first relation in \eqref{eqn:halflinkingorthogonality} to Appendix \ref{app:half-linking}. The second relation follows from the first one by noting that the generalized half-linking matrices $[{^{\mathcal{B}_1\mathcal{B}_2}}\mathbf{\Psi}^{ab}]_{*, *}$ and $[{^{\mathcal{B}_1\mathcal{B}_2}}\widetilde{\mathbf{\Psi}}^{ab}]_{*, *}$ with elements
\begin{eqnarray}
    [{^{\mathcal{B}_1\mathcal{B}_2}}\mathbf{\Psi}^{ab}]_{(\alpha \beta z w), (\mu x y)}\equiv {^{\mathcal{B}_1\mathcal{B}_2}}\Psi_{\alpha\beta(\mu x y)}^{(az)(bw)}\,, \quad [{^{\mathcal{B}_1\mathcal{B}_2}}\widetilde{\mathbf{\Psi}}^{ab}]_{(\mu x y), (\alpha \beta z w)}\equiv {^{\mathcal{B}_1\mathcal{B}_2}}\widetilde{\Psi}_{\alpha\beta(\mu x y)}^{(az)(bw)}\,
\end{eqnarray}
are square matrices,
where  $(\alpha \beta z w)$ and $(\mu x y)$ are the two sets of indices of the matrices. In other words, the size of these two index sets are equal,\footnote{When $a=b=1$, the left hand side counts the topological local operators in a 1+1d TQFT obtained by shrinking the sandwich setup whose bulk is $\mathrm{TV}_{\CC}$, and whose two boundary conditions are $\mathcal{B}_1$ and $\mathcal{B}_2$. The right hand side counts the number of simple topological boundary conditions of this 1+1d TQFT. It is known that the number of topological local operators equals the number of simple boundary conditions in 1+1d TQFTs \cite{Huang:2021zvu}. }
\begin{eqnarray}
    \sum_{\mu\in \mathrm{Irr}(Z(\CC))} \braket{\mu,a} \braket{\mu, b} = \sum_{\alpha, \beta\in \mathrm{Irr}(\mathcal{I})} (\widetilde{N}_L)_{b\alpha}^\beta (\widetilde{N}_R)_{\alpha a}^\beta\,.
\end{eqnarray}
The orthogonality relations \eqref{eqn:halflinkingorthogonality} then amount to the fact that the two generalized half-linking matrices defined in \eqref{eqn:halflinkingdefn} are inverse transposes of each other, 
\begin{eqnarray}
    [{^{\mathcal{B}_1\mathcal{B}_2}}\widetilde{\mathbf{\Psi}}^{ab}]^T = [{^{\mathcal{B}_1\mathcal{B}_2}}{\mathbf{\Psi}}^{ab}]^{-1}\,.
\end{eqnarray}
In the special case that $\mathcal{B}_1 = \mathcal{B}_2 = \mathcal{B}_{\mathrm{reg}}$, the orthogonality relations translate into nontrivial identities for the $\Omega$-symbols due to the relation \eqref{eq:psi_DD}.

\subsubsection{A boundary crossing relation}

In later sections, we frequently utilize the following ``boundary crossing relation'' for the 2+1d SymTFT $\mathrm{TV}_\mathcal{C}$,
\begin{align}\label{eq:collapsetube}
    \input{figures/collapsetube1.tikz}~= \sqrt{S_{11}} \sum_{\mu x y}{^{\mathcal{B}_1\mathcal{B}_2}}\Psi_{\alpha\beta(\mu x y)}^{(az)(bw)}~\input{figures/collapsetube2.tikz}~.
\end{align}
On the left-hand side, we have $\mathrm{TV}_\mathcal{C}$ in the bulk, and a hollow tube in the middle stretches between the two topological boundary conditions $\mathcal{B}_1$ and $\mathcal{B}_2$.
On the boundaries and on the hollow tube, various topological line defects and interfaces are present.
We then ``collapse'' the hollow tube so that it becomes a sum of line operators of $\mathrm{TV}_\mathcal{C}$ terminating on junctions on the two boundaries.
Such a configuration can then be expanded as in the right-hand side, and in Appendix \ref{app:half-linking}, we show that the coefficients are given by the generalized half-linking numbers.

The inverse of \eqref{eq:collapsetube} can be obtained using the orthogonality relation \eqref{eqn:halflinkingorthogonality}:
\begin{align}\label{eq:collapse2}
    \input{figures/collapsetube2.tikz}~=\frac{1}{\sqrt{S_{11}}} \sum_{\alpha\beta zw}{^{\mathcal{B}_1\mathcal{B}_2}}\widetilde{\Psi}_{\alpha\beta(\mu x y)}^{(a z)(b w)}~\input{figures/collapsetube1.tikz}\,.
\end{align}
A special case of these boundary crossing relations was discussed in \cite{Lin:2022dhv}, when $\mathcal{B}_1 = \mathcal{B}_2 = \mathcal{B}_{\mathrm{reg}}$ are both Dirichlet boundary conditions and $a=b=1$.

\section{Generalized tube algebras}\label{sec:tube}

Let $Q$ be a 1+1d quantum field theory with a (not-necessarily faithfully acting) symmetry category $\mathcal{C}$. An important problem is to understand how $\mathcal{C}$ acts on operators of diverse dimensions, and further how these operators organize into multiplets. 

This problem has been addressed \cite{Lin:2022dhv} in the case of bulk local operators and, more generally, twisted sector operators (i.e.\ local operators which live at the endpoints of topological lines). Such operators are acted on by the tube algebra $\mathrm{Tube}(\CC)$, whose structure we review in Section \ref{subsec:ordinarytube}.\footnote{When the spacetime dimension is higher than 1+1d, there are natural generalizations where the bulk (twisted sector) operators can be of higher dimensions \cite{
Bullivant:2019fmk,
Bhardwaj:2023ayw,  Bhardwaj:2023wzd, 
Bartsch:2022ytj, 
Bartsch:2023wvv, 
Bartsch:2022mpm, 
Bartsch:2023pzl
}.}
In Section \ref{subsec:interfacetube}, we write down a generalization of $\mathrm{Tube}(\mathcal{C})$ which acts on local operators sitting at junctions of conformal interfaces and boundary conditions.

\subsection{Ordinary tube algebra}\label{subsec:ordinarytube}

We start by reviewing the standard tube algebra, which encodes how the topological lines of $\mathcal{C}$ act on local operators and twisted sector operators of $Q$, following the treatment of \cite{Lin:2022dhv}.

Consider a twisted sector operator $\CO(x)$ attached to a topological line $a$. Given a pair of topological lines  $c,d\in \CC$, and topological point junctions $y\in\mathrm{Hom}_{\mathcal{C}}(a\otimes c,d)$ and $z\in\mathrm{Hom}_{\CC}(d,c \otimes b)$, one can obtain a \emph{lasso action}\footnote{In the literature, the two trivalent junctions $\bar{z},y$ are often combined into a single 4-way junction. 
We use the resolved junctions and keep track of the $\times$ marks. }
\begin{align} \label{eq:lasso}
    \mathsf{L}^{b,dyz}_{a,c}(\CO)
    = \input{figures/tubelassoresolved.tikz} \,
\end{align}
that maps  $\CO(x)$  to another twisted sector operator which is attached to a topological line $b$ which is generically different from $a$ \cite{Chang:2018iay}. The action involves shrinking the $c$ and $d$ lines and fusing the topological junctions $y, \bar{z}$ downwards onto $\CO$.  We call the operator implementing this lasso action a \emph{lasso operator}, and denote it as $\mathsf{L}^{b,dyz}_{a,c}$. By definition, $\mathsf{L}_{a,c}^{b,dyz}$ annihilates a twisted sector operator attached to a topological line $a'$ if $a'\neq a$.

\begin{figure}
\begin{center}
\begin{tikzpicture}
        \draw[thick, fill=black] (0+1, 0) circle (2pt) node[below] {$\mathcal{O}(x)$};
        \draw[thick, decoration = {markings, mark=at position -0.5 with {\arrow[scale=1]{stealth}}}, postaction=decorate] (0+1,0) to (0+1,2);
        \node[right] at (0+1,1) {$a$};
        \draw[thick, <->] (3,1) to (6,1);
        \node[above] at (4.5, 1) {state-operator};
        \node[below] at (4.5, 1) {correspondence};
        \node[right] at (7,1) {$\ket{\mathcal{O}} \in \mathcal{H}\Bigg($};
        \draw[thick] (9.88, 1) circle (25pt);
        \draw[thick, fill=black] (9, 1) circle (2pt) node[right] {$a$};
        \node[] at (10.95,1) {$\Bigg)$};
\end{tikzpicture}
\caption{State-operator correspondence: A local operator $\mathcal{O}(x)$ at the end of a topological line $a\in \CC$, is conformally equivalent to a state $|\mathcal{O}\rangle$ in the $S^1$ Hilbert space, decorated by an insertion of the topological line $a\in \mathcal{C}$ which we denote as $\CH_a$.
}\label{fig:twistedsectorop}
\end{center}
\end{figure}
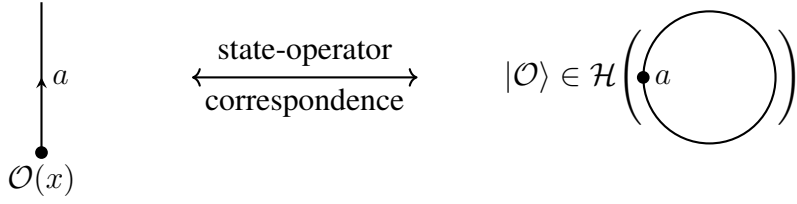

The lasso $\mathsf{L}_{a,c}^{b,dyz}$ can also be thought of as mapping states in the $a$-twisted $S^1$ Hilbert space (see the right of Figure \ref{fig:twistedsectorop}) to states in the $b$-twisted $S^1$ Hilbert space. We represent this operation diagrammatically as follows,
\begin{align}
    \mathsf{L}^{b,dyz}_{a,c}=
    ~\raisebox{-.35in}{\begin{tikzpicture}
    {
        \draw[thick, decoration = {markings, mark=at position 0.55 with {\arrow[scale=1.5]{stealth}}}, decoration = {markings, mark=at position 0.2 with {\arrow[scale=1.5]{stealth}},mark=at position 0.85 with {\arrow[scale=1.5]{stealth}}}, postaction=decorate] (0,0) to (0,2);
        \draw[thick, decoration = {markings, mark=at position 0.5 with {\arrow[scale=1.5]{stealth[reversed]}}}, postaction=decorate] (0,0.7) to (1, 1);
        \draw[thick, decoration = {markings, mark=at position 0.5 with {\arrow[scale=1.5]{stealth[reversed]}}}, postaction=decorate] (-1,1) to (0, 1.3);
        \node[] at (0,1.2) {$\times$};
        \node[] at (0,0.8) {$\times$};
        \node[below] at (-0.5, 1.6) {$c$};
        \node[right] at (0,0.3) {$a$};
        \node[left] at (0,1.05) {$d$};
        \node[right] at (0,1.75) {$b$};
        \node[] at (-0.2,0.7) {$y$};
        \node[] at (0.2,1.3) {$\bar{z}$};
        \node[] at (-1,1) {$\doubleslash$};
        \node[] at (1,1) {$\doubleslash$};
        }
\end{tikzpicture}}
\end{align}
where $\doubleslash$ means that the  points are periodically identified. We adopt the same symbol $\mathsf{L}^{b,dyz}_{a,c}$ to describe this operator which acts on states as opposed to on twisted sector operators;  again, $\mathsf{L}_{a,c}^{b,dyz}$ sends a state in $\CH_a$ to a state in $\CH_b$ and, by definition, annihilates all other states. In theories with a state/operator correspondence, a twisted sector operator $\CO$ at the end of a topological line $a\in \CC$ is equivalent to a state $\ket{\CO}$ in the $S^1$ Hilbert space $\CH_a$ decorated by an insertion of the same topological line $a\in \CC$ along the time direction, and the action of lassos on states is compatible with its action on operators.

The lasso operators can be composed, and the algebra formed by this composition is the \emph{tube algebra}, denoted as $\mathrm{Tube}(\CC)$. One notable feature is that, since lassos map between different twisted sectors, the algebra closes only on the direct sum of these twisted Hilbert spaces over all the topological lines in $\mathcal{C}$. In other words, the tube algebra acts on the \emph{extended Hilbert space},
\begin{align}\label{eq:extendedH}
    \mathcal{H}_{\mathcal{C}} \equiv \bigoplus_{a\in\mathrm{Irr}(\mathcal{C})}\mathcal{H}_a 
\end{align}
where $\mathrm{Irr}(\mathcal{C})$ is the set of simple lines in $\mathcal{C}$.

To find the algebra between the lasso operators explicitly, we consider the product $\mathsf{L}^{b',d'y'z'}_{a',c'} \times \mathsf{L}^{b,dyz}_{a,c}$. Graphically, we concatenate the legs labeled by $b$ and $a'$. Since both of them are simple objects in $\CC$, they can be concatenated (i.e.\ admit a non-trivial morphism between $b$ and $a'$) only when $b=a'$, as follows
\begin{eqnarray}\label{eq:LLlasso}
    \mathsf{L}^{b',d'y'z'}_{a',c'} \times \mathsf{L}^{b,dyz}_{a,c} = \delta_{a' b}\input{figures/tubemult1.tikz}\, .
\end{eqnarray}
Hence $\mathsf{L}^{b',d'y'z'}_{a',c'} \times \mathsf{L}^{b,dyz}_{a,c}$ is proportional to $\delta_{a' b}$. The explicit evaluation of the product requires using the $F$-symbols \eqref{eq:F_symbols}, completeness relations \eqref{eq:basis_bdy}, and junction isomorphisms \eqref{eq:junctionchangerelations} of $\mathcal{C}$. We relegate the derivation to  Appendix \ref{app:derivingtubealgebra}, and quote here only the final result:
\begin{align}\label{eq:tubealgebraLL}
\begin{split}
   &\mathsf{L}^{b',d'y'z'}_{a',c'} \times \mathsf{L}^{b,dyz}_{a,c} 
    =\delta_{a' b}\sum_{c'',d''\in\mathrm{Irr}(\mathcal{C})}\sum_{w=1}^{N_{cb}^d} \sum_{w'=1}^{N_{c'b'}^{d'}}\sum_{y''=1}^{N_{a c''}^{d''}}\sum_{w''=1}^{N_{\bar{c}''d''}^{b'}}\sum_{k=1}^{N_{\bar{c}d''}^{d'}}\sum_{l=1}^{N_{d c'}^{d''}}\sum_{i=1}^{N_{cc'}^{c''}} \sum_{\bar{z}'' =1}^{N_{c'' b'}^{d''}}\sqrt{\frac{\d_{c}\d_{c'}}{\d_{c''}}} \\ &[A_d^{cb}]_{\bar{z}w} [A_{d'}^{c'b'}]_{\bar{z}'w'} [A_{\bar{c}'' d''}^{b'}]_{w'' \bar{z}''}
     [C^{\bar c''}_{\bar c' \bar c}]_{j \bar \imath }[F_{\bar{c}d c'}^{d'}]^{-1}_{( b wy')(d''kl)} [F_{a c  c'}^{d''}]^{-1}_{(dyl)( c'' y''i)}[F_{\bar{c}'\bar{c}d''}^{b'}]_{(dw'k)(\bar{c}''iw'')}\mathsf{L}^{b',d''y''z''}_{a,c''} \,,
\end{split}
\end{align}
where the matrix $C^{\bar c''}_{\bar c \bar c'}$ is constructed out of $A$ and $B$ matrix symbols in \eqref{eq:junctionchangerelations} via $C^{\bar c''}_{\bar c \bar c'} = A_{\bar{c} \bar{c}'}^{\bar{c}''} \cdot B^{c \bar{c}''}_{\bar{c}'}\cdot A_{\bar{c}' c''}^{c}$.

\subsection{Incorporating boundaries and interfaces}\label{subsec:interfacetube}

We would like to generalize the ordinary tube algebra to settings involving boundaries and interfaces. More specifically, we aim to understand how symmetry lines in $\CC$ act on  local junction operators $\CO(x)$ sitting at the intersection of a collection of generically non-topological interfaces $\{I_i\}$, where $I_i$ interpolates between QFTs $Q_i$ and $Q_{i+1}$, with $Q_{n+1}=Q_1$. We assume the interfaces $I_i$ are simple in the sense that the only topological point operator they support is the identity, see Section \ref{sec:review}. 
We take the interfaces to be placed on straight lines. Since the lines are non-topological, the junction operator $\CO(x)$ generally depends on the angles $\theta_i$ at which $I_i$ and $I_{i+1}$ meet, however we suppress this data from our notation.\footnote{The angle dependence of the cusp between conformal defects has been studied recently in \cite{Cuomo:2024psk, Diatlyk:2024zkk}. }

\subsubsection{Defining the generalized tube algebra}

Suppose the theory $Q_i$ has a symmetry category $\CC_i$. A simple interface $I_i$ between $Q_i$ and $Q_{i+1}$ is a simple object of a $ (\mathcal{C}_i,\mathcal{C}_{i+1})$-bimodule category $\mathcal{I}_i$.
As explained in Section \ref{subsec:modulebimodule}, $\mathcal{I}_i$ can be thought of as the multiplet generated by acting on $I_i$ via parallel fusion by lines in $\mathcal{C}_i$ from one side, and lines in $\mathcal{C}_{i+1}$ from the other. The symmetry lines in $\CC_i$ act on  local junction operators via the following \emph{generalized lasso action},
\begin{align}\label{eq:generalizedlassoaction}
    \mathsf{H}_{\{I_i\}, \{c_i\}}^{\{J_i\}, \{K_i y_i z_i\}} (\CO) =\input{figures/junctionlasso.tikz}\, .
\end{align} 
This lasso operator maps a local junction operator $\CO$ attached to a collection of interfaces $\{I_i\}$, to another local junction operator attached to a generically different set of interfaces $\{J_i\}$. As with the ordinary tube algebra, this is achieved by shrinking the lines $\{c_i\}$ towards the junction. We denote the operator implementing the generalized lasso action by  $\mathsf{H}_{\{I_i\}, \{c_i\}}^{\{J_i\}, \{K_i y_i z_i\}}$.

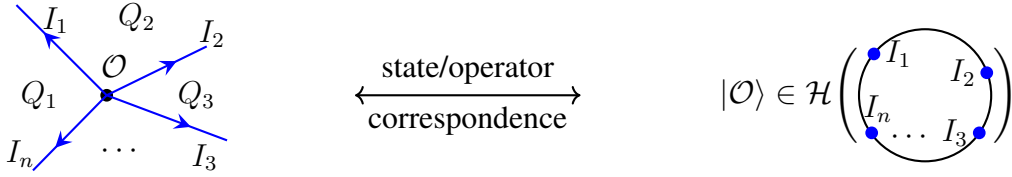
\begin{figure}
\begin{center}
\begin{tikzpicture}
	\def \X {1.1};
	\draw[thick, fill=black] (-1, 1) circle (2pt);
	\node[above] at (-.9, 1.1) {$\mathcal{O}$};
	\draw[decoration = {markings, mark=at position .7 with {\arrow[scale=1.5]{stealth}}}, postaction=decorate][thick,blue] (-1,1) to (-2*\X,2*\X);
	\draw[decoration = {markings, mark=at position .7 with {\arrow[scale=1.5]{stealth}}}, postaction=decorate] [thick,blue] (-1,1) to (0.3*\X,1.5*\X);
	\draw[decoration = {markings, mark=at position .7 with {\arrow[scale=1.5]{stealth}}}, postaction=decorate] [thick,blue] (-1,1) to (0.6,0.4);
	\draw[decoration = {markings, mark=at position .7 with {\arrow[scale=1.5]{stealth}}}, postaction=decorate] [thick,blue] (-1,1) to (-1.8*\X,0);
	\node[right] at (-2,2) {$I_1$};
	\node[above] at (0.4,1.5) {$I_2$};
	\node[below] at (0.3,0.5) {$I_3$};
	\node[above] at (-2.15, -.1) {$I_n$};
	\node[left] at (-1.5,1) {$Q_1$};
	\node[below] at (-0.8,0.5) {$\cdots$};
	\node[right] at (-0.2,1) {$Q_3$};
	\node[above] at (-0.6, 1.7) {$Q_2$};
	\draw[thick, <->] (3+0.3-1,1) to (6+0.3-1,1);
	\node[above] at (4.5+0.3-1, 1) {state/operator};
	\node[below] at (4.5+0.3-1, 1) {correspondence};
	\node[right] at (7,1) {$\ket{\mathcal{O}} \in \mathcal{H}\Bigg($};
	\draw[thick] (9.88, 1) circle (25pt);
	\draw[thick, blue,fill=blue] (9.17, 0.5) circle (2pt);
	\draw[thick,blue, fill=blue] (10.6, 0.5) circle (2pt);
	\draw[thick, blue,fill=blue] (10.7, 1.3) circle (2pt);
	\draw[thick, blue,fill=blue] (9.2, 1.55) circle (2pt);
	\node[above] at (9.25, 0.5) {$I_n$};
	\node[left] at (10.6, 0.5) {$I_3$};
	\node[left] at (10.7, 1.3) {$I_2$};
	\node[right] at (9.2, 1.55) {$I_1$};
	\node[above] at (9.7, 0.2) {$\cdots$};
	\node[] at (10.95,1) {$\Bigg)$};
\end{tikzpicture}
\caption{State-operator correspondence of a junction operator: A local  operator $\mathcal{O}(x)$ at the junction of interfaces  $\{I_i\}$. By the state/operator correspondence, $\mathcal{O}(x)$ may be viewed as a state $|\mathcal{O}\rangle$ in the $S^1$ Hilbert space, decorated by point-like insertions of the interfaces $I_1,\cdots, I_n$. }\label{fig:interfacejunction}
\end{center}
\end{figure}

In theories with a state/operator correspondence, a local junction operator $\CO$ at the common intersection point of interfaces $\{I_i\}$ is equivalent to a state $\ket{\CO}$ in the $S^1$ Hilbert space  $\CH_{\{I_i\}}$ decorated by point-like insertions of the $I_i$. See  Figure \ref{fig:interfacejunction}. By definition, the generalized lasso operator $\mathsf{H}_{\{I_i\}, \{c_i\}}^{\{J_i\}, \{K_i y_i z_i\}}$  maps the Hilbert space $\CH_{\{I_i\}}$ into the Hilbert space $\CH_{\{J_i\}}$, and annihilates states in other Hilbert spaces.

The generalized lasso operators can also be composed, and the algebra formed by this composition is the \emph{generalized tube algebra}, which we denote as $\mathrm{Tube}(\mathcal{I}_1| ... | \mathcal{I}_n)$. The explicit expression for this multiplication can also be worked out explicitly, in terms of the $\widetilde{F}_L$-, $\widetilde{F}_R$-, and $\widetilde{F}_M$-symbols of the bimodule categories $\mathcal{I}_i$, but we will not present the explicit formula here.  

As in the case of the ordinary tube algebra, 
the generalized tube algebra closes only on the direct sum of Hilbert spaces decorated by all possible combinations of interfaces $\{I_i\}$ in the bimodule categories $\{\mathcal{I}_i\}$. In other words, the generalized tube algebra acts on the \emph{extended Hilbert space}
\begin{align}\label{eq:extendedIntervalHilbert}
    \mathcal{H}_{\mathcal{I}_1\cdots \mathcal{I}_n} \equiv \bigoplus_{I_1\in\mathrm{Irr}(\mathcal{I}_1)}\cdots\bigoplus_{I_n\in\mathrm{Irr}(\mathcal{I}_n)}\mathcal{H}_{I_1 ... I_n}.
\end{align}

In the next section, we will develop a generalized SymTFT which establishes the representation theory of the tube algebras $\mathrm{Tube}(\mathcal{I}_1\vert\cdots\vert\mathcal{I}_n)$. We note that  $\mathrm{Tube}(\mathcal{I}_1\vert\cdots\vert\mathcal{I}_n)$ does not admit the structure of a Hopf algebra for arbitrary choices of the bimodule categories $\mathcal{I}_i$; correspondingly, the representation category $\mathrm{Rep}(\mathrm{Tube}(\mathcal{I}_1\vert\cdots\vert\mathcal{I}_n))$ will not be described by a fusion category in general. Nonetheless, as we will touch on briefly in the context of boundary tube algebras below, we expect that the Hopf structure is not completely gone, but rather replaced by something more general.

We end this section by unpacking a few special cases of this construction. One  is when the local junction operator is actually a twisted sector operator, in which case the generalized tube algebra reduces to the ordinary tube algebra discussed in Section \ref{subsec:ordinarytube}. Another case is when the local junction operator is a boundary changing operator between two boundary conditions, in which case the generalized tube algebra reduces to the \emph{boundary tube algebra}.

\subsubsection{Special case: ordinary tube algebras}

Topological line operators in a symmetry category $\mathcal{C}$ may be thought of as interfaces between a theory and itself. They transform in the \emph{regular} multiplet $\mathcal{I}_{\mathrm{reg}}$, which as a plain category is equivalent to $\mathcal{C}$. Thus, taking $n=1$ and $\mathcal{I}_1= \mathcal{I}_{\mathrm{reg}}$, the junction Hilbert spaces $\mathcal{H}_{I_1}$ become the twisted Hilbert spaces $\mathcal{H}_a$ in \eqref{eq:extendedH}, and the tube algebra $\mathrm{Tube}(\mathcal{I}_1)$ reduces to the standard tube algebra $\mathrm{Tube}(\mathcal{C})$ generated by  \eqref{eq:lasso}. It is in this sense that our construction can be thought of as a generalized tube algebra.

\subsubsection{Special case: boundary tube algebras}

Another interesting special case which will be important in subsequent sections is the case of boundary conditions, which may be thought of as interfaces between $Q_1$ the trivially gapped theory with trivial symmetry category $\mathcal{C}_1=\mathrm{Vec}$, and $Q_2=Q$ a quantum field theory  with symmetry category $\mathcal{C}_2\equiv\mathcal{C}$.

Let $\mathcal{B}_1$ and $\mathcal{B}_2$ be two left $\CC$-module categories, describing $\CC$-multiplets of boundary conditions.
We take $\mathcal{I}_1=\mathcal{B}_1^\vee$, thought of as a  $(\mathrm{Vec},\mathcal{C})$-bimodule category,  and $\mathcal{I}_2=\mathcal{B}_2$, thought of as a $(\mathcal{C},\mathrm{Vec})$-bimodule category, 
and study the space of local boundary-changing operators between $B_1\in\mathcal{B}_1^\vee$ and $B_2\in\mathcal{B}_2$, as well as the extended Hilbert space of boundary-changing  operators
\begin{align}
    \mathcal{H}_{\mathcal{B}_1^\vee\mathcal{B}_2} = \bigoplus_{B_1\in\mathcal{B}_1^\vee}\bigoplus_{B_2\in\mathcal{B}_2}\mathcal{H}_{B_1B_2}.
\end{align} 
In this case, we obtain a kind of \emph{boundary tube algebra} $\mathrm{Tube}(\mathcal{B}_1^\vee\vert \mathcal{B}_2)$, generated by \emph{boundary lasso operators} $\mathsf{H}^{C_1 C_2, y_1 y_2}_{B_1B_2, c}$ of the form,
\begin{align}\label{eq:Hlassoop}
	\begin{tikzpicture}
			\fill [gray, opacity=0.5] (0,-1.5) rectangle (1,1.5); 
			\node[left] at (-1.5, 0) {$\mathsf{H}^{C_1 C_2, y_1 y_2}_{B_1B_2, a}= $};
			\draw[thick, fill=black] (0, 0) circle (2pt);
			\draw[thick, decoration = {markings, mark=at position 0.75 with {\arrow[scale=1.5]{stealth}}}, postaction=decorate]  (0,1) -- (0,-1) arc(270:90:1) --cycle;
			\node[left] at (0, 0) {{$\mathcal{O}$}};
			\draw[thick,blue] (0, 1.5) -- (0,-1.5);
			\node[right] at (0, 1.5) {$C_2$};
			\node[right] at (0,-1.5) {$C_1$};
			\node[right] at (0, 0.5) {$B_2$};
			\node[right] at (0,-0.5) {$B_1$};
			\node[right] at (-0, 1) {$y_2$};
			\node[right] at (-0, -1) {$\bar{y}_1$};
			\node[] at (-0.8, -0.1) {$a$};
	\end{tikzpicture}
	~
\end{align}
where $a$ is a topological line operator in $\mathcal{C}$, and $y_2$ is a topological junction operator between boundaries $B_2,C_2\in\mathcal{B}_2$ on which $a$ ends (and similarly for $\bar{y}_1$). Note that we have changed conventions slightly compared to Equation \eqref{eq:generalizedlassoaction}.

We may also think of the boundary lasso operators as acting on the (extended) interval Hilbert space of a theory, and we diagrammatically represent this action as
\begin{align} \label{eq:Hlassostate}
	\mathsf{H}^{C_1 C_2, y_1 y_2}_{B_1B_2, a} = \quad \raisebox{-2.5em}
	{\begin{tikzpicture}
			\fill [gray, opacity=0.5] (0,-1) rectangle (-0.6,1); 
			\fill [gray, opacity=0.5] (2,-1) rectangle (2.6,1); 
			\draw[thick, decoration = {markings, mark=at position 0.5 with {\arrow[scale=1.5]{stealth}}}, postaction=decorate] (0,0) -- (2,0);
			\draw[thick,blue] (0,1) -- (0,-1);
			\draw[thick,blue] (2,1) -- (2,-1);
			\node[left] at (0,0.8) {$C_1$};
			\node[right] at (2,0.8) {$C_2$};
			\node[left] at (0,-0.8) {$B_1$};
			\node[right] at (2,-0.8) {$B_2$};
			\node[above] at (1,0) {$a$};
			\node[left] at (0,0) {$\bar{y}_1$};
			\node[right] at (2,0) {$y_2$};
	\end{tikzpicture}}\,.
\end{align}
In theories with a state/operator correspondence, the action of boundary lassos on boundary changing local operators is compatible with their action on interval states.

As with the standard tube algebra, the multiplication on $\mathrm{Tube}(\mathcal{B}_1^\vee\vert\mathcal{B}_2)$ can be determined from the structure of $\mathcal{B}_1$ and $\mathcal{B}_2$ as  $\mathcal{C}$-module categories. The multiplication of two boundary lasso operators is
\begin{equation}\label{eq:HHH}
\begin{split}
    &\mathsf{H}^{C'_1 C'_2, y'_1 y'_2}_{B'_1B'_2, a'} \times \mathsf{H}^{C_1 C_2, y_1 y_2}_{B_1B_2, a}
    =\delta_{B_1'C_1}\delta_{B_2'C_2}\sum_{a''\in\mathrm{Irr}(\CC)} \sum_{x=1}^{N_{a'a}^{a''}}\sum_{y_1''=1}^{}\sum_{y_2''=1}^{} \sqrt{\frac{\d_a\d_{a'}}{\d_{a''}}}  \\
     &\hspace{1in} [{^{\mathcal{B}_1}}\widetilde{F}_{\bar a'\bar a C_1'}^{B_1}]_{(C_1 y_1y_1')(\bar a'' xk_1)} [{^{\mathcal{B}_2}}\widetilde{F}_{a'aB_2}^{C_2'}]_{(C_2y_2'y_2)(a''xk_2)}\mathsf{H}_{B_1B_2,a''}^{C_1'C_2',y_1''y_2''}\, ,
\end{split}
\end{equation}
where here, we are using the $\widetilde{F}$-symbols of $\mathcal{B}_1$ and $\mathcal{B}_2$ thought of as \emph{left} $\CC$-module categories.

This algebra was studied in the context of lattice models for 2+1d gapped phases in \cite{2012CMaPh.313..351K} in the special case that $\mathcal{B}_1$ defines the same abstract left $\mathcal{C}$-module category as $\mathcal{B}_2$. (See also \cite{Barter_2022,Cordova:2024vsq,Copetti:2024onh,Cordova:2024iti,Copetti:2024dcz,Konechny:2024ixa,Inamura:2024jke}.) In this special case, they concluded that  $\mathrm{Tube}(\mathcal{B}_1^\vee\vert\mathcal{B}_2)$ naturally has the structure of a $C^\ast$ weak Hopf algebra. Thus, by Tannaka duality, we expect that $\mathrm{Rep}(\mathrm{Tube}(\mathcal{B}^\vee\vert\mathcal{B}))$ is a unitary fusion category; we will see that this expectation is borne out by the SymTFT. 

When $\mathcal{B}_1$ is not the same $\CC$-module category as $\mathcal{B}_2$, the boundary tube algebra is no longer Hopf, but it possesses some closely related structures. For example, the coalgebra structure of $\mathrm{Tube}(\mathcal{B}^\vee\vert\mathcal{B})$, 
\begin{align}\label{eqn:ordinarycoalgebra}
    \Delta: \mathrm{Tube}(\mathcal{B}^\vee\vert\mathcal{B}) \to \mathrm{Tube}(\mathcal{B}^\vee\vert\mathcal{B})\otimes \mathrm{Tube}(\mathcal{B}^\vee\vert\mathcal{B})
\end{align}
is replaced by a kind of ``collective'' coalgebra structure, 
\begin{align}\label{eqn:collectivecoalgebra}
    \Delta_{\mathcal{B}_1\mathcal{B}_2\mathcal{B}_3}:\mathrm{Tube}(\mathcal{B}_1^\vee\vert\mathcal{B}_3)\to \mathrm{Tube}(\mathcal{B}_1^\vee\vert\mathcal{B}_2)\otimes \mathrm{Tube}(\mathcal{B}_2^\vee\vert\mathcal{B}_3)
\end{align}
obtained by generalizing \cite[Equation 22]{2012CMaPh.313..351K} in the obvious way.
The consequence of this for the representation theory of the boundary tube algebras is that, while Equation \eqref{eqn:ordinarycoalgebra} allows one to take a tensor product of two representations of $\mathrm{Tube}(\mathcal{B}^\vee\vert\mathcal{B})$ to produce a third representation of $\mathrm{Tube}(\mathcal{B}^\vee\vert\mathcal{B})$, the collective coalgebra $\Delta_{\mathcal{B}_1\mathcal{B}_2\mathcal{B}_3}$ only allows one to take a tensor product of a representation of $\mathrm{Tube}(\mathcal{B}_1^\vee\vert\mathcal{B}_2)$ with one of $\mathrm{Tube}(\mathcal{B}_2^\vee\vert\mathcal{B}_3)$ to produce a representation of $\mathrm{Tube}(\mathcal{B}_1^\vee\vert\mathcal{B}_3)$. Of course, the existence of this structure is completely expected from the physics, simply because it is possible to take the OPE of boundary changing local operators in $\mathcal{H}_{\mathcal{B}_1^\vee\mathcal{B}_2}$ with those in $\mathcal{H}_{\mathcal{B}_2^\vee\mathcal{B}_3}$ to produce operators in $\mathcal{H}_{\mathcal{B}_1^\vee\mathcal{B}_3}$.

Similarly, the antipode of $\mathrm{Tube}(\mathcal{B}^\vee\vert\mathcal{B})$, 
\begin{align}
    S:\mathrm{Tube}(\mathcal{B}^\vee\vert\mathcal{B})\to \mathrm{Tube}(\mathcal{B}^\vee\vert\mathcal{B})
\end{align}
is replaced in general by a map
\begin{align}
    S:\mathrm{Tube}(\mathcal{B}_1^\vee\vert\mathcal{B}_2) \to \mathrm{Tube}(\mathcal{B}_2^\vee\vert\mathcal{B}_1)
\end{align}
again by generalizing \cite[Equation 24]{2012CMaPh.313..351K} in the obvious way. This means that the ``dual'' of a representation of $\mathrm{Tube}(\mathcal{B}_1^\vee\vert\mathcal{B}_2)$ is a representation of $\mathrm{Tube}(\mathcal{B}_2^\vee\vert\mathcal{B}_1)$.

We are also in a position to describe how the existence of a weakly symmetric boundary  is reflected in the structure of the boundary tube algebra. Recall \cite{Choi:2023xjw} that a boundary condition $B_{\mathrm{weak}}\in\mathcal{B}_1$ is said to be weakly symmetric with respect to $\mathcal{C}$ if every $a\in \mathcal{C}$ admits some topological junction on $B_{\mathrm{weak}}$. If there is a weakly symmetric boundary condition, one naturally obtains a subalgebra of $\mathrm{Tube}(\mathcal{B}_1^\vee\vert\mathcal{B}_2)$ which is generated by the subset of boundary lassos of the form 
\begin{align}
\mathsf{H}^{C_1 B_{\mathrm{weak}}, y_1 y_2}_{B_1B_{\mathrm{weak}}, a} = \quad \raisebox{-2.5em}
    {\begin{tikzpicture}
            \fill [gray, opacity=0.5] (0,-1) rectangle (-0.6,1); 
			\fill [gray, opacity=0.5] (2,-1) rectangle (2.6,1); 
    		\draw[thick, decoration = {markings, mark=at position 0.5 with {\arrow[scale=1.5]{stealth}}}, postaction=decorate] (0,0) -- (2,0);
    		\draw[blue,thick] (0,1) -- (0,-1);
    		\draw[blue,thick] (2,1) -- (2,-1);
    		\node[left] at (0,0.8) {$C_1$};
    		\node[right] at (2,0.8) {$B_{\mathrm{weak}}$};
    		\node[left] at (0,-0.8) {$B_1$};
    		\node[right] at (2,-0.8) {$B_{\mathrm{weak}}$};
    		\node[above] at (1,0) {$a$};
    		\node[left] at (0,0) {$\bar{y}_1$};
    		\node[right] at (2,0) {$y_2$};
    \end{tikzpicture}} \,.
\end{align} 
This subalgebra of $\mathrm{Tube}(\mathcal{B}_1^{\vee}\vert\mathcal{B}_2)$ naturally acts on the subspace 
\begin{align}
\mathcal{H}_{B_1 B_{\mathrm{weak}}}=\bigoplus_{B_1\in\mathrm{Irr}(\mathcal{B}_1^{\vee})}\mathcal{H}_{B_1B_{\mathrm{weak}}} \subset \mathcal{H}_{\mathcal{B}_1^{\vee},\mathcal{B}_2} \,.
\end{align}

Let us briefly mention some examples. First recall that, taking $\CC=\mathrm{Vec}_G$, the module categories of $\CC$ are labeled by pairs $(H,\psi)$ with $H<G$ a subgroup and $\psi\in H^2(H,\mathbb{C}^\times)$ a 2-cocycle. The module categories labeled by pairs with $H=G$ have only 1 simple object, and therefore describe symmetric boundary conditions. It turns out that, if $\mathcal{B}_1$ is taken to be the module category labeled by $(G,\psi_1)$, and $\mathcal{B}_2$ is the module category labeled by $(G,\psi_2)$, then the tube algebra is given by the twisted group algebra
\begin{align}
    \mathrm{Tube}(\mathcal{B}_1^\vee\vert\mathcal{B}_2) \cong \mathbb{C}^{\psi_1^{-1}\psi_2}[G].
\end{align}
In other words, if two $G$-symmetric boundaries transform in different $G$-multiplets, then their corresponding interval states (or boundary-changing local operators) transform projectively with respect to $G$, with 2-cocycle given by $\psi_1^{-1}\psi_2$. It is known that the tensor product of a representation of $\mathbb{C}^{\psi_1^{-1}\psi_2}[G]$ with a  representation of $\mathbb{C}^{\psi_2^{-1}\psi_3}[G]$ produces a  representation of $\mathbb{C}^{\psi_1^{-1}\psi_3}[G]$, and the dual of a representation of $\mathbb{C}^{\psi_1^{-1}\psi_2}[G]$ is a representation of $\mathbb{C}^{\psi_2^{-1}\psi_1}[G]$, which is consistent with the expectations established by our discussion on the ``collective'' coalgebra of boundary tube algebras.

As another example, recall that the only (indecomposable) module category of the Fibonacci unitary fusion category (described in more detail in Section \ref{sec:fib}) is the regular module category. Thus, simple boundary conditions in an irreducible $\mathrm{Fib}$-multiplet $\mathcal{B}_{\mathrm{reg}}$ can be labeled as $B_1$ and $B_W$, with $B_W$ guaranteed to be weakly symmetric, due to the fusion rule $W\otimes B_W\cong B_1\oplus B_W$. The subalgebra generated by the boundary lassos $\mathsf{H}_{B_WB_W,1}^{B_WB_W}$ and $\mathsf{H}_{B_WB_W,W}^{B_WB_W}$ was computed in \cite{Choi:2023xjw}, and takes the form 
\begin{align}
    \mathsf{H}_{B_WB_W,W}^{B_WB_W}\times \mathsf{H}_{B_WB_W,W}^{B_WB_W} = \mathsf{H}_{B_WB_W,1}^{B_WB_W}+\varphi^{-3/2}\mathsf{H}_{B_WB_W,W}^{B_WB_W}.
\end{align}
We refer to Section \ref{sec:fib} for further computations in $\mathrm{Tube}(\mathcal{B}_{\mathrm{reg}}^\vee\vert\mathcal{B}_{\mathrm{reg}})$.

Finally, we note that when $\mathcal{B}$ is a rank-1 module category of $\mathcal{C}$, the boundary tube algebra $\mathrm{Tube}(\mathcal{B}^\vee\vert\mathcal{B})$ admits a nice characterization in terms of Tannaka duality. Indeed, Tannaka duality asserts that every rank-1 $\mathcal{C}$-module category $\mathcal{B}$  defines a Hopf algebra $H_{\mathcal{B}}$ such that $\mathrm{Rep}(H_{\mathcal{B}})\cong \mathcal{C}$. We claim that 
\begin{align}
    \mathrm{Tube}(\mathcal{B}^\vee\vert\mathcal{B})\cong H_{\mathcal{B}}^\ast,
\end{align}
where $H_{\mathcal{B}}^\ast$ is the Hopf algebra dual to $H_{\mathcal{B}}$.

\section{The SymTFT of boundaries, interfaces, and junction operators}\label{sec:symTFT}

We are interested in understanding the representation theory of the generalized tube algebras defined in the previous section.  It is known \cite{Lin:2022dhv} that the irreducible representations (a.k.a.\ charges or multiplets) of the ordinary tube algebra, which acts on the local and twisted sector operators of a 1+1d QFT $Q$, are in one-to-one correspondence with simple anyonic excitations/topological line operators of the bulk SymTFT. A higher dimensional generalization of this fact was also recently discussed in \cite{Bhardwaj:2023ayw,  Bhardwaj:2023wzd, 
Bartsch:2022ytj, 
Bartsch:2023wvv, 
Bartsch:2022mpm, 
Bartsch:2023pzl}.
In this section, we develop an analogous three-dimensional SymTFT picture of boundaries, interfaces, and their junction operators to aid in the investigation of the representation theory of the generalized tube algebras $\mathrm{Tube}(\mathcal{I}_1\vert\cdots\vert\mathcal{I}_n)$.

\subsection{Basic setup of the SymTFT}\label{subsec:basicsetup}
\begin{figure}[t]
	\centering
	\raisebox{-73pt}{\begin{tikzpicture}
			\draw[dashed](2,0) -- (2,3) -- (3,4) -- (3,1) -- cycle;
			\draw (2,0) node[below]{$Q$};
   \draw (2.5,1.5) node[above]{$c$};
   \draw[preaction={draw=white,line width=6pt},decoration = {markings, mark=at position -0.45 with {\arrow[scale=1.5]{stealth}}}, postaction=decorate] [thick] (2,1.5) -- (3,2.5);
	\end{tikzpicture}}
	\quad $\leftrightharpoons$ ~
	\raisebox{-73pt}{\begin{tikzpicture}
			\draw (1.45,2.3) node[below]{\small  $\TV_\cC$};
			\draw[dashed] (0,0) -- (0,3) -- (1,4) -- (1,1) -- cycle;
			\draw (0,0) node[below]{$\mathcal{B}_{\mathrm{reg}}$};
			\draw[preaction={draw=white,line width=3pt}, dashed] (2,0) -- (2,3);
			\draw[dashed] (2,3) -- (3,4) -- (3,1) -- (2,0);
			\draw (2,0) node[below]{$\widetilde{Q}$};
   \draw[preaction={draw=white,line width=6pt},decoration = {markings, mark=at position -0.45 with {\arrow[scale=1.5]{stealth}}}, postaction=decorate] [thick] (0,1.5) -- (1,2.5);
   \draw  (.5,1.5) node[above] {$c$};
	\end{tikzpicture}}
	\caption{A 1+1d QFT $Q$ with symmetry $\CC$ is equivalent to a tuple $(\mathcal{B}_{\mathrm{reg}}, \mathrm{TV}_{\CC}, \widetilde{Q})$, where $\mathrm{TV}_{\CC}$ is the SymTFT and $\mathcal{B}_{\mathrm{reg}}$ and $\widetilde{Q}$ are its Dirichlet and physical boundary conditions, respectively. A symmetry line $c\in\mathcal{C}$ of $Q$ is supported on the Dirichlet boundary condition in the SymTFT picture.
	}
	\label{fig:sandwich}
\end{figure}
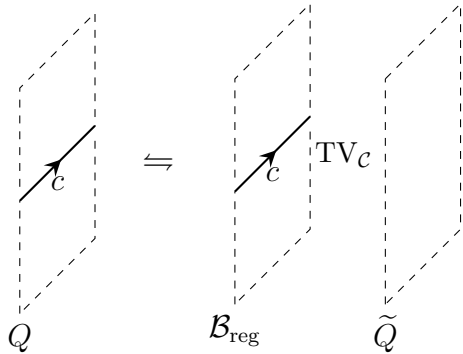

We begin by recalling the basic setup \cite{Gaiotto:2014kfa,Gaiotto:2020iye,Ji:2019jhk,Apruzzi:2021nmk,Freed:2022qnc, Kong:2015flk, Kong:2017hcw, Kong:2020cie, Kaidi:2022cpf, Antinucci:2022vyk}. The key idea is that any theory $Q$ with symmetry category $\CC$ can be equivalently represented by a triple
\begin{eqnarray}\label{eqn:basicSymTFT}
	Q \leftrightharpoons  (\mathcal{B}_{\mathrm{reg}}, \mathrm{TV}_{\mathcal{C}}, \widetilde{Q})
\end{eqnarray} 
as shown in Figure \ref{fig:sandwich}. 
This construction is colloquially known as the sandwich construction. 
We now explain the various ingredients arising in Equation \eqref{eqn:basicSymTFT}.

The SymTFT $\mathrm{TV}_{\CC}$ is the 2+1d Turaev-Viro TQFT \cite{Turaev:1992hq,Barrett:1993ab}, which can be informally thought of as a pure gauge theory for $\mathcal{C}$ \cite{Carqueville:2018sld,Kaidi:2021gbs}.\footnote{More precisely, in \cite{Kawagoe:2024tgv}, it has been put forward that the Levin-Wen string net model \cite{Levin:2004mi} based on a fusion category $\mathcal{C}$, which provides a Hamiltonian lattice formulation of the Turaev-Viro TQFT $\mathrm{TV}_{\CC}$, is a gauge theory where the gauge symmetry is the tube algebra $\mathrm{Tube}(\mathcal{C})$.}
Its anyonic excitations/topological line operators are described mathematically by the modular tensor category $Z(\mathcal{C})$, i.e.\ the Drinfeld center of $\mathcal{C}$, as discussed in Section \ref{subsec:halfbraiding}.

In the sandwich construction, the SymTFT $\mathrm{TV}_\CC$ is compactified on an interval with two boundary conditions imposed on the left and on the right, respectively. The left boundary is taken to be the ``Dirichlet" topological boundary. For reasons which are explained below, we label this boundary condition with the regular module category $\mathcal{B}_{\mathrm{reg}}$ of $\mathcal{C}$ (see \ref{subsec:halfbraiding} for a reminder on $\mathcal{C}$-module categories).

The dynamical degrees of freedom live on the right boundary of the SymTFT $\mathrm{TV}_{\mathcal{C}}$, which we denote $\widetilde{Q}$ and refer to as the \emph{physical boundary}.
This boundary condition has the property that the genuine local operators it hosts (i.e., its local operators which are not attached to any topological line) are the genuine local operators of $Q$ which are transparent to all the lines in $\mathcal{C}$.

Importantly, the original QFT $Q$ can be recovered by dimensionally reducing the sandwich construction along the interval direction down to 1+1d.
Conversely, any 1+1d QFT $Q$ with a $\mathcal{C}$ symmetry can be ``inflated' into a 2+1d picture as in Figure \ref{fig:sandwich}.

Such a 2+1d realization has the pleasant feature of separating out the universal kinematic aspects of the physics which follow from the $\mathcal{C}$ symmetry, from the aspects which are determined dynamically. Indeed, a reoccurring theme in this paper will be that the kinematic properties of $Q$ are associated with objects living in the bulk or near the topological boundary, while its dynamical properties are associated with physics near the physical boundary $\widetilde{Q}$. For example, in the SymTFT picture, the topological line defects implementing the $\mathcal{C}$ symmetry are supported on the topological boundary condition $\mathcal{B}_{\mathrm{reg}}$, as in Figure \ref{fig:sandwich}.

It is natural to ask what happens when $\mathcal{B}_{\mathrm{reg}}$ is replaced with some other topological boundary condition. Mathematically, topological boundary conditions of $\mathrm{TV}_{\mathcal{C}}$ are in one-to-one correspondence with any of the following three structures:
\begin{enumerate}[label=\arabic*)]
    \item Lagrangian algebras $L$ for $Z(\mathcal{C})$, up to isomorphism,
    \item module categories $\mathcal{B}$ for $\mathcal{C}$, up to $\mathcal{C}$-module equivalence, or
    \item gaugeable algebra objects $A$ of $\mathcal{C}$, up to Morita equivalence.
\end{enumerate}
For example, the Dirichlet boundary condition  in these three pictures corresponds to
\begin{enumerate}[label=\arabic*)]
\item the canonical Lagrangian algebra\footnote{See, e.g., Section 3.1 of \cite{bischoff2022computing}.} of $Z(\mathcal{C})$, 
\item the regular $\mathcal{C}$-module category $\mathcal{B}_{\mathrm{reg}}$, or 
\item the trivial algebra object $1\in\mathcal{C}$.
\end{enumerate}
In this paper, we mainly label topological boundary conditions of $\mathrm{TV}_{\mathcal{C}}$ using description 2), i.e.\ we label them by $\mathcal{C}$-module categories.

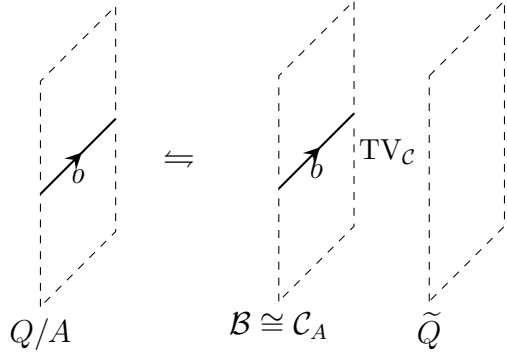
\begin{figure}[t]
	\centering
	\raisebox{-73pt}{\begin{tikzpicture}
			\draw[dashed](2,0) -- (2,3) -- (3,4) -- (3,1) -- cycle;
			\draw (2,0) node[below]{$Q/A$};
   \draw (2.5,1.5) node[above]{$b$};
   \draw[preaction={draw=white,line width=6pt},decoration = {markings, mark=at position -0.45 with {\arrow[scale=1.5]{stealth}}}, postaction=decorate] [thick] (2,1.5) -- (3,2.5);
	\end{tikzpicture}}
	\quad $\leftrightharpoons$ ~
	\raisebox{-73pt}{\begin{tikzpicture}
			\draw (1.45,2.3) node[below]{\small  $\TV_\cC$};
			\draw[dashed] (0,0) -- (0,3) -- (1,4) -- (1,1) -- cycle;
			\draw (0,0) node[below]{$\mathcal{B}\cong \CC_A$};
			\draw[preaction={draw=white,line width=3pt}, dashed] (2,0) -- (2,3);
			\draw[dashed] (2,3) -- (3,4) -- (3,1) -- (2,0);
			\draw (2,0) node[below]{$\widetilde{Q}$};
   \draw[preaction={draw=white,line width=6pt},decoration = {markings, mark=at position -0.45 with {\arrow[scale=1.5]{stealth}}}, postaction=decorate] [thick] (0,1.5) -- (1,2.5);
   \draw  (.5,1.5) node[above] {$b$};
	\end{tikzpicture}}
	\caption{The SymTFT picture for $Q/A$, the theory obtained by orbifolding the algebra object $A$ of $\mathcal{C}$. The Dirichlet boundary condition $\mathcal{B}_{\mathrm{reg}}$ is replaced by another topological boundary condition labeled by the left $\CC$-module category $\CB\cong \CC_A$. The dual symmetry $\mathcal{C}'$ of $Q/A$ is the category ${_A}\mathcal{C}_A$ of topological line operators supported on this topological boundary condition.
	}
	\label{fig:sandwichorbifold}
\end{figure}

On the other hand, it is known that the inequivalent ways of gauging a symmetry $\mathcal{C}$ in 1+1d are also labeled by algebra objects of $\mathcal{C}$, up to Morita equivalence. Using the equivalence of the descriptions  2) and 3) of topological boundaries, we deduce that swapping out the topological boundary condition described by the module category $\mathcal{B}_{\mathrm{reg}}$ for another module category $\mathcal{B} \cong \CC_A$ can be thought of as producing a SymTFT picture for the gauged theory $Q\big/A$. See Figure \ref{fig:sandwichorbifold}. 

The dual ``quantum'' symmetry $\mathcal{C}'$ of $Q\big/A$, which is guaranteed by the structure of the gauging procedure \cite{Bhardwaj:2017xup}, is exposed as the category of topological line operators supported on the topological boundary labeled by $\mathcal{B} \cong \CC_A$. This category $\mathcal{C}'$ can be described in three equivalent ways, corresponding to the three different pictures for topological boundary conditions of $\mathrm{TV}_{\mathcal{C}}$,
\begin{align}\label{eqn:dualsymmetrycategory}
    \mathcal{C}' = Z(\mathcal{C})_L \cong \mathrm{Fun}_{\mathcal{C}}(\mathcal{B},\mathcal{B})^{\mathrm{op}} \cong {_A}\mathcal{C}_A.
\end{align}
Here, $Z(\mathcal{C})_L$ is the category of right $L$-modules inside of $Z(\mathcal{C})$; $\mathrm{Fun}_{\mathcal{C}}(\mathcal{B},\mathcal{B})$ is the category of $\mathcal{C}$-module functors from $\mathcal{B}$ to itself; and ${_A}\mathcal{C}_A$ is the category of $A$-$A$-bimodules inside of $\mathcal{C}$.

For example, when $Q$ is a diagonal rational conformal field theory with chiral algebra $V$, and $\mathcal{C}$ is its category of Verlinde lines, then $\mathcal{C}$ is a modular tensor category, and it is known that the Drinfeld center is $Z(\mathcal{C})\cong \mathcal{C}\boxtimes \overline{\mathcal{C}}$. The physical boundary  $\widetilde{Q}$ is the boundary condition corresponding to the tensor product $V\otimes \overline{V}$ of the left- and right-moving chiral algebras; this comports with the fact that the Verlinde lines of a diagonal rational conformal field theory are precisely the topological lines of the theory which commute with $V\otimes \overline{V}$. By replacing the Dirichlet boundary $\mathcal{B}_{\mathrm{reg}}$ with another topological boundary $\mathcal{B}=\mathcal{C}_A$ labeled by a gaugeable algebra $A$ of $\mathcal{C}$, one obtains a rational conformal field theory corresponding to a different (potentially non-diagonal) modular invariant. This is in harmony with the results of \cite{Fuchs:2002cm}, where the authors showed that rational conformal field theories with chiral algebra $V$ are labeled by algebra objects of the category $\mathcal{C}\cong \mathrm{Rep}(V)$.

By ``unfolding'' this SymTFT picture along the topological boundary condition, the topological boundary of $Z(\mathcal{C})$ becomes a topological surface operator in $\mathcal{C}$, and one recovers the picture of Kapustin and Saulina \cite{Kapustin:2010if} which predates the SymTFT. In the case that the topological boundary condition is $\mathcal{B}_{\mathrm{reg}}$, upon unfolding, the surface operator of $\mathcal{C}$ which is obtained is the trivial surface.

As another example, when $Q$ is a generic CFT with a faithfully acting invertible symmetry category $\mathcal{C}=\mathrm{Vec}_G^\omega$, with $G$ a finite group and $\omega\in H^3(G,\mathbb{C}^\times)$ the anomaly, then $Z(\mathcal{C})=\mathcal{D}_\omega(G)$ is a Dijkgraaf--Witten theory \cite{Dijkgraaf:1989pz}. Its topological boundary conditions are known to correspond to pairs $(H,\psi)$ up to conjugation, where $H$ is a subgroup of $G$ on which the restriction of $\omega$ is trivial, and $\psi\in H^2(H,\mathbb{C}^\times)$. 
Replacing the Dirichlet boundary condition with the boundary described by $(H,\psi)$ corresponds to gauging the non-anomalous subgroup $H$ with discrete torsion given by $\psi$.

\subsection{Twisted sector operators}\label{subsec:twistedsectorlocalops}

Let us turn to the 2+1d description of twisted sector operators of $Q$, elaborating on the discussion in \cite{Lin:2022dhv}. 
In the SymTFT picture, an operator $\mathcal{O}$ in $\mathcal{H}_{a}$ can be decomposed into a triple 
\begin{align}\label{eqn:decomptwistedlocalop}
    \mathcal{O}\leftrightharpoons(x, \mu, \widetilde{\mathcal{O}}), \qquad x\in W^\mu_a, \quad \mu\in Z(\mathcal{C}), \quad \widetilde{\mathcal{O}}\in \mathcal{V}_\mu.
\end{align}
Here, $\mu\in Z(\mathcal{C})$ is an anyon of the bulk SymTFT, which stretches between the two boundary conditions $\mathcal{B}_{\mathrm{reg}}$ and $\widetilde{Q}$. On the physical boundary $\widetilde{Q}$, the line $\mu$ terminates on a (not necessarily topological) operator $\widetilde{\mathcal{O}}$, and we use $\mathcal{V}_\mu$ to denote the Hilbert space of such operators. On the Dirichlet boundary, $\mu$ ends on a topological junction operator $x$ which is connected to the line $a\in\mathcal{C}$ supported on the topological boundary $\mathcal{B}_{\mathrm{reg}}$, and we similarly use the notation $W_a^\mu$ to denote the (finite-dimensional) Hilbert space of such junction operators. See Figure \ref{fig:defect.operators}.

Mathematically, the space $W_a^\mu$ to which $x$ belongs is
\begin{align}\label{eqn:tuberepspaces}
    W^\mu_a\equiv \mathrm{Hom}_{\mathcal{C}}(F(\mu),a),
\end{align} where $F:Z(\mathcal{C})\to\mathcal{C}$ is the forgetful functor which describes the fate of a bulk anyon when it is pushed onto the Dirichlet topological boundary, as explained in Section \ref{subsec:halfbraiding}.\footnote{In particular, an operator $\mathcal{O}\in\mathcal{H}_a$ can only be decomposed into a triple involving a bulk anyon $\mu$ if $F(\mu)= a\oplus \cdots$.}  Physically, $W^\mu_a$ can be reinterpreted by a version of the state/operator correspondence as the Hilbert space of $\mathrm{TV}_{\mathcal{C}}$ on the disk $D^2$ with various decorations: $\mathcal{B}_{\mathrm{reg}}$ is imposed as the boundary condition of the disk, $\mu$ forms a point-like puncture at the origin of $D^2$, and $a$ pierces the boundary at a point. See the right of Figure \ref{fig:twistedsectorhilbertspaces}.

Similarly, if $Q$ is a conformal field theory, then by the state/operator correspondence, $\mathcal{V}_\mu$ can be thought of as the disk Hilbert space of $\mathrm{TV}_{\mathcal{C}}$ with $\mu$ puncturing the origin and with the physical boundary $\widetilde{Q}$ imposed on $\partial D^2$. See the left of Figure \ref{fig:twistedsectorhilbertspaces}. We emphasize that, in general, $\mathcal{V}_{1}$ is the space of $\mathcal{C}$-neutral untwisted local operators of $Q$. 

To return to our reoccurring example, suppose that $Q$ is a rational conformal field theory with chiral algebra $V$, and take  $\mathcal{C}$ to be the category of Verlinde lines. In this situation, $\mathcal{C}$ is modular, so the Drinfeld center is $Z(\mathcal{C})\cong \mathcal{C}\boxtimes \overline{\mathcal{C}}$ and the forgetful functor is $F(a,b)=a\otimes b$. This implies that the junction Hilbert space $W^{(b,c)}_a$ has dimension $N_{bc}^a$, with $N_{bc}^a$ the fusion coefficients of $\mathcal{C}$. On the other hand, one has that $\mathcal{V}_{(a,b)}=V_a\otimes \overline{V_b}$, where $V_a$ is the irreducible module of the chiral algebra $V$ labeled by $a\in\mathcal{C}$.

\subsection{A Schur-Weyl duality from the extended Hilbert space}

From \eqref{eqn:decomptwistedlocalop} and the discussion in the previous subsection, it follows that the extended Hilbert space $\mathcal{H}_{\mathcal{C}}$ admits a decomposition of the form \cite{Lin:2022dhv}
\begin{align}\label{eqn:schurweyldecomp}
    \mathcal{H}_{\mathcal{C}} \cong \bigoplus_{\mu\in\mathrm{Irr}(Z(\mathcal{C}))} W^\mu \otimes \mathcal{V}_\mu, \hspace{.5in} W^\mu\equiv \bigoplus_{a\in\mathrm{Irr}(\mathcal{C})}W^\mu_a.
\end{align}
For example, in the case of diagonal rational conformal field theories with chiral algebra $V$, this decomposition reduces to 
\begin{align}
    \mathcal{H}_a \cong \bigoplus_{(b,c)\in\mathrm{Irr}(\CC\boxtimes\overline{\CC})}N_{bc}^a V_b \otimes \overline{V_c}.
\end{align}

Let us understand the decomposition in \eqref{eqn:schurweyldecomp} representation-theoretically. Recall that $\mathrm{Tube}(\mathcal{C})$ acts on the extended Hilbert space $\mathcal{H}_{\mathcal{C}}$. The lasso diagrams of the tube algebra live on the topological boundary condition $\mathcal{B}_{\mathrm{reg}}$, and are therefore incapable of altering $\mu$ or $\widetilde{\mathcal{O}}$ in the decomposition $\mathcal{O}\leftrightharpoons (x, \mu, \widetilde{\mathcal{O}})$. In particular, the tube algebra acts only on the topological junction operator $x$, in a way that depends only on $\mu$ and not on $\widetilde{\mathcal{O}}$. In fact, it turns out that $\mathrm{Tube}(\mathcal{C})$ acts irreducibly on the spaces $W^\mu$, and that all irreducible representations of $\mathrm{Tube}(\mathcal{C})$ arise in this way \cite{Lin:2022dhv}. 
To further illustrate this, in Section \ref{sec:matrix}, we explicitly compute the matrix elements of the lasso operators in $\mathrm{Tube}(\mathcal{C})$ within the irreducible representations $W^\mu$.

On the other hand, by restriction, the extended Hilbert $\mathcal{H}_{\mathcal{C}}$ is also a module of $\mathcal{V}_{1}$, the subalgebra of $\mathcal{C}$-neutral genuine local operators of $Q$.  In fact, it is not too difficult to see that the $\mathcal{V}_\mu$ are representations of $\mathcal{V}_{1}$. Indeed, simply note that the OPE of an operator of the form $\mathcal{O}_1\leftrightharpoons(1, 1, \widetilde{\mathcal{O}}_1)$ with an operator of the form $\mathcal{O}_2\leftrightharpoons(x, \mu, \widetilde{\mathcal{O}}_2)$ will only produce operators of the form $\mathcal{O}_3\leftrightharpoons (x, \mu, \widetilde{\mathcal{O}}_3)$; in other words, the action of $\mathcal{V}_{1}$ operates only within the spaces $\mathcal{V}_\mu$. Again, it turns out that the $\mathcal{V}_\mu$ are irreducible, and that, at least in conformal field theories with faithfully acting symmetry categories, every irreducible representation of $\mathcal{V}_{1}$ takes this form. Indeed, for example, in the case of diagonal rational conformal field theories, the spaces $\mathcal{V}_{(a,b)}=V_a\otimes \overline{V_b}$ are precisely all of the irreducible modules of $\mathcal{V}_{1}=V\otimes\overline{V}$.

Finally, note that the action of $\mathcal{V}_{1}$ commutes with the action of $\mathrm{Tube}(\mathcal{C})$. The SymTFT makes this plain to see, since $\mathrm{Tube}(\mathcal{C})$ operates entirely on the topological boundary $\mathcal{B}_{\mathrm{reg}}$, while $\mathcal{V}_{1}$ operates entirely on the physical boundary $\widetilde{Q}$. Thus, the decomposition in \eqref{eqn:schurweyldecomp} is a decomposition of $\mathcal{H}_{\mathcal{C}}$ into irreducible  $(\mathrm{Tube}(\mathcal{C}),\mathcal{V}_{1})$-modules. In fact, at least in CFTs with $\mathcal{C}$ acting faithfully, this decomposition furnishes a kind of Schur--Weyl duality, in the sense that it gives a one-to-one correspondence between irreducible representations of $\mathrm{Tube}(\mathcal{C})$ and irreducible representations of $\mathcal{V}_{1}$.\footnote{This correspondence was used extensively in recent studies of {symmetry topological order}, {topological holography}, and gapless phases and quantum phase transitions \cite{Inamura:2023ldn, Chatterjee:2022jll, Chatterjee:2022tyg, Ji:2021esj, Kong:2020cie, Ji:2019jhk, Ji:2019ebr, Ji:2019eqo}.} (Note that the representation categories of both $\mathrm{Tube}(\mathcal{C})$ and $\mathcal{V}_{1}$ are equivalent to $Z(\mathcal{C})$.) 

We note in passing a connection to symmetry/subalgebra duality. In \cite{Rayhaun:2023pgc}, it was emphasized that, in the context of chiral conformal field theories (i.e.\ CFTs with only left-moving degrees of freedom), there is a correspondence between symmetry categories $\mathcal{C}$ of $Q$ and conformal subalgebras of $\mathcal{H}$. This correspondence assigns to $\mathcal{C}$ precisely the subalgebra of $\mathcal{C}$-neutral operators, i.e.\ $\mathcal{V}_{1}$; conversely, given a conformal subalgebra $W\subset\mathcal{H}$ (of ``finite index''), one may reconstruct a finite symmetry category of $Q$  with the property that $\mathcal{V}_{1}\cong W$, using SymTFT arguments. The same ideas likely apply to general 1+1d CFTs: that is, in 
a general 1+1d CFT, $\mathcal{V}_{1}$ can be thought of as the subalgebra assigned to $\mathcal{C}$ by symmetry/subalgebra duality; conversely, one expects that if one is able to identify a suitable finite index conformal subalgebra of the Hilbert space $\mathcal{H}$, then this allows one to define a symmetry of $Q$. 

The SymTFT perspective on the extended Hilbert space makes it straightforward to determine the fate of (twisted sector) local operators after gauging. Indeed, as we saw earlier, gauging the QFT $Q$ by an algebra object $A$ of $\mathcal{C}$ corresponds to replacing the Dirichlet topological boundary condition $\mathcal{B}_{\mathrm{reg}}$ by another topological boundary condition $\mathcal{B}=\mathcal{C}_A$. The topological line operators supported on the boundary $\mathcal{B}$ are described by the fusion category $\mathcal{C}'={_A}\mathcal{C}_A$ (cf.\ \eqref{eqn:dualsymmetrycategory}), and there is a corresponding forgetful functor $F' :Z(\mathcal{C}) \to \mathcal{C}'$ which describes what happens when a bulk line is pushed onto the boundary labeled by $\mathcal{B}$. We may then form the spaces $(W')^\mu_a\equiv \mathrm{Hom}_{\mathcal{C}'}(F'(\mu),a)$, where $a\in \mathcal{C}'$ and $\mu\in Z(\mathcal{C})$, from which the extended Hilbert space of the gauged theory $Q/A$ immediately follows,
\begin{align}
    \mathcal{H}'_{\mathcal{C}'} \cong \bigoplus_{\mu\in \mathrm{Irr}(Z(\mathcal{C}))}(W')^\mu\otimes \mathcal{V}_\mu, \ \ \ \ (W')^\mu \equiv \bigoplus_{a\in\mathrm{Irr}(\mathcal{C}^\ast)}(W')^\mu_a.
\end{align}
In other words, to form the extended Hilbert space $\mathcal{H}'_{\mathcal{C}'}$ of an orbifolded theory $Q/A$, one simply replaces the representation spaces $W^\mu$ with the spaces $(W')^\mu$, keeping the spaces $\mathcal{V}_\mu$ the same as in the original theory. See, e.g.,\ \cite{Fuchs:2002cm} for a description of the spaces $(W')^\mu_a$ in the case of rational conformal field theories.

Finally, we note that, thinking of the new boundary condition as a Lagrangian algebra $L$ of $Z(\mathcal{C})$ (instead of as an algebra $A$ of $\mathcal{C}$), the dual category is $\mathcal{C}'\cong Z(\mathcal{C})_L$. Thus, each line $a\in\mathcal{C}'$ of the dual symmetry may then be thought of as an $L$-module, and may accordingly be decomposed into simple objects of $Z(\mathcal{C})$. The $(W')^\mu_a$ can be interpreted as the multiplicity spaces of this decomposition, i.e.\ 
\begin{align}
    a \cong \bigoplus_{\mu\in \mathrm{Irr}(Z(\mathcal{C}))}\dim\left((W')^\mu_a\right) ~\mu. 
\end{align}

\subsection{Boundaries and interfaces}\label{subsec:interfaces}

We move on to the SymTFT representation of extended objects, like boundaries and interfaces. Because a boundary may be thought of as an interface between a quantum field theory and a trivially gapped theory, we can focus on interfaces without loss of generality. 

We consider an interface $I$ between two 1+1d quantum field theories $Q_1$ and $Q_2$, with symmetry categories $\mathcal{C}_1$ and $\mathcal{C}_2$, respectively. When we blow this configuration up into 2+1d, far away from the interface in the direction of $Q_1$ or $Q_2$, we expect to find the standard SymTFT setup: 
\begin{eqnarray}
    Q_1 \leftrightharpoons  (\mathcal{B}_{\mathrm{reg},1}, \mathrm{TV}_{\mathcal{C}_1}, \widetilde{Q}_1), \qquad Q_2 \leftrightharpoons  (\mathcal{B}_{\mathrm{reg},2}, \mathrm{TV}_{\mathcal{C}_2}, \widetilde{Q}_2)
\end{eqnarray}
as explained in Section \ref{subsec:basicsetup}.
Near the interface, there must then be suitable objects in the SymTFT which interpolate between these two configurations.

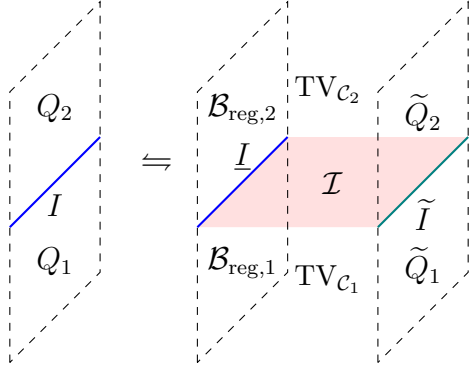
\begin{figure}[t]
	\centering
	\raisebox{-73pt}{\begin{tikzpicture}[scale=1.2]
			\draw[dashed](2,0) -- (2,3) -- (3,4) -- (3,1) -- cycle;
			\draw (2.5,1.4) node[below]{$Q_1$};
   \draw (2.5,3.1) node[below]{$Q_2$};
   \draw (2.5,1.5) node[above]{$I$};
   \draw [thick, blue] (2,1.5) -- (3,2.5);
	\end{tikzpicture}}
	\quad $\leftrightharpoons$ ~
	\raisebox{-73pt}{\begin{tikzpicture}[scale=1.2]
 \fill[fill=red!40,opacity=.3] (0,1.5)  -- (2,1.5)  -- (3,2.5) -- (1,2.5) -- cycle;
			\draw (1.45,1.2) node[below]{\small  $\TV_{\mathcal{C}_1}$};
   \draw (1.45,3.3) node[below]{\small  $\TV_{\mathcal{C}_2}$};
			\draw[dashed] (0,0) -- (0,3) -- (1,4) -- (1,1) -- cycle;
			\draw (.5,1.4) node[below]{$\mathcal{B}_{\mathrm{reg},1}$};
   \draw (.5,3.1) node[below]{$\mathcal{B}_{\mathrm{reg},2}$};
			\draw[ dashed] (2,0) -- (2,3);
			\draw[dashed] (2,3) -- (3,4) -- (3,1) -- (2,0);
			\draw (2.5,1.4) node[below]{$\widetilde{Q}_1$};
   \draw (2.5,3.1) node[below]{$\widetilde{Q}_2$};
   \draw [style=boundarylines] (0,1.5) -- (1,2.5);
   \draw [thick,color=teal] (2,1.5) -- (3,2.5);
   \draw (1.5,1.75) node[above] {$\mathcal{I}$}; 
   \draw  (.5,2) node[above] {$\underline{I}$};
   \draw  (2.5,1.35) node[above] {$\widetilde{I}$};
	\end{tikzpicture}}
	\caption{The SymTFT interpretation of a line interface between two QFTs $Q_1$ and $Q_2$ with symmetry categories  $\CC_1$ and $\CC_2$, respectively.
	}
	\label{fig:interfacesandwich}
\end{figure}

Our proposal is that, in analogy with the case of twisted sector local operators, $I$ decomposes into a triple of the form 
\begin{align}
    I\leftrightharpoons(\underline{I},\mathcal{I},\widetilde{I}), \qquad \underline{I}\in \mathrm{Int}_{\mathcal{I}}(\mathcal{B}_{\mathrm{reg},1},\mathcal{B}_{\mathrm{reg},2}), \quad \mathcal{I}\in \mathrm{Bimod}(\mathcal{C}_1,\mathcal{C}_2), \quad  \widetilde{I}\in \mathrm{Int}_{\mathcal{I}}(\widetilde{Q}_1,\widetilde{Q}_2)
\end{align} 
whose components are depicted in  Figure \ref{fig:interfacesandwich}, and which we now explain.

First, $\mathcal{I}$ is a two-dimensional \emph{topological interface} between the  SymTFTs $\mathrm{TV}_{\mathcal{C}_1}$ and $\mathrm{TV}_{\mathcal{C}_2}$. To state which interface exactly, recall that, mathematically, topological interfaces between $\mathrm{TV}_{\mathcal{C}_1}$ and $\mathrm{TV}_{\mathcal{C}_2}$ are in one-to-one correspondence with $(\mathcal{C}_1,\mathcal{C}_2)$-bimodule categories.\footnote{This can be seen by folding: surfaces between $\mathrm{TV}_{\mathcal{C}_1}$ and $\mathrm{TV}_{\mathcal{C}_2}$ are the same as boundaries in $\mathrm{TV}_{\mathcal{C}_1}\otimes \overline{\mathrm{TV}_{\mathcal{C}_2}}$. As a modular tensor category, this latter TQFT is represented by $Z(\mathcal{C}_1\boxtimes \mathcal{C}_2^{\mathrm{op}})$, and so its boundaries are $\mathcal{C}_1\boxtimes \mathcal{C}_2^{\mathrm{op}}$-module categories, which are by definition the same as $(\mathcal{C}_1,\mathcal{C}_2)$-bimodule categories.} On the other hand, by performing parallel fusion of the lines in $\mathcal{C}_1$ and $\mathcal{C}_2$ onto the interface $I$, we generate a $(\mathcal{C}_1,\mathcal{C}_2)$-multiplet $\mathcal{I}'$ of interfaces, which we saw in Section \ref{subsec:modulebimodule} also carries the structure of a $(\mathcal{C}_1,\mathcal{C}_2)$-bimodule category, i.e.\ $\mathcal{I}'\in \mathrm{Bimod}(\mathcal{C}_1,\mathcal{C}_2)$. (Recall that the NIM-rep under which the interface $I$ transforms is part of the data of this bimodule category $\mathcal{I}'$.)  One of central claims of this section is that these two $(\mathcal{C}_1,\mathcal{C}_2)$-bimodule categories coincide.

\begin{claim}
    Suppose that $I$ is a line interface between two 1+1d quantum field theories $Q_1$ and $Q_2$ with symmetry categories $\mathcal{C}_1$ and $\mathcal{C}_2$, respectively, and further that $I$ belongs to a $(\mathcal{C}_1,\mathcal{C}_2)$-bimodule category $\mathcal{I}'$ of such interfaces. The two-dimensional topological interface $\mathcal{I}$ between  $\TV_{\CC_1}$ and $\TV_{\CC_2}$ arising in the SymTFT construction of $I$, as in Figure \ref{fig:interfacesandwich}, coincides with $\mathcal{I}'$ when both are thought of as abstract $(\CC_1,\CC_2)$-bimodule categories. That is, $\mathcal{I}\cong \mathcal{I}'$.
\end{claim}
Therefore, we henceforth abusively conflate these two structures --- multiplets of line interfaces between 1+1d QFTs and topological interfaces between the corresponding bulk SymTFTs --- and label the surface $\mathcal{I}$ arising in the SymTFT construction of $I$ by the $(\mathcal{C}_1,\mathcal{C}_2)$ multiplet  of interfaces to which $I$ belongs. This prescription is in harmony with the idea that multiplets of $n$-dimensional objects in a QFT should be labeled by suitable $(n+1)$-dimensional objects in the bulk SymTFT.

This surface $\mathcal{I}$ must  terminate on suitable junctions $\widetilde{I}$ and $\underline{I}$ on the ``loaves'' of the sandwich.  More specifically, $\widetilde{I}$ is a (not necessarily topological) line interface between the physical boundaries $\widetilde{Q}_1$ and $\widetilde{Q}_2$ on which $\mathcal{I}$ terminates; we denote the category of such interfaces as $\mathrm{Int}_{\mathcal{I}}(\widetilde{Q}_1,\widetilde{Q}_2)$. Similarly, $\underline{I}$ is a \emph{topological line junction} between the topological boundaries $\mathcal{B}_{\mathrm{reg},1}$ and $\mathcal{B}_{\mathrm{reg},2}$ on which $\mathcal{I}$ terminates; we denote the category of such interfaces as $\mathrm{Int}_{\mathcal{I}}(\mathcal{B}_{\mathrm{reg},1},\mathcal{B}_{\mathrm{reg},2})$.

One may ask what happens when one manipulates various objects in the SymTFT. For example, we may consider swapping out the topological line junction $\underline{I}$ for another one. It turns out that that the simple line junctions which can sit at the intersection of $\mathcal{B}_{\mathrm{reg},1}$, $\mathcal{B}_{\mathrm{reg},2}$, and $\mathcal{I}$ are precisely the simple objects of the category $\mathcal{I}$. That is, there is an equivalence 
\begin{align}
    \mathrm{Int}_{\mathcal{I}}(\mathcal{B}_{\mathrm{reg},1},\mathcal{B}_{\mathrm{reg},2})\cong\mathcal{I}
\end{align} 
of $(\mathcal{C}_1,\mathcal{C}_2)$-bimodule categories. (As a sanity check on this claim, note that when $\mathcal{C}_1\cong \mathcal{C}_2$ and $\mathcal{I}=\mathcal{I}_{\mathrm{reg}}$ is chosen to be the trivial surface, then this formula reproduces the familiar fact that topological line operators on the Dirichlet boundary of $\TV_{\mathcal{C}}$ are described by $\mathcal{C}$.)
As we noted in the previous paragraph, $\mathcal{I}$ can be thought of as the multiplet of interfaces  generated by acting with symmetries on $I$. Thus, we arrive at the conclusion that swapping out $\underline{I}\in\mathrm{Int}_{\mathcal{I}}(\mathcal{B}_{\mathrm{reg},1},\mathcal{B}_{\mathrm{reg},2})$ for another choice of line junction $\underline{I}'\in\mathrm{Int}_{\mathcal{I}}(\mathcal{B}_{\mathrm{reg},1},\mathcal{B}_{\mathrm{reg},2})$ corresponds in 1+1d to toggling from $I\in\mathcal{I}$ to another interface $I'\in\mathcal{I}$ in the same multiplet as $I$. 

More dramatically, we may consider swapping out the Dirichlet boundaries $\mathcal{B}_{\mathrm{reg},i}$  for other topological boundary conditions described by module categories $\mathcal{B}_i\cong\mathcal{C}_{A_i}$. We saw earlier that this has the effect of gauging the 1+1d QFTs we start with, i.e.\ $Q_1$ is replaced with $Q_1\big/A_1$, and similarly $Q_2$ is replaced with $Q_2\big/A_2$. After gauging, the original multiplet of interfaces $\mathcal{I}$ will go over to a new multiplet of interfaces $A_1\big\backslash\mathcal{I}\big/A_2$, which transforms as a bimodule category with respect to the dual symmetries
\begin{align}
    \mathcal{C}_i' \cong {_{A_i}}(\mathcal{C}_i)_{A_i} \cong \mathrm{Fun}_{\mathcal{C}_i}(\mathcal{B}_i,\mathcal{B}_i)^{\mathrm{op}}, \ \ \ \ (i=1,2)
\end{align}
of $Q_1\big/A_1$ and $Q_2\big/A_2$, respectively. In purely 1+1-dimensional terms, we would say that this new multiplet consists of the (generally non-simple) interfaces in $\mathcal{I}$ on which the gauging meshes $A_1$ and $A_2$ can topologically end. The condition of $A_1$ and $A_2$ having a topological junction with an interface in the multiplet $\mathcal{I}$ means that that interface should form an $A_1$-$A_2$-bimodule (cf.\ e.g.\ \cite{Huang:2021zvu} for the analogous statement describing how boundary conditions behave under orbifolding, and \cite{Collier:2021ngi} for the special case of orbifolds of invertible symmetries). Thus, we recover the mathematical fact that 
\begin{align}
    A_1\big\backslash\mathcal{I}\big/A_2 \cong {_{A_1}}\mathcal{I}_{A_2}
\end{align}
where ${_{A_1}}\mathcal{I}_{A_2}$ is the category of $A_1$-$A_2$ bimodules inside of $\mathcal{I}$.
This is manifestly an $({_{A_1}}(\mathcal{C}_1)_{A_1},{_{A_2}}(\mathcal{C}_2)_{A_2})$-bimodule category, with e.g.\ the left ${_{A_1}}(\mathcal{C}_1)_{A_1}$ action coming from tensor product over $A_1$.\footnote{See e.g.\ \cite{Bhardwaj:2017xup} for a discussion of what ``tensor product over $A_1$'' means.}

In 2+1d terms, we see that $A_1\big\backslash\mathcal{I}\big/A_2$ consists of triples $(\underline{I},\mathcal{I},\widetilde{I})$, where $\underline{I}$ is now allowed to vary over objects of $\mathrm{Int}_{\mathcal{I}}(\mathcal{B}_1,\mathcal{B}_2)$, i.e.\ topological line interfaces between $\mathcal{B}_1$ and $\mathcal{B}_2$ on which $\mathcal{I}$ terminates. To determine this category, we can fuse the interface $\mathcal{I}$ onto the boundary $\mathcal{B}_2$ of $\mathrm{TV}_{\mathcal{C}_2}$ to obtain a boundary $\mathcal{I}\boxtimes_{\mathcal{C}_2}\mathcal{B}_2$ of $\mathrm{TV}_{\mathcal{C}_1}$.\footnote{See e.g.\ \cite{etingof2010fusion} for a mathematical discussion of tensor product of module categories.} Then, we are tasked with determining the category of topological line interfaces which can interpolate between the boundaries $\mathcal{B}_1$ and $\mathcal{I}\boxtimes_{\mathcal{C}_2}\mathcal{B}_2$ of $\mathrm{TV}_{\mathcal{C}_1}$. Mathematically, it is known \cite{2012CMaPh.313..351K} that this is given by 
\begin{align}
    A_1\big\backslash\mathcal{I}\big/A_2\cong \mathrm{Int}_{\mathcal{I}}(\mathcal{B}_1,\mathcal{B}_2)\cong  \mathrm{Fun}_{\mathcal{C}_1}(\mathcal{B}_1,\mathcal{I}\boxtimes_{\mathcal{C}_2}\mathcal{B}_2) .
\end{align}
This is then manifestly a $(\mathrm{Fun}_{\mathcal{C}_1}(\mathcal{B}_1,\mathcal{B}_1)^{\mathrm{op}},\mathrm{Fun}_{\mathcal{C}_2}(\mathcal{B}_2,\mathcal{B}_2)^{\mathrm{op}})$-bimodule category, with e.g.\ the left action of $\mathrm{Fun}_{\mathcal{C}_1}(\mathcal{B}_1,\mathcal{B}_1)^{\mathrm{op}}$ coming from the composition of functors. Of course, both the 1+1d description and the 2+1d description agree with one another.

Let us consider a special case of this construction that we have already encountered. Consider a 1+1d QFT $Q$ with a fusion category symmetry $\mathcal{C}$. The lines in $\mathcal{C}$ may be thought of as interfaces between $Q$ and itself. We know that, when blown up into 3d, any line operator in $\mathcal{C}$ may be thought of as living on the topological Dirichlet boundary condition $\mathcal{B}_{\mathrm{reg}}$ of $\mathrm{TV}_{\mathcal{C}}$. This means that the triples $(\underline{I},\mathcal{I},\widetilde{I})$ corresponding to lines $I\in \mathcal{C}$ have $\mathcal{I}\cong \mathcal{I}_{\mathrm{reg}}$ as the trivial surface operator of $\mathrm{TV}_{\mathcal{C}}$, and $\widetilde{I}$ as the trivial line operator on the physical boundary $\widetilde{Q}$. Meanwhile, $\underline{I}$ toggles over elements of the $(\mathcal{C},\mathcal{C})$-bimodule category $\mathcal{I}_{\mathrm{reg}}$, whose objects are in one-to-one correspondence with objects of $\mathcal{C}$. Thus, Figure \ref{fig:interfacesandwich} reduces to Figure \ref{fig:sandwich} in this special case.

Another example is the category of half-space gauging interfaces. Indeed, let $Q_1$ be a 1+1d QFT with symmetry category $\mathcal{C}_1=\mathcal{C}$, and let $Q_2 = Q_1\big/A$ be the orbifold of $Q_1$ by a gaugeable algebra $A$ of $\mathcal{C}$, which has symmetry category $\mathcal{C}_2={_A}\mathcal{C}_A$. One can consider the SymTFT representation of the interfaces between $Q_1$ and $Q_2$ obtained by starting with $Q_1$ and gauging $A$ in half of spacetime, varying over different boundary conditions for the 1+1d gauge fields at the location of the interface. Then, the triples $(\underline{I},\mathcal{I},\widetilde{I})$ corresponding to such interfaces have $\mathcal{I}\cong \mathcal{C}_A$, which can be thought of as a $(\mathcal{C}_1,\mathcal{C}_2)=(\mathcal{C},{_A}\mathcal{C}_A)$-bimodule category, and $\underline{I}\in \mathcal{C}_A$. Moreover  $\widetilde{I}$ is the trivial line operator since the gauging interface is topological. 

\subsection{Symmetric boundaries, anomalies, and magnetic boundaries of SymTFT}\label{subsec:symmetricboundariesanomalies}

Let us specialize to the case of boundaries, and discuss the relationship between anomalies of $\mathcal{C}$, and the possibility of a 1+1d QFT having a $\mathcal{C}$-symmetric simple boundary condition.\footnote{For the case of ordinary symmetries, the relation between anomalies and boundary conditions has been extensively studied in the past \cite{Wang:2013yta,Han:2017hdv,Jensen:2017eof,Numasawa:2017crf,Smith:2020rru,Smith:2020nuf,Thorngren:2020yht,Tong:2021phe,Li:2022drc,Zeng:2022grc,Wang:2022ucy}.}
As we will see, our SymTFT picture affords useful alternative ways of looking at familiar facts. For the purposes of this discussion, our definition of what it means for $\mathcal{C}$ to be non-anomalous is that it should be compatible with a 1+1d trivially-gapped phase. Our definition of what it means for a simple boundary condition to be $\mathcal{C}$-symmetric is that it should be invariant under parallel fusion by all of the lines in $\mathcal{C}$; this happens if and only if the boundary transforms in a module category of $\mathcal{C}$ with exactly one simple object.\footnote{The definition of symmetric boundary condition which we employ here coincides with the notion of \emph{strongly-symmetric} boundary condition defined in \cite{Choi:2023xjw}.}

\subsubsection{Anomaly free $\cong$ kinematic compatibility with simple symmetric boundary condition}

There is an equivalent reformulation of this definition of anomaly. For example, it is known that $\mathcal{C}$ is non-anomalous (i.e. compatible with a trivially gapped phase) if and only if $\mathcal{C}$ admits a fiber functor $F$, i.e.\ a tensor functor $F:\mathcal{C}\to \mathrm{Vec}$. The intuition behind this is that $\mathcal{C}$ is non-anomalous if and only if there is no kinematical obstruction to the existence of  an RG flow from a UV theory with symmetry $\mathcal{C}$ to the trivially gapped theory with trivial symmetry category $\mathrm{Vec}$; on general grounds, in any RG flow we expect a mapping from the symmetry in the UV to the symmetry in the IR, which in the present situation must be implemented by a fiber functor. 

On the other hand, the existence of a fiber functor is equivalent to $\mathcal{C}$ admitting a module category with one simple object. Indeed, if $F:\mathcal{C}\to\mathcal{D}$ is a tensor functor of fusion categories, then any $\mathcal{D}$-module category $\mathcal{B}$ may be ``pulled back'' to a $\mathcal{C}$-module category using $F$. Thus, the unique rank-1 module category of $\mathrm{Vec}$ can be pulled back to a rank-1 module category of $\mathcal{C}$ if $\mathcal{C}$ admits a fiber functor. The converse turns out to be true as well. From this reformulation, it straightforwardly follows that $\mathcal{C}$ is non-anomalous if and only if it is kinematically compatible with the existence of a simple symmetric boundary condition. 

We could reach the same conclusion from the perspective of the SymTFT. In analogy with the case of interfaces, a  boundary condition $B$  decomposes into a triple $(\underline{B},\mathcal{B}, \widetilde{B})$, where $\mathcal{B}$ is a topological boundary condition of $\mathrm{TV}_{\mathcal{C}}$ (which we represent as a $\mathcal{C}$-module category), $\underline{B}\in\mathcal{B}$ is a topological line junction between $\mathcal{B}_{\mathrm{reg}}$ and $\mathcal{B}$, and $\widetilde{B}$ is a physical line junction between $\widetilde{Q}$ and $\mathcal{B}$. See Figure \ref{fig:bdysymtft}.

Kinematical compatibility with a symmetric boundary can then be rephrased as the existence of a topological boundary $\mathcal{B}$ which admits only one topological interface $\underline{B}$ with the Dirichlet boundary $\mathcal{B}_{\mathrm{reg}}$, since the number of such interfaces is the number of simple boundaries of $Q$ in the $\mathcal{C}$-multiplet to which $B$ belongs.
In particular, a boundary $B$ is symmetric if and only if the topological boundary $\mathcal{B}$ in the decomposition $(\underline{B},\mathcal{B},\widetilde{B})$ corresponds mathematically to a rank-1 module category, which, as we saw earlier, can exist only if $\mathcal{C}$ is non-anomalous.  See also  \cite{Putrov:2024uor} for a discussion on the relationship between symmetric boundaries and anomalies from the SymTFT perspective.

\subsubsection{Anomaly free $\cong$ magnetic boundary condition of SymTFT}

\begin{figure}[t]
	\centering
	\raisebox{-73pt}{\begin{tikzpicture}
			\draw[dashed](2,0) -- (2,3) -- (3,4) -- (3,1) -- cycle;
			\draw (2,0) node[below]{$Q_{\mathcal{B}}$};
	\end{tikzpicture}}
	\quad $\leftrightharpoons$ ~
	\raisebox{-73pt}{\begin{tikzpicture}
			\draw (1.45,2.3) node[below]{\small  $\TV_\cC$};
			\draw[dashed] (0,0) -- (0,3) -- (1,4) -- (1,1) -- cycle;
			\draw (0,0) node[below]{$\mathcal{B}_{\mathrm{reg}}$};
			\draw[preaction={draw=white,line width=3pt}, dashed] (2,0) -- (2,3);
			\draw[dashed] (2,3) -- (3,4) -- (3,1) -- (2,0);
			\draw (2,0) node[below]{$\mathcal{B}$};
	\end{tikzpicture}}
	\caption{The definition of the theory $Q_{\mathcal{B}}$.
	}
	\label{fig:QB}
\end{figure}
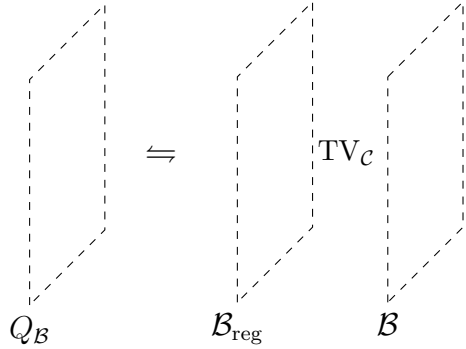

There is another reformulation of what it means for $\mathcal{C}$ to be non-anomalous, which is that the SymTFT $\mathrm{TV}_{\mathcal{C}}$ should admit a magnetic boundary condition \cite{Zhang:2023wlu}. By definition, a boundary condition $\mathcal{B}$ of $\mathrm{TV}_{\mathcal{C}}$  is said to be magnetic if the only bulk anyon which is condensed both on the Dirichlet boundary $\mathcal{B}_{\mathrm{reg}}$ and on $\mathcal{B}$ is the trivial anyon. Indeed, given such a magnetic $\mathcal{B}$, one can dimensionally reduce the SymTFT $\mathrm{TV}_{\mathcal{C}}$ on an interval with $\mathcal{B}_{\mathrm{reg}}$ and $\mathcal{B}$ imposed at both ends; let us call this theory $Q_{\mathcal{B}}$, depicted in Figure \ref{fig:QB}. This resulting 1+1d TFT has $\mathcal{C}$-symmetry, simply because the Dirichlet boundary hosts line operators described by $\mathcal{C}$. Moreover the theory has only one genuine local operator, since by the discussion in Section \ref{subsec:twistedsectorlocalops}, such operators correspond in 2+1d to triples $(x,\mu,y)$, where $x$ is a junction between $\mu$ and the trivial line on $\mathcal{B}_{\mathrm{reg}}$, and $y$ is a junction between $\mu$ and the trivial line on $\mathcal{B}$; by the assumption that $\mathcal{B}$ is magnetic, only the identity line $\mu=1$ admits such junctions $x$ and $y$, and the junctions are unique up to rescaling. By the state/operator correspondence, the theory has only one state on $S^1$, and is therefore a  $\mathcal{C}$-SPT in 1+1d, and in particular trivially gapped. Hence $\mathcal{C}$ is non-anomalous. The yoga of the SymTFT allows one to show the reverse as well. 

\begin{figure}
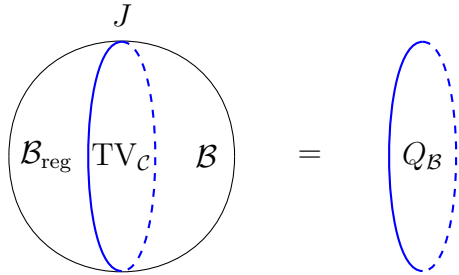

    \ctikzfig{figures/QBsolidball}\vspace{-.75cm}
    \caption{The topological line interfaces $J$ between the $\mathcal{B}_{\mathrm{reg}}$ and $\mathcal{B}$ boundary conditions of $\mathrm{TV}_{\mathcal{C}}$ correspond to boundary conditions of $Q_{\mathcal{B}}$.}\label{fig:QBdisk}
\end{figure}

Recall that from the perspective of the SymTFT, a 1+1d QFT $Q$ with $\mathcal{C}$-symmetry being compatible with a $\mathcal{C}$-symmetric boundary is the same as the existence of a topological boundary $\mathcal{B}$ of $\mathrm{TV}_{\mathcal{C}}$ which admits only a single topological line interface with the Dirichlet boundary condition $\mathcal{B}_{\mathrm{reg}}$. We now show that this happens if and only if $\mathcal{B}$ is magnetic, which in turn is equivalent to $\mathcal{C}$ being non-anomalous by the discussion of the previous paragraph. There are two arguments for this. 

The first is to note that the category of interfaces between $\mathcal{B}_{\mathrm{reg}}$ and $\mathcal{B}$ is the same as the category of boundary conditions in $Q_{\mathcal{B}}$. Indeed, one can start with $\mathrm{TV}_{\mathcal{C}}$ on a solid 3-dimensional ball, with $\mathcal{B}_{\mathrm{reg}}$ imposed on one hemisphere, $\mathcal{B}$ imposed on the other, and a topological interface $J$ at the equator. By squashing this ball down onto a disk, one obtains the TFT $Q_{\mathcal{B}}$ on a disk, and $J$ descends to a boundary condition, as in Figure \ref{fig:QBdisk}. The reverse is true as well. Thus, $\mathcal{B}$ and $\mathcal{B}_{\mathrm{reg}}$ admit a unique topological interface between them if and only if $Q_{\mathcal{B}}$ admits a unique simple boundary condition, which happens if and only if it is a $\mathcal{C}$-symmetric SPT, which we saw happens if and only if $\mathcal{B}$ is a magnetic boundary.

\begin{figure}
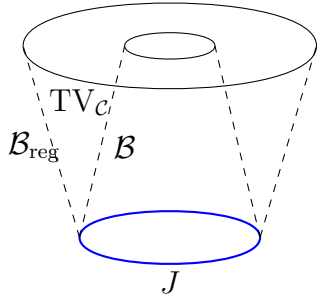

    \ctikzfig{figures/statelineopcorrespondence}\vspace{-.5cm}
    \caption{There is a correspondence between 1) topological line interfaces $J$ between the $\mathcal{B}_{\mathrm{reg}}$ and $\mathcal{B}$ topological boundary conditions of $\mathrm{TV}_{\mathcal{C}}$, and 2) states of $\mathrm{TV}_{\mathcal{C}}$ on $I\times S^1$ with $\mathcal{B}_{\mathrm{reg}}$ and $\mathcal{B}$ imposed at the two ends.}\label{fig:statelineopcorrespondence}
\end{figure}

The second argument uses a kind of state/operator correspondence between 1) the topological line interfaces between $\mathcal{B}_{\mathrm{reg}}$ and $\mathcal{B}$, and 2) states of $\mathrm{TV}_{\mathcal{C}}$ on its $I\times S^1$ Hilbert space, where $\mathcal{B}_{\mathrm{reg}}$ and $\mathcal{B}$ are imposed at the two ends. Indeed, the operator to state map can be seen, for example, by performing the Euclidean path integral described in Figure \ref{fig:statelineopcorrespondence}. On the other hand, by dimensional reduction on the interval, the dimension of this Hilbert space is the same as the dimension of the $S^1$ Hilbert space of $Q_{\mathcal{B}}$, which is equal to $1$ if and only if $\mathcal{B}$ is magnetic. 

Before moving on, we pause to emphasize that, although one might be tempted to conclude that the SymTFT shows that every 1+1d QFT $Q$ with a non-anomalous symmetry $\mathcal{C}$ admits a $\mathcal{C}$-symmetric boundary condition, the discussion so far has only been about the kinematical compatibility of $Q$ with such a boundary. The question of whether such a boundary is actually \emph{dynamically} realized is equivalent to the question of whether $\mathrm{TV}_{\mathcal{C}}$ admits a magnetic boundary condition $\mathcal{B}$ (or equivalently, a boundary condition corresponding to a rank-1 module category) which has a suitable junction $\widetilde{B}$ with the physical boundary $\widetilde{Q}$.\footnote{Really, one is interested if the junction $\widetilde{B}$ is such that the triple $(J,\mathcal{B},\widetilde{B})$ leads to a simple boundary $B$. We will see in the next subsubsection that this happens if and only if, of all the line operators supported on the topological boundary $\mathcal{B}$, only the identity line admits a topological end-point on both $\widetilde{B}$ and $J$.} In fact, it is known that not every 1+1d CFT admits a simple $\mathcal{C}$-symmetric conformal boundary condition (see e.g.\ \cite{Choi:2023xjw} for a minimal model example).

\subsection{Local junction operators}\label{subsec:junctionops}

Finally, we study the SymTFT perspective on local operators which sit at the junction of a collection of interfaces between 1+1d QFTs with symmetry. We use the same conventions as in Section \ref{subsec:interfacetube}. Namely, we assume that we are given theories $Q_i$ with symmetry categories $\mathcal{C}_i$, and interfaces $I_i$ between $Q_i$ and $Q_{i+1}$ that transform in multiplets  $\mathcal{I}_i$, where $i=1,\cdots,n$. Each $\mathcal{I}_i$ possesses the structure of a $(\mathcal{C}_i,\mathcal{C}_{i+1})$-bimodule category.

\subsubsection{Local junction operators and representation of generalized tube algebra from SymTFT}

\begin{figure}
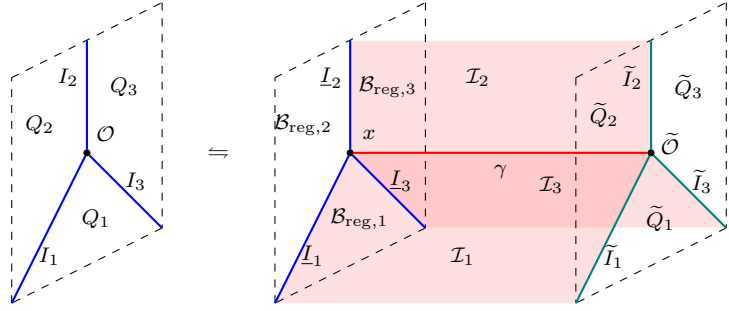

\begin{center}
    \ctikzfig{figures/junctionsymTFT}
    \caption{The SymTFT picture of local junction operators, illustrated in the special case that $n=3$.}\label{fig:junctionsymTFT}
\end{center}
\end{figure}

Following the pattern of the previous subsections, the three-dimensional formulation of this setup has several components, which we illustrate in Figure \ref{fig:junctionsymTFT}.
\begin{enumerate}[label=\arabic*)]
    \item Each theory $Q_i$ is blown up into a triple,
    \begin{align}
        Q_i\leftrightharpoons(\mathcal{B}_{\mathrm{reg},i},\mathrm{TV}_{\mathcal{C}_i},\widetilde{Q}_i)
    \end{align}
    where $\widetilde{Q}_i$ and $\mathcal{B}_{\mathrm{reg},i}$ are the physical and Dirichlet boundaries, respectively, of $\mathrm{TV}_{\mathcal{C}_i}$. This has been discussed in Section \ref{subsec:basicsetup}.
    \item Each interface $I_i$ is inflated into a triple, 
    \begin{align}
        I_i\leftrightharpoons (\underline{I}_i,\mathcal{I}_i,\widetilde{I}_i)
    \end{align}
    where $\mathcal{I}_i\in\mathrm{Bimod}(\mathcal{C}_i,\mathcal{C}_{i+1})$ is a topological surface between $\mathrm{TV}_{\mathcal{C}_i}$ and $\mathrm{TV}_{\mathcal{C}_{i+1}}$ described by the $(\mathcal{C}_i,\mathcal{C}_{i+1})$-multiplet of $I_i$. Further, $\underline{I}_i\in \mathrm{Int}_{\mathcal{I}_i}(\mathcal{B}_{\mathrm{reg},i},\mathcal{B}_{\mathrm{reg},i+1})\cong \mathcal{I}_i$ is a topological line junction between $\mathcal{B}_{\mathrm{reg},i}$ and $\mathcal{B}_{\mathrm{reg},i+1}$ on which $\mathcal{I}_i$ terminates, and $\widetilde{I}_i$ is a physical line interface  between $\widetilde{Q}_i$ and $\widetilde{Q}_{i+1}$ on which $\mathcal{I}_i$ terminates. This has been discussed in Section \ref{subsec:interfaces}.
    \item Finally, each operator $\mathcal{O}\in\mathcal{H}_{I_1\cdots I_n}$ decomposes, as in the case of genuine local operators, into a triple 
\begin{align}\label{eqn:junctiontriple}
    \mathcal{O}\leftrightharpoons (x,\gamma,\widetilde{\mathcal{O}}), \qquad  x \in W^\gamma_{\underline{I}_1\cdots\underline{I}_n}, \quad  \gamma \in \mathrm{Jun}(\mathcal{I}_1,\cdots,\mathcal{I}_n), \quad \widetilde{\mathcal{O}}\in \mathcal{V}_\gamma^{\widetilde{I}_1\cdots\widetilde{I}_n}.
\end{align}
Here, $\gamma$ is a topological line junction between the surfaces $\mathcal{I}_1,\cdots,\mathcal{I}_n$. The object $\widetilde{\mathcal{O}}$ is a (not necessarily topological) local operator which sits at the junction of the physical line interfaces $\widetilde{I}_i$, and similarly $x$ is a topological local operator which sits at the junction of the topological line interfaces $\underline{I}_i$. 
\end{enumerate}  

In particular, as a result of \eqref{eqn:junctiontriple}, we have the following decompositions, which generalize \eqref{eqn:schurweyldecomp},
\begin{align}
\begin{split}
    \mathcal{H}_{I_1\cdots I_n}& \cong \bigoplus_{\gamma\in\mathrm{Jun}(\mathcal{I}_1,\cdots,\mathcal{I}_n)}\mathcal{V}_\gamma^{\widetilde{I}_1 \cdots \widetilde{I}_n}\otimes W^\gamma_{\underline{I}_1\cdots\underline{I}_n} \\
    \mathcal{H}_{\mathcal{I}_1\cdots\mathcal{I}_n}&\cong \bigoplus_{\gamma\in\mathrm{Jun}(\mathcal{I}_1,\cdots,\mathcal{I}_n)}\mathcal{V}_\gamma^{\widetilde{I}_1\cdots\widetilde{I}_n}\otimes W^\gamma, 
\end{split}
\end{align}
where we have defined the space 
\begin{align}
    W^\gamma \equiv \bigoplus_{\underline{I}_1\in\mathrm{Irr}(\mathcal{I}_1)}\cdots \bigoplus_{\underline{I}_n\in\mathrm{Irr}(\mathcal{I}_n)}W^\gamma_{\underline{I}_1\cdots \underline{I}_n}.
\end{align}
In theories with a state/operator correspondence, $\mathcal{V}_\gamma^{\widetilde{I}_1\cdots\widetilde{I}_n}$ can be thought of as a particular $D^2$ Hilbert space of the SymTFT, decorated by various defects. For example, the topological surface $\mathcal{I}_i$ appears as a line defect which stretches from a point on the boundary $\partial D^2$ where it terminates on the junction $\widetilde{I}_i$, to the origin where it terminates on the defect $\gamma$. The disk is thus partitioned by the $\mathcal{I}_i$ into $n$ pizza slices, where the $i$th slice is in the $\mathrm{TV}_{\mathcal{C}_i}$ phase and has the physical boundary $\widetilde{Q}_i$ imposed at the crust, as in the left of Figure \ref{fig:junctionophilbertspaces}.

The Hilbert space $W^\gamma_{\underline{I}_1\cdots\underline{I}_n}$ can be obtained similarly, but with the physical boundaries $\widetilde{Q}_i$ replaced with the Dirichlet boundaries $\mathcal{B}_{\mathrm{reg},i}$, and with the junctions $\widetilde{I}_i$ replaced with $\underline{I}_i$, as in the right of Figure \ref{fig:junctionophilbertspaces}. Note that these Hilbert spaces generalize those described in Figure \ref{fig:twistedsectorhilbertspaces}. Following logic identical to that of Section \ref{subsec:twistedsectorlocalops}, the tube algebra $\mathrm{Tube}(\mathcal{I}_1\vert \cdots \vert \mathcal{I}_n)$ acts irreducibly on the spaces $W^\gamma$, and we expect that all irreducible representations can be obtained  in this manner.

\begin{figure}
    \vspace{-.5cm}\ctikzfig{figures/VmutildeI}\vspace{-.5cm}
    \caption{Left: the Hilbert space $\mathcal{V}_\gamma^{\widetilde{I}_1\cdots\widetilde{I}_n}$. Right: the Hilbert space $W^\gamma_{\underline{I}_1\cdots\underline{I}_n}$.}\label{fig:junctionophilbertspaces}
\end{figure}

\begin{figure}
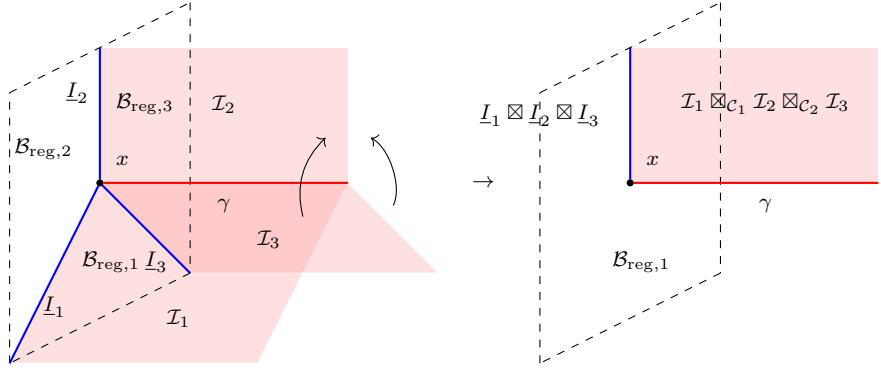

    \begin{center}
        \ctikzfig{figures/swingsurfaces}
        \caption{Swinging the surfaces $\mathcal{I}_1,\cdots,\mathcal{I}_n$ around $\gamma$ and fusing them onto each other produces a surface $\mathcal{I}=\mathcal{I}_1\boxtimes_{\mathcal{C}_1}\cdots\boxtimes_{\mathcal{C}_{n-1}}\mathcal{I}_n$ in $\mathrm{TV}_{\mathcal{C}_1}$, and the line junctions $\underline{I}_1,\cdots,\underline{I}_n$ fuse to form a junction $\underline{I}_1\boxtimes\cdots\boxtimes \underline{I}_n$. Then $\gamma$ can be thought of as a line operator which bounds the surface $\mathcal{I}_1\boxtimes_{\mathcal{C}_1}\cdots\boxtimes_{\mathcal{C}_{n-1}}\mathcal{I}_n$.}\label{fig:swingsurfaces}
    \end{center}
\end{figure}

Let us describe where everything lives in more mathematical detail. We begin with the line junction $\gamma$. By swinging the surfaces $\mathcal{I}_1,\cdots,\mathcal{I}_n$ around  $\gamma$ and fusing them all together, we obtain a surface $\mathcal{I}\equiv \mathcal{I}_1\boxtimes_{\mathcal{C}_2} \mathcal{I}_2 \boxtimes_{\mathcal{C}_3}\cdots \boxtimes_{\mathcal{C}_{n}}\mathcal{I}_n$ in $\mathrm{TV}_{\mathcal{C}_1}$ (i.e.\ a $(\mathcal{C}_1,\mathcal{C}_1)$-bimodule category) and $\gamma$ descends to a topological line on its boundary, as in Figure \ref{fig:swingsurfaces}.\footnote{Alternatively, we could have fused them in a different order to obtain that $\gamma$ can be thought of as a line operator at the boundary of the surface $\mathcal{I}\equiv \mathcal{I}_i\boxtimes_{\mathcal{C}_{i+1}}\mathcal{I}_{i+1}\boxtimes_{\mathcal{C}_{i+2}}\cdots \boxtimes_{\mathcal{C}_{i-1}}\mathcal{I}_{i-1}$ in $\mathrm{TV}_{\mathcal{C}_i}$ for any $i=1,\cdots, n$.} The notation $\boxtimes_{\mathcal{C}_i}$ stands for ``tensor product over $\mathcal{C}_i$'' and is a kind of categorical generalization of tensor product of modules over an algebra; we refer readers to Section 3 of \cite{etingof2010fusion} for the relevant mathematical background.
Lines at the boundary of a surface $\mathcal{I}$ can be thought of as topological line interfaces between the trivial surface $\mathcal{I}_{\mathrm{reg},1}$ and the surface $\mathcal{I}$, which are captured by the category 
\begin{align}\label{eqn:interfacetuberepresentations}
    \gamma\in\mathrm{Jun}(\mathcal{I}_1,\cdots,\mathcal{I}_n)\equiv Z_{\mathcal{C}_1}(\mathcal{I}),
\end{align} 
where $Z_{\mathcal{C}_1}(\mathcal{I})$ is the category $\mathrm{Fun}_{(\mathcal{C}_1,\mathcal{C}_1)}(\mathcal{I}_{\mathrm{reg},1},\mathcal{I})$ of $(\mathcal{C}_1,\mathcal{C}_1)$ bimodule functors from $\mathcal{I}_{\mathrm{reg},1}$ to $\mathcal{I}$. Thus, we learn that the category of representations of $\mathrm{Tube}(\mathcal{I}_1\vert\cdots\vert\mathcal{I}_n)$ can be identified as 
\begin{align}\label{eq:centerbimodle}
    \mathrm{Rep}(\mathrm{Tube}(\mathcal{I}_1\vert\cdots\vert\mathcal{I}_n)) \cong Z_{\mathcal{C}_1}(\mathcal{I}_1\boxtimes_{\CC_2}\mathcal{I}_2\boxtimes_{\CC_2}\cdots\boxtimes_{\CC_n} \mathcal{I}_n).
\end{align}

Another way to determine the representations of $\mathrm{Tube}(\mathcal{I}_1\vert\cdots\vert\mathcal{I}_n)$ is as follows. It is known that $(\mathcal{C}_1,\mathcal{C}_1)$-bimodule categories (and hence, topological surfaces of $\mathrm{TV}_{\mathcal{C}_1}$) are in canonical one-to-one correspondence with $Z(\mathcal{C}_1)$-module categories, see e.g.\ Corollary 2.2 of \cite{Ostrik:2002ohv}. Moreover, the objects of a $Z(\mathcal{C}_1)$-module category can be thought of precisely as line operators which can live at the boundary of the corresponding surface. Thus, if we call $\mathcal{S}$ the $Z(\mathcal{C}_1)$-module category associated to the $(\mathcal{C}_1,\mathcal{C}_1)$-bimodule category $\mathcal{I}_1\boxtimes_{\mathcal{C}_{2}}\cdots \boxtimes_{\mathcal{C}_{n}}\mathcal{I}_{n}$ via this correspondence, then we learn that $\gamma$ is an object of $\mathcal{S}$. 

We turn next to the description of the topological point junctions $x$. When we fuse the topological interfaces $\mathcal{I}_1,\cdots,\mathcal{I}_n$ together, the boundary junctions $\underline{I}_1,\cdots,\underline{I}_n$ go along for the ride. In particular, after fusion, the surface $\mathcal{I}$ terminates on a line junction $\underline{I}\equiv \underline{I}_1\boxtimes\cdots\boxtimes \underline{I}_n\in\mathcal{I}$ supported on the Dirichlet boundary $\mathcal{B}_{\mathrm{reg},1}$. We are interested in the space of point junctions between $\gamma$ and $\underline{I}$. For this, we require a generalization of the forgetful functor, one which describes the result of bringing a line operator $\gamma$ which is attached to a topological surface $\mathcal{I}$ close to the Dirichlet boundary,
\begin{align}
    \input{figures/generalizedforgetful.tikz}
\end{align}
In this case, we claim that the generalized forgetful functor is given by 
\begin{align}
\begin{split}
F_{\mathcal{I}}:\mathrm{Jun}(\mathcal{I}_1,\cdots,\mathcal{I}_n):=\mathrm{Fun}_{(\mathcal{C}_1,\mathcal{C}_1)}(\mathcal{I}_{\mathrm{reg},1},\mathcal{I}) &\to \mathcal{I}=:\mathcal{I}_1\boxtimes_{\mathcal{C}_2}\cdots\boxtimes_{\mathcal{C}_n}\mathcal{I}_n \\
\gamma &\mapsto \gamma(1),
\end{split}
\end{align}
where $1$ is the identity object in $\mathcal{I}_{\mathrm{reg},1}$, and $\gamma(1)$ means ``evaluate the functor $\gamma$ on the identity object of $\mathcal{I}_{\mathrm{reg},1}$''.\footnote{Suppose the Dirichlet boundary has a topological line $a\in\mathcal{C}_1$ inserted on it. As one might expect, $\gamma(a)$ describes the result of fusing $\gamma$ onto the Dirichlet boundary on top of this boundary $a$ line.}
It follows that the space of topological point junctions $x$ between $\gamma$ and $\underline{I}$ is given by $W^\gamma_{\underline{I}_1\cdots\underline{I}_n}\equiv \mathrm{Hom}_{\mathcal{I}}(F_{\mathcal{I}}(\gamma),\underline{I})$.

\subsubsection{Special case: twisted sector operators}

As a sanity check on this mathematical formalism, let us briefly describe how the results of Section \ref{subsec:twistedsectorlocalops} are recovered as a special case. 

In that context, we were interested in twisted sector local operators, i.e.\ local operators which live at the endpoints of topological lines belonging to a fusion category $\mathcal{C}_1$. As discussed earlier, topological lines may be thought of as interfaces between a theory and itself. In particular, we may take $n=1$ and $\mathcal{I}_1=\mathcal{C}_1$ (where $\mathcal{C}_1$ may be thought of as a $(\mathcal{C}_1,\mathcal{C}_1)$-bimodule category in the obvious way, describing the trivial surface of $\mathrm{TV}_{\mathcal{C}_1}$), in which case $\mathrm{Tube}(\mathcal{I}_1\vert\cdots\vert\mathcal{I}_n)$ reduces to the ordinary tube algebra $\mathrm{Tube}(\mathcal{C}_1)$. Equation  \eqref{eqn:interfacetuberepresentations} asserts that the irreducible representations of $\mathrm{Tube}(\mathcal{C}_1)$ are in correspondence with the simple objects of $\mathrm{Fun}_{(\mathcal{C}_1,\mathcal{C}_1)}(\mathcal{C}_1,\mathcal{C}_1)$, since $\mathcal{I}=\mathcal{I}_1=\mathcal{C}_1$. Indeed, we claim that this is consistent with Equation \eqref{eqn:Rep(Tube(C))} because $\mathrm{Fun}_{(\mathcal{C}_1,\mathcal{C}_1)}(\mathcal{C}_1,\mathcal{C}_1)\cong Z(\mathcal{C}_1)$, with the equivalence given by 
\begin{align}
\begin{split}
    Z(\mathcal{C}_1)&\to \mathrm{Fun}_{(\mathcal{C}_1,\mathcal{C}_1)}(\mathcal{C}_1,\mathcal{C}_1) \\
    \gamma &\mapsto (a\mapsto F(\gamma)\otimes a)
\end{split}
\end{align}
where $F:Z(\mathcal{C}_1)\to\mathcal{C}_1$ is the standard forgetful functor.  In other words, $Z_{\mathcal{C}_1}(\mathcal{C}_1)$ reduces to the standard Drinfeld center $Z(\mathcal{C}_1)$ of $\CC_1$, as the notation suggests.

We can also confirm that the representation spaces $W^\gamma$ reduce to those in \eqref{eqn:tuberepspaces}. This follows simply by noticing that the generalized forgetful functor $F_{\mathcal{I}}:\mathrm{Fun}_{(\mathcal{C}_1,\mathcal{C}_1)}(\mathcal{C}_1,\mathcal{I})\to\mathcal{I}$ reduces to the standard forgetful functor $F:Z(\mathcal{C}_1)\to\mathcal{C}_1$ in the case that $n=1$ and $\mathcal{C}_1=\mathcal{I}$. Indeed, we already saw above that $Z(\mathcal{C}_1)\cong \mathrm{Fun}_{(\mathcal{C}_1,\mathcal{C}_1)}(\mathcal{C}_1,\mathcal{C}_1)$, and moreover we furnished the explicit equivalence. Using this, it follows straightforwardly from the definitions that $F_{\mathcal{C}_1}(\gamma)=F(\gamma)$. 

\subsubsection{Special case: boundary-changing local operators}

It will also be useful to treat the special case of boundary-changing local operators, where we can provide alternative formulations of some of the ingredients which enter.

\begin{figure}[t]
	\centering
 {\scriptsize
    \raisebox{-73pt}{\begin{tikzpicture}
			\draw [color=blue, thick, decoration = {markings, mark=at position 0.5 with {\arrow[scale=1]{stealth[reversed]}}}, postaction=decorate] (2,3) -- (2.25,3.25) node[left]{\color{black}$B_1$} -- (2.5,3.5);
			\draw [color=blue, thick, decoration = {markings, mark=at position 0.5 with {\arrow[scale=1]{stealth[reversed]}}}, postaction=decorate] (2.5,3.5) -- (2.75,3.75) node[left]{\color{black}$B_2$} -- (3,4);
			\draw [fill=blue] (2.5,3.5) circle (0.04) node [below] {{\color{black}$\cO$}};
			\draw[dashed](3,4) -- (3,1) -- (2,0) -- (2,3);
			\draw (2,0) node[below]{$Q$};
	\end{tikzpicture}}
    \quad $\leftrightharpoons$ ~
	\raisebox{-73pt}{\begin{tikzpicture}
             \fill[fill=red!40,opacity=.3] (0-1,3)  -- (3-1,3)  -- (4-1,4) -- (1-1,4) -- cycle;
			\draw (1.35-.5,0.8) node[below]{\small $\TV_\cC$};
			\draw[color=black, dashed] (1-1,4) -- (1-1,1) -- (0-1,0) -- (0-1,3);
			\draw [color=blue, thick, decoration = {markings, mark=at position 0.5 with {\arrow[scale=1]{stealth[reversed]}}}, postaction=decorate] (0-1,3) -- (0.25-1,3.25) node[left]{\color{black}$\underline{B}_1$} -- (0.5-1,3.5);
			\draw [color=blue, thick, decoration = {markings, mark=at position 0.5 with {\arrow[scale=1]{stealth[reversed]}}}, postaction=decorate] (0.5-1,3.5) -- (0.75-1,3.75) node[left]{\color{black}$\underline{B}_2$} -- (1-1,4);
			\draw (-1,0) node[below]{$\mathcal{B}_{\mathrm{reg}}$};
			\draw [color=blue, thick, decoration = {markings, mark=at position 0.6 with {\arrow[scale=1]{stealth}}}, postaction=decorate] (2.5,3.5) -- (1.5-.25,3.5) node[below]{\color{black}$\alpha$} -- (0.5-1,3.5);
   \draw (.5,3.55) node[below]  {$\mathcal{B}_1$};
   \draw (1,4.1) node[below]  {$\mathcal{B}_2$};
			\draw [fill=black] (0.5-1,3.5) circle (0.04) node [below] {$x$};
			\draw[color=black, preaction={draw=white,line width=3pt}, dashed] (2,0) -- (2,3);
			\draw[color=black, dashed] (3,4) -- (3,1) -- (2,0);
			\draw (2,0) node[below]{ $\widetilde{Q}$};
			\draw[color=black, dashed] (0-1,3) -- (2,3);
			\draw [color=teal, thick, decoration = {markings, mark=at position 0.5 with {\arrow[scale=1]{stealth[reversed]}}}, postaction=decorate] (2,3) -- (2.25,3.25) node[left]{\color{black} $\widetilde{B}_1$} -- (2.5,3.5);
			\draw [color=teal, thick, decoration = {markings, mark=at position 0.5 with {\arrow[scale=1]{stealth[reversed]}}}, postaction=decorate] (2.5,3.5) -- (2.75,3.75) node[left]{\color{black} $\widetilde{B}_2$} -- (3,4);
			\draw[color=black, dashed ] (1-1,4) -- (3,4);
			\draw [fill=black] (2.5,3.5) circle (0.04) node [below] {$\widetilde{\cO}$};
	\end{tikzpicture}}
 }
	\caption{
 The SymTFT picture of boundary-changing local operators.
	}
	\label{fig:bdydefect.operators}
\end{figure}
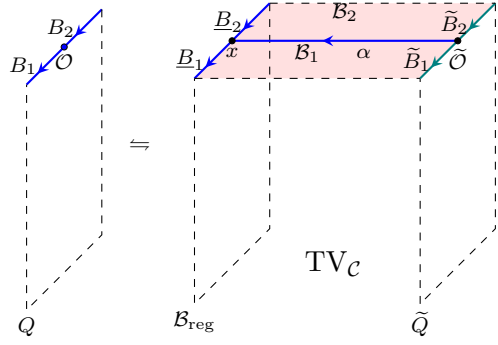

Suppose we have a QFT $Q$ with symmetry category $\mathcal{C}$, which admits two left $\CC$-module categories $\mathcal{B}_1$ and $\mathcal{B}_2$ of boundary conditions.
We wish to study boundary local operators $\mathcal{O}(x)$ of $Q$ which interpolate between two boundary conditions $B_1\in\mathcal{B}_1^\vee$ and $B_2\in\mathcal{B}_2$. Such operators belong to an extended Hilbert space which we write as
\begin{align}
    \mathcal{H}_{\mathcal{B}_1^\vee\mathcal{B}_2} = \bigoplus_{B_1\in\mathcal{B}_1^\vee}\bigoplus_{B_2\in\mathcal{B}_2} \mathcal{H}_{B_1B_2}
\end{align}
where $\mathcal{H}_{B_1B_2}$ is the Hilbert space of boundary-changing local operators between the boundary conditions $B_1$ and $B_2$.
The SymTFT picture of such operators is depicted in Figure \ref{fig:bdydefect.operators}, though we emphasize that it is simply a special case of Figure \ref{fig:junctionsymTFT}. For completeness, we spell out the ingredients explicitly once more.

\begin{figure}
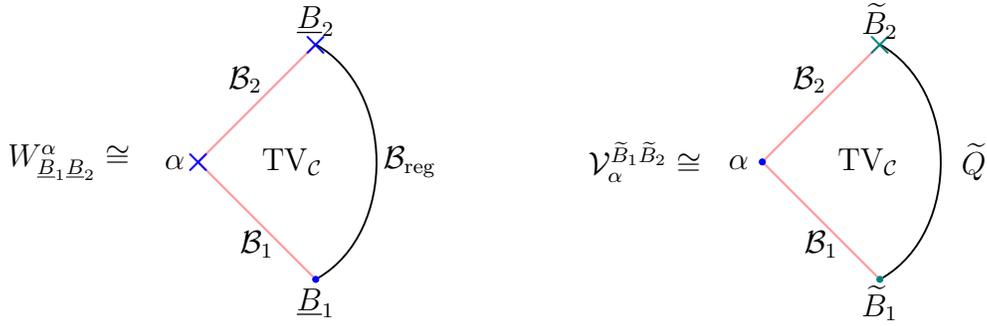

\ctikzfig{figures/boundarySymTFTHilbertspaces}
\caption{The description $W^\alpha_{\underline{B}_1\underline{B}_2}$ and $\mathcal{V}_\alpha^{\widetilde{B}_1\widetilde{B}_2}$ as Hilbert spaces of states of the SymTFT.}\label{fig:SymTFTboundaryHilbertspaces}
\end{figure}

\begin{enumerate}[label=\arabic*)]
\item As before, $Q$ can be blown up into a 2+1d sandwich, with the bread and the plant-based meat alternative forming the following triple,
\begin{align}
    Q\leftrightharpoons (\mathcal{B}_{\mathrm{reg}},\mathrm{TV}_{\mathcal{C}},\widetilde{Q}).
\end{align}
\item The two boundary conditions $B_1$ and $B_2$ of $Q$ expand as 
\begin{align}
    B_i \leftrightharpoons (\underline{B}_i,\mathcal{B}_i,\widetilde{B}_i),
\end{align}
where $i=1,2$.
Here, $\mathcal{B}_i$ is a topological boundary condition of $\mathrm{TV}_{\mathcal{C}}$, which can be described either by the abstract $\mathcal{C}$-module category $\mathcal{B}_i$, or by an algebra object $A_i$ of $\mathcal{C}$ such that $\mathcal{B}_i\cong \mathcal{C}_{A_i}$. This topological boundary meets at a (not necessarily topological) line junction $\widetilde{B}_i$ with the physical boundary $\widetilde{Q}$, and at a topological line junction $\underline{B}_i$ with the Dirichlet boundary $\mathcal{B}_{\mathrm{reg}}$. We recall that $\underline{B}_i$ can be thought of as an object of the category $\mathcal{B}_i$ because the category of line interfaces between the Dirichlet boundary condition and the $\mathcal{B}_i$ boundary condition is equivalent to $\mathcal{B}_i$. In particular, we may also think of $\underline{B}_i$ as a right $A_i$-module, using $\mathcal{B}_i\cong \mathcal{C}_{A_i}$.
\item Finally, each operator $\mathcal{O}\in \mathcal{H}_{B_1B_2}$ decomposes as 
\begin{align}\label{eqn:tripleboundaryops}
    \mathcal{O}\leftrightharpoons (x,\alpha,\widetilde{\mathcal{O}}), \qquad  x \in W^\alpha_{\underline{B}_1\underline{B}_2}, \quad \alpha\in \mathrm{Jun}(\mathcal{B}_1,\mathcal{B}_2),\quad \widetilde{\mathcal{O}}\in\mathcal{V}_\alpha^{\widetilde{B}_1\widetilde{B}_2}.
\end{align}
Here, $\alpha$ is a topological line interface which interpolates between the two topological boundaries $\mathcal{B}_1$ and $\mathcal{B}_2$ of $\mathrm{TV}_{\mathcal{C}}$. The local operator $\widetilde{\mathcal{O}}$ sits at the junction between the line interfaces $\widetilde{B}_1$, $\alpha$, and $\widetilde{B}_2$, and similarly $x$ sits at the junction between the line interfaces $\underline{B}_1$, $\alpha$, and $\underline{B}_2$.
\end{enumerate}
The implication of \eqref{eqn:tripleboundaryops} is that the extended Hilbert space $\mathcal{H}_{\mathcal{B}_1^\vee\mathcal{B}_2}$ of boundary-changing local operators between boundary conditions in $\mathcal{B}_1^\vee$ and  boundary conditions in $\mathcal{B}_2$ admits a decomposition of the form 
\begin{align}\label{eqn:extendedboundaryhilbertspacedecomposition}
    \mathcal{H}_{\mathcal{B}_1^\vee\mathcal{B}_2} = \bigoplus_{\alpha\in\mathrm{Jun}(\mathcal{B}_1,\mathcal{B}_2)} W^\alpha \otimes \mathcal{V}_\alpha^{\widetilde{B}_1\widetilde{B}_2} \,, \ \ \ \ W^\alpha\equiv \bigoplus_{B_1\in\mathcal{B}_1^\vee}\bigoplus_{B_2\in\mathcal{B}_2}W^\alpha_{\underline{B}_1\underline{B}_2}.
\end{align}
As a special case of Figure \ref{fig:junctionophilbertspaces}, we note that, in theories with a state/operator correspondence, the spaces $\mathcal{V}_\alpha^{\widetilde{B}_1\widetilde{B}_2}$ and $W^\alpha_{\underline{B}_1\underline{B}_2}$ can be described as Hilbert spaces of states of the SymTFT. See Figure \ref{fig:SymTFTboundaryHilbertspaces}.

It is known that the category $\mathrm{Jun}(\mathcal{B}_1,\mathcal{B}_2)$ to which $\alpha$ belongs can be described mathematically in the following two ways 
\begin{align}\label{eqn:twodescriptionsjunction}
   \mathrm{Jun}(\mathcal{B}_1,\mathcal{B}_2)= {_{A_1}}\mathcal{C}_{A_2}\cong \mathrm{Fun}_{\mathcal{C}}(\mathcal{B}_1,\mathcal{B}_2) \,.
\end{align}
The mathematical interpretation of this category is that it is the representation category of the boundary tube algebra, $\mathrm{Tube}(\mathcal{B}_1^\vee\vert \mathcal{B}_2)$. We note that, when $\mathcal{B}_1\cong \mathcal{B}_2$, the representation category actually admits the structure of a fusion category, in which case one should replace $\mathrm{Fun}(\mathcal{B}_1,\mathcal{B}_1)$ with $\mathrm{Fun}(\mathcal{B}_1,\mathcal{B}_1)^{\mathrm{op}}$; one way to see that one obtains a fusion category is that ${_A}\mathcal{C}_A$ is precisely the dual symmetry category which arises if one gauges $A$ in a theory with $\mathcal{C}$ symmetry. This is in accordance with the fact that $\mathrm{Tube}(\mathcal{B}_1^\vee\vert\mathcal{B}_2)$ is a weak $C^\ast$ Hopf algebra when $\mathcal{B}_1\cong\mathcal{B}_2$ (see Section \ref{subsec:interfacetube}), since it is known that the representation categories of weak $C^\ast$ Hopf algebras are fusion by a version of Tannaka duality. 

One interesting feature we draw attention to, which holds even away from the special case that $\mathcal{B}_1\cong\mathcal{B}_2$, is that both of the descriptions in \eqref{eqn:twodescriptionsjunction} imply a natural notion of ``tensor product'' of representations of boundary tube algebras. Indeed, given a representation $\alpha_1$ of $\mathrm{Tube}(\mathcal{B}_1^\vee\vert\mathcal{B}_2)$ and a representation $\alpha_2$ of $\mathrm{Tube}(\mathcal{B}_2^\vee\vert\mathcal{B}_3)$, we claim that one can form a representation $\alpha_1\otimes \alpha_2$ of $\mathrm{Tube}(\mathcal{B}_1^\vee\vert \mathcal{B}_3)$. If we think of $\alpha_1$ as an $A_1$-$A_2$ bimodule and $\alpha_2$ as an $A_2$-$A_3$ bimodule, then  $\alpha_1\otimes \alpha_2$ is the $A_1$-$A_3$ bimodule obtained by tensoring over $A_2$, see e.g.\ \cite{Bhardwaj:2017xup} for a description. Alternatively, if we think of $\alpha_1$ as a functor $\mathcal{B}_1\to\mathcal{B}_2$ and $\alpha_2$ as a functor $\mathcal{B}_2\to \mathcal{B}_3$, then $\alpha_1\otimes \alpha_2$ is the functor $\mathcal{B}_1\to \mathcal{B}_3$ obtained by composition. Physically, this tensor product is describing the transformation properties with respect to $\mathrm{Tube}(\mathcal{B}_1^\vee\vert\mathcal{B}_3)$ of boundary-changing local operators which arise in the OPE $\mathcal{O}_1\times \mathcal{O}_2$, where $\mathcal{O}_1\in\mathcal{H}_{\mathcal{B}_1^\vee\mathcal{B}_2}$ and $\mathcal{O}_2\in\mathcal{H}_{\mathcal{B}_2^\vee\mathcal{B}_3}$. The existence of this tensor product could have been anticipated from the ``collective co-algebra'' structure on boundary tube algebras  discussed in Section \ref{subsec:interfacetube}. 

Appealing to the descriptions in \eqref{eqn:twodescriptionsjunction} also makes it easier to define the components $W^\alpha_{\underline{B}_1\underline{B}_2}$ of the representation spaces $W^\alpha$. Indeed, recall that we may think of $\underline{B}_1$ either as an object of the category $\mathcal{B}_1$, or as a right $A_1$-module, and similarly for $\underline{B}_2$. If we think about $\underline{B}_1$ and $\underline{B}_2$ in the former way, and think of $\alpha$ as a functor $\mathcal{B}_1\to\mathcal{B}_2$, then we may define 
\begin{align}
    W^\alpha_{\underline{B}_1\underline{B}_2}=\mathrm{Hom}_{\mathcal{B}_2}(\alpha(\underline{B}_1),\underline{B}_2),
\end{align} 
i.e.\ the dimension of $W^\alpha_{\underline{B}_1\underline{B}_2}$ is the number of times $\underline{B}_2$ appears in the decomposition of $\alpha(\underline{B}_1)$ into simple objects of $\mathcal{B}_2$. On the other hand, if we think of $\underline{B}_1$ and $\underline{B}_2$ as $A_1$ and $A_2$ modules, respectively, and $\alpha$ as an $A_1$-$A_2$ bimodule, then we may similarly define 
\begin{align}
    W^\alpha_{\underline{B}_1\underline{B}_2} = \mathrm{Hom}_{\mathcal{C}_{A_2}}(\underline{B}_1\otimes_{A_1}\alpha,\underline{B}_2)
\end{align}
where again $\underline{B}_1\otimes_{A_1}\alpha$ refers to the tensor product over $A_1$. Again, the dimension of $W^\alpha_{\underline{B}_1\underline{B}_2}$ is equal to the number of times which $\underline{B}_2$ arises in the decomposition of $\underline{B}_1\otimes_{A_1}\alpha$ into simple right $A_2$-modules. 

Finally, we note that, when $\mathcal{B}_1=\mathcal{B}_2$, the discussion below \eqref{eqn:schurweyldecomp} generalizes to the present situation. Namely,  $\mathrm{Jun}(\mathcal{B}_1,\mathcal{B}_2)\cong\mathrm{Rep}(\mathrm{Tube}(\mathcal{B}_1^\vee\vert\mathcal{B}_2))$ is a fusion category so that there is a trivial representation of $\mathrm{Tube}(\mathcal{B}_1^\vee\vert\mathcal{B}_2)$, and the space $\mathcal{V}_{1}^{\widetilde{B}_1\widetilde{B}_2}$ defines an operator subalgebra of $\mathcal{H}_{\mathcal{B}_1^\vee\mathcal{B}_2}$. The spaces $\mathcal{V}_{\mu}^{\widetilde{B}_1\widetilde{B}_2}$ then define irreducible representations of $\mathcal{V}_{1}^{\widetilde{B}_1\widetilde{B}_2}$, and \eqref{eqn:extendedboundaryhilbertspacedecomposition} becomes a kind of Schur-Weyl decomposition of the extended Hilbert space into irreducible $(\mathcal{V}_{1}^{\widetilde{B}_1\widetilde{B}_2}, \mathrm{Tube}(\mathcal{B}_1^\vee\vert\mathcal{B}_2))$ modules.

\section{Partition functions of the SymTFT} \label{sec:matrix}

The SymTFT picture developed in Section \ref{sec:symTFT} allows us to effectively compute the matrix elements of generalized tube algebras in a given representation.
In particular, they are obtained by performing the path integral of SymTFT on various geometries in the presence of bulk and boundary topological line insertions, and they can often be computed explicitly in terms of the category-theoretic data associated with the symmetry fusion category $\mathcal{C}$, which are discussed in Section \ref{sec:review}.

Below, we compute the matrix elements of ordinary and boundary tube algebra representations, which will be used in later sections.
The general case involving interfaces can be treated similarly.
Furthermore, we discuss the ``quantum characters'' of tube algebra representations, which are again given by certain SymTFT partition functions.
In particular, we show that the characters for a boundary tube algebra satisfy a version of Verlinde formula.

\subsection{Matrix elements of generalized tube algebras}

\subsubsection{Ordinary tube algebra}

Recall that the irreducible representations of the tube algebra $\mathrm{Tube}(\mathcal{C})$ of a fusion category $\mathcal{C}$ are labeled by the bulk simple topological lines $\mu \in \mathrm{Irr}(Z(\mathcal{C}))$ \cite{Lin:2022dhv}.
As reviewed in Section \ref{subsec:twistedsectorlocalops}, the representation spaces are given by $W^\mu \equiv \bigoplus_{a \in \mathrm{Irr}(\mathcal{C})} W_a^\mu$, where $W_a^\mu \equiv \mathrm{Hom}_{\mathcal{C}} (F (\mu),a)$. 
We now show that the matrix elements in these representations are given by half-braiding $\Omega$-symbols \eqref{eq:half-braiding}. 

Recall that, as shown in Figure \ref{fig:twistedsectorhilbertspaces}, the vector space $W_a^\mu$ is the Hilbert space of the SymTFT defined on a disk $D^2$, where we impose the Dirichlet boundary condition $\mathcal{B}_{\mathrm{reg}}$ on the boundary $\partial D^2$, with $\mu$ and $a$ punctures inside and on the boundary of the disk, respectively.
Basis states inside this Hilbert space are prepared by the following path integral of the SymTFT on a solid hemisphere:
\begin{equation} \label{eq:x_state}
    \ket{x,\mu,a} = \input{figures/Wmustates.tikz} \in W_a^\mu \,.
\end{equation}
We define the corresponding ``bra'' states by the path integral
\begin{equation} \label{eq:x_state_bra}
    \bra{x,\mu,a} = \input{figures/Wmustatesbra.tikz} \in \left( W_a^\mu \right)^* \,.
\end{equation}

Now, consider a lasso operator $\mathsf{L}_{a,c}^{b, dyz} \in \mathrm{Tube}(\mathcal{C})$ given in \eqref{eq:lasso}.
Its matrix elements in an irreducible representation $W^\mu$ labeled by the bulk topological line $\mu$ can be explicitly computed as follows.
First, note that there is a natural inner product between the above ket and bra states which can be defined by a path integral of the SymTFT on a solid three-dimensional ball,\footnote{We always normalize the SymTFT partition functions on a solid three-dimensional ball with a topological boundary condition imposed to be 1, when there are no topological lines inserted both in the bulk and on the boundary, by tuning the boundary Euler counterterm.}
\begin{align} \label{eq:xinner}
    \langle x', \mu', a'|x, \mu, a\rangle = \delta_{aa'} \delta_{\mu \mu'}\input{figures/Wmuinnerproduct.tikz}=\sqrt{\d_\mu \d_a}\delta_{aa'}\delta_{\mu \mu'}\delta_{xx'}\,,
\end{align}
where we have used the orthogonality relation \eqref{eq:basis_bulk_to_bdy} to shrink the topological lines to a point. 
Using this inner product, we obtain
\begin{align}
    \langle x', \mu, a'|\mathsf{L}_{a,c}^{b,dyz}|x, \mu, a\rangle &=\delta_{a'b}~\input{figures/tubematrixelement.tikz}=\delta_{a'b} \sqrt{\d_c\d_d\d_\mu} [\Omega^d_{c\mu}]_{(axy)(bx'z)}
\end{align}
and therefore conclude that 
\begin{align} \label{eq:bulk_tube_matrix_elem}
    \mathsf{L}_{a,c}^{b, dyz}|x, \mu, a\rangle = \sqrt{\frac{\d_c\d_d}{\d_b}} \sum_{x' =1}^{\langle \mu , b \rangle} [\Omega^d_{c\mu}]_{(axy)(bx'z)}|x', \mu, b\rangle \,.
\end{align}
Above, we deformed the boundary lines in the path integral until they became concentrated at the south pole, used the definition of the generalized half-linking numbers \eqref{eqn:halflinkingdefn}, and then finally the relation to the $\Omega$-symbols given in \eqref{eq:psi_DD}.

We note that, thanks to \eqref{eq:ABF} and \eqref{eqn:modularhalfbraiding}, when the fusion category $\mathcal{C}$ admits the structure of an MTC, \eqref{eq:bulk_tube_matrix_elem} gives an explicit expression for the matrix elements of the tube algebra in a given representation in terms of the $F$- and $R$-symbols of $\mathcal{C}$, which are known for a variety of categories (see e.g.\ \cite{rowell2009classification}).

\subsubsection{Boundary tube algebra}

Let us repeat the kind of analysis we carried out for $\mathrm{Tube}(\mathcal{C})$ in the case of the boundary tube algebra $\mathrm{Tube}(\mathcal{B}_1^\vee \vert \mathcal{B}_2)$. 
We recall that in this case, the irreducible representations are labeled by topological line interfaces $\alpha$ between the $\mathcal{B}_1$ and $\mathcal{B}_2$ topological boundary conditions of the SymTFT $\mathrm{TV}_{\mathcal{C}}$.

As discussed in Section \ref{sec:symTFT}, the representation spaces for the boundary tube algebra are $W^\alpha=\bigoplus_{\underline{B}_1 \underline{B}_2} W^\alpha_{\underline{B}_1 \underline{B}_2}$.
Physically, $W^\alpha_{\underline{B}_1 \underline{B}_2}$ is the disk Hilbert space of the SymTFT where the boundary of the disk is decorated as in Figure \ref{fig:SymTFTboundaryHilbertspaces}.
Similar to before, we prepare basis states inside $W^\alpha_{\underline{B}_1 \underline{B}_2}$ by the following path integral of the SymTFT on a solid hemisphere:
\begin{equation} \label{eq:y_state}
    \ket{y,\alpha,\underline{B}_1 \underline{B}_2} = ~~\scalebox{0.8}{\input{figures/Walphastates.tikz}} ~~ \in W^\alpha_{\underline{B}_1 \underline{B}_2} \,.
\end{equation}
On the boundary of the solid hemisphere, we have the three topological line interfaces $\alpha$, $\underline{B}_1$, and $\underline{B}_2$, dividing the boundary into three topological boundary conditions $\mathcal{B}_1$, $\mathcal{B}_2$, and $\mathcal{B}_{\mathrm{reg}}$.
The three interfaces meet at the south pole where we have the topological junction operator $y$.
The corresponding bra states are defined by the path integral
\begin{equation} \label{eq:y_state_bra}
    \bra{y,\alpha,\underline{B}_1 \underline{B}_2} = ~~\scalebox{0.8}{\raisebox{-45pt}{\input{figures/Walphastatesbra.tikz}}}~~ \in \left( W^\alpha_{\underline{B}_1 \underline{B}_2}\right)^* \,.
\end{equation}
Their inner product is given by
\begin{align} \label{eq:yinner}
\begin{split}
    \langle y',\alpha',\underline{B}'_1 \underline{B}'_2|y,\alpha,\underline{B}_1 \underline{B}_2\rangle &= \delta_{\underline{B}_1 \underline{B}'_1} \delta_{\underline{B}_2 \underline{B}'_2} \delta_{\alpha \alpha'}~~\scalebox{0.8}{\raisebox{-45pt}{\input{figures/Walphainnerproduct.tikz}}}\\ &=\sqrt{\qd_{\underline{B}_1} \qd_{\underline{B}_2} \qd_\alpha}\delta_{\underline{B}_1 \underline{B}'_1} \delta_{\underline{B}_2 \underline{B}'_2} \delta_{\alpha \alpha'} \delta_{yy'}\,,
\end{split}
\end{align}
where we have used \eqref{eq:interfaces_comp_ortho}.

The matrix elements for the boundary lasso operator \eqref{eq:Hlassostate} are then given by the following solid ball partition function of the SymTFT:
\begin{equation} \label{eq:bdymatrix}
    \langle y',\alpha,\underline{B}'_1 \underline{B}'_2| \mathsf{H}^{C_1 C_2, z_1 z_2}_{B_1B_2, a}|y,\alpha,\underline{B}_1 \underline{B}_2\rangle =  \delta_{\underline{B}'_1 \underline{C}_1} \delta_{\underline{B}'_2 \underline{C}_2} ~~\raisebox{-45pt}{\input{figures/Walphamatrix.tikz}}~~ \,.
\end{equation}
We now proceed to explicitly compute \eqref{eq:bdymatrix}. 
For simplicity, let us focus on the case that the two boundary multiplets $\mathcal{B}_1$ and $\mathcal{B}_2$ are (kinematically) equivalent, $\mathcal{B}_1\cong \mathcal{B}_2$ (though we will go back to the more general case when we compute characters in the next subsection).
In such a case, $\alpha$ is a topological line in the fusion category $\mathcal{C}' \equiv \mathrm{Fun}_{\mathcal{C}}(\mathcal{B}_1,\mathcal{B}_1)^{\mathrm{op}}$ (see Section \ref{subsec:modulebimodule}), and furthermore, topological interfaces $\underline{B}_i$ are now objects in a $(\mathcal{C},\mathcal{C}')$-bimodule category.
The matrix element of the boundary tube algebra \eqref{eq:bdymatrix} is then determined by the (inverse) $\widetilde{F}_M$-symbols defined in \eqref{eq:middle}.
Specifically, we have
\begin{equation}
    \langle y',\alpha,\underline{B}'_1 \underline{B}'_2| \mathsf{H}^{C_1 C_2, z_1 z_2}_{B_1B_2, a}|y,\alpha,\underline{B}_1 \underline{B}_2\rangle = \delta_{\underline{B}'_1 \underline{C}_1} \delta_{\underline{B}'_2 \underline{C}_2}\sqrt{\qd_a \qd_{\underline{C}_1} \qd_{\underline{C}_2} \qd_\alpha} \left[ 
        ( \widetilde{F}_M )_{ a \underline{C}_2 \alpha }^{\underline{B}_1}
    \right]^{-1}_{(\underline{B}_2 z_2 y)(\underline{C}_1 z_1 y')} \,.
\end{equation}
This is obtained by ``crossing'' the $z_2$ junction through the $y$ junction using the middle associator, and then shrinking the resulting network of interfaces to a point, on the right-hand side of \eqref{eq:bdymatrix}.
Combined with \eqref{eq:yinner}, this also implies
\begin{equation}
    \mathsf{H}^{C_1 C_2, z_1 z_2}_{B_1B_2, a}|y,\alpha,\underline{B}_1 \underline{B}_2\rangle = \sum_{y'=1}^{(\widetilde{N}_R)_{\underline{C}_2 \alpha}^{\underline{C}_1} }
    \sqrt{\qd_{a}}
    \left[ 
        ( \widetilde{F}_M )_{ a \underline{C}_2 \alpha }^{\underline{B}_1}
    \right]^{-1}_{(\underline{B}_2 z_2 y)(\underline{C}_1 z_1 y')} |y',\alpha,\underline{C}_1 \underline{C}_2\rangle \,.
\end{equation}

\subsection{Quantum characters and a generalized Verlinde formula}\label{subsec:quantumcharacters}

Intuition from the representation theory of finite groups tells us that there should be a kind of ``character theory'' for generalized tube algebras.
Below, we discuss characters of ordinary and boundary tube algebras, which appear in later sections when we discuss symmetry-resolved torus and annulus partition functions of a 1+1d CFT.
In particular, characters for boundary tube algebras and the associated generalized Verlinde formula play an important role in the study of twisted sector boundary states, which we discuss in Section \ref{sec:boundariesclosedstring}.

\subsubsection{Ordinary tube algebra}

We first discuss the characters of an ordinary tube algebra.
Given an element $\mathsf{L} \in \mathrm{Tube}(\mathcal{C})$, and an irreducible representation $W^\mu = \bigoplus_{a \in \mathrm{Irr}(\cal C)} W_a^\mu$, we define the character of $\mathsf{L}$ in the representation $W^\mu$ as the trace of $\mathsf{L}$ over $W^\mu$.
In particular, the character for a lasso operator $\mathsf{L}_{a,c}^{b,dyz}$ given in \eqref{eq:lasso} is
\begin{equation} \label{eq:character_trace}
    \left[\chi_\mu \right]_{a}^{c,dyz} \equiv \mathrm{Tr}_{W_a^\mu} \left( \mathsf{L}_{a,c}^{a,dyz} \right) \,,
\end{equation}
which we call the ``$a$-twisted quantum character'' of the representation $W^\mu$ of $\mathrm{Tube}(\mathcal{C})$. 
Recalling that $W_a^\mu$ is the Hilbert space of $\mathrm{TV}_{\mathcal{C}}$ on a disk $D^2$ with Dirichlet boundary condition $\mathcal{B}_{\mathrm{reg}}$ imposed on $\partial D^2$, with $\mu$ puncturing the origin of $D^2$ and $a$ puncturing a point on $\partial D^2$, we can obtain an expression for the $a$-twisted quantum character as a path integral of the SymTFT on a solid torus with various insertions,
\begin{equation} \label{eq:character}
    \left[\chi_\mu \right]_{a}^{c,dyz} 
    =~\raisebox{-45pt}{
    \begin{tikzpicture}
\draw[thick, decoration = {markings, mark=at position -0.3 with {\arrow[scale=1.5]{stealth}}}, postaction=decorate] (1.5,0) -- (2.025,0.975);
\draw[thick, decoration = {markings, mark=at position -0.3 with {\arrow[scale=1.5]{stealth}}}, postaction=decorate]  (0.975,2.025) -- (0,1.5);
\draw[thick, decoration = {markings, mark=at position -0.3 with {\arrow[scale=1.5]{stealth[reversed]}}}, postaction=decorate] (2.025,0.975) -- (3,1.5);
\draw[thick, decoration = {markings, mark=at position -0.3 with {\arrow[scale=1.5]{stealth}}}, postaction=decorate] (0.975,2.025) -- (1.5,3);
\draw[thick, decoration = {markings, mark=at position -0.3 with {\arrow[scale=1.5]{stealth}}}, postaction=decorate] (2.025,0.975) -- (0.975,2.025);
\draw [red, thick, preaction={draw=white,line width=6pt}, decoration = {markings, mark=at position .8 with {\arrow[scale=1.5]{stealth}}}, postaction=decorate] (2.3,3) to (2.3,0);
\draw[] (1.7,0.9) node[] {$y$};
\draw[] (0.9,1.7) node[] {$\bar{z}$};
\draw[] (2.75,1.7) node[] {$c$};
\draw[] (1.3,.15) node[] {$a$};
\draw[] (1.5,1.9) node[] {$d$};
\draw[] (2.7,0.3) node[] {$\mu$};
\draw[] (0.3,0.3) node[] {$\mathcal{B}_{\mathrm{reg}}$};
\draw[] (1.9,1.05) node[] {$+$};
\draw[] (1.05,1.95) node[] {$+$};
\draw[dashed] (0,0) rectangle (3,3);
\end{tikzpicture}} \,.
\end{equation}
Here, the dashed square on the right-hand side represents a solid torus with the Dirichlet boundary condition $\mathcal{B}_{\mathrm{reg}}$ imposed on the boundary.
The vertical direction corresponds to the non-contractible cycle of the solid torus, around which the bulk $\mu$ line wraps.
On the boundary, we have a network of boundary topological lines corresponding to the action of the tube algebra element $\mathsf{L}_{a,c}^{a,dyz}$.

To compute \eqref{eq:character}, we push the bulk $\mu$ line onto the boundary by first using \eqref{eq:half-braiding_2} and then \eqref{eq:basis_bulk_to_bdy}.
After that, we shrink the network of boundary topological lines to a point.\footnote{We also use the fact that $W_a^1$ is empty unless $a=1$.}
We obtain
\begin{equation} \label{eq:tube_character}
    \left[\chi_\mu \right]_{a}^{c,dyz} = \sqrt{\frac{\d_c \d_d}{\d_a}} \sum_{x =1}^{\langle \mu , a \rangle} [\Omega^d_{c\mu}]_{(axy)(axz)} \,,
\end{equation}
which is consistent with the explicit matrix elements in \eqref{eq:bulk_tube_matrix_elem}.

\subsubsection{Boundary tube algebra}

We now discuss the characters of the boundary tube algebra $\mathrm{Tube}(\mathcal{B}_1^\vee \vert \mathcal{B}_2)$.
Similar to before, we define the character as the trace of a boundary tube algebra element over the representation space $W^{\alpha} = \bigoplus_{\underline{B}_1 \underline{B}_2} W^\alpha_{\underline{B}_1 \underline{B}_2}$.
In particular, given the boundary tube algebra generator $\mathsf{H}_{B_1 B_2,a}^{B_1 B_2, y_1 y_2}$ defined in \eqref{eq:Hlassostate}, we define
\begin{equation} \label{eq:character_bdy}
    \left[ \chi_{\alpha} \right]_{\underline{B}_1 \underline{B}_2}^{ay_1 y_2} \equiv \mathrm{Tr}_{W^\alpha_{\underline{B}_1 \underline{B}_2}} \left( \mathsf{H}_{B_1 B_2,a}^{B_1 B_2, y_1 y_2} \right) \,.
\end{equation}
We call it the ``quantum character'' of $\mathsf{H}_{B_1 B_2,a}^{B_1 B_2, y_1 y_2} \in \mathrm{Tube}(\mathcal{B}_1^\vee \vert \mathcal{B}_2)$ in the representation $W^\alpha$.
 
Similar to the ordinary tube algebra case, the fact that the representation space $W^\alpha_{\underline{B}_1 \underline{B}_2}$ is a $D^2$ Hilbert space of the SymTFT decorated with various topological line interfaces on the boundary $\partial D^2$, as shown in Figure \ref{fig:SymTFTboundaryHilbertspaces}, allows us to represent the character geometrically as a partition function of the SymTFT, which can in turn be explicitly computed.
In particular, the character \eqref{eq:character_bdy} is given by the following solid torus partition function of the SymTFT:
\begin{align} \label{eq:bdy_char_solid_torus}
\begin{split}
    \left[ \chi_{\alpha} \right]_{\underline{B}_1 \underline{B}_2}^{ay_1 y_2} = \input{figures/characterverlinde.tikz} \,.
\end{split}
\end{align}
On the boundary of the solid torus, we have three topological line interfaces $\alpha$, $\underline{B}_1$, and $\underline{B}_2$ wrapping around the nontrivial cycle of the solid torus, and they divide the boundary  into three regions, with boundary conditions $\mathcal{B}_1$, $\mathcal{B}_2$, and $\mathcal{B}_{\mathrm{reg}}$ imposed.
There is a topological line $a \in \mathcal{C}$ on the Dirichlet boundary $\mathcal{B}_{\mathrm{reg}}$, stretched between $\mathcal{B}_1$ and $\mathcal{B}_2$, representing the action of $\mathsf{H}_{B_1 B_2,a}^{B_1 B_2, y_1 y_2}$ on the disk Hilbert space $W^\alpha_{\underline{B}_1 \underline{B}_2}$ of the SymTFT.

We have drawn the solid torus in \eqref{eq:bdy_char_solid_torus} in a suggestive way so that we can readily apply the ``collapsing tube'' formula in \eqref{eq:collapsetube}.
We obtain
\begin{equation}
    \left[ \chi_{\alpha} \right]_{\underline{B}_1 \underline{B}_2}^{ay_1 y_2} = \sqrt{S_{11}} \sum_{\mu x y}  {^{\mathcal{B}_1\mathcal{B}_2}}\Psi_{\alpha\alpha (\mu xy)}^{11}~~~\input{figures/characterverlinde2.tikz} \,,
\end{equation}
where now the character is reduced to a solid ball partition function of the SymTFT decorated by various bulk and boundary topological lines and interfaces.
We then proceed by applying \eqref{eqn:halflinkingid1} and \eqref{eqn:halflinkingid1_tilde} to push the bulk $\mu$ line to the boundary, and then shrink the resulting network of topological lines and interfaces on the boundary of the solid ball to a point.
This gives us
\begin{equation} \label{eq:verlinde}
    \left[ \chi_{\alpha} \right]_{\underline{B}_1 \underline{B}_2}^{ay_1 y_2}
    = \sqrt{\qd_a} \sum_{\mu x y y'} \frac{
    {^{\mathcal{B}_1\mathcal{B}_{\mathrm{reg}}}}\widetilde{\Psi}_{\underline{B}_1 \underline{B}_1 (\mu xy')}^{1(ay_1)}    {^{\mathcal{B}_1\mathcal{B}_2}}\Psi_{\alpha\alpha (\mu xy)}^{11}
    {^{\mathcal{B}_2\mathcal{B}_{\mathrm{reg}}}}\Psi_{\underline{B}_2 \underline{B}_2 (\mu yy')}^{1(ay_2)}    }{\sqrt{S_{1 \mu}}} \,.
\end{equation}
That is, the boundary tube algebra characters are determined by the generalized half-linking numbers \eqref{eqn:halflinkingdefn}.

As a special case, consider what happens when one takes $a=1$ to be the trivial topological line.
Then, the character reduces to the dimension of the representation space $W^\alpha_{\underline{B}_1 \underline{B}_2}$ (which is essentially a fusion coefficient), and \eqref{eq:verlinde} becomes
\begin{equation} \label{eq:verlinde_trivial_a}
    \mathrm{dim}_{\mathbb{C}} ( W^\alpha_{\underline{B}_1 \underline{B}_2} )
    = \sum_{\mu x y y'} \frac{
    {^{\mathcal{B}_1\mathcal{B}_{\mathrm{reg}}}}\widetilde{\Psi}_{\underline{B}_1 \underline{B}_1 (\mu xy')}^{11}    {^{\mathcal{B}_1\mathcal{B}_2}}\Psi_{\alpha\alpha (\mu xy)}^{11}
    {^{\mathcal{B}_2\mathcal{B}_{\mathrm{reg}}}}\Psi_{\underline{B}_2 \underline{B}_2 (\mu yy')}^{11}    }{\sqrt{S_{1 \mu}}} \,.
\end{equation}
We refer to \eqref{eq:verlinde} as a generalized Verlinde formula (where $a$ is not necessarily trivial).
The ordinary Verlinde formula as well as some of its generalizations are reproduced as special cases of \eqref{eq:verlinde_trivial_a}, which we now briefly comment on.

Consider the special case that $\mathcal{C}$ is an MTC, and take $\mathcal{B}_1 = \mathcal{B}_2 = \mathcal{B}_{\mathrm{reg}}$.
For instance, this is the setup relevant for Cardy boundary conditions in diagonal RCFTs.
In this case, $\mathrm{dim}_{\mathbb{C}} ( W^\alpha_{\underline{B}_1 \underline{B}_2} )$ becomes a fusion coefficient of $\mathcal{C}$.
Furthermore, on the right-hand side of \eqref{eq:verlinde_trivial_a}, the summation over $\mu$ runs over $\mu = (a, \bar{a})$, where $a \in \mathrm{Irr}(\mathcal{C})$, $\sqrt{S_{1 (a,\bar{a})}} = \mathbb{S}_{1a}$, and the summation over the junctions $x$, $y$, $y'$ is trivial.
Using \eqref{eq:psi_S}, the generalized Verlinde formula \eqref{eq:verlinde_trivial_a} (with trivial $a=1$) then reduces to 
\begin{equation}
    N_{\underline{B}_2 \alpha}^{\underline{B}_1} = \sum_{a} \frac{\mathbb{S}^*_{\underline{B}_1 a} \mathbb{S}_{\alpha a} \mathbb{S}_{\underline{B}_2 a}}{\mathbb{S}_{1a}} \,,
\end{equation}
which is the standard Verlinde formula \cite{Verlinde:1988sn}.

On the other hand, if one takes $\mathcal{B}_1 = \mathcal{B}_{\mathrm{reg}}$ to be the regular module category, while allowing $\mathcal{B}_2 =\mathcal{B}$ to be an arbitrary $\mathcal{C}$-module category, then \eqref{eq:verlinde_trivial_a} becomes a Verlinde formula for NIM-rep coefficients of the module category $\mathcal{B}$ over the fusion category $\mathcal{C}$.
Here, $\mathcal{C}$ does not have to be an MTC. 
This generalizes a result of \cite{Gaberdiel:2002qa}, where such a Verlinde formula for NIM-rep coefficients was obtained for an MTC $\mathcal{C}$.
See also \cite{Lin:2022dhv,Shen:2019wop}.

We note that we may consider the ``dual'' character,
\begin{align}\label{eq:dualbdycharacter_solidtorus}
\begin{split}
    \left[ \widetilde{\chi}_{\alpha} \right]_{\underline{B}_1 \underline{B}_2}^{ay_1 y_2} = \input{figures/characterverlindedual.tikz} \,,
\end{split}
\end{align}
where we flip the orientation of the topological line $a$, compared to \eqref{eq:character_bdy}.
One may compute the dual character similarly as before, by first applying the collapsing tube formula in \eqref{eq:collapse3}, and then using \eqref{eqn:halflinkingid1} and \eqref{eqn:halflinkingid1_tilde}.
We obtain
\begin{equation} \label{eq:verlinde_dual}
    \left[ \widetilde{\chi}_{\alpha} \right]_{\underline{B}_1 \underline{B}_2}^{ay_1 y_2}
    = \sqrt{\qd_a} \sum_{\mu x y y'} \frac{
    {^{\mathcal{B}_1\mathcal{B}_{\mathrm{reg}}}}\Psi_{\underline{B}_1 \underline{B}_1 (\mu xy')}^{1(ay_1)}    {^{\mathcal{B}_1\mathcal{B}_2}}\widetilde{\Psi}_{\alpha\alpha (\mu xy)}^{11}
    {^{\mathcal{B}_2\mathcal{B}_{\mathrm{reg}}}}\widetilde{\Psi}_{\underline{B}_2 \underline{B}_2 (\mu yy')}^{1(ay_2)}    }{\sqrt{S_{1 \mu}}} \,.
\end{equation}
When we take $a=1$ to be the identity line, this also gives an alternative expression for the dimension of the representation space $W^\alpha_{\underline{B}_1 \underline{B}_2}$,
\begin{equation} 
    \mathrm{dim}_{\mathbb{C}} ( W^\alpha_{\underline{B}_1 \underline{B}_2} )
    = \sum_{\mu x y y'} \frac{
    {^{\mathcal{B}_1\mathcal{B}_{\mathrm{reg}}}}\Psi_{\underline{B}_1 \underline{B}_1 (\mu xy')}^{11}    {^{\mathcal{B}_1\mathcal{B}_2}}\widetilde{\Psi}_{\alpha\alpha (\mu xy)}^{11}
    {^{\mathcal{B}_2\mathcal{B}_{\mathrm{reg}}}}\widetilde{\Psi}_{\underline{B}_2 \underline{B}_2 (\mu yy')}^{11}    }{\sqrt{S_{1 \mu}}} \,.
\end{equation}

Finally, we record here an orthogonality relation satisfied by $\chi_\alpha$ and $\widetilde{\chi}_\alpha$, whose proof we postpone to Section \ref{sec:annulus}, 
\begin{align} \label{eq:ortho_character}
    \frac{1}{\dim(\CC)^2}\sum_{\underline{B}_1,\underline{B}_2}\sum_{a,y_1,y_2} [\widetilde{\chi}_\alpha]^{ay_1y_2}_{\underline{B}_1\underline{B}_2} [\chi_\beta]^{ay_1y_2}_{\underline{B}_1\underline{B}_2}=\delta_{\alpha\beta},
\end{align}
where $\dim(\CC)^2=\sum_{a\in\mathrm{Irr}(\CC)} \d_a^2$. This formula vastly generalizes the familiar orthogonality relation satisfied by the standard characters of a finite group. In particular, it reduces to the standard orthogonality relation, 
\begin{align}
    \frac{1}{|G|}\sum_{a\in G} \chi_\alpha(a)^\ast \chi_\beta(a) = \delta_{\alpha\beta}
\end{align}
in the case that $\CC=\mathrm{Vec}_G$, and $\mathcal{B}_1$ and $\mathcal{B}_2$ are both taken to be the canonical $\mathrm{Vec}_G$-module category with 1 simple object. Indeed, under these conditions, the summation over $\underline{B}_1$, $\underline{B}_2$, $y_1$, and $y_2$ trivializes, $\alpha$ is an object of $\mathrm{Rep}(G)$, and the summation over $a$ is a summation over the elements of $G$. See \cite{Bridgeman:2022gdx} for related results.

\section{Symmetry-resolved torus partition functions}\label{sec:torus}

In this section, we apply some of the technology developed in previous sections to define symmetry-resolved torus partition functions of 1+1d CFTs with $\mathcal{C}$ symmetry. 
We start by introducing the \emph{symmetry} and \emph{representation} basis torus partition functions and discussing some of their properties.
In Sections \ref{sec:annulus} and \ref{sec:boundariesclosedstring}, we discuss the case of annulus partition functions and twisted sector boundary states, where the story is enriched by the existence of two different channels in which one may work: the open and closed string channels.

\subsection{The two bases}\label{subsec:torustwobases}
Consider a 1+1d CFT $Q$ which has a fusion category $\mathcal{C}$ as a symmetry. 
As explained in Section \ref{sec:symTFT}, the irreducible representations $W^\mu$ of the tube algebra $\mathrm{Tube}(\mathcal{C})$ under which (twisted sector) local operators of $Q$ transform are labeled by simple line operators of $Z(\mathcal{C})$. 
In particular, from \eqref{eqn:schurweyldecomp}, we see that the ``multiplicity'' with which $W^\mu$ appears in the decomposition of the extended Hilbert space $\mathcal{H}_{\mathcal{C}}$ is $\mathcal{V}_\mu$, which is the Hilbert space of the SymTFT $\mathrm{TV}_{\mathcal{C}}$ on a disk $D^2$ with the anyon $\mu$ puncturing the center, and with the physical boundary condition $\widetilde{Q}$ imposed on $\partial D^2$.\footnote{Of course, $\mathcal{V}_\mu$ is an infinite-dimensional Hilbert space, however it is graded by conformal dimension, $\mathcal{V}_\mu = \bigoplus_{h,\bar h} \mathcal{V}_{\mu,h,\bar{h}}$, with each component $\mathcal{V}_{\mu,h,\bar{h}}$ being finite dimensional if the theory under consideration is compact. So one might more usefully say that $\dim(\mathcal{V}_{\mu,h,\bar{h}})$ counts the number of multiplets of states in $\mathcal{H}_{\mathcal{C}}$ with conformal dimensions $(h,\bar{h})$ transforming in the irreducible representation $W^\mu$ of $\mathrm{Tube}(\mathcal{C})$.}

We define the corresponding $D^2\times S^1$ partition function of $\mathrm{TV}_{\mathcal{C}}$ as the symmetry-resolved torus partition functions of $Q$ in the ``representation basis,'' 
\begin{align} \label{eq:rep_torus}
    \mathbf{Z}_\mu(\tau) \equiv \mathrm{Tr}_{\mathcal{V}_\mu}q^{L_0-c/24}\bar{q}^{\bar{L}_0-\bar{c}/24} \ \ \ q=e^{2\pi i \tau}\, ,
\end{align}
For example, consider the case that $Q$ is a diagonal rational conformal field theory with chiral algebra $V$, and the symmetry category $\mathcal{C} = \mathrm{Rep}(V)$ of its Verlinde lines \cite{Verlinde:1988sn,Petkova:2000ip}.
We write the irreducible representations of the chiral algebra as $V_a$, and the corresponding characters as 
\begin{align}
    \mathrm{ch}_a(q) = \mathrm{Tr}_{V_a}q^{L_0-c/24}.
\end{align}
Then, $\mathcal{C}$ is a modular tensor category \cite{Moore:1988qv}, so that $Z(\mathcal{C})\cong \mathcal{C}\boxtimes\overline{\mathcal{C}}$, and we can express the bulk topological lines as pairs $\mu=(a,b)$. The representation basis partition functions are then 
\begin{align}
    \mathbf{Z}_{(a,b)}(\tau) = \mathrm{ch}_a(q) \overline{\mathrm{ch}_b(q)}.
\end{align}

On the other hand, we may consider torus partition functions in the ``symmetry basis.'' 
That is, we may study the torus partition function in the presence of topological line defects wrapping around the two nontrivial cycles of the torus, which in the case of invertible symmetries corresponds to turning on a classical background gauge field for the global symmetry.
To this end, consider the $S^1$ Hilbert space $\mathcal{H}_a$ of the CFT twisted by a simple topological line $a$.
On this twisted Hilbert space, we have the action of lasso operators $\mathsf{L}^{a,dyz}_{a,c}$.
We define the symmetry-resolved torus partition functions in the ``symmetry basis'' to be
\begin{align} \label{eq:torus_symm}
    Z_a^{c,dyz}(\tau) \equiv \mathrm{Tr}_{\mathcal{H}_a} 
    \left( \mathsf{L}^{a,dyz}_{a,c} q^{L_0-c/24}\bar{q}^{\bar{L}_0-\bar{c}/24} \right)  =   \raisebox{-25pt}{
\begin{tikzpicture}
\draw[] (0,0) rectangle (2,2);
\draw[thick, decoration = {markings, mark=at position -0.3 with {\arrow[scale=1.5]{stealth}}}, postaction=decorate] (1,0) -- (1.35,1-.35);
\draw[thick, decoration = {markings, mark=at position -0.3 with {\arrow[scale=1.5]{stealth}}}, postaction=decorate]  (1-.35,1.35) -- (0,1);
\draw[thick, decoration = {markings, mark=at position -0.3 with {\arrow[scale=1.5]{stealth}}}, postaction=decorate] (1.35,1-.35) -- (1-.35,1+.35);
\draw[thick, decoration = {markings, mark=at position -0.3 with {\arrow[scale=1.5]{stealth}}}, postaction=decorate] (1-.35,1+.35) -- (1,2);
\draw[thick, decoration = {markings, mark=at position -0.3 with {\arrow[scale=1.5]{stealth}}}, postaction=decorate] (2,1) -- (1.35,1-.35);
\draw[] (1.5,0.5) node[] {$y$};
\draw[] (0.5,1.5) node[] {$\bar{z}$};
\draw[] (1.85,1.45) node[below] {$c$};
\draw[] (.6,.15) node[right] {$a$};
\draw[] (.6,.85) node[right] {$d$};
\draw[] (0.93,0.75) node[right] {$+$};
\draw[] (0.41,1.25) node[right] {$+$};
\end{tikzpicture}} \,.
\end{align}
Below, we first discuss a few properties of the symmetry and representation torus partition functions, and then derive the explicit basis transformation formulas relating the two.

\subsection{Modular transformations}\label{subsec:modulartransforms}

It is known that, under modular transformations, the representation basis elements transform into each other according to the modular data of the Drinfeld center $Z(\mathcal{C})$ of $\mathcal{C}$, i.e.\ 
\begin{align}\label{eqn:repbasismodulartransformtorus}
    \mathbf{Z}_\mu(-1/\tau) = \sum_\nu S_{\mu\nu} \mathbf{Z}_\nu(\tau)  \,, \ \ \ \mathbf{Z}_\mu(\tau+1) = e^{2\pi i (\theta_\mu - (c-\bar{c})/24)}\mathbf{Z}_\mu(\tau) \,,
\end{align}
where $S_{\mu\nu}$ is the S-matrix of $Z(\mathcal{C})$ and $\theta_\mu$ is the twist of the anyon $\mu$.

On the other hand, in the symmetry basis, one may derive the identities
\begin{align}\label{eqn:modtranssymbasis}
\begin{split}
    Z_a^{c,dyz}(-1/\tau) &= \sum_{b\in\mathrm{Irr}(\mathcal{C})}\sum_{z'=1}^{N_{ca}^d}\sum_{i,i'=1}^{N_{ab}^c}\sum_{j=1}^{N_{ba}^c}
    [A_{ab}^c]_{i\bar\imath'} [B_d^{ca}]_{\bar{z}z'} \left[F_{ac\bar{a}}^c \right]^{-1}_{(dyz')(bij)}
    Z_c^{\bar{a},bji}(\tau) \,, \\
    Z_a^{c,dyz}(\tau+1) &= e^{-2\pi i (c-\bar c)/24}\sum_{b\in\mathrm{Irr}(\mathcal{C})} \sum_{z'=1}^{N_{ca}^d} \sum_{i=1}^{N_{ab}^c} \sum_{j,j'=1}^{N_{ba}^c} [B_{c\bar{a}}^b]_{j \bar\jmath'} [B_d^{ca}]_{\bar{z}z'} \left[ F_{ac\bar{a}}^c \right]^{-1}_{(dyz')(bij)} Z_a^{b,cij'}(\tau)\,.
\end{split}
\end{align}
Such transformation rules were considered, for instance, in \cite{Perez-Lona:2023djo} for the special case where all the fusion coefficients $N_{ab}^c$ are either 0 or 1, and we derived \eqref{eqn:modtranssymbasis} similarly.
Analogous results also appeared recently in \cite{Perez-Lona:2024sds}.

\subsection{Linear dependencies}\label{subsec:toruslineardependencies}

On kinematic grounds alone, there are not generally any linear dependencies between the different representation basis partition functions $\mathbf{Z}_\mu(\tau)$.
However, in any given CFT, the dynamics might lead to different linear combinations of these basis elements vanishing. For example, suppose that $Q$ is a diagonal rational conformal field theory whose associated modular tensor category $\mathcal{C}$ has an object $a$ which is not self-dual, $a\neq \bar{a}$. In this case, it is known that the corresponding characters of the chiral algebra are the same, $\mathrm{ch}_a(q) = \mathrm{ch}_{\bar a}(q)$, and hence the representation basis partition functions have linear dependencies. The simplest example is the $SU(3)_1$ Wess-Zumino-Witten model, where the modular tensor category is $\mathrm{Vec}_{\mathbb{Z}_3}$ when thought of as a fusion category.

On the other hand, on kinematic grounds alone, the partition functions in the symmetry basis generically have linear dependencies. 
This is a generalization of the familiar fact from the representation theory of finite groups, which says that the character $\chi_R(g)$ of a representation $R$ depends only on the conjugacy class of $g$. In the case of general fusion categories, we can discover analogous linear dependencies by nucleating a simple line on the torus, and then fusing it onto the existing network of lines. Schematically, the manipulations look as follows,
\begin{align} \label{eqn:toruslineardependency}
     \qd_b Z_{a}^{c;(dyz)}(\tau)
= ~\raisebox{-25pt}{\begin{tikzpicture}
		\draw[] (0,0) rectangle (2,2);
		\draw[thick, decoration = {markings, mark=at position -0.3 with {\arrow[scale=1]{stealth}}}, postaction=decorate] (1,0) -- (1.35,1-.35);
		\draw[thick, decoration = {markings, mark=at position -0.3 with {\arrow[scale=1]{stealth}}}, postaction=decorate]  (1-.35,1.35) -- (0,1);
		\draw[thick, decoration = {markings, mark=at position -0.3 with {\arrow[scale=1]{stealth}}}, postaction=decorate] (1.35,1-.35) -- (1-.35,1+.35);
		\draw[thick, decoration = {markings, mark=at position -0.3 with {\arrow[scale=1]{stealth}}}, postaction=decorate] (1-.35,1+.35) -- (1,2);
		\draw[thick, decoration = {markings, mark=at position -0.3 with {\arrow[scale=1]{stealth}}}, postaction=decorate] (2,1) -- (1.35,1-.35);
		\draw[] (1.5,0.5) node[] {$y$};
		\draw[] (0.5,1.5) node[] {$\bar{z}$};
		\draw[] (1.9,0.95) node[below] {$c$};
		\draw[] (.65,.15) node[right] {$a$};
		\draw[] (.7,.8) node[right] {$d$};
		\draw[thick, decoration = {markings, mark=at position -0.3 with {\arrow[scale=1]{stealth}}}, postaction=decorate] (1.5, 1.5) circle (6pt);
  \draw[] (1.1,1.5) node[] {$b$};
  \draw[] (0.93,0.75) node[right] {$+$};
\draw[] (0.41,1.25) node[right] {$+$};
\end{tikzpicture}}~
= ~\raisebox{-25pt}{\begin{tikzpicture}
		\draw[] (0,0) rectangle (2,2);
		\draw[thick, decoration = {markings, mark=at position -0.3 with {\arrow[scale=1]{stealth}}}, postaction=decorate] (1,0) -- (1.35,1-.35);
		\draw[thick, decoration = {markings, mark=at position -0.3 with {\arrow[scale=1]{stealth}}}, postaction=decorate]  (1-.35,1.35) -- (0,1);
		\draw[thick, decoration = {markings, mark=at position -0.3 with {\arrow[scale=1]{stealth}}}, postaction=decorate] (1.35,1-.35) -- (1-.35,1+.35);
		\draw[thick, decoration = {markings, mark=at position -0.3 with {\arrow[scale=1]{stealth}}}, postaction=decorate] (1-.35,1+.35) -- (1,2);
		\draw[thick, decoration = {markings, mark=at position -0.3 with {\arrow[scale=1]{stealth}}}, postaction=decorate] (2,1) -- (1.35,1-.35);
		\draw[thick, decoration = {markings, mark=at position 0.6 with {\arrow[scale=1]{stealth[reversed]}}}, postaction=decorate, bend left, looseness=1.25] (0.7,2) to (0, 1.3);
		\node[] at (.35, 1.6) {};
		\draw[thick, decoration = {markings, mark=at position 0.6 with {\arrow[scale=1]{stealth}}}, postaction=decorate, in=-135, out=-135, looseness=2.50] (1.3,2) to (2, 1.3);
		\node[] at (1.3,1.3) {$b$};
		\draw[thick, decoration = {markings, mark=at position 0.6 with {\arrow[scale=1]{stealth[reversed]}}}, postaction=decorate, in=45, out=45, looseness=2.50] (0,.7) to (.7, 0);
		\node[] at (.7,.7) {};
		\draw[thick, decoration = {markings, mark=at position 0.6 with {\arrow[scale=1]{stealth[reversed]}}}, postaction=decorate, bend left, looseness=1.25] (1.3,0) to (2, 0.7);
		\node[above] at (1.5, 0) {};
        \draw[] (0.93,0.75) node[right] {$+$};
\draw[] (0.41,1.25) node[right] {$+$};
\end{tikzpicture}}\,.
\end{align}
In the last picture, we have suppressed the labels of some of the topological lines and junctions. One may then locally fuse lines which are parallel to each other using completeness relations, and then simplify the results using various $F$-moves.
One then finds that the right-hand side of \eqref{eqn:toruslineardependency} is expressed as a linear combination of the symmetry basis torus partition functions.\footnote{The calculation is similar to the calculation of the plaquette terms in the Hamiltonian of string-net models \cite{Levin:2004mi}.}
The linear relation among the symmetry basis partition functions that we get this way is not always nontrivial, but it demonstrates that generically there exist linear relations among them.
In \cite{Robbins:2024tqf}, such a computation was explicitly carried out for the $\mathrm{Rep}(S_3)$ fusion category, where \eqref{eqn:toruslineardependency} gives rise to a nontrivial linear relation between the symmetry basis torus partition functions.

\subsection{Change of basis}\label{subsec:toruschangeofbasis}

Symmetry-resolved torus partition functions in the two bases are related by a linear transformation.
First, recall that the $a$-twisted Hilbert space $\mathcal{H}_a$ of the QFT $Q$ decomposes as \cite{Lin:2022dhv}
\begin{equation}
    \mathcal{H}_a = \bigoplus_{\mu \in \mathrm{Irr}(Z(\mathcal{C}))} W_a^\mu \otimes \mathcal{V}_\mu \,,
\end{equation}
which we reviewed in Section \ref{sec:symTFT}.
As discussed there, the tube algebra acts on the space $W_a^\mu$, while leaving $\mathcal{V}_\mu$ intact.
On the other hand, the Virasoro algebra of the CFT acts only on $\mathcal{V}_\mu$, but not on $W_a^\mu$.
Therefore, we find that the symmetry basis partition function in \eqref{eq:torus_symm} can be written as a linear combiniation of the representation basis ones \eqref{eq:rep_torus} as follows:
\begin{align}\label{eq:symtoreptorus}
\begin{split}
     Z_a^{c,dyz}(\tau) &= \sum_{\mu \in \mathrm{Irr}(Z(\mathcal{C}))} \left( \mathrm{Tr}_{W_a^\mu}  \mathsf{L}^{a,dyz}_{a,c}  \right) \left( 
        \mathrm{Tr}_{\mathcal{V}_\mu} q^{L_0-c/24}\bar{q}^{\bar{L}_0-\bar{c}/24}
     \right) = \sum_{\mu \in \mathrm{Irr}(Z(\mathcal{C}))} [\chi_\mu]_a^{c,dyz} \mathbf{Z}_\mu (\tau) \\
     &= \sqrt{\frac{\d_c \d_d}{\d_a}} \sum_{\mu \in \mathrm{Irr}(Z(\mathcal{C}))} \sum_{x =1}^{\langle \mu , a \rangle} [\Omega^d_{c\mu}]_{(axy)(axz)} \mathbf{Z}_\mu (\tau) \,,
\end{split}
\end{align}
where we see that such a basis change is governed by the quantum characters $[\chi_\mu]_a^{c,dyz}$ of the tube algebra (and in turn by $\Omega$-symbols), which are defined and computed in Section \ref{sec:matrix}. See \cite{Shen:2019wop,Gaiotto:2020iye,Lin:2022dhv} for similar ideas.

Conversely, we may express the representation basis partition function $\mathbf{Z}_\mu(\tau)$ as a linear combination of symmetry basis ones as follows.
In \cite{Lin:2022dhv}, it was shown that the operator
\begin{equation} \label{eq:bulk_projector}
    P^\mu_a =\sum_{\nu \in \mathrm{Irr}(Z(\mathcal{C}))}  \overline{S}_{\mu\nu} S_{\mu 1} \raisebox{-2em}{
\begin{tikzpicture}
\draw[thick, decoration = {markings, mark=at position -0.3 with {\arrow[scale=1.5]{stealth}}}, postaction=decorate] (1,0) -- (1,2);
\draw[red, thick, preaction={draw=white,line width=6pt}, decoration = {markings, mark=at position -0.3 with {\arrow[scale=1.5]{stealth}}}, postaction=decorate] (2,1) -- (0,1);
\draw[] (1,1.7) node[right] {$a$};
\draw[red] (1.7,1) node[below] {$\nu$};
\draw[] (0,1) node[] {$\doubleslash$};
\draw[] (2,1) node[] {$\doubleslash$};
\end{tikzpicture}}
\end{equation}
acting on the twisted Hilbert space $\mathcal{H}_a $ is the projection operator onto the subspace $\mathcal{H}_a^\mu \equiv  W_a^\mu \otimes \mathcal{V}_\mu \subset \mathcal{H}_a$ which transforms under the fixed representation $\mu$ of the tube algebra.
Here, the horizontal $\nu$ line on the right-hand side indicates the $F(\nu) \in \mathcal{C}$ topological line acting on the twisted Hilbert space $\mathcal{H}_a$ with the choice of the 4-fold topological junction at the middle given by the half-braiding morphism \eqref{eq:hb_morphism} between $F(\nu)$ and $a$.

\begin{figure}[t!]
\centering
    \input{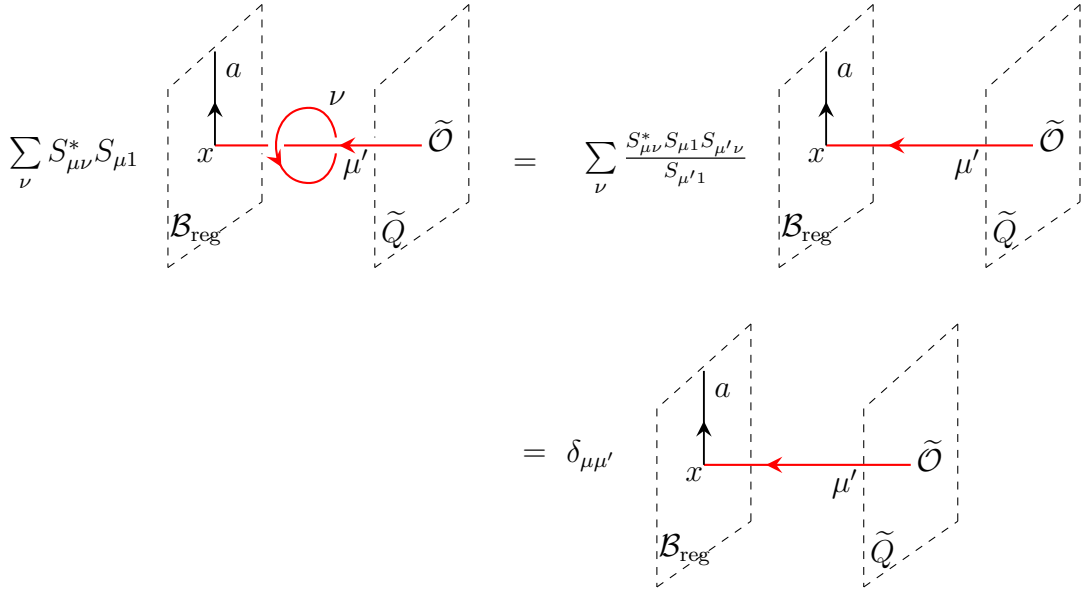}\vspace{.5cm}
\caption{The 2+1d SymTFT description of the projector $P^\mu_a$.}\label{fig:proj_bulk}
\end{figure}

The fact that \eqref{eq:bulk_projector} acts as a projector can be understood from the SymTFT picture.
Suppose that an operator $\mathcal{O}$ belongs to  $\mathcal{H}_a^{\mu'}$.
In the 2+1d SymTFT picture, this operator corresponds to a triple $(x,\mu',\widetilde{\mathcal{O}})$ as in \eqref{eqn:decomptwistedlocalop} and in Figure \ref{fig:defect.operators}.
The action of $P^\mu_a$ is shown in Figure \ref{fig:proj_bulk}, and the operator $\mathcal{O} \leftrightharpoons (x,\mu',\widetilde{\mathcal{O}})$ is mapped to
\begin{equation}
    P^\mu_a : \mathcal{O} \mapsto \delta_{\mu \mu'} \mathcal{O} \,,
\end{equation}
as desired.

Using the projector \eqref{eq:bulk_projector} and the $\Omega$-symbols in \eqref{eq:half-braiding_2} for the half-braiding morphism, we obtain
\begin{align} \label{eq:Hamu}
\begin{split}
    &~~~~~~\mathrm{Tr}_{\mathcal{H}_a^{\mu}} \left( q^{L_0 -c/24} \bar{q}^{\bar{L}_0 - \bar{c}/24} \right)  
    = \sum_{\nu \in \mathrm{Irr}(Z(\mathcal{C}))}  S^*_{\mu\nu} S_{\mu 1} \raisebox{-25pt}{
\begin{tikzpicture}
\draw[thick, decoration = {markings, mark=at position -0.3 with {\arrow[scale=1.5]{stealth}}}, postaction=decorate] (1,0) -- (1,2);
\draw[red, thick, preaction={draw=white,line width=6pt}, decoration = {markings, mark=at position -0.3 with {\arrow[scale=1.5]{stealth}}}, postaction=decorate] (2,1) -- (0,1);
\draw[red] (1.7,1) node[below] {$\nu$};
\draw[] (1,1.7) node[right] {$a$};
\draw[] (0,0) rectangle (2,2);
\end{tikzpicture}} \\
&=  \sum_{\nu \in \mathrm{Irr}(Z(\mathcal{C}))}  \sum_{c,d \in \mathrm{Irr}(\mathcal{C})} \sum_{x=1}^{\langle \nu, c \rangle} \sum_{y=1}^{N_{ac}^d} \sum_{z=1}^{N_{ca}^d}  
S^*_{\mu\nu} S_{\mu 1} \sqrt{\frac{\qd_d}{\qd_a \qd_c}} \left[ \Omega_{a \nu}^d \right]_{(dxz)(dxy)}
\raisebox{-25pt}{
\begin{tikzpicture}
\draw[] (0,0) rectangle (2,2);
\draw[thick, decoration = {markings, mark=at position -0.3 with {\arrow[scale=1.5]{stealth}}}, postaction=decorate] (1,0) -- (1.35,1-.35);
\draw[thick, decoration = {markings, mark=at position -0.3 with {\arrow[scale=1.5]{stealth}}}, postaction=decorate]  (1-.35,1.35) -- (0,1);
\draw[thick, decoration = {markings, mark=at position -0.3 with {\arrow[scale=1.5]{stealth}}}, postaction=decorate] (1.35,1-.35) -- (1-.35,1+.35);
\draw[thick, decoration = {markings, mark=at position -0.3 with {\arrow[scale=1.5]{stealth}}}, postaction=decorate] (1-.35,1+.35) -- (1,2);
\draw[thick, decoration = {markings, mark=at position -0.3 with {\arrow[scale=1.5]{stealth}}}, postaction=decorate] (2,1) -- (1.35,1-.35);
\draw[] (1.5,0.5) node[] {$y$};
\draw[] (0.5,1.5) node[] {$\bar{z}$};
\draw[] (1.85,1.45) node[below] {$c$};
\draw[] (.6,.15) node[right] {$a$};
\draw[] (.6,.85) node[right] {$d$};
\draw[] (0.93,0.75) node[right] {$+$};
\draw[] (0.41,1.25) node[right] {$+$};
\end{tikzpicture}}
\\
&= \sum_{\nu \in \mathrm{Irr}(Z(\mathcal{C}))}  \sum_{c,d \in \mathrm{Irr}(\mathcal{C})} \sum_{x=1}^{\langle \nu, c \rangle} \sum_{y=1}^{N_{ac}^d} \sum_{z=1}^{N_{ca}^d}  
S^*_{\mu\nu} S_{\mu 1} \sqrt{\frac{\qd_d}{\qd_a \qd_c}} \left[ \Omega_{a \nu}^d \right]_{(dxz)(dxy)} Z_a^{c;dyz}(\tau) \\
&= \langle \mu , a \rangle \mathbf{Z}_\mu(\tau) \,,
\end{split}
\end{align}
where the last equality follows from the fact that $\mathcal{H}_a^\mu = W_a^\mu \otimes \mathcal{V}_\mu$ and $\mathrm{dim}_{\mathbb{C}} (W_a^\mu) = \langle \mu , a \rangle$.
Next, by taking the quantum dimensions of both sides of Equation \eqref{eq:forgetmu}, we have 
\begin{equation} \label{eq:dmu}
\qd_\mu = \sum_{a \in \mathrm{Irr}(\mathcal{C})} \langle \mu , a \rangle \qd_a \,.
\end{equation}
Combining \eqref{eq:Hamu} and \eqref{eq:dmu}, we obtain
\begin{align} \label{eq:symm_to_rep}
\begin{split}
    &~~~~~\mathbf{Z}_\mu(\tau) =   \sum_{a \in \mathrm{Irr}(\mathcal{C})} \frac{\qd_a}{\qd_\mu} \mathrm{Tr}_{\mathcal{H}_a^{\mu}} \left( q^{L_0 -c/24} \bar{q}^{\bar{L}_0 - \bar{c}/24} \right) \\
    &=   \sum_{a,c,d \in \mathrm{Irr}(\mathcal{C})} \sum_{y=1}^{N_{ac}^d} \sum_{z=1}^{N_{ca}^d}
    \sqrt{\frac{\qd_a \qd_d}{\qd_c}}
     \left( \sum_{\nu \in \mathrm{Irr}(Z(\mathcal{C}))} \sum_{x=1}^{\langle \nu, c\rangle}
    S^*_{\mu\nu} S_{11}  \left[ \Omega_{a \nu}^d \right]_{(dxz)(dxy)} \right) Z_a^{c,dyz}(\tau) \,.
\end{split}
\end{align}
This is the desired expression for the representation basis partition function $\mathbf{Z}_\mu(\tau)$ as a linear combination of the symmetry basis partition functions $Z_a^{c,dyz}(\tau)$.
Using the modular transformation property of $\mathbf{Z}_\mu(\tau)$, one may equivalently write
\begin{equation} \label{eq:symm_to_rep_2}
    \mathbf{Z}_\mu(-1/\tau) =  S_{11}\sum_{a,c,d \in \mathrm{Irr}(\mathcal{C})}\sum_{y=1}^{N_{ac}^d} \sum_{z=1}^{N_{ca}^d} \sum_{x=1}^{\langle \mu, c \rangle}
    \sqrt{\frac{\qd_a \qd_d}{\qd_c}}  \left[ \Omega_{a \nu}^d \right]_{(dxz)(dxy)}  Z_a^{c,dyz}(\tau) \,.
\end{equation}
We remark that one reproduces the result of \cite{Lin:2022dhv} by taking the high-temperature limit of \eqref{eq:symm_to_rep}, which we do not explicitly write here.

Before moving on, we note that the formalism described thus far implies that the representation basis partition functions $\mathbf{Z}_\mu(\tau)$ for a QFT $Q$ are the same for all of its orbifolds $Q/A$, where $A$ is a gaugeable algebra object of $\mathcal{C}$ \cite{Fuchs:2002cm,Bhardwaj:2017xup}. 
Furthermore, the computation of the symmetry basis partition functions of $Q/A$ can be reduced to the problem of computing the characters
\begin{equation} 
    \left[\chi_\mu^A \right]_{a}^{c,dyz} 
    =~\raisebox{-45pt}{
    \begin{tikzpicture}
\draw[thick, decoration = {markings, mark=at position -0.3 with {\arrow[scale=1.5]{stealth}}}, postaction=decorate] (1.5,0) -- (2.025,0.975);
\draw[thick, decoration = {markings, mark=at position -0.3 with {\arrow[scale=1.5]{stealth}}}, postaction=decorate]  (0.975,2.025) -- (0,1.5);
\draw[thick, decoration = {markings, mark=at position -0.3 with {\arrow[scale=1.5]{stealth[reversed]}}}, postaction=decorate] (2.025,0.975) -- (3,1.5);
\draw[thick, decoration = {markings, mark=at position -0.3 with {\arrow[scale=1.5]{stealth}}}, postaction=decorate] (0.975,2.025) -- (1.5,3);
\draw[thick, decoration = {markings, mark=at position -0.3 with {\arrow[scale=1.5]{stealth}}}, postaction=decorate] (2.025,0.975) -- (0.975,2.025);
\draw [red, thick, preaction={draw=white,line width=6pt}, decoration = {markings, mark=at position .8 with {\arrow[scale=1.5]{stealth}}}, postaction=decorate] (2.3,3) to (2.3,0);
\draw[] (1.7,0.9) node[] {$y$};
\draw[] (0.9,1.7) node[] {$\bar{z}$};
\draw[] (2.75,1.7) node[] {$c$};
\draw[] (1.3,.15) node[] {$a$};
\draw[] (1.5,1.9) node[] {$d$};
\draw[] (2.7,0.3) node[] {$\mu$};
\draw[] (0.3,0.3) node[] {$\mathcal{C}_A$};
\draw[] (1.9,1.05) node[] {$+$};
\draw[] (1.05,1.95) node[] {$+$};
\draw[dashed] (0,0) rectangle (3,3);
\end{tikzpicture}} \,,
\end{equation}
which is represented by a solid torus partition function of the SymTFT.
The above character differs from the one in \eqref{eq:character} in that the boundary of the solid torus is now taken to be the one labeled by the module category $\mathcal{C}_A$, and $a,c,d,\bar{z},y$ are taken to be lines and topological point junctions in the dual symmetry category ${_A}\mathcal{C}_A$.

The characters $\chi^A_\mu$  are essentially given by the $\Omega$-symbols for the dual symmetry category ${_A}\mathcal{C}_A$, similar to \eqref{eq:tube_character}. In the case that these are known, this gives a streamlined prescription to perform generalized orbifolds on a torus (see \cite{Diatlyk:2023fwf,Choi:2023vgk,Perez-Lona:2023djo,Perez-Lona:2024sds} for recent examples of generalized orbifolds).
Namely, the symmetry-basis torus partition functions of the orbifolded theory $Q/A$ are given by
\begin{equation}
     {{}^{Q/A}}Z_a^{c,dyz}(\tau)= \sum_{\mu \in \mathrm{Irr}(Z(\mathcal{C}))} [\chi_\mu^A]_a^{c,dyz} \mathbf{Z}_\mu (\tau)  \,.
\end{equation}
By re-expressing the representation basis partition functions on the right-hand side in terms of the symmetry basis ones of the original theory $Q$ using \eqref{eq:symm_to_rep}, we may further deduce that
\begin{align} \label{eq:torus_generalized_orbifolds}
\begin{split}
&~~~{{}^{Q/A}}Z_{a'}^{c',d'y'z'}(\tau)\\
&= S_{11}\sum_{\mu,\nu \in \mathrm{Irr}(Z(\mathcal{C}))}  \sum_{a,c,d \in \mathrm{Irr}(\mathcal{C})} \sum_{x=1}^{\langle \nu, c\rangle}\sum_{y=1}^{N_{ac}^d} \sum_{z=1}^{N_{ca}^d}  \sqrt{\frac{\qd_a \qd_d}{\qd_c}} [\chi_\mu^A]_{a'}^{c',d'y'z'}
    S^*_{\mu\nu}   \left[ \Omega_{a \nu}^d \right]_{(dxz)(dxy)} {{}^{Q}}Z_a^{c,dyz}(\tau) \,,
\end{split}
\end{align}
which expresses the torus partition functions of the (generalized) orbifold theory $Q/A$, decorated with topological line insertions, as linear combinations of those of the original theory $Q$.
This may be thought of as the non-invertible generalization of the standard formula for finite group orbifolds \cite[Equation (8.17)]{Ginsparg:1988ui}.

\section{Symmetry-resolved annulus partition functions}\label{sec:annulus}

Let $Q$ again be a 1+1d CFT with a fusion category symmetry $\mathcal{C}$, and let $\mathcal{B}_1$ and $\mathcal{B}_2$ be two (possibly distinct) multiplets of conformal boundary conditions, which define left module categories over $\mathcal{C}$.
We use $A_1$ and $A_2$ to denote the algebra objects of $\mathcal{C}$ for which $\mathcal{B}_i \cong \mathcal{C}_{A_i}$. 
Using the case of $T^2$ partition functions as a guide, we move on to the analysis of $S^1\times I$ (i.e.\ conformally an annulus) partition functions, with boundary conditions from $\mathcal{B}_1$ and $\mathcal{B}_2$ imposed at the two ends of the interval $I$. 
Many of the statements made in Section \ref{sec:torus} in the context of the torus have close analogs in the open string channel, though as we will see, there are several new features arising in the closed string channel which make the story  richer. 
This will be discussed in Section \ref{sec:boundariesclosedstring}.
Much of what we say here can likely be extended to the case of $T^2$ partition functions decorated by conformal interfaces, but we stick to annuli for simplicity.

\subsection{Open string channel bases}

Just as for the torus, there are two natural bases in the open string channel: the symmetry basis and the representation basis. We begin with the latter.

Recall that the natural symmetry structure which acts on the (extended) interval Hilbert space is the boundary tube algebra, $\mathrm{Tube}(\mathcal{B}_1^\vee\vert\mathcal{B}_2)$. 
The irreducible representations of $\mathrm{Tube}(\mathcal{B}_1^\vee\vert\mathcal{B}_2)$ are labeled by simple objects $\alpha$ of the category $\mathrm{Fun}_{\mathcal{C}}(\mathcal{B}_1,\mathcal{B}_2) \cong {_{A_1}}\mathcal{C}_{A_2}$ (cf.\ \eqref{eqn:twodescriptionsjunction}). 
More physically, the choice of an $\alpha$ describes a topological line interface which interpolates between the two topological boundary conditions $\mathcal{B}_1$ and $\mathcal{B}_2$ of the SymTFT $\mathrm{TV}_{\mathcal{C}}$. Given such an $\alpha$, one may form the $D^2$ Hilbert space $\mathcal{V}_\alpha^{\widetilde{B}_1\widetilde{B}_2}$ of the SymTFT described in Section \ref{subsec:junctionops} (see Figure \ref{fig:SymTFTboundaryHilbertspaces}), and perform a graded trace over it,
\begin{align} \label{eq:rep_annulus}
    \mathbf{Z}_\alpha (\delta) \equiv \mathrm{Tr}_{\mathcal{V}_\alpha^{\widetilde{B}_1\widetilde{B}_2}}q^{L_0- c/24}\,, \ \ \ \ q=e^{- \pi \delta} \,.
\end{align}
This graded trace can be recast as a path integral of the SymTFT on $D^2\times S^1$, i.e.\ on the solid torus, where $D^2$ is decorated as above, and $S^1$ is the Euclidean time direction.
We call \eqref{eq:rep_annulus} the \emph{representation-basis} annulus partition function, in the open string channel.

In the case of a diagonal rational conformal field theory whose chiral algebra has representation category $\mathcal{C}$, the Cardy boundary conditions transform in the regular module category of $\mathcal{C}$, i.e.\ $\mathcal{B}_1\cong\mathcal{B}_2\cong \mathcal{B}_{\mathrm{reg}}$, and both $A_1$ and $A_2$ are Morita trivial. 
The irreducible representations of the boundary tube algebra are therefore labeled by objects of ${_{1}}\mathcal{C}_{1} \cong \mathcal{C}$. 
In particular, one has that 
\begin{align} \label{eq:rep_char}
    \mathbf{Z}_\alpha(\delta) = \mathrm{ch}_\alpha(q), \ \ \ \alpha\in\mathcal{C}
\end{align}
where $\mathrm{ch}_\alpha(q)$ are the irreducible characters of the chiral algebra of the rational conformal field theory.

We may also contemplate formulating the symmetry-basis partition functions in the open string channel. Given boundary conditions $B_1\in\mathcal{B}_1$ and $B_2\in\mathcal{B}_2$, and a symmetry line $a\in\mathcal{C}$ admitting topological point junctions $\bar{z}_1$ and $z_2$ on $B_1$ and $B_2$, respectively, one obtains an operator $\mathsf{H}_{B_1 B_2,a}^{B_1 B_2, z_1 z_2}$, defined in \eqref{eq:Hlassostate}, which acts on the interval Hilbert space $\mathcal{H}_{B_1 B_2}$. 
We define 
\begin{align}\label{eqn:openstringsymmetrybasis}
    Z_{B_1B_2}^{az_1z_2}(\delta) \equiv  \mathrm{Tr}_{\mathcal{H}_{B_1B_2}}\left(\mathsf{H}_{B_1 B_2,a}^{B_1 B_2, z_1 z_2} q^{L_0- c/24}\right) = 
    \raisebox{-2.5em}
	{\begin{tikzpicture}
			\fill [gray, opacity=0.5] (0,-1) rectangle (-0.6,1); 
			\fill [gray, opacity=0.5] (2,-1) rectangle (2.6,1); 
			\draw[thick, decoration = {markings, mark=at position 0.5 with {\arrow[scale=1.5]{stealth}}}, postaction=decorate] (0,0) -- (2,0);
			\draw[thick, blue] (0,1) -- (0,-1);
			\draw[thick,blue] (2,1) -- (2,-1);
			\node[left] at (0,0.8) {$B_1$};
			\node[right] at (2,0.8) {$B_2$};
			\node[left] at (0,-0.8) {$B_1$};
			\node[right] at (2,-0.8) {$B_2$};
			\node[above] at (1,0) {$a$};
			\node[left] at (0,0) {$\bar{z}_1$};
			\node[right] at (2,0) {$z_2$};
            \draw[dashed] (0,-1) -- (2,-1);
            \draw[dashed] (0,1) -- (2,1);
            \node[] at (1,1) {$\doubleslash$};
            \node[] at (1,-1) {$\doubleslash$};
	\end{tikzpicture}}
    \,,
\end{align}
and refer to them as the \emph{symmetry-basis} annulus partition functions in the open string channel.

Note that we do not need to consider the case that an additional symmetry line wraps around the non-trivial cycle of the annulus, because such a symmetry line can always be pushed onto either the left or the right boundary, and therefore the corresponding partition function can always be expressed as a linear combination of the symmetry basis partition functions of the form appearing in Equation \eqref{eqn:openstringsymmetrybasis}.

\subsection{Linear dependencies}\label{subsec:lineardependencies}
Just as in the case of the torus, the representation basis partition functions are generally independent from one another, though again the dynamics of any given theory might render certain combinations of them vanishing. An easy example is again the $SU(3)_1$ WZW model, taking $\mathcal{B}_1=\mathcal{B}_2$ to be the multiplet of Cardy states. Then, the representation basis partition functions are simply the characters of the corresponding chiral algebra, and  it is known that the two non-vacuum characters of the $SU(3)_1$ current algebra are equal to one another. 

On the other hand, the symmetry basis partition functions generally possess kinematic relationships. They are invariant under ``conjugation,'' which we can see by nucleating a line $b$ and using completeness relations to fuse it onto the boundaries,
\begin{align}
\begin{split}
   \d_b Z_{B_1B_2}^{ax_1x_2}(\delta)&=\input{figures/annuluslineardependencies1.tikz}=\input{figures/annuluslineardependencies2.tikz} \\
   &=\sum_{B_1',B_2'}\sum_{y_1,y_1',y_2,y_2'}\frac{1}{\d_b}\sqrt{\frac{\d_{B_1'}\d_{B_2'}}{\d_{B_1}\d_{B_2}}} [\widetilde{A}_{B_2'}^{bB_2}]_{\bar y_2y_2'}[\widetilde{B}_{B_1\bar b}^{B_1'}]_{y_1y_1'}~\input{figures/annuluslineardependencies3.tikz} \,,
\end{split}
\end{align}
where $\widetilde{A}$ and $\widetilde{B}$ are boundary versions of the $A$- and $B$-symbols defined in Equation \eqref{eq:ABF}, 
\begin{align}
    [\widetilde{A}_{B_2'}^{bB_2}]_{\bar y_2 y_2'} = \sqrt{\frac{\d_a \d_{B'}}{\d_B}} [{^{\mathcal{B}_2}}\widetilde{F}_{\bar a a B'}^{B'}]^{-1}_{1,(By_2'y_2)}\,, \ \ \ \ [\widetilde{B}_{B_1a}^{B_1'}]_{y_1\bar y_1'} = \sqrt{\frac{\d_b \d_{B_1}}{\d_{B_1'}}}[{^{\mathcal{B}_1^\vee}}\widetilde{F}_{B_1 a\bar a}^{B_1}]_{(B_1'y_1y_1')1}\,.
\end{align}
From this, we can deduce that the symmetry-basis annulus partition functions satisfy a ``class function'' property, which says that 
\begin{align}\label{eqn:annulusconjugation}
\begin{split}
    &\mathrm{Tr}_{\mathcal{H}_{B_1B_2}}\left(\mathsf{H}_{B_1B_2,a}^{B_1B_2,x_1x_2}q^{L_0-c/24}\right) =\sum_{B_1',B_2'}\sum_{y_1,y_1',y_2,y_2'}  \frac{1}{\d_b}\sqrt{\frac{\d_{B_1'}\d_{B_2'}}{\d_{B_1}\d_{B_2}}} [\widetilde{A}_{B_2'}^{bB_2}]_{\bar y_2y_2'}[\widetilde{B}_{B_1\bar b}^{B_1'}]_{y_1\bar y_1'} \\
    & \hspace{.4in} \mathrm{Tr}_{\mathcal{H}_{B_1'B_2'}}\left(\mathsf{H}_{B_1'B_2',\bar b}^{B_1B_2,y_1y_2'}\mathsf{H}_{B_1B_2,a}^{B_1B_2,x_1x_2}\mathsf{H}_{B_1B_2,b}^{B_1'B_2',y_1'y_2}q^{L_0-c/24}\right).
\end{split}
\end{align}
For example, in a theory with $\mathrm{Vec}_G$ symmetry and a $G$-symmetric boundary condition $B$, Equation \eqref{eqn:annulusconjugation} reduces to the statement that the symmetry basis annulus partition functions are ``class functions'' on $G$, i.e.\ $Z^a_{BB}(\delta) = Z^{b^{-1}ab}_{BB}(\delta)$. The characters of the tube algebra defined in Section \ref{subsec:quantumcharacters} obey the same class function property.

If one would like, one may also use the multiplication rules on the boundary lassos, Equation \eqref{eq:HHH}, to further reduce the right-hand side of Equation \eqref{eqn:annulusconjugation} to a linear combination of symmetry-basis annulus partition functions. The relations \eqref{eqn:annulusconjugation} are not always nontrivial, but they show that the symmetry basis partition functions are generically linearly dependent.

\subsection{Change of basis}

We now discuss the relationship between the two bases of annulus partition functions, and derive the explicit change of basis formula between them.

Recall the decomposition of the extended interval Hilbert space \eqref{eqn:extendedboundaryhilbertspacedecomposition}.
When we impose the two boundary conditions $B_1$ and $B_2$ on two ends of the interval, the Hilbert space then decomposes as
\begin{equation}
    \mathcal{H}_{B_1 B_2} = \bigoplus_{\alpha}  W_{\underline{B}_1 \underline{B}_2}^\alpha \otimes \mathcal{V}_{\alpha}^{\widetilde{B}_1 \widetilde{B}_2} \,.
\end{equation}
Consider the annulus partition function in the symmetry basis \eqref{eqn:openstringsymmetrybasis}.
The symmetry operator $\mathsf{H}_{B_1 B_2,a}^{B_1 B_2, z_1 z_2}$ acts only on the space $W_{\underline{B}_1 \underline{B}_2}^\alpha$, and leaves $\mathcal{V}_{\alpha}^{\widetilde{B}_1 \widetilde{B}_2}$ untouched.
On the other hand, the Virasoro algebra acts only on $\mathcal{V}_{\alpha}^{\widetilde{B}_1 \widetilde{B}_2}$, but not on $W_{\underline{B}_1 \underline{B}_2}^\alpha$.

Therefore, by taking the trace in \eqref{eqn:openstringsymmetrybasis}, we get 
\begin{align} \label{eq:annulus_rep_to_symm}
\begin{split}
    Z^{az_1z_2}_{B_1B_2}(\delta) &= \sum_{\alpha } 
    \left( \mathrm{Tr}_{W_{\underline{B}_1 \underline{B}_2}^\alpha} \mathsf{H}_{B_1 B_2,a}^{B_1 B_2, z_1 z_2} \right)
    \left(\mathrm{Tr}_{\mathcal{V}_{\alpha}^{\widetilde{B}_1 \widetilde{B}_2}} q^{L_0 - c/24}\right) \\
    &= \sum_{\alpha} \left[ \chi_\alpha \right]_{\underline{B}_1 \underline{B}_2}^{a z_1 z_2} \mathbf{Z}_\alpha (\delta)\,,
\end{split}
\end{align}
where $\left[ \chi_\alpha \right]_{B_1 B_2}^{a z_1 z_2}$ is the boundary tube algebra character, which is determined by the generalized half-linking numbers through the generalized Verlinde formula \eqref{eq:verlinde}, as discussed in Section \ref{sec:matrix}.
This gives an explicit basis transformation formula from the representation basis to the symmetry basis.

We may also consider going in the other direction, namely, from the symmetry basis to the representation basis.
In order to do this, we define a projection operator
\begin{align}
P^\alpha_{B_1B_2}:\mathcal{H}_{B_1B_2}\to \mathcal{H}^\alpha_{B_1B_2}    =  W_{\underline{B}_1 \underline{B}_2}^\alpha \otimes \mathcal{V}_{\alpha}^{\widetilde{B}_1 \widetilde{B}_2} \,,
\end{align}
which has the property that it annihilates a boundary local operator $\mathcal{O}\leftrightharpoons (z,\beta,\widetilde{\mathcal{O}})$ if it transforms in a representation $\beta \neq \alpha$ of the boundary tube algebra, and otherwise  it acts as identity. 
Given such an operator, we have 
\begin{align} \label{eq:projector_trace}
   \mathrm{Tr}_{\mathcal{H}_{B_1 B_2}}\left(P^\alpha_{B_1B_2}q^{L_0-c/24}\right) = \dim_{\mathbb{C}}(W^\alpha_{\underline{B}_1 \underline{B}_2}) \mathbf{Z}_\alpha (\delta) \,,
\end{align}
since $\mathcal{H}^\alpha_{B_1B_2}= W_{\underline{B}_1 \underline{B}_2}^\alpha \otimes \mathcal{V}_{\alpha}^{\widetilde{B}_1 \widetilde{B}_2}$.
Now, we use the fact that $\dim_{\mathbb{C}}(W^\alpha_{\underline{B}_1 \underline{B}_2})$ is the fusion coefficient appearing in the fusion of topological line interfaces, $\underline{B}_2 \otimes \alpha = \bigoplus_{\underline{B}_1} \dim_{\mathbb{C}}(W^\alpha_{\underline{B}_1 \underline{B}_2}) \underline{B}_1$.
By taking the quantum dimension of both sides of the fusion algebra, multiplying both sides by $\qd_{\underline{B}_2}$, and then finally summing over $\underline{B}_2$, we obtain
\begin{align}
    \qd_\alpha S_{11}^{-1} = \sum_{{B}_1  {B}_2} \d_{\underline{B}_1} \d_{\underline{B}_2} \dim(W_{\underline{B}_1 \underline{B}_2}^\alpha) \,,
\end{align}
where we used $S_{11}^{-1} = \sum_{\underline{B}_2} \qd_{\underline{B}_2}^2$.
Combined with \eqref{eq:projector_trace}, it follows that
\begin{align}
    \mathbf{Z}_\alpha (\delta)=\frac{S_{11}}{\qd_\alpha }\sum_{{B}_1  {B}_2} \qd_{\underline{B}_1}\qd_{\underline{B}_2} \mathrm{Tr}_{\mathcal{H}_{B_1B_2}}\left(P^{\alpha}_{B_1B_2}q^{L_0-c/24}\right) \,.
\end{align}

An explicit expression for the projector $P^\alpha_{B_1 B_2}$ can be derived as follows.
First, consider the following configuration of the 2+1d SymTFT:
\begin{align} \label{eq:muhalflink}
\begin{split}
    &\input{annulusprojector1.tikz} \,.
    \end{split}
\end{align}
Here, $\mathcal{O}\leftrightharpoons(\underline{\cO}, \beta, \widetilde{\mathcal{O}})$ is a boundary-changing local operator between two conformal boundary conditions $B_1$ and $B_2$, which transforms under the irreducible representation of the boundary tube algebra labeled by the topological line interface $\beta$ in the SymTFT picture.
That is, $\mathcal{O} \in \mathcal{H}_{B_1 B_2}^{\beta}$.

If we push the bulk $\mu$ line, which is half-linked with the interface $\beta$, onto the Dirichlet boundary condition $\mathcal{B}_{\mathrm{reg}}$ of the SymTFT, it reduces to a linear combination of the boundary tube algebra generators \eqref{eq:Hlassostate}.
In particular, we get
\begin{align} \label{eq:pushing_mu}
\begin{split}
    &\sum_{a x'}\sqrt{\frac{\qd_a}{\qd_\mu}}~\input{annulusprojector2.tikz}
    \hspace{.5in} \\
    &= \sum_{a x' z_1 z_2}\frac{\sqrt{\qd_a}    {^{\mathcal{B}_1\mathcal{B}_{\mathrm{reg}}}}\Psi^{1(az_1)}_{\underline{B}_1 \underline{B}_1(\mu x x')}    {^{\mathcal{B}_2\mathcal{B}_{\mathrm{reg}}}}\widetilde{\Psi}^{1(a z_2)}_{\underline{B}_2 \underline{B}_2(\mu y x')}   }{\sqrt{\qd_\mu } \qd_{\underline{B}_1} \qd_{\underline{B}_2} S_{11} }~\input{annulusprojector3.tikz} \,,
\end{split}
\end{align}
where we have used the completeness relation in \eqref{eq:basis_bulk_to_bdy} as well as Equations \eqref{eqn:halflinkingid1} and \eqref{eqn:halflinkingid1_tilde} which are derived in Appendix \ref{app:half-linking}.
The final SymTFT configuration on the right-hand side represents an element of the boundary tube algebra, and in particular, each summand is proportional to $\mathsf{H}_{B_1 B_2,a}^{B_1 B_2, z_1 z_2}$.

Next, by combining (a special case of) \eqref{eqn:halflinkingid1} and the orthgonality relation \eqref{eqn:halflinkingorthogonality} for the generalized half-linking numbers, we obtain
\begin{equation} \label{eq:delta_projector}
    \sum_{\mu x y} \sqrt{S_{11}\qd_\alpha^2} {^{\mathcal{B}_1\mathcal{B}_2}}\widetilde{\Psi}_{\alpha\alpha(\mu xy)}^{11}
     \raisebox{-19pt}{
\begin{tikzpicture}
\draw[color=blue, thick, decoration = {markings, mark=at position -0.5 with {\arrow[scale=1.5]{stealth[reversed]}}}, postaction=decorate] (-1.5,0.5) -- (1,0.5);
\draw[] (-1.5,0.5) node[left] {$\beta$};
\draw[color=red, preaction={draw=white,line width=4pt}, thick, decoration = {markings, mark=at position 0.7 with {\arrow[scale=1.5]{stealth[reversed]}}}, postaction=decorate] (0,0) arc [start angle=-90, end angle=90, x radius=0.3, y radius=0.5];
\draw[] (0.4,0.7) node[above] {$\mu$};
\draw [] (0.1,0) node[left]{\small  $y$};
\draw [] (0.1,1) node[left]{\small  $\bar{x}$};
\draw[] (-1.0,0.7) node[above] {$\mathcal{B}_1$};
\draw[] (-1.0,0.3) node[below] {$\mathcal{B}_2$};
\end{tikzpicture}}
= \delta_{\alpha \beta}
\raisebox{-19pt}{
\begin{tikzpicture}
\draw[color=blue, thick, decoration = {markings, mark=at position -0.5 with {\arrow[scale=1.5]{stealth[reversed]}}}, postaction=decorate] (-1.5,0.5) -- (1,0.5);
\draw[] (-1.5,0.5) node[left] {$\alpha$};
\draw[] (-1.0,0.7) node[above] {$\mathcal{B}_1$};
\draw[] (-1.0,0.3) node[below] {$\mathcal{B}_2$};
\end{tikzpicture}} \,.
\end{equation}
Together with \eqref{eq:pushing_mu}, we deduce that the following operator acts as the projector:
\begin{equation} \label{eq:projector_annulus}
    P^{\alpha}_{B_1 B_2} = \sum_{\substack{\mu xy \\ ax' z_1 z_2} } \sqrt{\frac{\qd_a}{\qd_\mu S_{11}}} \frac{\qd_\alpha}{\qd_{\underline{B}_1} \qd_{\underline{B}_2}}  
{^{\mathcal{B}_1\mathcal{B}_2}}\widetilde{\Psi}_{\alpha\alpha(\mu xy)}^{11} {^{\mathcal{B}_1 \mathcal{B}_{\mathrm{reg}}}}\Psi_{\underline{B}_1 \underline{B}_1 (\mu x x')}^{1(az_1)}
{^{\mathcal{B}_2 \mathcal{B}_{\mathrm{reg}}}}\widetilde{\Psi}_{\underline{B}_2 \underline{B}_2 (\mu yx')}^{1(az_2)} 
\mathsf{H}^{B_1 B_2, z_1 z_2}_{B_1 B_2, a} \,.
\end{equation}
The intuition is as follows (analogous to the torus case in Figure \ref{fig:proj_bulk}). 
The projector operator $P^{\alpha}_{B_1 B_2}$ is given by a particularly chosen linear combination of the boundary tube algebra generators \eqref{eq:Hlassostate} such that in the SymTFT picture, it can be lifted to a sum of bulk topological lines half-linked with the topological interface $\beta$ as in \eqref{eq:muhalflink}, which in turn implements the projection onto the chosen representation sector labeled by the topological interface $\alpha$ thanks to \eqref{eq:delta_projector}. This projector will be crucial in deriving the non-invertible symmetry-resolved Affleck-Ludwig-Cardy formula as well as the symmetry-resolved entanglement entropy in the companion paper \cite{Choi:2024wfm}. 

Using the projector \eqref{eq:projector_annulus}, we arrive at the desired change-of-basis formula,
\begin{align}\label{eq:changebasislast}
\begin{split}
    \mathbf{Z}_\alpha(\delta) &= \sqrt{S_{11}}\sum_{B_1,B_2}
     \sum_{\substack{\mu xy \\ ax' z_1 z_2} } \sqrt{\frac{\qd_a}{\qd_\mu }}   
{^{\mathcal{B}_1\mathcal{B}_2}}\widetilde{\Psi}_{\alpha\alpha(\mu xy)}^{11} {^{\mathcal{B}_1 \mathcal{B}_{\mathrm{reg}}}}\Psi_{\underline{B}_1 \underline{B}_1 (\mu x x')}^{1(az_1)}
{^{\mathcal{B}_2 \mathcal{B}_{\mathrm{reg}}}}\widetilde{\Psi}_{\underline{B}_2 \underline{B}_2 (\mu yx')}^{1(az_2)} 
    Z_{B_1 B_2}^{a z_1 z_2}(\delta) \\
    &= S_{11} \sum_{B_1,B_2}
     \sum_{\substack{ a  z_1 z_2} }  
 \left[ \widetilde{\chi}_{\alpha} \right]_{\underline{B}_1 \underline{B}_2}^{az_1 z_2} Z_{B_1 B_2}^{a z_1 z_2}(\delta) \,,
\end{split}
\end{align}
which expresses the representation basis annulus partition functions as a linear combination of the symmetry basis ones.
We see that the coefficients are given by the dual characters \eqref{eq:dualbdycharacter_solidtorus} for the boundary tube algebra.

Finally note that, by taking the special case $Z_{B_1B_2}^{az_1z_2}(\delta) = [\chi_\beta]_{\underline{B}_1\underline{B}_2}^{az_1z_2}$, in which case $\mathbf{Z}_\alpha=\delta_{\alpha\beta}$, we recover the orthogonality relation advertised in Equation \eqref{eq:ortho_character}.

\section{Boundaries in the closed string channel}\label{sec:boundariesclosedstring}

In a 1+1d CFT $Q$, a conformal boundary condition $B$ defines a corresponding (non-normalizable) \emph{boundary state} $|B\rangle$ in the $S^1$ Hilbert space of the theory \cite{Cardy:2004hm}. 
Indeed, this can be seen, for example, by taking time to run perpendicularly to the boundary condition. 
These boundary states can then be used as ingredients in computing other observables: for example, their overlaps define ``closed string presentations'' of annulus partition functions of $Q$. 

It is interesting to ask how this discussion can be refined by the incorporation of symmetries. 
For example, if a topological line $a$ in a symmetry category $\mathcal{C}$ admits a topological junction $z$ on a boundary condition $B$,\footnote{Such a topological junction exists if and only if the (untwisted) boundary state satisfies $a \ket{B} = \ket{B} + \cdots$ under the action of the topological line $a$. In this situation, \cite{Choi:2023xjw} refers to $B$ as a ``weakly symmetric'' boundary.} then one expects that there should be a  boundary state $|B\rangle_{a,z}$ corresponding to this configuration, now living in the $a$-twisted $S^1$ Hilbert space $\mathcal{H}_a$ of $Q$. See Figure \ref{fig:twistedboundarystatedefn}.

Below, we discuss how such twisted sector boundary states (twisted boundary states for short) can actually be computed.
To explain this, we will associate to any $\mathcal{C}$-multiplet of conformal boundary conditions of $Q$ a collection of generalized Ishibashi states, which reduce to the standard Ishibashi states when $Q$ is a diagonal rational conformal field theory and $\mathcal{C}$ is its category of Verlinde lines. 
In analogy with the standard Ishibashi states, we will show how all the twisted boundary states of $Q$ as well as its orbifolds can be obtained as linear combinations of these generalized Ishibashi states.

Twisted boundary states have been discussed in the context of RCFTs in the past. 
See, for instance \cite{Gaberdiel:2002qa,Fuchs:1997kt,Fuchs:2000cm,Fuchs:2004dz,Collier:2021ngi}, and also \cite{Fukusumi:2021zme, Ebisu:2021acm} for $\mathbb{Z}_2$ symmetry in the context of fermionization. 
However, our discussion below does not rely on the rationality of the underlying theory $Q$, nor the existence of an extended chiral algebra.
Furthermore, our general results reduce to the known ones when $Q$ is rational.
We heavily rely on technologies developed in earlier sections, and in particular the SymTFT understanding of boundary conditions and boundary-changing local operators plays a central role.

\subsection{Half-Ishibashi states}\label{subsec:generalizedishibashi}

Suppose that $Q$ is a 1+1d conformal field theory, and that $\mathcal{B}$ is a $\mathcal{C}$-multiplet of conformal boundary conditions of $Q$, so that $\mathcal{B}$ has the structure of a $\mathcal{C}$-module category. 
Recall that any boundary condition $B\in\mathcal{B}$ can be decomposed into a triple $B\leftrightharpoons (\underline{B},\mathcal{B},\widetilde{B})$ as in Figure \ref{fig:bdydefect.operators}, where $\widetilde{B}$ is a conformal line interface between the topological boundary condition of $\mathrm{TV}_{\mathcal{C}}$ defined by $\mathcal{B}$ and the physical boundary condition $\widetilde{Q}$ of $\mathrm{TV}_{\mathcal{C}}$ defined by $Q$. 
This physical line interface $\widetilde{B}$ is common to all the boundary conditions $B$ in the multiplet $\mathcal{B}$, and it will be a crucial ingredient for defining our generalized Ishibashi states.

We define a \emph{half}-Ishibashi state as the state in the Hilbert space $\mathcal{V}_{\bar{\mu}}$ of the SymTFT $\mathrm{TV}_{\mathcal{C}}$ on a disk $D^2$ punctured by a bulk topological line $\mu$ (past-oriented) with the physical boundary condition $\widetilde{Q}$ imposed on the boundary $\partial D^2$ of the disk, which is prepared by the following path integral over a ``solid cone'':
\begin{align} \label{eq:half_Ishi}
    |y;\mu,\mathcal{B} \rrangle \equiv ~\input{figures/halfishibashistate.tikz}~~~ \in \mathcal{V}_{\bar{\mu}} \,.
\end{align}
Here, $\mathcal{B}$ is the topological boundary condition of the SymTFT (imposed on the boundary of the solid cone) given by the module category that the boundary condition $B$ belongs to, and $y$ is a choice of a topological junction on the boundary $\mathcal{B}$ on which the bulk topological line $\mu$ can end.
If one were to stop performing the path integral just shy of the top of this cone, then this would define a state in the disk Hilbert space $\mathcal{H}(D^2_{\mathcal{B},\bar{\mu}})$ of $\mathrm{TV}_{\mathcal{C}}$ punctured by $\mu$ and with $\mathcal{B}$ imposed at the boundary of the disk. One then applies $\widetilde{B}$ to this state to obtain a state in $\mathcal{V}_{\bar{\mu}}$: indeed, because $\widetilde{B}$ is a line interface between the $\mathrm{TV}_{\mathcal{C}}$-boundaries $\mathcal{B}$ and $\widetilde{Q}$, it also defines an operator $\widetilde{B}:\mathcal{H}(D^2_{\mathcal{B},\bar{\mu}})\to\mathcal{V}_{\bar{\mu}}$.
Since $\widetilde{B}$ is a conformal interface between a topological boundary condition $\mathcal{B}$ and a conformal boundary condition $\widetilde{Q}$ of the SymTFT $\mathrm{TV}_{\mathcal{C}}$, under the action of the Virasoro algebra on $\mathcal{V}_{\bar{\mu}}$, the half-Ishibashi state satisfies the usual condition that
\begin{equation}
    \left( L_n - \bar{L}_{-n} \right)|y;\mu,\mathcal{B} \rrangle = 0  ~~~\text{for all $n$} \,,
\end{equation}
where $L_n$ and $\bar{L}_{n}$ are the left- and right-moving Virasoro generators, respectively.
We define the ``bra'' half-Ishibashi state similarly by the path integral
\begin{align} \label{eq:half_Ishi_bra}
    \llangle y;\mu,\mathcal{B} | \equiv ~\input{figures/halfishibashistate_bra.tikz}~~~ \in \mathcal{V}_{\bar{\mu}}^* \,.
\end{align}

The reason we refer to $|\mu;y, \mathcal{B} \rrangle$ as a ``half'' Ishibashi state is that it is not quite yet a state in the extended $S^1$ Hilbert space $\mathcal{H}_{\mathcal{C}} =\bigoplus_{a \in \mathrm{Irr}(\mathcal{C})} \mathcal{H}_a$ of $Q$. 
However, when it is combined with a vector in $W^{\bar{\mu}} = \bigoplus_{a \in \mathrm{Irr}(\mathcal{C})} W_a^{\bar{\mu}}$,
it produces a state in the (extended) Hilbert space of the CFT $Q$, due to the Schur-Weyl decomposition \eqref{eqn:schurweyldecomp} of $\mathcal{H}_{\mathcal{C}}$.
In particular, we consider the state of the form
\begin{align} \label{eq:full_Ishi}
    |\bar{x}, \bar{\mu}, \bar{a}\rangle\otimes |y;\mu, \mathcal{B} \rrangle \in W_{\bar{a}}^{\bar{\mu}}\otimes \mathcal{V}_{\bar{\mu}} \subset \mathcal{H}_{\bar{a}} \,,
\end{align}
which \emph{does} live in the extended Hilbert space $\mathcal{H}_{\mathcal{C}}$, where the state $|\bar{x}, \bar{\mu}, \bar{a}\rangle \in W^{\bar{\mu}}$ is prepared by the path integral of the SymTFT shown in \eqref{eq:x_state}.
We refer to \eqref{eq:full_Ishi} as a generalized \emph{full} Ishibashi state, or just Ishibashi state for short.
When $a=1$, we call the Ishibashi state untwisted, and if $a \neq 1$, we call it $a$-twisted.\footnote{It is our choice of convention that an $a$-twisted Ishibashi state \eqref{eq:full_Ishi} lives in the $\bar{a}$-twisted Hilbert space $\mathcal{H}_{\bar{a}}$, as opposed to the $a$-twisted Hilbert space $\mathcal{H}_{a}$.
Similarly, later we define an $a$-twisted boundary condition such that the corresponding twisted boundary state also lives in $\mathcal{H}_{\bar{a}}$ instead of $\mathcal{H}_{a}$. With such a convention, a bulk operator which is in the $a$-twisted sector can have a nontrivial 1-point function in the presence of an $a$-twisted boundary condition, say, on a disk, where the topological line $a$ runs from the bulk operator to the boundary.\label{footnote:abar}}

Similar to ordinary Ishibashi states, our half-Ishibashi states \eqref{eq:full_Ishi} generally have an infinite norm and are therefore not normalizable.
Physically, this is because when we attempt take an inner product of two half-Ishibashi states, there is a short-distance divergence coming from the two copies of the conformal interface $\widetilde{B}$, which are coincident. 
However, as is standard for ordinary Ishibashi states, we can consider the inner product between two half-Ishibashi states by first time-evolving one of them by a finite amount $1/\delta$:
\begin{equation} \label{eq:Ishi_solid_ball}
    \llangle y_1;\mu,  \mathcal{B}_1 | \tilde q^{\frac12 (L_0 + \bar{L}_0 - c/12)} | y_2;\mu, \mathcal{B}_2 \rrangle = \raisebox{-2.4em}{~\input{figures/Ishi_overlap.tikz}} \,, ~~~ \tilde{q} = e^{-4\pi \delta} \,.
\end{equation}
The SymTFT partition function on the right-hand side can be explicitly evaluated, using the (inverse of the) ``collapsing tube'' formula in \eqref{eq:collapse2}.
We relegate the detailed computation to Appendix \ref{app:boundary}.
The result is
\begin{equation} \label{eq:Ishi_overlap}
    \llangle y_1;\mu, \mathcal{B}_1 | \tilde{q}^{\frac12 (L_0 + \bar{L}_0 - c/12)} | y_2;\mu, \mathcal{B}_2 \rrangle = \frac{1}{\sqrt{S_{11}}} \sum_{\alpha\in \mathrm{Irr}(\mathcal{I})} {^{\mathcal{B}_1\mathcal{B}_2}}\Psi_{\alpha\alpha(\mu y_1 y_2)}^{11} \mathbf{Z}_\alpha (\delta) \,,
\end{equation}
where $\mathcal{I}$ is the multiplet of topological line interfaces between the two topological boundary conditions $\mathcal{B}_1$ and $\mathcal{B}_2$ of the SymTFT.
We see that the closed string channel overlap between the half-Ishibashi states is given by a linear combination of the representation basis annulus partition functions $\mathbf{Z}_\alpha (\delta)$, defined in \eqref{eq:rep_annulus}, in the open string channel, where the coefficients are given by the generalized half-linking numbers.

\subsubsection{Special case: diagonal rational conformal field theories}
As an illustration, let us describe how this goes in a diagonal RCFT, although our discussions apply to general CFTs which are not necessarily rational.
Denote the extended chiral algebra as $V$, and take $\mathcal{C}$ to be the category of Verlinde lines. 
Since $\mathcal{C}$ is a modular tensor category, $Z(\mathcal{C})\cong \mathcal{C}\boxtimes\overline{\mathcal{C}}$, and we can label bulk lines as $\mu=(a,b)$. In this situation, the Cardy states live in a multiplet described by the regular module category of $\mathcal{C}$, so that $\mathcal{B}\cong \mathcal{B}_{\mathrm{reg}}$ and the boundary states are labeled by objects in $\mathcal{C}$ (equivalently, by primary operators of $Q$).
The standard (untwisted) Ishibashi states $|a\rrangle_{\mathrm{Ish}}$ of $Q$ are known to be also labeled by elements $a\in\mathrm{Irr}(\mathcal{C})$ as well \cite{Cardy:2004hm}.

In our formalism, to produce a generalized Ishibashi state, we require a bulk line $\mu$ which is capable of terminating topologically on the Dirichlet boundary condition given by the regular module category $\mathcal{B}_{\mathrm{reg}}$. 
Since the bulk-to-boundary forgetful functor in this situation is given by $F(a,b)=a\otimes b$ as discussed around \eqref{eq:MTC_ffunctor}, a line $\mu = (a,b)$ can terminate topologically on the Dirichlet boundary if and only if $b=\bar{a}$, in which case the junction $\ket{\bar{x},(\bar{a},a),1}$ is unique.
Similarly, the choice of the junction $y$ in \eqref{eq:half_Ishi} is also unique for a given $\mu = (a,\bar{a})$.
Our normalization convention for the generalized Ishibashi states is such that they are related to the standard Ishibashi states in the case of diagonal RCFTs by
\begin{align} \label{eq:Ishi_dRCFT}
    |a\rrangle_{\mathrm{Ish}}= \frac{S_{11}^{1/4}}{ \sqrt{\qd_a}}\ket{\bar{x},(\bar{a},a),1}\otimes |y;(a,\bar{a}), \mathcal{B}_{\mathrm{reg}}\rrangle \in W^{(\bar{a},a)}_1 \otimes\mathcal{V}_{(\bar{a},a)} = V_{\bar{a}} \otimes\overline{V_a} \,.
\end{align}
In fact, from the overlap of half-Ishibashi states \eqref{eq:Ishi_overlap}, and also $\braket{\bar{x}_a | \bar{x}_b} = \delta_{ab} \qd_a$, we reproduce
\begin{equation}
   {}_{\mathrm{Ish}}\llangle a |  \tilde q^{\frac12 (L_0 + \bar{L}_0 - c/12)}    |b\rrangle_{\mathrm{Ish}} = \delta_{ab} \mathrm{ch}_a (\tilde q) \,,
\end{equation}
where we use the fact that the generalized half-linking numbers appearing in \eqref{eq:Ishi_overlap} in this case reduce to the $S$-matrix of $\mathcal{C}$ as in \eqref{eq:psi_S}, and that the representation basis annulus partition functions are given by the chiral algebra characters as in \eqref{eq:rep_char}.
Similarly, $c$-twisted Ishibashi states are of the form 
\begin{align}
    |\bar{x},(\bar{a},a),\bar{c} \rangle\otimes| y; (a,\bar{a}), \mathcal{B}_{\mathrm{reg}}\rrangle \in W^{(\bar{a},a)}_{\bar{c}}\otimes\mathcal{V}_{(\bar{a},a)}=N_{\bar{a}a}^c V_{\bar{a}}\otimes \overline{V_a}
\end{align}
where we require that $N_{\bar{a}a}^c>0$.

\subsection{Twisted sector boundary states}\label{subsec:twistedboundarystates}

Having laid out the basic theory of generalized Ishibashi states, let us describe how one may take linear combinations of them to obtain the true (twisted sector) boundary states. 

Consider the problem of finding the boundary state $|B\rangle_{a,z}$ corresponding to a boundary condition $B\leftrightharpoons (\underline{B},\mathcal{B}, \widetilde{B})$ hosting a junction $z$ on which a topological line $a\in\mathcal{C}$ topologically terminates.  
Our convention is such that such a twisted boundary state belongs to the $\bar{a}$-twisted Hilbert space (albeit non-normalizable), i.e.\ $|B\rangle_{a,z}\in\mathcal{H}_{\bar{a}}$ (see Footnote \ref{footnote:abar}).
To obtain the twisted boundary state as a linear combination of generalized Ishibashi states \eqref{eq:full_Ishi}, we consider the following Euclidean configuration of the 1+1d CFT $Q$ as well as the corresponding configuration of the 2+1d SymTFT:
\begin{align} \label{eq:twisted_bdy_Euclidean}
    \input{figures/twistedboundarystate0.tikz}~\leftrightharpoons~\input{figures/twistedboundarystate1.tikz}=\sqrt{S_{11}}\sum_{\mu xy} {^{\mathcal{B}\mathcal{B}_{\mathrm{reg}}}}\Psi_{\underline{B}\underline{B}(\mu y x)}^{1(az)}~\input{figures/twistedboundarystate2.tikz} \,.
\end{align}
To the left, we begin by putting the theory $Q$ on a plane, with a small disk $D^2$ excised around the origin.
At the boundary of $D^2$, we impose the boundary condition $B$.
A topological line $a$ terminates on $B$ at a topological junction $z$.
Upon a conformal transformation to a (half-infinite) cylinder, this Euclidean configuration defines the twisted boundary state $|B\rangle_{a,z}$.

We then blow up this 1+1d configuration into the corresponding 2+1d SymTFT configuration, following the discussions in Section \ref{sec:symTFT}.
The excised disk on the 1+1d side maps to a hollow tube stretched between two boundary conditions, $\mathcal{B}_{\mathrm{reg}}$ and $\widetilde{Q}$, of the SymTFT.
On the hollow tube itself, the topological boundary condition $\mathcal{B}$ of the SymTFT, which labels the multiplet of boundary conditions that $B$ belongs to, is imposed.
In between $\mathcal{B}_{\mathrm{reg}}$ and $\mathcal{B}$, there is a boundary-changing topological line interface $\underline{B}$, on which the topological line $a \in \mathcal{C}$ from the $\mathcal{B}_{\mathrm{reg}}$ ends, forming the $z$ junction.
We moved the interface $\underline{B}$ to the middle of the hollow tube without any cost, since $\underline{B}$, as well as $a$ and the junction $z$, are topological.
Finally, on the other end of the hollow tube, we have the conformal line interface $\widetilde{B}$ between the two boundary conditions $\mathcal{B}$ and $\widetilde{Q}$.

To connect this SymTFT configuration to the earlier discussions on generalized Ishibashi states, we now apply the ``collapsing tube'' formula \eqref{eq:collapsetube}, and reach the right-hand side of \eqref{eq:twisted_bdy_Euclidean}.
Each summand contains a bulk topological line $\mu$ stretched between the two topological boundary conditions $\mathcal{B}_{\mathrm{reg}}$ and $\mathcal{B}$, ending on the topological junctions $\bar{x}$ and $y$, respectively, which results from collapsing the hollow tube.
The conformal line interface $\widetilde{B}$ remains and still divides $\mathcal{B}$ and $\widetilde{Q}$.

Now, on the right-hand side of \eqref{eq:twisted_bdy_Euclidean}, notice that, upon appropriate state-operator maps, the $\bar{x}$ junction maps to the state $|\bar{x}, \bar{\mu}, \bar{a}\rangle \in W_{\bar{a}}^{\bar{\mu}}$, whereas the $y$ junction, when time-evolved just past $\widetilde{B}$, precisely maps to the half-Ishibashi state $|\mu;y, \mathcal{B} \rrangle \in \mathcal{V}_{\bar{\mu}}$.
Therefore, this Euclidean configuration defines a state which is proportional to $|\bar{x}, \bar{\mu}, \bar{a}\rangle \otimes |\mu;y, \mathcal{B} \rrangle$ (but not exactly the same on the nose), which is precisely the generalized Ishibashi state \eqref{eq:full_Ishi}.
The exact proportionality constant is $S_{1\mu}^{-1/2}$, which we derive in Appendix \ref{app:boundary}.\footnote{Intuitively, the reason why the Euclidean configuration on the right-hand side of \eqref{eq:twisted_bdy_Euclidean} corresponds to a state which differs from the naive $|\bar{x}, \bar{\mu}, \bar{a}\rangle \otimes |\mu;y, \mathcal{B} \rrangle$ by an overall normalization is because the tensor product state $|\bar{x}, \bar{\mu}, \bar{a}\rangle \otimes |\mu;y, \mathcal{B} \rrangle$ is prepared by the Euclidean path integral of the SymTFT on two disconnected solid cones, each hosting a $\mu$ line (recall \eqref{eq:x_state} and \eqref{eq:half_Ishi}), whereas the Euclidean configuration on the right-hand side of \eqref{eq:twisted_bdy_Euclidean} is on a connected manifold with a single $\mu$ line. Similar issues arise when one considers the decomposition of the (twisted sector) local operators \eqref{eqn:decomptwistedlocalop}. See \cite{Felder:1999cv} for a related discussion in the context of RCFTs.}

Combining all the ingredients, we arrive at an explicit expression for the twisted boundary state in terms of the generalized Ishibashi states:
\begin{equation} \label{eq:twisted_bdy_state}
    \ket{B}_{a,z} = \sum_{\mu x y} \sqrt{\frac{S_{11}}{S_{1\mu}}} {^{\mathcal{B}\mathcal{B}_{\mathrm{reg}}}}\Psi_{\underline{B}\underline{B}(\mu y x)}^{1(az)} |\bar{x}, \bar{\mu}, \bar{a}\rangle \otimes |y;\mu, \mathcal{B} \rrangle \,.
\end{equation}
That is, a twisted boundary state is obtained as a linear combination of generalized Ishibashi states, where the coefficients are given by the generalized half-linking numbers \eqref{eqn:halflinkingdefn}.
The result applies to an arbitrary $\mathcal{C}$-multiplet of conformal boundary conditions of an arbitrary  1+1d CFT, without relying on an extended chiral algebra or the rationality of the theory.\footnote{We do assume that the CFT as well as the boundary conditions are compact, in the sense that the circle and interval Hilbert spaces have discrete spectra.}
The formula \eqref{eq:twisted_bdy_state} allows us to reduce various computations involving twisted boundary states essentially to an exercise in category theory and TQFT.
Similarly, one can also obtain the ``bra'' twisted boundary state:\footnote{We \emph{define} the bra boundary state $\prescript{}{a,z}{\bra{B}}$ such that the topological line $a$ emanates from the boundary, starting at a topological junction $\bar{z}$. On the other hand, for the ket boundary $\ket{B}_{a,z}$, the topological line $a$ terminates on the boundary at a topological junction $z$. Depending on the choice of junction vectors $z$ and $\bar{z}$ (see \eqref{eq:basis_bdy}), $\widetilde{\Psi}$'s that appear in the expansion of $\prescript{}{a,z}{\bra{B}}$ are not necessarily complex conjugates of $\Psi$'s that appear in the expansion of $\ket{B}_{a,z}$.}
\begin{equation} \label{eq:twisted_bdy_state_bra}
    \prescript{}{a,z}{\bra{B}} = \sum_{\mu x y} \sqrt{\frac{S_{11}}{S_{1\mu}}} {^{\mathcal{B}\mathcal{B}_{\mathrm{reg}}}}\widetilde{\Psi}_{\underline{B}\underline{B}(\mu y x)}^{1(az)} \langle\bar{x}, \bar{\mu}, \bar{a}| \otimes \llangle y;\mu, \mathcal{B} | \,.
\end{equation}

As a sanity check, consider the boundary state \eqref{eq:twisted_bdy_state} in the special case of a diagonal RCFT, where we take $\mathcal{C}$ to be the category of its Verlinde lines, and furthermore restrict to the untwisted case $a=1$.
The boundary conditions preserving (half of) the chiral algebra form the regular module category $\mathcal{B}_{\mathrm{reg}}$ over $\mathcal{C}$.
We denote the simple boundary conditions as $B_a, B_b, \cdots$, where $a,b, \cdots$ label the primaries as well as the Verlinde lines.
As was discussed, in this special case, the generalized Ishibashi states reduce to the standard Ishibashi states up to an overall normalization as in \eqref{eq:Ishi_dRCFT}, and the half-linking numbers appearing in \eqref{eq:twisted_bdy_state} reduce to the S-matrix elements of $\mathcal{C}$ as in \eqref{eq:psi_S}.
Combining these observations, we obtain
\begin{equation} \label{eq:cardy_bdy}
    \ket{B_a} = \sum_b \frac{\mathbb{S}_{ab}}{\sqrt{\mathbb{S}_{1b}}} |b\rrangle_{\mathrm{Ish}} \,,
\end{equation}
which reproduces the standard result \cite{Cardy:2004hm}.

Finally, let $A$ be a gaugeable algebra object of $\mathcal{C}$, and consider boundary states in the gauged theory $Q/A$ \cite{Fuchs:2002cm,Bhardwaj:2017xup,Komargodski:2020mxz,Huang:2021zvu}, where boundary states may now be in a twisted sector for the dual symmetry $_A \mathcal{C}_A$.
In the SymTFT picture for twisted sector boundary conditions in \eqref{eq:twisted_bdy_Euclidean}, this corresponds to replacing the Dirichlet boundary $\mathcal{B}_{\mathrm{reg}}$ on the left by a new topological boundary condition $\mathcal{C}_A$, given by the category of $A$-modules in $\mathcal{C}$.
On the other hand, the configuration near the physical boundary condition $\widetilde{Q}$ is unaltered.
In particular, the half-Ishibashi states are shared among the theory $Q$ and all of its possible orbifolds $Q/A$ by an algebra object in $\mathcal{C}$.
Any twisted sector boundary state of the gauged theory is then given by an expression similar to \eqref{eq:twisted_bdy_state}, but now the states $\ket{\bar{x},\bar{\mu},\bar{a}}$ are replaced by the ones prepared by the path integral \eqref{eq:x_state} with the Dirichlet boundary condition replaced by $\mathcal{C}_A$, and $a$ now a topological line in $_A \mathcal{C}_A$.
The multiplet $\mathcal{B}$ of boundary conditions in $Q$ reorganizes into a new multiplet $_A \mathcal{B}$ in the theory $Q/A$, transforming under the dual symmetry $_A \mathcal{C}_A$, where $_A \mathcal{B}$ denotes the category left $A$-modules in $\mathcal{B}$ \cite{Lin:2021udi}.\footnote{The objects of $_A \mathcal{B}$ correspond to (not necessarily simple) boundary conditions of $Q$ on which the mesh of the algebra object $A$ can consistently end, hence defining boundary conditions in the gauged theory $Q/A$.}
In the SymTFT picture, the topological line interface $\underline{B}$ as well as the generalized half-linking numbers are then also appropriately modified, which we do not explicitly write here. See also \cite{KRW} for an alternative approach to boundaries of an orbifolded theory which centers around the algebra object implementing the orbifold.

\subsection{Twisted Cardy condition}

A fundamental consistency condition that a conformal boundary condition in a 1+1d CFT must satisfy is the Cardy condition \cite{Cardy:2004hm}.
It requires that the open and closed string channel interpretations of an annulus (or cylinder) partition function are compatible with each other.
In the case of twisted boundary states, the Cardy condition is enriched by the presence of a topological line defect streched across the interval in between the two boundary conditions  that the theory is quantized on.
Specifically, one requires
\begin{equation} \label{eq:twisted_Cardy}
    \prescript{}{a,z_1}{\bra{B_1}} \tilde{q}^{\frac12 (L_0 + \bar{L}_0 - c/12)} \ket{B_2}_{a,z_2} \stackrel{!}{=} \mathrm{Tr}_{\mathcal{H}_{B_1 B_2}} \left( 
        \mathsf{H}_{B_1 B_2 ,a}^{B_1 B_2, z_1 z_2} q^{L_0 -c/24}
    \right) \equiv Z_{B_1 B_2}^{a z_1 z_2} (\delta) \,,
\end{equation}
where the right-hand side is the symmetry basis annulus partition function in the open string channel, defined in \eqref{eqn:openstringsymmetrybasis}.
Here, $\mathsf{H}_{B_1 B_2 ,a}^{B_1 B_2, z_1 z_2}$ is the boundary tube algebra element defined in \eqref{eq:Hlassostate}, acting on the interval Hilbert space $\mathcal{H}_{B_1 B_2}$, and we have $q=e^{-\pi \delta}$, $\tilde{q}=e^{-4\pi/\delta}$.

Using the overlap between the half-Ishibashi states given in  \eqref{eq:Ishi_overlap}, and the expansion of twisted boundary states in terms of generalized Ishibashi states as in \eqref{eq:twisted_bdy_state} and \eqref{eq:twisted_bdy_state_bra}, we can compute the overlap between the twisted boundary states and verify that the twisted Cardy condition \eqref{eq:twisted_Cardy} is in fact satisfied by our boundary states.
We obtain
\begin{align}\label{eq:twisted_Cardy2}
\begin{split}
    &~~~~ \prescript{}{a,z_1}{\bra{B_1}} \tilde{q}^{\frac12(L_0 + \bar{L}_0 - c/12)} \ket{B_2}_{a,z_2} \\ &= \sum_\alpha \left( 
        \sqrt{\qd_a} \sum_{\mu x y y'} \frac{
    {^{\mathcal{B}_1\mathcal{B}_{\mathrm{reg}}}}\widetilde{\Psi}_{\underline{B}_1 \underline{B}_1 (\mu yx)}^{1(az_1)}    {^{\mathcal{B}_1\mathcal{B}_2}}\Psi_{\alpha\alpha (\mu yy')}^{11}
    {^{\mathcal{B}_2\mathcal{B}_{\mathrm{reg}}}}\Psi_{\underline{B}_2 \underline{B}_2 (\mu y' x)}^{1(az_2)}    }{\sqrt{S_{1 \mu}}}
    \right) \mathbf{Z}_\alpha (\delta) \\
    &= \sum_\alpha \left[ \chi_\alpha \right]_{\underline{B}_1 \underline{B}_2}^{a z_1 z_2} \mathbf{Z}_\alpha (\delta) \\
    &= Z_{B_1 B_2}^{a z_1 z_2} (\delta) \,,
\end{split}
\end{align}
as desired.
In the second to the last equality, we have used the generalized Verlinde formula \eqref{eq:verlinde} satisfied by the boundary tube algebra characters, and in the last equality, we have used the basis transformation formula \eqref{eq:annulus_rep_to_symm} for the annulus partition functions, from the representation basis to the symmetry basis.

The fact that we need to use the generalized Verlinde formula to verify the twisted Cardy condition resembles that one needs to use the ordinary Verlinde formula to verify the ordinary Cardy condition for the Cardy boundary states in diagonal RCFTs.

\subsection{Open-closed duality in the representation basis}

The twisted Cardy condition \eqref{eq:twisted_Cardy} is an instance of the \emph{open-closed} duality for annulus partition functions, which results from one's freedom to choose either the circle or the interval factor of the annulus as the Euclidean time direction. (It is analogous to the modular covariance property of torus partition functions.)
More precisely, the twisted Cardy condition represents the open-closed duality for the symmetry-basis annulus partition functions.

Below, we discuss a version of open-closed duality for the representation basis annulus partition functions, where the story is enriched by the fact that the open string channel partition function of a theory is related in the closed string channel to an overlap between boundary states in an orbifolded theory, as we now explain.
Such an observation was also made recently in \cite{Kusuki:2023bsp} for the case of ordinary finite group symmetries.

\begin{figure}
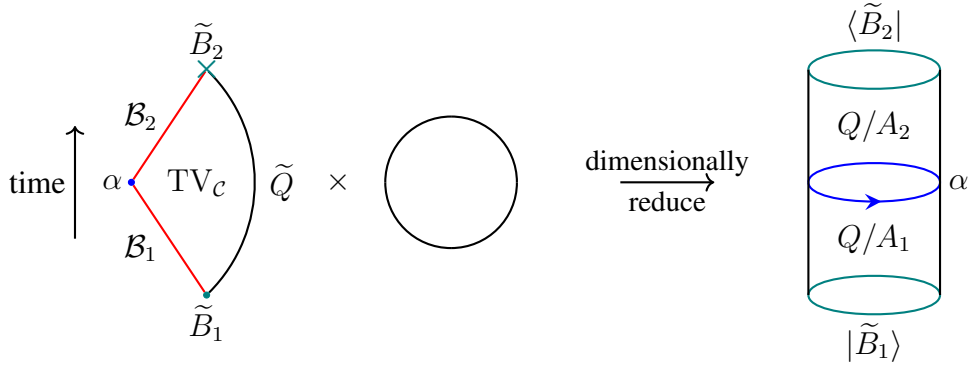

    \ctikzfig{figures/boundaryrepbasis}
    \caption{Representation basis annulus partition functions $\mathbf{Z}_\alpha(\delta)$ in the closed string channel, before and after dimensional reduction on the interval direction orthogonal to time.}\label{fig:closedstringrepbasis}
\end{figure}

The representation basis partition functions are obtained by placing the SymTFT on a background which is topologically $D^2\times S^1$, as in Figure \ref{fig:closedstringrepbasis} (also recall Figure \ref{fig:SymTFTboundaryHilbertspaces}). Instead of taking the circle direction to correspond to time, we can choose time so that it runs orthogonally to the $S^1$, as depicted in Figure \ref{fig:closedstringrepbasis}. Then one encounters, at a generic spatial slice, an $S^1\times I$ topology, where $\widetilde{Q}$ is imposed at one end of the interval, and a topological boundary condition (either $\mathcal{B}_1$ or $\mathcal{B}_2$) is imposed at the other. 

Recall one of the basic features of the SymTFT reviewed in Section \ref{sec:symTFT}: any 1+1d theory $Q$ with $\mathcal{C}$ symmetry can be expanded into a triple $(\CB_{\mathrm{reg}},\mathrm{TV}_{\mathcal{C}}, \widetilde{Q})$. 
If one replaces the Dirichlet boundary condition $\CB_{\mathrm{reg}}$ with another topological boundary condition, say, $\mathcal{B}_1 \cong \mathcal{C}_{A_1}$, then upon dimensional reduction back to 1+1d along the interval direction, one obtains the orbifolded theory $Q/A_1$. Similarly, by dimensionally reducing the configuration on the left of Figure \ref{fig:closedstringrepbasis} along the interval direction orthogonal to time, one obtains a 1+1d theory on the annulus, which in the lower half is $Q/A_1$, and in the upper half is $Q/A_2$. In the middle there is a topological line interface $\alpha$ which interpolates between these two phases, and can be thought of in particular as an operator mapping from the $S^1$ Hilbert space of $Q/A_1$ to that of $Q/A_2$. Another way to think of $\alpha$ is as a half-space gauging interface between $Q/A_1$ and $Q/A_2$.

Let us understand the boundary conditions $|\widetilde{B}_1\rangle$ and $|\widetilde{B}_2\rangle$, focusing for ease of exposition on $|\widetilde{B}_1\rangle$. By definition, it is the boundary condition of $Q/A_1$ obtained by folding along the $\widetilde{B}_1$ line interface between the $\mathcal{B}_1$ and $\widetilde{Q}$ boundaries of the SymTFT. Mathematically it can be described as follows. Recall that if a theory $Q$ with $\mathcal{C}$ symmetry admits a multiplet $\mathcal{B}$ of boundary conditions, then the orbifolded theory $Q/A$ admits a dual multiplet $\mathcal{B}/A$ of boundary conditions. This dual multiplet consists of boundaries of the original multiplet on which the mesh $A$ can consistently end, which mathematically means that $\mathcal{B}/A$ is the category of $A$-modules inside of $\mathcal{B}$. Applying this to the situation at hand, we learn that the boundary multiplet of $Q/A_1$ dual to $\mathcal{B}_1$ is ${_{A_1}}\mathcal{B}_1\cong {_{A_1}}\mathcal{C}_{A_1}$. Note that ${_{A_1}}\mathcal{C}_{A_1}$ is a fusion category; we claim that $|\widetilde{B}_1\rangle$ is the boundary state corresponding to the identity object of this fusion category. 

The line interface $\alpha$, which can be thought of as an $A_2$-$A_1$ bimodule, can be pushed down onto the $|\widetilde{B}_1\rangle$ boundary. It then produces a boundary condition $\alpha | \widetilde{B}_1 \rangle_{Q/{A_1}} = |\alpha\otimes_{A_1}\widetilde{B}_1\rangle_{Q/{A_2}}$ of $Q/A_2$. This is an object of the boundary multiplet ${_{A_2}}\mathcal{B}_1={_{A_2}}\mathcal{C}_{A_1}$ dual to $\mathcal{B}_1$ in the orbifolded theory $Q/A_2$. Since $|\widetilde{B}_1\rangle$ corresponds to the identity object of ${_{A_1}}\mathcal{C}_{A_1}$, it follows that $|\alpha\otimes_{A_1}\widetilde{B}_1\rangle$ simply corresponds to $\alpha$, thought of as an object of ${_{A_2}}\mathcal{C}_{A_1}$.

In total, we obtain the following expression:
\begin{align}
    \mathbf{Z}_\alpha(\delta) = {_{Q/A_2}}\langle\widetilde{B}_2| e^{-H_{\mathrm{cl}}/\delta}\alpha |\widetilde{B}_1\rangle_{Q/A_1} \,.
\end{align}
This generalizes Equation (4.4) of \cite{Kusuki:2023bsp} to the case of non-invertible symmetries.

These expressions are particularly powerful in the context of diagonal RCFTs. Indeed, as we saw earlier, all of the boundary states arising in these open-closed duality equations --- $|\alpha\otimes_{A_1} \widetilde{B}_1\rangle$, $|\widetilde{B}_2\rangle$, $|B_i\rangle_{a,x_i}$, etc. --- can be determined as linear combinations of Ishibashi states involving the half-linking/half-braiding numbers. We have given a formula for the half-braiding numbers of a modular tensor category in \eqref{eqn:modularhalfbraiding}, and moreover the inner products of Ishibashi states are known. Thus, one has complete control over annulus partition functions in this large class of theories from a variety of different angles.

\section{An example: Fibonacci symmetry}\label{sec:fib}

The technology we have developed throughout this paper becomes particularly powerful when $\CC$ is a modular tensor category, and when the boundary conditions under consideration transform in the regular $\CC$-module category $\mathcal{B}_{\mathrm{reg}}$. One situation where this occurs (though not the only one) is when $Q$ is a diagonal rational conformal field theory, $\CC$ is the category of Verlinde lines of the maximally extended chiral algebra, and the boundary conditions are taken to be the Cardy boundary conditions. Also, any theory with a Fibonacci symmetry will satisfy these conditions: we consider this case in detail here to showcase some of our formulae. 

\subsection{Basic data}
\label{subsec:basicdataFib}

The basic category theoretic data of $\mathrm{Fib}$ can be found e.g.\ in \cite{Bonderson:2007ci}, which we review below. 
Recall that $\mathrm{Fib}$ is the unique unitary non-invertible fusion category of rank 2. We write its simple lines as $1$ and $W$: they satisfy the fusion rule $W^2=1\oplus W$. The quantum dimensions of its simple lines are $\d_{1}=1$ and $\d_W=\varphi$ where $\varphi = \frac{1+\sqrt{5}}{2}$ is the golden ratio. Its only non-trivial $F$-symbols are given by 
\begin{align}
    F_{WWW}^W = \left( \begin{array}{rr} \varphi^{-1} & \varphi^{-1/2} \\ \varphi^{-1/2} & -\varphi^{-1}\end{array}\right).
\end{align}
It is also a modular tensor category, with non-trivial $R$-symbols given by $R_{WW}^{1}=e^{-4\pi i/5}$ and $R_{WW}^W=e^{3\pi i/5}$. Its modular S-matrix and twists are 
\begin{align}
    S = \frac{1}{\sqrt{2+\varphi}} 
    \left( \begin{array}{rr} 1 & \varphi \\ \varphi & -1\end{array}\right), \ \ \theta_1 = 1, \ \ \theta_W = e^{4\pi i/5} \,.
\end{align}

\subsection{CFTs with Fibonacci symmetry}\label{subsec:Fibtheories}

There are a plethora of 1+1d conformal field theories which possess Fibonacci symmetry to which the formulae of this section apply. Perhaps the simplest examples are the $(G_2)_1$ and $(F_4)_1$ Wess--Zumino--Witten models, for which $\mathrm{Fib}$ arises as the category of Verlinde lines, i.e.\ the category of topological line operators which commute with the maximal chiral algebra. In fact, all rational conformal field theories with $c<25$ and with $\mathrm{Fib}$ as the category of Verlinde lines were classified in \cite{Mukhi:2022bte}; furthermore, many 1+1d conformal field theories for which $\mathrm{Fib}$ arises as a \emph{subcategory} of the category of Verlinde lines were found in \cite{Rayhaun:2023pgc}. Other examples include the tricritical Ising model \cite{Chang:2018iay}, the chiral $(E_8)_1$ WZW model \cite{Rayhaun:2023pgc}, and the chiral monster CFT $V^\natural$ \cite{Fosbinder-Elkins:2024hff,MRb}.

\subsection{Boundary tube algebra}

The general expression for the multiplication of boundary lasso operators is given by \eqref{eq:HHH}. The only indecomposable module category of $\mathrm{Fib}$ is the regular module category. Hence the $\widetilde{F}$-symbols reduces to the $F$-symbols (or their inverses) collected in Section \ref{subsec:basicdataFib}. 

Let us look at some special cases of this multiplication. Feeding the data of the $\mathrm{Fib}$ category into this formula, we get the multiplication rule 
\begin{eqnarray}
     \mathsf{H}_{B_aB_b,W}^{B_W B_W}\times \mathsf{H}_{B_W B_W,W}^{B_aB_b} = \varphi \sum_{c=1,W}\frac{[F^W_{WWW}]^{-1}_{a c}[F^W_{WWW}]_{b c}}{\sqrt{\d_c}} \mathsf{H}_{B_WB_W,c}^{B_WB_W}\,.
\end{eqnarray}
Unpacking this a bit more, when $(a,b)=(1,1)$, we find
\begin{eqnarray}
     \mathsf{H}_{B_1B_1,W}^{B_WB_W}\times \mathsf{H}_{B_WB_W,W}^{B_1B_1} = \varphi^{-1} \mathsf{H}_{B_WB_W,1}^{B_WB_W}  + \varphi^{-1/2} \mathsf{H}_{B_WB_W,W}^{B_WB_W} \,.
\end{eqnarray}
When $(a,b)=(W,1)$ or $(1,W)$,  we obtain
\begin{eqnarray}
     \mathsf{H}_{B_1B_W,W}^{B_WB_W}\times \mathsf{H}_{B_WB_W,W}^{B_1B_W} = \mathsf{H}_{B_WB_1,W}^{B_WB_W} \times \mathsf{H}_{B_WB_W,W}^{B_WB_1} = \varphi^{-1/2} \mathsf{H}_{B_WB_W,1}^{B_WB_W}  - \varphi^{-3/2} \mathsf{H}_{B_WB_W,W}^{B_WB_W} \,,
\end{eqnarray}
and finally when $(a,b)=(W,W)$, one finds
\begin{eqnarray}
    \mathsf{H}_{B_WB_W,W}^{B_WB_W} \times \mathsf{H}_{B_WB_W,W}^{B_WB_W} = \mathsf{H}_{B_WB_W,1}^{B_WB_W}  + \varphi^{-3/2} \mathsf{H}_{B_WB_W,W}^{B_WB_W} \,.
\end{eqnarray}
The last case appeared in \cite{Choi:2023xjw}. In contrast to the fusion of topological lines in the bulk, where the fusion coefficients are non-negative integers, the fusion rules of topological lines in the presence of boundaries  are in general non-integral. This is of course because the lines are terminating on local topological junction operators, which can be rescaled by arbitrary complex numbers.

\subsection{Torus partition functions}

Next, let us consider  the symmetry-resolved torus partition functions of a theory with Fibonacci symmetry. For each of the theories listed in Section \ref{subsec:Fibtheories}, it is straightforward to write down the partition functions in the representation basis. 

For example, when $Q$ is a diagonal rational conformal field theory with $\mathrm{Fib}$ as its Verlinde lines, one has that 
\begin{align}
    \mathbf{Z}_{(a,b)}(\tau) = \mathrm{ch}_{a}(q) \overline{\mathrm{ch}_{b}(q)}, \ \ \ q=e^{2\pi i \tau}
\end{align}
where the $\mathrm{ch}_a(q)$ are the characters of the chiral algebra of $Q$, with $a\in \{1,W\}$. (See \cite{Mukhi:2022bte,Rayhaun:2023pgc} for detailed information on these characters.) 

Similarly, in the case that $Q$ is a chiral conformal field theory, like $(E_8)_1$ or the monster CFT $V^\natural$, the subalgebra $\mathcal{V}_{1}$ of $\mathrm{Fib}$-neutral operators will form a chiral algebra in and of itself, and the representation basis partition functions are precisely equal to the characters of the irreducible modules $\mathcal{V}_{(a,b)}$ of $\mathcal{V}_{1}$. For example, in the case of the chiral $(E_8)_1$ WZW model, it is known that the subalgebra $\mathcal{V}_{1}$ of operators which commute with $\mathrm{Fib}$ is isomorphic to the chiral algebra of $(G_2)_1\otimes (F_4)_1$, and hence
\begin{align}
    \mathbf{Z}_{(a,b)}(\tau) = \mathrm{ch}_a^{(G_2)_1}(q) \mathrm{ch}_b^{(F_4)_1}(q)\,.
\end{align}
Similarly, in the case of the monster CFT $V^\natural$, the characters of the subalgebra $\mathcal{V}_{1}$ were worked out in \cite{MRb}, using some results of \cite{Bae:2020pvv,Rayhaun:2023pgc}.

With the representation basis partition functions in hand, one may then ask for the torus partition functions in the symmetry basis. We will derive the representation $\to$ symmetry basis transformation in two ways and check the two methods against each other. The five consistent symmetry basis partition functions are 
\begin{align}
\begin{split}
    &\hspace{.5in}Z(\tau)\equiv Z_{1}^{1,1}(\tau) = \input{figures/Z111.tikz}\ , \ \ \ Z_{W}^{1,W}(\tau) = \input{figures/ZW1W.tikz}\ ,  \\
    &Z_{1}^{W,W}(\tau) = \input{figures/Z1WW.tikz}\ , \ \ \ Z_W^{W,1}(\tau) = \input{figures/ZWW1.tikz}\ , \ \ \ Z_{W}^{W,W}(\tau) = \input{figures/ZWWW.tikz}\ .
\end{split}
\end{align}
The first two partition functions simply count the genuine local operators and the $W$-twisted sector local operators, respectively. To determine these, we may use the fact that 
\begin{equation}
\begin{split}
    Z(\tau) &= \sum_{(a,b)\in Z(\mathrm{Fib})} \dim(W^{(a,b)}_{1}) \mathbf{Z}_{(a,b)}(\tau),\\
    Z_W^{1,W}(\tau) &= \sum_{(a,b)\in Z(\mathrm{Fib})}\dim(W_W^{(a,b)})\mathbf{Z}_{(a,b)}(\tau)
\end{split}
\end{equation}
where $\dim(W_c^{(a,b)})$ is the number of ways that the bulk line $(a,b)\in Z(\mathrm{Fib})$ of the SymTFT may form a junction with the line $c$ on the boundary. (We hope that the overloading of the symbol ``$W$'' will not cause confusion: it both refers to the junction space in Equation \eqref{eqn:tuberepspaces}, and to the non-trivial simple line of $\mathrm{Fib}$.) Using the fact that the bulk-to-boundary map is $F(a,b)=a\otimes b$, we find that $(1,1)$ and $(W,W)$ can each form a single junction with $1$ on the boundary, while $(W,1)$, $(1,W)$, and $(W,W)$ can each form a single junction with $W$ on the boundary. Thus, 
\begin{equation}\label{eqn:ZZW1Wrep2sym}
\begin{split}
    Z(\tau) &= \mathbf{Z}_{(1,1)}(\tau)+\mathbf{Z}_{(W,W)}(\tau), \\Z_{W}^{1,W}(\tau) &= \mathbf{Z}_{(W,1)}(\tau) + \mathbf{Z}_{(1,W)}(\tau) + \mathbf{Z}_{(W,W)}(\tau). 
\end{split}
\end{equation}

To calculate $Z_{1}^{W,W}(\tau)$ and $Z_W^{W,1}(\tau)$, we may use the fact that these can be obtained from $Z_{W}^{1,W}(\tau)$ via modular transformations. Indeed, from \eqref{eqn:modtranssymbasis}, one may deduce that 
\begin{align}
    Z_{1}^{W,W}(\tau)=Z_{W}^{1,W}(-1/\tau), \ \ \ Z_W^{W,1}(\tau) =Z_W^{1,W}(\tau-1).
\end{align}
On the other hand, the partition functions in the representation basis transform according to the modular data of the $Z(\mathrm{Fib})$ modular tensor category, as in Equation \eqref{eqn:repbasismodulartransformtorus}, and so we find that 
\begin{align}\label{eqn:Z1WWZWW1rep2sym}
\begin{split}
    Z_{1}^{W,W}(\tau) &= \varphi \mathbf{Z}_{(1,1)}(\tau) - \varphi^{-1}\mathbf{Z}_{(W,W)}(\tau) \\
    Z_W^{W,1}(\tau) &= e^{4 \pi i/5}\mathbf{Z}_{(1,W)}(\tau) + e^{-4\pi i/5}\mathbf{Z}_{(W,1)}(\tau) + \mathbf{Z}_{(W,W)}(\tau)\,.
\end{split}
\end{align}

Finally, the last symmetry basis partition function $Z_W^{W,W}(\tau)$ can be computed using the fact that there are linear dependencies which come from ``conjugation'' by the $W$-line, \eqref{eqn:toruslineardependency}. Indeed, taking $b=W$ in this equation, one finds that 
\begin{align}\label{eqn:ZWWrep2sym}
\begin{split}
    Z_W^{W,W}(\tau) &= \sqrt{2+\sqrt{5}}\left(\varphi Z(\tau)-Z_{1}^{W,W}(\tau)-Z_{W}^{1,W}(\tau)-Z_W^{W,1}(\tau)\right)  \\
    &= (e^{\pi i/5}-1)\varphi^{3/2} \mathbf{Z}_{(W,1)}(\tau)-(e^{4\pi i/5}+1)\varphi^{3/2}\mathbf{Z}_{(1,W)}(\tau)  
     + (\sqrt{5}-2)^{1/2}\mathbf{Z}_{(W,W)}(\tau)
\end{split}
\end{align}
so that $Z_W^{W,W}(\tau)$ can be determined from the other 4 symmetry-basis partition functions. 

On the other hand, since $\mathrm{Fib}$ is itself a modular tensor category, its half-braiding is straight-forwardly determined via \eqref{eqn:modularhalfbraiding}. Then, one may check that Equations \eqref{eqn:ZZW1Wrep2sym}, \eqref{eqn:Z1WWZWW1rep2sym}, and \eqref{eqn:ZWWrep2sym} all follow from the character-theoretic change-of-basis formula in \eqref{eq:symtoreptorus}.

For completeness, we record the modular transformation equations in the symmetry basis, \eqref{eqn:modtranssymbasis},
\begin{align}
\begin{split}
    \left(\begin{array}{c} Z \\ Z_W^{1,W} \\ Z_{1}^{W,W} \\ Z_{W}^{W,1} \\ Z_W^{W,W}\end{array}\right)(-1/\tau)  &= \left(\begin{array}{ccccc} 1 & 0 & 0 & 0 & 0 \\ 
    0 & 0 & 1 & 0 & 0 \\ 0 & 1 & 0 & 0 & 0  \\ 0 & 0 &0 & \varphi^{-1} & \varphi^{-1/2} \\ 0 & 0 & 0 &\varphi^{-1/2} & -\varphi^{-1}\end{array}\right)\left(\begin{array}{c} Z \\ Z_W^{1,W} \\ Z_{1}^{W,W} \\ Z_{W}^{W,1} \\ Z_W^{W,W}\end{array}\right)(\tau)  \\
     \left(\begin{array}{c} Z \\ Z_W^{1,W} \\ Z_{1}^{W,W} \\ Z_{W}^{W,1} \\ Z_W^{W,W}\end{array}\right)(\tau+1)  &= e^{2\pi i (c-\bar c)/24} \left(\begin{array}{ccccc} 1 & 0 & 0 & 0 & 0 \\ 
    0 & 0 & 0 & \varphi^{-1} & \varphi^{-1/2} \\ 0 & 0 & 1 & 0 & 0  \\ 0 & 1 &0 & 0 & 0 \\ 0 & 0 & 0 &\varphi^{-1/2} & -\varphi^{-1}\end{array}\right)\left(\begin{array}{c} Z \\ Z_W^{1,W} \\ Z_{1}^{W,W} \\ Z_{W}^{W,1} \\ Z_W^{W,W}\end{array}\right)(\tau) 
\end{split}
\end{align}
and note that they are easily checked to be consistent with the modular transformation properties of the representation basis, \eqref{eqn:repbasismodulartransformtorus}.

\subsection{Annulus partition functions and twisted boundary states}

Next, we move on to the symmetry-resolved annulus partition functions. The only indecomposable module category of $\mathrm{Fib}$ is the regular module category, i.e.\ $\mathrm{Fib}$ itself. Accordingly, there is only one kind of multiplet of boundary conditions possible in a theory with Fibonacci symmetry. Let us call the two boundaries in such a multiplet $B_{1}$ and $B_W$. The representations of the boundary tube algebra in this case are also labeled by objects $\alpha\in\{1,W\}.$

The chiral CFTs with Fibonacci symmetry, such as $(E_8)_1$ and the monster $V^\natural$, cannot admit boundary conditions for the simple reason that they have a gravitational anomaly. However the rest of the theories in Section \ref{subsec:Fibtheories} have a plethora of boundary conditions. For example, the analysis below applies most straightforwardly to the Cardy boundary conditions of the rational conformal field theories with $\mathrm{Fib}$ as their category of Verlinde lines, such as the $(G_2)_1$ WZW model. In this case, one has that the representation basis annulus partition functions are given by the characters of the chiral algebra, $\mathbf{Z}_\alpha(\delta) = \mathrm{ch}_\alpha(q)$  with $\alpha \in \{1,W\}$ and $q=e^{-\pi \delta}.$

There are five  annulus partition functions in the symmetry basis. The first four are of the form $Z_{B_1B_2}(\delta)$ with $B_1,B_2\in \{B_{1}, B_W\}$; they correspond to placing the theory on a cylinder with boundaries $B_1$ and $B_2$ imposed at the two ends, without the additional insertion of any topological line operators. The fifth symmetry basis partition function is $Z_{B_WB_W}^W(\delta)$, which corresponds to placing the theory on an annulus with the boundary $B_W$ imposed at both ends, and with the bulk topological line $W$ stretching from one boundary to the other. 

From the SymTFT, one finds that the first four symmetry basis partition functions are related to the representation basis partition functions via the formula 
\begin{align}
    Z_{B_1B_2}(\delta) = \sum_{\alpha\in \{1,W\}}\widetilde{N}_{B_1\alpha}^{B_2}\mathbf{Z}_\alpha(\delta)
\end{align}
where $\widetilde{N}_{B_1a}^{B_2}$ are the fusion coefficients of the regular $\mathrm{Fib}$-module category, i.e.\ $\widetilde{N}_{B_1a}^{B_2}$ are just the standard fusion coefficients of the Fibonacci category. In particular, this implies that 
\begin{align}
    Z_{B_{1}B_{1}}(\delta) = \mathbf{Z}_{1}(\delta), \ \ \ Z_{B_{1}B_W}(\delta) = Z_{B_WB_{1}}(\delta) = \mathbf{Z}_W(\delta), \ \ \ Z_{B_WB_W}(\delta) = \mathbf{Z}_{1}(\delta) +\mathbf{Z}_W(\delta).
\end{align}
The fifth partition function can be derived in a variety of ways. One approach is to use the linear dependency equation, \eqref{eqn:annulusconjugation}, from which one concludes that 
\begin{align}
    Z_{B_WB_W}^W(\delta) = \varphi^{1/2}\mathbf{Z}_{1}(\delta) -\varphi^{-1/2} \mathbf{Z}_W(\delta).
\end{align}
One can check that the same result can be derived using the character theory of the boundary tube algebra. In particular, one can use \eqref{eq:annulus_rep_to_symm} to express $Z^W_{B_WB_W}(\delta)$ in terms of representation basis partition functions with coefficients given by the characters $\chi_\alpha$ defined in Equation \eqref{eq:character_bdy}. Then, one may use the generalized Verlinde formula of Equation \eqref{eq:verlinde} to express the characters in terms of generalized half-linking numbers. Next, one can use Equation \eqref{eq:psi_DD} to express the generalized half-linking numbers in terms of the $\Omega$-symbols of $\mathrm{Fib}$, and finally Equation \eqref{eqn:modularhalfbraiding} to express the $\Omega$-symbols in terms of the $F$-symbols and $R$-matrices of $\mathrm{Fib}$, which we reported at the beginning of this section.

\section{Applications}\label{sec:apps}

In this section, we telegraphically sketch a variety of applications that flow from the technology that we have developed throughout this paper. We hope to return to some of these in more detail in future work.

\subsection{Selection rules on bulk/boundary/interface correlators}\label{subsec:correlators}

The most basic application of global symmetries is that they imply selection rules on correlation functions. In \cite{Lin:2022dhv}, the authors described how to systematically derive selection rules on correlation functions of bulk local operators in a 1+1d quantum field theory with $\mathcal{C}$ symmetry from the SymTFT. We extend this analysis so that it applies to more general correlators of bulk/boundary/interface local operators.

We consider two conformal field theories $Q_1$ and $Q_2$ with symmetry categories $\mathcal{C}_1$ and $\mathcal{C}_2$, which are separated by a conformal interface $I$. A CFT with boundary is recovered as a special case, corresponding to taking $Q_2$ to be the trivially gapped theory and $\mathcal{C}_2=\mathrm{Vec}$. We consider Euclidean correlation functions of the form
\begin{align}\label{eq:bulkinterfacecorrelator}
   \langle \mathcal{O}^{(1)}_1(x_1)\cdots\mathcal{O}^{(1)}_n(x_n)\mathcal{O}^{(2)}_1(y_1)\cdots \mathcal{O}^{(2)}_m(y_m) \psi_1(t_1)\cdots \psi_l(t_l)\rangle
\end{align}
where the $\mathcal{O}^{(1)}_i$ are local operators of $Q_1$, the $\mathcal{O}^{(2)}_i$ are local operators of $Q_2$, and the $\psi_i$ are local operators sitting on the interface $I$. We ask under what conditions such a correlator is forced to vanish. 

By repeatedly using the bulk-bulk and bulk-interface operator product expansion (OPE), the correlator can be reduced to a sum of 1-point functions of operators supported on the interface $I$. However, by conformal invariance, the only operator supported on the interface which has a non-vanishing 1-point function (on the sphere) is the identity operator. Thus, it immediately follows that the correlator vanishes if we can demonstrate that the identity operator cannot appear after successive applications of the OPE. 

\begin{figure}
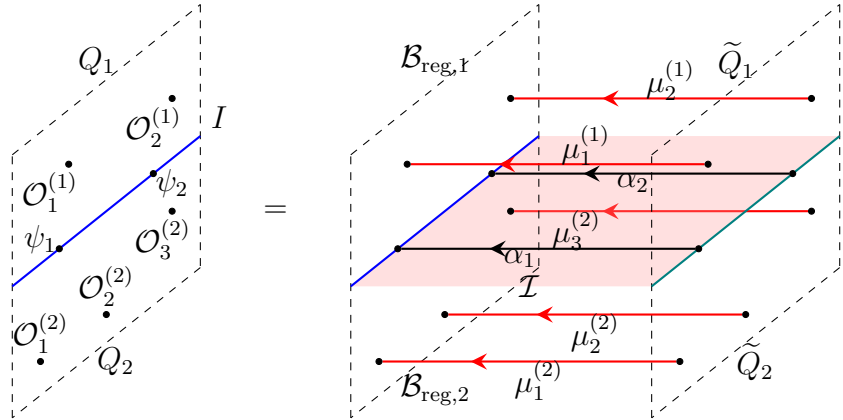

    \ctikzfig{figures/bulkinterfacecorrelator}
    \caption{A bulk/interface correlation function, blown up into 2+1d using the SymTFT.}\label{fig:bulkinterfacecorrelator}
\end{figure}

Determining when the identity operator appears can be systematically addressed from the perspective of the SymTFT. Blowing up to three dimensions, we find that $I$ expands into a topological interface $\mathcal{I}$ between $\mathrm{TV}_{\mathcal{C}_1}$ and $\mathrm{TV}_{\mathcal{C}_2}$. Further, each $\mathcal{O}^{(1)}_i$ expands to a topological line $\mu^{(1)}_i$ of $\mathrm{TV}_{\mathcal{C}_1}$,  each $\mathcal{O}^{(2)}_i$ expands to a topological line $\mu^{(2)}_i$ of $\mathrm{TV}_{\mathcal{C}_2}$, and each $\psi_i$ is expands to a topological line $\alpha_i$ supported on the topological interface $\mathcal{I}$. Mathematically, the $\mu^{(k)}_i$ belong to the modular tensor categories $Z(\mathcal{C}_k)$, and the $\alpha_i$ belong to the fusion category $\mathrm{Fun}_{(\mathcal{C}_1,\mathcal{C}_2)}(\mathcal{I},\mathcal{I})$ of $(\mathcal{C}_1,\mathcal{C}_2)$-bimodule functors from $\mathcal{I}$ to itself. See Figure \ref{fig:bulkinterfacecorrelator}. 

Now, if we take the bulk-bulk OPE of $\mathcal{O}_i^{(k)}$ with $\mathcal{O}_j^{(k)}$, we see that the operators which appear must, in the SymTFT, be attached to a line of $\mathrm{TV}_{\mathcal{C}_k}$ appearing in the decomposition of $\mu^{(k)}_i\otimes \mu^{(k)}_j$ into simples. Likewise, if we expand a bulk operator $\mathcal{O}^{(k)}_i$ into a sum of operators supported on the interface $I$, then the operators which appear must, in the SymTFT, be attached to a topological line arising in the decomposition of $F_k(\mu_i^{(k)})$ into simples, where $F_k$ is the bulk-interface map which describes the result of pushing a bulk line in $\mathrm{TV}_{\mathcal{C}_k}$ onto the topological interface $\mathcal{I}$.\footnote{Note that the bulk-to-interface maps $F_k$ can be understood as more conventional bulk-to-boundary forgetful functors by folding along $\mathcal{I}$.} Using the fact that the identity operator on the interface is attached to the identity line $1$ in the fusion category $\mathrm{Fun}_{(\mathcal{C}_1,\mathcal{C}_2)}(\mathcal{I},\mathcal{I})$, we deduce the following selection rule.  \\

\begin{claim}
    \parbox{\textwidth}{
    The correlator in \eqref{eq:bulkinterfacecorrelator} vanishes unless the identity line $1$ appears in the decomposition of $F_1(\mu_1^{(1)}\otimes\cdots\otimes \mu_n^{(1)})\otimes F_2(\mu_1^{(2)}\otimes\cdots\otimes \mu_m^{(2)})\otimes \alpha_1\otimes\cdots\otimes\alpha_l$ into simple line operators on the topological interface $\mathcal{I}$.
    }%
\end{claim}
We note that it does not matter what order any of the $\mu_i^{(k)}$ appear in this equation because the bulk-to-interface maps are ``central functors'' (see e.g.\ \cite{Kong:2022cpy} for the definition), which in particular means that $F_k(\mu_i^{(k)})\otimes\alpha_j \cong \alpha_j \otimes F_k(\mu_i^{(k)})$. Physically this is clear, because one can always lift the line $F_k(\mu_i^{(k)})$ into the bulk, and then push it back onto the interface on the other side of $\alpha_j$ to change the order of fusion. 

We note also that the same argument carries through mutatis mutandis in the case that the local operators are attached to topological line operators in $\mathcal{C}_1$ and $\mathcal{C}_2$.

As an example of this selection rule in action, consider correlation functions of a diagonal rational conformal field theory in the presence of a Cardy boundary. Call $\mathcal{C}$ the category of Verlinde lines. If a bulk local operator $\mathcal{O}_i$ is a descendant of the primary labeled by $a_i\in\mathcal{C}$, then in the SymTFT picture, it expands into the bulk line $(a_i,a_i^\ast)\in Z(\mathcal{C})\cong \mathcal{C}\boxtimes \overline{\mathcal{C}}$. On the other hand, if $\psi_j$ is a local operator on a Cardy boundary labeled by $b\in\mathcal{C}$, then $\psi_j$, when viewed from the perspective of the SymTFT, must extend to a topological line $\alpha_j\in\mathcal{C}$ on the Dirichlet boundary satisfying $N_{b\alpha_j}^b>0$, where $N_{ab}^c$ are the fusion rules of $\mathcal{C}$. Using the fact that the bulk-to-boundary map is $F(a,b)=a\otimes b$ for the Dirichlet boundary of $\mathcal{C}\boxtimes \overline{\mathcal{C}}$, we derive a selection rule governing bulk/boundary correlators $\langle \mathcal{O}_1(x_1)\cdots\mathcal{O}_n(x_n)\psi_1(t_1)\cdots \psi_l(t_l)\rangle$ which asserts that this correlator vanishes unless 
\begin{align}
    1\subset a_i\otimes a_i^\ast\cdots a_n\otimes a_n^\ast \otimes \alpha_1\otimes\cdots\otimes \alpha_l \in\mathcal{C} .
\end{align}
Of course in this situation, the same selection rule could have been derived by the standard analysis of fusion rules of the rational conformal field theory, but we have reproduced it from the perspective of the SymTFT.

\subsection{Degeneracy and SPT transitions}\label{subsec:degeneracy}

Another physically interesting application of internal global symmetries is that they can often guarantee a degeneracy of states in various Hilbert spaces. A famous example of this is Kramers doubling in certain time-reversal symmetric systems.

As another well-known example, consider a 1+1d QFT with an invertible global symmetry $\mathcal{C}=\mathrm{Vec}_G^\omega$, with $\omega\in H^3(G,\mathbb{C}^\times)$ capturing the anomaly. For each group element $g\in G$, define a 2-cocycle $\psi_g\in H^2(C_G(g),\mathbb{C}^\times)$ as follows, 
\begin{align}\label{eq:psig}
\psi_g(h,h')=\omega(g,h,h')\omega(h,h',g^{hh'})\omega(h,g^h,h')^\ast,
\end{align}
see e.g.\ \cite{Evans:2018qgz}. Here $C_G(g)$ is the centralizer of $g$ in $G$.  The formula \eqref{eq:psig} is the slant product of $\omega$ with respect to $g$. 
Recall that the simple anyons of the SymTFT $Z(\mathrm{Vec}_G^\omega)$ are labeled by pairs $\mu=([g],\rho)$, with $[g]$ a conjugacy class of $G$ and $\rho$ an irreducible representation of $C_G(g)$ transforming projectively with respect to the 2-cocycle $\psi_g$. The lowest energy subspace of the twisted sector Hilbert space $\mathcal{H}_h$ must decompose into a direct sum of the spaces $W^{([g],\rho)}_{h}$, by virtue of \eqref{eqn:schurweyldecomp}. So in particular, if these spaces are either trivial or have $\dim(W^{([g],\rho)}_h)>1$, then it is guaranteed that the ground space is degenerate. 

Recalling that the bulk-to-boundary map is $F([g],\rho) = \dim(\rho) \bigoplus_{h\in [g]}h$, and that $W^{([g],\rho)}_h = \mathrm{Hom}_{\mathcal{C}}(F([g],\rho),h)$,  we learn that $\dim(W^{[g],\rho}_h)=\delta_{[g][h]}\dim(\rho)$. Indeed, $W^{([g],\rho)}_h$ can essentially be thought of as the projective representation $\rho$ of $C_G(g)$ with 2-cocycle $\psi_g$.  If the anomaly is trivial, $\omega=1$, then the twisted sector Hilbert space $\mathcal{H}_h$ transforms under a linear representation of $C_G(g)$; in particular, the trivial representation $\rho=1$ is always possible, and so e.g.\ the lowest energy state(s) in $\mathcal{H}_h$ may not come with any degeneracy. However, if the anomaly is non-trivial, $\omega\neq 1$, then the twisted sector might transform projectively under $C_G(h)$, depending on whether $\psi_h$ is cohomologically trivial or not, and since the projective irreducible representations of a finite group always have dimension greater than 1, we conclude that the ground space must be degenerate. 

To summarize, the twisted sectors with respect to an anomalous invertible symmetry are generally degenerate. As one can imagine, analogous considerations apply when the symmetry is non-invertible.

We derive similar degeneracy constraints on the Hilbert space $\mathcal{H}_{B_1B_2}$ of  local operators which sit at the junction of two $\mathcal{C}$-symmetric boundaries $B_1$ and $B_2$. Equivalently, by the state/operator correspondence, this is the space of states on the interval with $B_1$ and $B_2$ imposed at the two ends.  Indeed, suppose that $\mathcal{C}$ is non-anomalous (see the discussion in Section \ref{subsec:symmetricboundariesanomalies}), and further that it has two inequivalent left $\mathcal{C}$-module categories $\mathcal{B}_1$ and $\mathcal{B}_2$ with $\mathrm{rank}(\mathcal{B}_i)=1$. Then we have the following.\footnote{A related statement in the context of interface modes between distinct non-invertible SPT phases represented by matrix product states recently appeared in \cite{Inamura:2024jke}. See also \cite{Seifnashri:2024dsd} for the case of SPT phases protected by the $\mathrm{Rep}(D_8)$ symmetry.}
\begin{claim}
    \parbox{\textwidth}{
    Suppose that a 1+1d QFT $Q$ admits two $\mathcal{C}$-symmetric boundary conditions $B_1$ and $B_2$  belonging to multiplets which are inequivalent as $\mathcal{C}$-module categories. Then every energy level of the interval Hilbert space $\mathcal{H}_{B_1B_2}$ is guaranteed to be degenerate.
    }%
\end{claim}

To see why, we analyze the irreducible representations $W^\alpha=W^\alpha_{\underline{B}_1\underline{B}_2}$ of the boundary tube algebra $\mathrm{Tube}(\mathcal{B}_1^\vee\vert\mathcal{B}_2)$ into which the Hilbert space $\mathcal{H}_{B_1B_2}$ must decompose. Here, $\underline{B}_i$ is the unique topological line interface between the Dirichlet boundary condition of $\mathrm{TV}_{\mathcal{C}}$ described by $\mathcal{B}_{\mathrm{reg}}$, and the topological boundary described by $\mathcal{B}_i$. Let $A_1$ and $A_2$ be algebra objects of $\mathcal{C}$ such that $\mathcal{B}_i = \mathcal{C}_{A_i}$. Recall that the label $\alpha$ of the irreducible representation can be thought of as a simple object of ${_{A_1}}\mathcal{C}_{A_2}$, or more physically as a topological line interface between the two topological boundary conditions of $\mathrm{TV}_{\mathcal{C}}$ defined by $\mathcal{B}_1$ and $\mathcal{B}_2$. Furthermore, the dimension of the corresponding representation, $\dim(W^\alpha)$, is defined by the equation
\begin{align}
    \underline{B}_1 \otimes_{A_1} \alpha = \dim(W^\alpha) \underline{B}_2
\end{align}
where $\otimes_{A_1}$ refers to the tensor product of $A_1$-modules. In other words, the dimension of $W^\alpha$ agrees with the quantum dimension $\d_\alpha$ of the line interface $\alpha$. Similarly, we have that 
\begin{align}
    \underline{B}_2\otimes_{A_2}\bar\alpha = \dim(W^{\bar\alpha}) \underline{B}_1
\end{align}
where $\bar\alpha$ is the orientation-reversal of the interface $\alpha$. The interface $\bar\alpha$ is described mathematically by an $A_2$-$A_1$ bimodule, and it labels an irreducible representation of $\mathrm{Tube}(\mathcal{B}_2^\vee\vert\mathcal{B}_1)$. Thus, $\dim(W^{\bar\alpha})=\d_{\bar\alpha}=\d_\alpha$. 

To constrain the dimension $\d_\alpha$, we note that one can think of $\alpha\otimes_{A_2}\bar\alpha$ (the tensor product of the representations $\alpha$ and $\bar\alpha$, which produces a representation of $\mathrm{Tube}(\mathcal{B}_1^\vee\vert\mathcal{B}_1)$) as a gaugeable algebra in the fusion category ${_{A_1}}\mathcal{C}_{A_1}$. Moreover, it is an algebra which, when gauged, gives ${_{A_2}}\mathcal{C}_{A_2}$ as the dual symmetry. Because $A_1$ and $A_2$ are assumed to be Morita inequivalent, this algebra object $\alpha\otimes_{A_2}\bar\alpha$ must be non-trivial, and hence have a quantum dimension greater than 1. Therefore, it follows that $\dim(W^\alpha)^2=\d_\alpha^2 = \d_{ \alpha\otimes_{A_2}\bar \alpha} >1$.

\paragraph{Example: invertible symmetry} The simplest example of this phenomenon in action is a theory with a non-anomalous invertible symmetry $\mathcal{C}=\mathrm{Vec}_G$ for which $H^2(G,\mathbb{C}^\times)$ is non-trivial. Rank-1 module categories (and therefore, multiplet types of $G$-symmetric boundaries) are labeled by classes $\psi \in H^2(G,\mathbb{C}^\times)$. Suppose that a  theory admits two $G$-symmetric boundaries $B_1$ and $B_2$ with 2-cocycles $\psi_1$ and $\psi_2$, respectively. Then $\mathrm{Tube}(\mathcal{B}_1^\vee\vert \mathcal{B}_2)$ is the twisted group algebra $\mathbb{C}^{\psi_1^{-1}\psi_2}[G]$, and hence $\mathcal{H}_{B_1B_2}$ transforms as a projective representation of $G$ with $2$-cocycle $\psi_1^{-1}\psi_2$. Because projective representations of a group have dimension greater than 1, the ground space must be degenerate. 

\paragraph{Example: $\mathrm{Rep}(D_8)$} As another example, consider $\mathcal{C}=\mathrm{Rep}(D_8)$. This category is known to have 3 inequivalent module categories of rank-1, i.e.\ 3 different kinds of multiplets of symmetric boundary conditions \cite{Seifnashri:2024dsd, Thorngren:2019iar, Cordova:2023bja, Inamura:2021wuo, Bhardwaj:2024qrf, tambara2000representations}. Suppose that a theory admits a pair of symmetric boundary conditions $B_1$ and $B_2$ belonging to two multiplets $\mathcal{B}_1$ and $\mathcal{B}_2$ which are inequivalent as $\mathrm{Rep}(D_8)$ module categories. Then by the logic above, the corresponding interval Hilbert space $\mathcal{H}_{B_1B_2}$ must be degenerate at every energy level. 

There is a slick way to compute the minimum degeneracy without doing any computations. We can simply note that $\mathrm{Tube}(\mathcal{B}_1^\vee\vert\mathcal{B}_2)$ is an $8$-dimensional $C^\ast$-algebra. Since finite-dimensional $C^\ast$-algebras are semisimple, the Wedderburn-Artin theorem allows us to conclude that it must be a direct sum of matrix algebras. The only possibilities in dimension $8$ are $8\cdot M_{1}(\mathbb{C})$, $4\cdot M_1(\mathbb{C})\oplus M_2(\mathbb{C})$, and $2\cdot M_2(\mathbb{C})$. The last option is the only one which does not admit any irreducible representations of dimension $1$, and in fact it only admits 2-dimensional representations. Thus, the entire spectrum of $\mathcal{H}_{B_1B_2}$ is doubled at each energy level. 

\medskip

Before moving on, we briefly note that there is an interesting application of this degeneracy phenomenon, in the spirit of \cite{Cordova:2022lms}, which treated the case of invertible symmetries. Indeed, suppose one is interested in constraining the central charge of any CFT which is capable of mediating a phase transition between two gapped $\mathcal{C}$-symmetric SPT phases. By definition, such a CFT admits two relevant $\mathcal{C}$-symmetric deformations which trivially gap the theory, and hence, by activating these deformations in half of spacetime, it comes equipped with two distinguished $\mathcal{C}$-symmetric ``RG boundary conditions.'' 

Recall from the discussion in Section \ref{subsec:symmetricboundariesanomalies} that $\mathcal{C}$-symmetric SPT phases in 1+1d are labeled by rank-1 $\mathcal{C}$-module categories, as are multiplet types of $\mathcal{C}$-symmetric boundaries. One then has the expectation that a $\mathcal{C}$-symmetric SPT phase to which a CFT can be deformed, and the multiplet type of the corresponding RG boundary, precisely agree as abstract $\mathcal{C}$-module categories.

Now, suppose that a CFT which transitions between two SPTs labeled by $\mathcal{C}$-module categories $\mathcal{B}_1$ and $\mathcal{B}_2$ is rational and diagonal. Call the RG boundary conditions $B_1$ and $B_2$. In this situation, \cite{Cordova:2022lms} derived an inequality of the form 
\begin{align}
    \log(d)\leq \frac{\pi}{8}c_{\mathrm{eff}}, \ \ \ c_{\mathrm{eff}}\to\infty,
\end{align}
where $d$ is the ground state degeneracy of the interval Hilbert space $\mathcal{H}_{B_1B_2}$ of the CFT with the two RG boundary conditions imposed, and $c_{\mathrm{eff}}=c+8h_{\mathrm{min}}$ with $h_{\mathrm{min}}$ the minimum scaling dimension of boundary local operators in $\mathcal{H}_{B_1B_2}$. (See Equation (1.13) of \cite{Cordova:2022lms} for a more complicated bound which holds away from large $c_{\mathrm{eff}}$.) Thus, one sees that any non-trivial lower bound on degeneracies $d$ in the interval Hilbert space translate to lower bounds on the effective central charge $c_{\mathrm{eff}}$ of an SPT-transitioning CFT. As we have just seen, the representation theory of boundary tube algebras precisely furnishes such lower bounds on $d$.

\subsection{Interface fusion}\label{subsec:interfacefusion}

Another application of our SymTFT picture is that it constrains the fusion of conformal interfaces. 

While the fusion of topological defects of arbitrary codimension is by now a standard consideration in the field, it is less appreciated that conformal defects may also be fused with each other in a sensible way, and admit a well-defined operator algebra. Part of the subtlety is that, just as with local operators, the limit in which two conformal defects are brought close to each other is in general singular. However, conformal invariance tightly constrains the possible singularities that may arise: in the special case of an interface $I$ between CFTs $Q_1$ and $Q_2$ and an interface $I'$ between CFTs $Q_2$ and $Q_3$, their fusion may be rendered finite using just a single self-energy counterterm \cite{Bachas:2007td},
\begin{align}
    (I\circ I')(\gamma) \equiv \lim_{\epsilon\to 0} e^{2\pi a/\epsilon}I(\gamma_\epsilon) I'(\gamma)\end{align}
where $\gamma$ is a 1-dimensional line, and $\gamma_\epsilon$ is the translation of $\gamma$ by $\epsilon$ in the direction orthogonal to its worldline. Remarkably, $I\circ I'$ is a direct sum/linear combination of simple conformal interfaces with non-negative integer coefficients,
\begin{align}\label{eqn:interfacefusion}
    I\circ I' = \bigoplus_{I''}N_{II'}^{I''}I'', \ \ \ \ \ N_{I,I'}^{I''}\in \mathbb{Z}^{\geq 0}.
\end{align}
See \cite{Diatlyk:2024zkk} for a more general discussion which applies to conformal defects of arbitrary codimension inside of conformal field theories with arbitrary dimension.

\begin{figure}
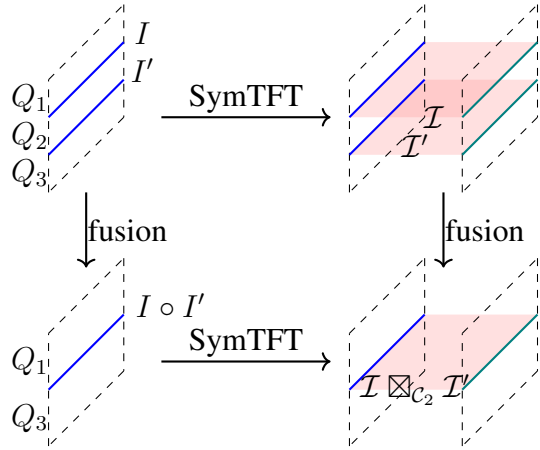

    \ctikzfig{figures/interfacefusion}
    \caption{The $(\mathcal{C}_1,\mathcal{C}_3)$-multiplet (i.e.\ $(\mathcal{C}_1,\mathcal{C}_3)$-bimodule category) to which the fusion $I\circ I'$ of two conformal interfaces $I$ and $I'$ belongs is represented by the topological interface $\mathcal{I}\boxtimes_{\mathcal{C}_2}\mathcal{I}'$ in the SymTFT. }\label{fig:interfacefusion}
\end{figure}

Global symmetries can be used to constrain which interfaces $I''$ arise in the fusion product of \eqref{eqn:interfacefusion}. In other words, global symmetries provide selection rules on the fusion coefficients $N_{II'}^{I''}$. Indeed, assume that $Q_i$ has a symmetry category $\mathcal{C}_i$.  To describe the resulting selection rules, recall that, in the SymTFT picture, $I$ blows up into  a topological interface $\mathcal{I}$ between the SymTFTs $\mathrm{TV}_{\mathcal{C}_1}$ and $\mathrm{TV}_{\mathcal{C}_2}$ of $Q_1$ and $Q_2$, respectively. (Mathematically, $\mathcal{I}$ is a $(\mathcal{C}_1,\mathcal{C}_2)$ bimodule category.)  Likewise, $I'$ expands into a topological interface $\mathcal{I}'$ between $\mathrm{TV}_{\mathcal{C}_2}$ and $\mathrm{TV}_{\mathcal{C}_3}$. Whatever line $I''$ appears in \eqref{eqn:interfacefusion}, it must be attached to a topological interface $\mathcal{I}''$ between $\mathrm{TV}_{\mathcal{C}_1}$ and $\mathrm{TV}_{\mathcal{C}_3}$ which arises in the decomposition of $\mathcal{I}\boxtimes_{\mathcal{C}_2}\mathcal{I}'$, the fusion of $\mathcal{I}$ with $\mathcal{I}'$, into indecomposable $(\mathcal{C}_1,\mathcal{C}_3)$ bimodule categories. See Figure \ref{fig:interfacefusion}. We summarize this as follows.\\

\begin{claim}
    \parbox{\textwidth}{
    The fusion coefficient $N_{II'}^{I''}$ vanishes unless the topological interface $\mathcal{I}''$ between $\mathrm{TV}_{\mathcal{C}_1}$ and $\mathrm{TV}_{\mathcal{C}_3}$, which describes the $(\mathcal{C}_1,\mathcal{C}_3)$ multiplet structure of $I''$, arises in the decomposition of   $\mathcal{I}\boxtimes_{\mathcal{C}_2} \mathcal{I}'$ into indecomposable interfaces. 
    }%
\end{claim}

It would be interesting to verify these selection rules in the context of $c=1$ compact boson CFTs, where a variety of conformal interfaces are known \cite{Bachas:2007td}. To this end, we remark that the fusion of topological interfaces is under good mathematical control in the SymTFTs corresponding to non-anomalous Abelian finite group symmetry categories (see e.g.\ Proposition 3.19 of \cite{etingof2010fusion}).

\subsection{Line operator RG flows}

Another important application of global symmetries is to RG flows. The prototype for this  is the following situation:  if a (say 1+1d) quantum field theory $Q$ in the UV admits a relevant deformation which preserves an anomalous symmetry, then the theory in the IR cannot be trivially gapped: it must either be gapless, or else a topological field theory with degenerate vacua. In 1+1d, it is known that a fusion category is anomalous if it does not admit any rank-1 module categories.

One may ask if there are analogous constraints on RG flows of line operators. In particular, consider starting with a bulk 1+1d CFT with symmetry category $\mathcal{C}$ in the presence of a line operator $I$ which belongs to a $\mathcal{C}$-multiplet $\mathcal{I}$ and decomposes into a triple $I\leftrightharpoons(\underline{I},\mathcal{I},\widetilde{I})$ in the SymTFT, as in Figure \ref{fig:interfacesandwich}. We then imagine deforming the theory by integrating a local operator  $\mathcal{O}$ over the line, and asking what the fate of $I$ is in the IR \cite{Wu:2023ezm, Aharony:2023amq, Aharony:2022ntz, Cuomo:2022xgw, Cuomo:2021kfm, Cuomo:2021rkm, Nagar:2024mjz, Giombi:2022vnz, Graham:2003nc, Konechny:2012wm, Konechny:2020jym, Konechny:2019wff, Andrei:2018die, Konechny:2015qla, Kormos:2009sk}. 

Now, $\mathcal{O}$ itself decomposes into a triple $(x, \alpha, \widetilde{\mathcal{O}})$ in the SymTFT, where $\alpha$ is a topological line operator supported on the topological surface $\mathcal{I}$ (which is a special case of the Figure \ref{fig:junctionsymTFT}). Generally, if the operator $\mathcal{O}$ transforms in a non-trivial representation of $\mathrm{Tube}(\mathcal{I}^\vee\vert\mathcal{I})$, i.e.\ if $\alpha$ is not the identity line, then deforming by $\mathcal{O}$ will modify the $(\mathcal{C},\mathcal{C})$-bimodule structure of the multiplet of interfaces to which $I$ belongs; this is because when $\mathcal{O}$ is integrated along the worldline of $I$, the topological line  $\alpha$ is correspondingly  smeared along the surface $\mathcal{I}$ in the bulk SymTFT which encodes this $(\mathcal{C},\mathcal{C})$-bimodule structure. On the other hand, if $\alpha$ is trivial, so that $\mathcal{O}$ is a symmetry-preserving deformation, then one observes in the SymTFT that integrating $\mathcal{O}$ along $I$ amounts to integrating just $\widetilde{\mathcal{O}}$ along $\widetilde{I}$, without touching the bulk surface $\mathcal{I}$. Thus, the $(\mathcal{C},\mathcal{C})$-bimodule structure is matched in the UV and in the IR, in a manner spiritually similar to the matching of 't Hooft anomalies in bulk RG flows. 

This observations of the previous paragraph, coupled with the fact that the topological lines in $\mathcal{C}$ (including the trivial line) transform in the regular $(\mathcal{C},\mathcal{C})$ multiplet, leads to the following. \\

\begin{claim}
    \parbox{\textwidth}{
    Suppose $Q$ is a 1+1d CFT with a symmetry category $\mathcal{C}$ and a line operator $I$ which is deformed by a $\mathcal{C}$-preserving relevant local operator $\mathcal{O}$ supported on $I$. If $I$ does not transform in the ``regular multiplet'' (i.e.\ if the bulk surface operator $\mathcal{I}$ in the SymTFT is non-trivial), then $I$ cannot be completely screened in the IR. That is, $I$ flows either to a non-trivial conformal interface, or a non-trivial topological line operator.
    }%
\end{claim}

Although we have specialized to two dimensions, it is clear that the logic employed generalizes to defects of arbitrary codimension in QFTs of arbitrary dimension.

\section*{Acknowledgements}

We thank Andrea Antinucci, Giovanni Galati, Hongliang Jiang, Ryohei Kobayashi, Zohar Komargodski, Justin Kulp, Rajath Radhakrishnan, Daniel Ranard, Daniel Robbins, Shinsei Ryu, Sahand Seifnashri, Shu-Heng Shao, Nikita Sopenko, Zhengdi Sun, and Fei Yan for useful discussions. YC gratefully acknowledges funding provided by the Roger Dashen Member Fund and the Fund for Memberships in Natural Sciences at the Institute for Advanced Study. BR gratefully acknowledges NSF grant PHY-2210533 and DOE grant DE-SC0009988. YZ gratefully acknowledges NSFC grant No.12505093 and the start-up fund from UCAS.

\appendix

\section{More on generalized half-linking numbers} \label{app:half-linking}

\subsection{Orthogonality}

Here, we prove the first orthogonality relation in \eqref{eqn:halflinkingorthogonality} satisfied by the generalized half-linking numbers \eqref{eqn:halflinkingdefn}, following the analogous derivation in \cite[Appendix C]{Lin:2022dhv}.
We first note that
\begin{align}\label{eqn:halflinkingid1}
    \input{figures/halflinkid1lhs.tikz} = \sum_{\beta \in \mathrm{Irr}(\mathcal{I})} \sum_{z=1}^{(\widetilde{N}_{R})_{\alpha a}^{\beta}}  \sum_{w=1}^{(\widetilde{N}_L)_{b\alpha}^{\beta}}\frac{{^{\mathcal{B}_1\mathcal{B}_2}}\Psi_{\alpha\beta(\mu x y)}^{(az)(bw)}}{\sqrt{S_{11} \qd_\alpha^2}} ~~ \input{figures/halflinkid1rhs.tikz} \,,
\end{align}
which generalizes Equation (C.5) of \cite{Lin:2022dhv}, and can be proven by expanding the right-hand side in unknown coefficients and then plugging into the defining equation for the generalized half-linking numbers, \eqref{eqn:halflinkingdefn}. 
Here $\mathcal{I}$ is the multiplet of topological line  interfaces between $\mathcal{B}_1$ and $\mathcal{B}_2$ topological boundary conditions.
One can similarly derive
\begin{align}\label{eqn:halflinkingid1_tilde}
    \input{figures/halflinkid1lhs_tilde.tikz} = \sum_{\beta \in \mathrm{Irr}(\mathcal{I})} \sum_{z=1}^{(\widetilde{N}_{R})_{\alpha a}^{\beta}}  \sum_{w=1}^{(\widetilde{N}_L)_{b\alpha}^{\beta}}\frac{{^{\mathcal{B}_1\mathcal{B}_2}}\widetilde{\Psi}_{\alpha\beta(\mu x y)}^{(az)(bw)}}{\sqrt{S_{11} \qd_\alpha^2}} ~~ \input{figures/halflinkid1rhs_tilde.tikz} \,.
\end{align}

Using \eqref{eqn:halflinkingid1}, we obtain
\begin{align}\label{eqn:halflinkinglemma1}
\input{figures/halflinkinglemma2.tikz}=\sum_{\beta \in \mathrm{Irr}(\mathcal{I})} \sum_{z=1}^{(\widetilde{N}_{R})_{\alpha a}^{\beta}}  \sum_{w=1}^{(\widetilde{N}_L)_{b\alpha}^{\beta}} \frac{\sqrt{\qd_a \qd_b}}{S_{11}\qd_\alpha}  {^{\mathcal{B}_1\mathcal{B}_2}}\Psi_{\alpha\beta(\mu x y)}^{(az)(bw)}{^{\mathcal{B}_1\mathcal{B}_2}}\widetilde{\Psi}_{\alpha\beta(\mu' x' y')}^{(az)(bw)} \,.
\end{align}
To derive \eqref{eqn:halflinkinglemma1}, one first uses \eqref{eqn:halflinkingid1} locally around the $\mu$ line, and then use the definition of $\widetilde{\Psi}$ in \eqref{eqn:halflinkingdefn}.
On the other hand, we can compute this configuration, summed over $\alpha$, in another way. First, note that arguments which are nearly identical to the ones given in \cite{Lin:2022dhv} can be used to show that
\begin{align}
   \sum_{\alpha \in \mathrm{Irr}(\mathcal{I})} \d_\alpha ~\input{figures/halflinkinglemma1.tikz}\propto \delta_{\mu,1} \,.
\end{align}
Using this, we find that 
\begin{align}\label{eqn:halflinkinglemma2}
\begin{split}
    &\sum_{\alpha\in \mathrm{Irr}(\mathcal{I})} \qd_\alpha ~ \input{figures/halflinkinglemma2.tikz} = \sum_{\alpha\in \mathrm{Irr}(\mathcal{I})} \sum_{\rho\in Z(\mathcal{C})}\sum_{i=1}^{N_{\bar{\mu}\mu'}^{\rho}} \qd_\alpha \sqrt{\frac{\qd_\rho}{\qd_\mu \qd_{\mu'}}}~\input{figures/halflinkinglemma3.tikz} \\&= \sum_{\alpha\in \mathrm{Irr}(\mathcal{I})} \frac{\qd_\alpha}{\qd_\mu } \delta_{\mu\mu'} ~\input{figures/halflinkinglemma4.tikz} 
    = \sqrt{\qd_a \qd_b}\delta_{\mu\mu'}\delta_{xx'}\delta_{yy'}\sum_{\alpha\in \mathrm{Irr}(\mathcal{I})} \qd_\alpha^2 
    = S_{11}^{-1}  \sqrt{\qd_a \qd_b}\delta_{\mu\mu'}\delta_{xx'}\delta_{yy'} \,.
\end{split}
\end{align}
The first line of \eqref{eqn:halflinkingorthogonality} then follows by combining \eqref{eqn:halflinkinglemma1} and \eqref{eqn:halflinkinglemma2}.

\subsection{2+1d boundary crossing relations}

Here, we derive the boundary crossing relation \eqref{eq:collapsetube} and discuss its variants.
We first expand the right-hand side in terms of unknown coefficients ${^{\mathcal{B}_1\mathcal{B}_2}}A^{(az)(bw)}_{\alpha\beta(\mu x y)}$,
\begin{align}
    \input{figures/collapsetube1.tikz}~=\sum_{\mu x y}{^{\mathcal{B}_1\mathcal{B}_2}}A^{(az)(bw)}_{\alpha\beta(\mu x y)}~\input{figures/collapsetube2.tikz}~.
\end{align}
To compute ${^{\mathcal{B}_1\mathcal{B}_2}}A^{(az)(bw)}_{\alpha\beta(\mu x y)}$, we consider the following partition function of $\mathrm{TV}_{\mathcal{C}}$ on a solid torus:\
\begin{align}
Z\equiv \input{figures/collapsetubeproof1.tikz} \,.
\end{align}
We may compute $Z$ in two ways. 
On the one hand, we can collapse the tube at the core using \eqref{eq:collapsetube}, and use the definition of the generalized half-linking numbers to find that\footnote{We normalize the SymTFT partition function on a solid ball $B^3$ with a topological boundary condition imposed on the boundary $\partial B^3$ as 1, by tuning the boundary Euler counterterm, in the absence of any topological lines.} 
\begin{align} \label{eq:Z1}
    Z=\sqrt{\frac{\qd_a \qd_b}{S_{11}}}\sum_{\mu xy}{^{\mathcal{B}_1\mathcal{B}_2}}A_{\alpha\beta(\mu x y)}^{(az)(bw)} ~{^{\mathcal{B}_1\mathcal{B}_2}}\widetilde{\Psi}_{\alpha'\beta'(\mu x y)}^{(a z') (b w')} \,.
\end{align}
To compute $Z$ a second way, we use the fact that the $D^2$ disk Hilbert space of $\mathrm{TV}_{\mathcal{C}}$, twisted by an arbitrary simle topological line $c$ on either boundary condition $\mathcal{B}_1$ or $\mathcal{B}_2$, is empty unless $c$ is the identity line, in which case it is 1-dimensional.\footnote{Such a Hilbert space is in 1-to-1 correspondence with the space of topological point operators on which the line $c$ can terminate by the state-operator map, and hence empty if $c \neq 1$.
If $c = 1$, there is a unique state (up to overall normalization) corresponding to the identity operator. Recall that the topological boundaries $\mathcal{B}_i$ are always assumed to be simple.}
By bringing the loop formed out of the line interfaces $\beta'$ and $\alpha'$ close to the loop formed out of $\beta$ and $\alpha$ and fusing them together, we find that we must have $\beta'=\beta$ and $\alpha'=\alpha$ in order to have a non-vanishing partition function, because this is the only situation in which there is a contribution to the partition function for which the identity line is wrapping the non-contractible cycle. 
From there, one may use \eqref{eq:interfaces_comp_ortho} repeatedly to obtain
\begin{align} \label{eq:Z2}
Z=\sqrt{\d_a\d_b}\delta_{\alpha\alpha'}\delta_{\beta\beta'}\delta_{zz'}\delta_{ww'} \,.
\end{align}
Equating \eqref{eq:Z1} and \eqref{eq:Z2}, it follows from the orthgonality relation \eqref{eqn:halflinkingorthogonality} that
\begin{align}
    {^{\mathcal{B}_1\mathcal{B}_2}}A_{\alpha\beta(\mu x y)}^{(az)(bw)} = \sqrt{S_{11}}{^{\mathcal{B}_1\mathcal{B}_2}}\Psi_{\alpha\beta(\mu x y)}^{(az)(bw)} \,,
\end{align}
giving us \eqref{eq:collapsetube}, as desired.

The following variant of \eqref{eq:collapsetube}, with slight changes in junctions and orientations of lines, can also be derived similarly:
\begin{align} \label{eq:collapse3}
    \input{figures/collapsetube3.tikz}~=\sqrt{S_{11}}\sum_{\mu x y}{^{\mathcal{B}_2\mathcal{B}_1}}\widetilde{\Psi}_{\beta\alpha(\mu y x)}^{(bw)(az)}~\input{figures/collapsetube4.tikz}~.
\end{align}
The corresponding inverse relation is
\begin{align} \label{eq:collapse4}
    \input{figures/collapsetube4.tikz}~=\frac{1}{\sqrt{S_{11}}}\sum_{\mu x y}{^{\mathcal{B}_2\mathcal{B}_1}}\Psi_{\beta\alpha(\mu y x)}^{(bw)(az)}~\input{figures/collapsetube3.tikz}~.
\end{align}

\section{Explicit derivation of tube algebra}
\label{app:derivingtubealgebra}

In this Appendix, we lay out the details for deriving \eqref{eq:tubealgebraLL}. As already explained in the main text, the product of two lasso operators amounts to evaluating the figure in \eqref{eq:LLlasso}. To proceed, we first move $\times$ marks appropriately using \eqref{eq:AABB}, perform an $F$-move using \eqref{eq:F_symbols}, and then use the completeness relation \eqref{eq:basis_bdy} on $c, c'$ lines. 
\begin{align}
\begin{split}
   &\mathsf{L}^{b',d'y'z'}_{a',c'} \times \mathsf{L}^{b,dyz}_{a,c} \\ &= \delta_{a' b}
   \sum_{w=1}^{N_{cb}^d} \sum_{w'=1}^{N_{c'b'}^{d'}}
   [A_d^{cb}]_{\bar{z}w} [A_{d'}^{c'b'}]_{\bar{z}'w'}
   \input{figures/tubemult1.tikz} \\
   &=  \delta_{a' b}\sum_{d''\in\mathrm{Irr}(\mathcal{C})} \sum_{w=1}^{N_{cb}^d} \sum_{w'=1}^{N_{c'b'}^{d'}} \sum_{k=1}^{N_{\bar{c}d''}^{d'}}\sum_{l=1}^{N_{d c'}^{d''}} [A_d^{cb}]_{\bar{z}w} [A_{d'}^{c'b'}]_{\bar{z}'w'} [F_{\bar{c}d c'}^{d'}]^{-1}_{( b wy')(d''kl)}\input{figures/tubemult2.tikz} \\[-.3in]
    &=\delta_{a' b}\sum_{c'',d''\in\mathrm{Irr}(\mathcal{C})}\sum_{w=1}^{N_{cb}^d} \sum_{w'=1}^{N_{c'b'}^{d'}}\sum_{k=1}^{N_{\bar{c}d''}^{d'}}\sum_{l=1}^{N_{d c'}^{d''}}\sum_{i=1}^{N_{c c'}^{c''}}\sqrt{\frac{\d_{c''}}{\d_c \d_{c'}}} [A_d^{cb}]_{\bar{z}w} [A_{d'}^{c'b'}]_{\bar{z}'w'}[F_{\bar{c}d c'}^{d'}]^{-1}_{( b wy')(d''kl)}~\input{figures/tubemult3.tikz} \,.
\end{split}
\end{align}
We further apply another $F$-move to resolve the two triangle islands into two bubbles, i.e. 
\begin{align}\label{eq:LLapp}
\begin{split}
    &\input{figures/tubemult3.tikz} = \sum_{c''_1, c''_2\in \mathrm{Irr} (\CC)}  \sum_{j=1}^{N_{c c'}^{c''_1}} \sum_{w''=1}^{N_{\bar{c}''_1 d''}^{b'}} \sum_{y''=1}^{N_{a c''_2}^{d''}} \sum_{i'=1}^{N_{c c'}^{c''_2}} [F_{\bar{c}'\bar{c} d''}^{b'}]_{(d'w'k)(\bar{c}'''jw'')} [F_{a c c'}^{d''}]_{(dyl)(c''_2y'' i')}
    \input{figures/tubemult4.tikz} \,.
\end{split}
\end{align}
The bubble on the right can be shrunk using \eqref{eq:basis_bdy}. However, shrinking the bubble on the left needs caution, because despite the $\times$ marks are on external legs $c''$ and $c''_2$  of the bubble, we need one of the two junctions in the bubble to be in the dual junction basis, i.e. with a bar. Changing the junction to a dual junction can be achieved by using a sequence of moves listed in \eqref{eq:junctionchangerelations}. 
Concretely, we have
\begin{eqnarray}\label{eq:tornado}
	\input{figures/tornado.tikz}  =  \sum_{j'=1}^{N_{c c'}^{c''_1}} [C_{\bar{c}' \bar{c}}^{\bar{c}'''}]_{j \bar\jmath'}    \input{figures/tornado2.tikz}    
\end{eqnarray}
where 
\begin{eqnarray}\label{eq:CCdef}
    C^{\bar c''}_{\bar c \bar c'} = A_{\bar{c} \bar{c}'}^{\bar{c}''} \cdot B^{c \bar{c}''}_{\bar{c}'}\cdot A_{\bar{c}' c''}^{c} \,.
\end{eqnarray}
Substituting \eqref{eq:tornado} and \eqref{eq:CCdef} into \eqref{eq:LLapp}, and after a final rearrangement of an $x$ mark using \eqref{eq:AABB}, we find
\begin{align}\label{eq:tubealgebraLLapp}
\begin{split}
   &\mathsf{L}^{b',d'y'z'}_{a',c'} \times \mathsf{L}^{b,dyz}_{a,c} 
    =\delta_{a' b}\sum_{c'',d''\in\mathrm{Irr}(\mathcal{C})}\sum_{w=1}^{N_{cb}^d} \sum_{w'=1}^{N_{c'b'}^{d'}}\sum_{y''=1}^{N_{a c''}^{d''}}\sum_{w''=1}^{N_{\bar{c}''d''}^{b'}}\sum_{k=1}^{N_{\bar{c}d''}^{d'}}\sum_{l=1}^{N_{d c'}^{d''}}\sum_{i=1}^{N_{cc'}^{c''}} \sum_{\bar{z}'' =1}^{N_{c'' b'}^{d''}}\sqrt{\frac{\d_{c}\d_{c'}}{\d_{c''}}} \\ &[A_d^{cb}]_{\bar{z}w} [A_{d'}^{c'b'}]_{\bar{z}'w'} [A_{\bar{c}'' d''}^{b'}]_{w'' \bar{z}''} 
     %\hspace{.4in} 
     [C^{\bar c''}_{\bar c' \bar c}]_{j \bar \imath }[F_{\bar{c}d c'}^{d'}]^{-1}_{( b wy')(d''kl)} [F_{a c  c'}^{d''}]^{-1}_{(dyl)( c'' y''i)}[F_{\bar{c}'\bar{c}d''}^{b'}]_{(dw'k)(\bar{c}''iw'')}\mathsf{L}^{b',d''y''z''}_{a,c''}
\end{split}
\end{align}
which is \eqref{eq:tubealgebraLL}, as desired.

\section{More on generalized Ishibashi states} \label{app:boundary}

\subsection{Overlap between half-Ishibashi states}

Here, we discuss how the overlap between half-Ishibashi states in Equation \eqref{eq:Ishi_overlap} is derived.
First, recall that such an overlap is given by the partition function of the SymTFT on a solid ball, decorated by bulk and boundary lines and interfaces, as shown in \eqref{eq:Ishi_solid_ball}.
We use the inverse of a ``collapsing tube'' formula given in \eqref{eq:collapse4} to turn this into a sum over solid torus partition functions of the SymTFT.
That is,
\begin{equation} \label{eq:Ishi_solid_torus}
    \llangle y_1; \mu, \mathcal{B}_1 | \tilde{q}^{\frac{1}{2}(L_0 + \bar{L}_0 - c/12)} | y_2;\mu, \mathcal{B}_2 \rrangle = \frac{1}{\sqrt{S}_{11}}  \sum_{\alpha\in \mathrm{Irr}(\mathcal{I})} {^{\mathcal{B}_1\mathcal{B}_2}}\Psi_{\alpha\alpha(\mu y_1 y_2)}^{11} \raisebox{-2.4em}{~\input{figures/Ishi_torus.tikz}} \,.
\end{equation}
We then recognize the solid torus SymTFT partition functions on the right-hand side as the representation basis annulus partition functions $\mathbf{Z}_\alpha (1/\delta)$ defined in \eqref{eq:rep_annulus}, in the dual open string channel.
This gives us the desired Equation \eqref{eq:Ishi_overlap}.

\subsection{Generalized Ishibashi states}

Here, we discuss the subtle normalization difference of the state represented by the Euclidean configuration of the SymTFT on the right-hand side of \eqref{eq:twisted_bdy_Euclidean}, which we denote as $\ket{X} \in W_{\bar{a}}^{\bar{\mu}}\otimes \mathcal{V}_{\bar{\mu}}$, compared to the naive guess $|\bar{x}, \bar{\mu}, \bar{a}\rangle \otimes |y;\mu, \mathcal{B} \rrangle$.
A parallel discussion in the context of RCFTs can be found in \cite{Felder:1999cv}.

From the SymTFT picture, it is clear that the state $\ket{X}$ is proportional to $|\bar{x}, \bar{\mu}, \bar{a}\rangle \otimes |y;\mu, \mathcal{B} \rrangle$.
That is,
\begin{equation}
    \ket{X} = t |\bar{x}, \bar{\mu}, \bar{a}\rangle \otimes |y;\mu, \mathcal{B} \rrangle \,,
\end{equation}
where $t$ is an as-yet-unknown proportionality constant, which we assume to be real.
We consider the (regularized) norm of the state $\ket{X}$,
which is represented by a SymTFT partition function,
\begin{align} \label{eq:Xoverlap}
\begin{split}
    \langle X | \tilde{q}^{\frac{1}{2}(L_0 + \bar{L}_0 - c/12)} | X \rangle &= t^2 \langle \bar{x}, \bar{\mu}, \bar{a} | \bar{x}, \bar{\mu}, \bar{a} \rangle
    \llangle y;\mu, \mathcal{B} | \tilde{q}^{\frac{1}{2}(L_0 + \bar{L}_0 - c/12)} | y;\mu, \mathcal{B} \rrangle \\
    &= \raisebox{-2.4em}{~\scalebox{0.8}{\input{figures/norm1.tikz}}} \,.
\end{split}
\end{align}
Here, we have the SymTFT on a manifold which is topologically $S^2 \times I$, where $I$ is an interval.
There are two boundary components, inner and outer $S^2$.
On the inner $S^2$ boundary, we have the Dirichlet boundary condition $\mathcal{B}_{\mathrm{reg}}$ imposed, and a bulk line $\mu$ meets at two topological junctions $x$ and $\bar{x}$ with the line $a$.
On the outer $S^2$ boundary, we have the familiar configuration which appears in the overlap of half-Ishibashi states \eqref{eq:Ishi_solid_ball}.
This is to be contrasted with the corresponding norm of the product state $|\bar{x}, \bar{\mu}, \bar{a}\rangle \otimes |y;\mu, \mathcal{B} \rrangle$ is computed by the SymTFT partition function on two disconnected solid balls, with appropriate decorations.

We compute the SymTFT partition function in \eqref{eq:Xoverlap} as follows:
\begin{align} \label{eq:norm}
\begin{split}
   \raisebox{-1.5em}{~ \scalebox{0.8}{\input{figures/norm11.tikz}}} &= \frac{\sqrt{\qd_a}}{S_{11}\sqrt{\qd_\mu}} \raisebox{-2.4em} {~\input{figures/norm2.tikz}}\\ &= \frac{\sqrt{\qd_a}}{S_{11}\sqrt{\qd_\mu}} \llangle y;\mu, \mathcal{B} | \tilde{q}^{\frac{1}{2}(L_0 + \bar{L}_0 - c/12)} | y;\mu, \mathcal{B} \rrangle \,,
\end{split}
\end{align}
where in the first step, we have collapsed the inner hollow ball, which results in the factor of $\sqrt{\qd_a}/(S_{11}\sqrt{\qd_\mu})$ in our conventions (cf.\ \cite{Felder:1999cv}).
Recall that we have $\langle \bar{x}, \bar{\mu}, \bar{a} | \bar{x}, \bar{\mu}, \bar{a} \rangle = \sqrt{\qd_\mu \qd_a}$ as explained in \eqref{eq:xinner}. 
Comparing the above equation with the first equality of \eqref{eq:Xoverlap}, we conclude that $t = (S_{11} \qd_\mu)^{-1/2} = S_{1\mu}^{-1/2}$.
That is,
\begin{equation}
    \ket{X} = S_{1\mu}^{-1/2} |\bar{x}, \bar{\mu}, \bar{a}\rangle \otimes |y;\mu, \mathcal{B} \rrangle \,.
\end{equation}
This then leads to the correctly normalized twisted sector boundary states given in \eqref{eq:twisted_bdy_state}.

\bibliographystyle{JHEP}
\bibliography{ref}

\providecommand{\href}[2]{#2}\begingroup\raggedright\begin{thebibliography}{100}

\bibitem{Poland:2018epd}
D.~Poland, S.~Rychkov, and A.~Vichi, {\it {The Conformal Bootstrap: Theory, Numerical Techniques, and Applications}},  {\em Rev. Mod. Phys.} {\bf 91} (2019) 015002, [\href{http://arxiv.org/abs/1805.04405}{{\tt arXiv:1805.04405}}].

\bibitem{Billo:2016cpy}
M.~Bill\`o, V.~Gon\c{c}alves, E.~Lauria, and M.~Meineri, {\it {Defects in conformal field theory}},  {\em JHEP} {\bf 04} (2016) 091, [\href{http://arxiv.org/abs/1601.02883}{{\tt arXiv:1601.02883}}].

\bibitem{Chang:2018iay}
C.-M. Chang, Y.-H. Lin, S.-H. Shao, Y.~Wang, and X.~Yin, {\it {Topological Defect Lines and Renormalization Group Flows in Two Dimensions}},  {\em JHEP} {\bf 01} (2019) 026, [\href{http://arxiv.org/abs/1802.04445}{{\tt arXiv:1802.04445}}].

\bibitem{Thorngren:2019iar}
R.~Thorngren and Y.~Wang, {\it {Fusion Category Symmetry I: Anomaly In-Flow and Gapped Phases}},  \href{http://arxiv.org/abs/1912.02817}{{\tt arXiv:1912.02817}}.

\bibitem{Thorngren:2021yso}
R.~Thorngren and Y.~Wang, {\it {Fusion Category Symmetry II: Categoriosities at $c$ = 1 and Beyond}},  \href{http://arxiv.org/abs/2106.12577}{{\tt arXiv:2106.12577}}.

\bibitem{Komargodski:2020mxz}
Z.~Komargodski, K.~Ohmori, K.~Roumpedakis, and S.~Seifnashri, {\it {Symmetries and strings of adjoint QCD$_{2}$}},  {\em JHEP} {\bf 03} (2021) 103, [\href{http://arxiv.org/abs/2008.07567}{{\tt arXiv:2008.07567}}].

\bibitem{Kong:2020cie}
L.~Kong, T.~Lan, X.-G. Wen, Z.-H. Zhang, and H.~Zheng, {\it {Algebraic higher symmetry and categorical symmetry -- a holographic and entanglement view of symmetry}},  {\em Phys. Rev. Res.} {\bf 2} (2020), no.~4 043086, [\href{http://arxiv.org/abs/2005.14178}{{\tt arXiv:2005.14178}}].

\bibitem{Verlinde:1988sn}
E.~P. Verlinde, {\it {Fusion Rules and Modular Transformations in 2D Conformal Field Theory}},  {\em Nucl. Phys. B} {\bf 300} (1988) 360--376.

\bibitem{Moore:1988qv}
G.~W. Moore and N.~Seiberg, {\it {Classical and Quantum Conformal Field Theory}},  {\em Commun. Math. Phys.} {\bf 123} (1989) 177.

\bibitem{Bhardwaj:2017xup}
L.~Bhardwaj and Y.~Tachikawa, {\it {On finite symmetries and their gauging in two dimensions}},  {\em JHEP} {\bf 03} (2018) 189, [\href{http://arxiv.org/abs/1704.02330}{{\tt arXiv:1704.02330}}].

\bibitem{Carqueville:2012dk}
N.~Carqueville and I.~Runkel, {\it {Orbifold completion of defect bicategories}},  {\em Quantum Topol.} {\bf 7} (2016), no.~2 203--279, [\href{http://arxiv.org/abs/1210.6363}{{\tt arXiv:1210.6363}}].

\bibitem{Brunner:2013xna}
I.~Brunner, N.~Carqueville, and D.~Plencner, {\it {A quick guide to defect orbifolds}},  {\em Proc. Symp. Pure Math.} {\bf 88} (2014) 231--242, [\href{http://arxiv.org/abs/1310.0062}{{\tt arXiv:1310.0062}}].

\bibitem{Lin:2019hks}
Y.-H. Lin and S.-H. Shao, {\it {Duality Defect of the Monster CFT}},  {\em J. Phys. A} {\bf 54} (2021), no.~6 065201, [\href{http://arxiv.org/abs/1911.00042}{{\tt arXiv:1911.00042}}].

\bibitem{Ji:2019ugf}
W.~Ji, S.-H. Shao, and X.-G. Wen, {\it {Topological Transition on the Conformal Manifold}},  {\em Phys. Rev. Res.} {\bf 2} (2020), no.~3 033317, [\href{http://arxiv.org/abs/1909.01425}{{\tt arXiv:1909.01425}}].

\bibitem{Fuchs:2002cm}
J.~Fuchs, I.~Runkel, and C.~Schweigert, {\it {TFT construction of RCFT correlators 1. Partition functions}},  {\em Nucl. Phys. B} {\bf 646} (2002) 353--497, [\href{http://arxiv.org/abs/hep-th/0204148}{{\tt hep-th/0204148}}].

\bibitem{Frohlich:2006ch}
J.~Frohlich, J.~Fuchs, I.~Runkel, and C.~Schweigert, {\it {Duality and defects in rational conformal field theory}},  {\em Nucl. Phys. B} {\bf 763} (2007) 354--430, [\href{http://arxiv.org/abs/hep-th/0607247}{{\tt hep-th/0607247}}].

\bibitem{Frohlich:2009gb}
J.~Frohlich, J.~Fuchs, I.~Runkel, and C.~Schweigert, {\it {Defect Lines, Dualities and Generalised Orbifolds}},  in {\em {16th International Congress on Mathematical Physics}}, pp.~608--613, 2010.
\newblock \href{http://arxiv.org/abs/0909.5013}{{\tt arXiv:0909.5013}}.

\bibitem{Gaiotto:2014kfa}
D.~Gaiotto, A.~Kapustin, N.~Seiberg, and B.~Willett, {\it {Generalized Global Symmetries}},  {\em JHEP} {\bf 02} (2015) 172, [\href{http://arxiv.org/abs/1412.5148}{{\tt arXiv:1412.5148}}].

\bibitem{etingof2015tensor}
P.~Etingof, S.~Gelaki, D.~Nikshych, and V.~Ostrik, {\em Tensor Categories}.
\newblock Mathematical surveys and monographs. American Mathematical Society, 2015.

\bibitem{Choi:2024wfm}
Y.~Choi, B.~C. Rayhaun, and Y.~Zheng, {\it {A Non-Invertible Symmetry-Resolved Affleck-Ludwig-Cardy Formula and Entanglement Entropy from the Boundary Tube Algebra}},  \href{http://arxiv.org/abs/2409.02806}{{\tt arXiv:2409.02806}}.

\bibitem{Lin:2022dhv}
Y.-H. Lin, M.~Okada, S.~Seifnashri, and Y.~Tachikawa, {\it {Asymptotic density of states in 2d CFTs with non-invertible symmetries}},  \href{http://arxiv.org/abs/2208.05495}{{\tt arXiv:2208.05495}}.

\bibitem{Huang:2021zvu}
T.-C. Huang, Y.-H. Lin, and S.~Seifnashri, {\it {Construction of two-dimensional topological field theories with non-invertible symmetries}},  {\em JHEP} {\bf 12} (2021) 028, [\href{http://arxiv.org/abs/2110.02958}{{\tt arXiv:2110.02958}}].

\bibitem{Choi:2023xjw}
Y.~Choi, B.~C. Rayhaun, Y.~Sanghavi, and S.-H. Shao, {\it {Remarks on boundaries, anomalies, and noninvertible symmetries}},  {\em Phys. Rev. D} {\bf 108} (2023), no.~12 125005, [\href{http://arxiv.org/abs/2305.09713}{{\tt arXiv:2305.09713}}].

\bibitem{Diatlyk:2023fwf}
O.~Diatlyk, C.~Luo, Y.~Wang, and Q.~Weller, {\it {Gauging non-invertible symmetries: topological interfaces and generalized orbifold groupoid in 2d QFT}},  {\em JHEP} {\bf 03} (2024) 127, [\href{http://arxiv.org/abs/2311.17044}{{\tt arXiv:2311.17044}}].

\bibitem{Cordova:2024vsq}
C.~Cordova, D.~Garc\'\i{}a-Sep\'ulveda, and N.~Holfester, {\it {Particle-soliton degeneracies from spontaneously broken non-invertible symmetry}},  {\em JHEP} {\bf 07} (2024) 154, [\href{http://arxiv.org/abs/2403.08883}{{\tt arXiv:2403.08883}}].

\bibitem{Cordova:2024iti}
C.~Cordova, N.~Holfester, and K.~Ohmori, {\it {Representation Theory of Solitons}},  \href{http://arxiv.org/abs/2408.11045}{{\tt arXiv:2408.11045}}.

\bibitem{Copetti:2024dcz}
C.~Copetti, L.~Cordova, and S.~Komatsu, {\it {S-Matrix Bootstrap and Non-Invertible Symmetries}},  \href{http://arxiv.org/abs/2408.13132}{{\tt arXiv:2408.13132}}.

\bibitem{Inamura:2024jke}
K.~Inamura and S.~Ohyama, {\it {1+1d SPT phases with fusion category symmetry: interface modes and non-abelian Thouless pump}},  \href{http://arxiv.org/abs/2408.15960}{{\tt arXiv:2408.15960}}.

\bibitem{Levin:2004mi}
M.~A. Levin and X.-G. Wen, {\it {String net condensation: A Physical mechanism for topological phases}},  {\em Phys. Rev. B} {\bf 71} (2005) 045110, [\href{http://arxiv.org/abs/cond-mat/0404617}{{\tt cond-mat/0404617}}].

\bibitem{Bridgeman:2019wyu}
J.~C. Bridgeman and D.~Barter, {\it {Computing data for Levin-Wen with defects}},  {\em Quantum} {\bf 4} (2020) 277, [\href{http://arxiv.org/abs/1907.06692}{{\tt arXiv:1907.06692}}].

\bibitem{jia2024weak}
Z.~Jia, S.~Tan, and D.~Kaszlikowski, {\it Weak hopf symmetry and tube algebra of the generalized multifusion string-net model},  {\em arXiv preprint arXiv:2403.04446} (2024).

\bibitem{2012CMaPh.313..351K}
A.~{Kitaev} and L.~{Kong}, {\it {Models for Gapped Boundaries and Domain Walls}},  {\em Communications in Mathematical Physics} {\bf 313} (July, 2012) 351--373, [\href{http://arxiv.org/abs/1104.5047}{{\tt arXiv:1104.5047}}].

\bibitem{Barter_2022}
D.~Barter, J.~Bridgeman, and R.~Wolf, {\it Computing associators of endomorphism fusion categories},  {\em {SciPost} Physics} {\bf 13} (aug, 2022).

\bibitem{Copetti:2024onh}
C.~Copetti, {\it {Defect Charges, Gapped Boundary Conditions, and the Symmetry TFT}},  \href{http://arxiv.org/abs/2408.01490}{{\tt arXiv:2408.01490}}.

\bibitem{Konechny:2024ixa}
A.~Konechny and V.~Vergioglou, {\it {On fusing matrices associated with conformal boundary conditions}},  \href{http://arxiv.org/abs/2405.10189}{{\tt arXiv:2405.10189}}.

\bibitem{Bridgeman:2022gdx}
J.~C. Bridgeman, L.~Lootens, and F.~Verstraete, {\it {Invertible Bimodule Categories and Generalized Schur Orthogonality}},  {\em Commun. Math. Phys.} {\bf 402} (2023), no.~3 2691--2714, [\href{http://arxiv.org/abs/2211.01947}{{\tt arXiv:2211.01947}}].

\bibitem{Barter:2018hjs}
D.~Barter, J.~C. Bridgeman, and C.~Jones, {\it {Domain Walls in Topological Phases and the Brauer\textendash{}Picard Ring for ${{\rm Vec} (\mathbb{Z}/p\mathbb{Z})}$}},  {\em Commun. Math. Phys.} {\bf 369} (2019), no.~3 1167--1185, [\href{http://arxiv.org/abs/1806.01279}{{\tt arXiv:1806.01279}}].

\bibitem{Koide:2023rqd}
M.~Koide, Y.~Nagoya, and S.~Yamaguchi, {\it {Non-invertible symmetries and boundaries in four dimensions}},  \href{http://arxiv.org/abs/2304.01550}{{\tt arXiv:2304.01550}}.

\bibitem{Koide:2021zxj}
M.~Koide, Y.~Nagoya, and S.~Yamaguchi, {\it {Non-invertible topological defects in 4-dimensional $\mathbb {Z}_2$ pure lattice gauge theory}},  {\em PTEP} {\bf 2022} (2022), no.~1 013B03, [\href{http://arxiv.org/abs/2109.05992}{{\tt arXiv:2109.05992}}].

\bibitem{Gaiotto:2020iye}
D.~Gaiotto and J.~Kulp, {\it {Orbifold groupoids}},  {\em JHEP} {\bf 02} (2021) 132, [\href{http://arxiv.org/abs/2008.05960}{{\tt arXiv:2008.05960}}].

\bibitem{Ji:2019jhk}
W.~Ji and X.-G. Wen, {\it {Categorical symmetry and noninvertible anomaly in symmetry-breaking and topological phase transitions}},  {\em Phys. Rev. Res.} {\bf 2} (2020), no.~3 033417, [\href{http://arxiv.org/abs/1912.13492}{{\tt arXiv:1912.13492}}].

\bibitem{Apruzzi:2021nmk}
F.~Apruzzi, F.~Bonetti, I.~G. Etxebarria, S.~S. Hosseini, and S.~Schafer-Nameki, {\it {Symmetry TFTs from String Theory}},  \href{http://arxiv.org/abs/2112.02092}{{\tt arXiv:2112.02092}}.

\bibitem{Freed:2022qnc}
D.~S. Freed, G.~W. Moore, and C.~Teleman, {\it {Topological symmetry in quantum field theory}},  \href{http://arxiv.org/abs/2209.07471}{{\tt arXiv:2209.07471}}.

\bibitem{Kong:2015flk}
L.~Kong, X.-G. Wen, and H.~Zheng, {\it {Boundary-bulk relation for topological orders as the functor mapping higher categories to their centers}},  \href{http://arxiv.org/abs/1502.01690}{{\tt arXiv:1502.01690}}.

\bibitem{Kong:2017hcw}
L.~Kong, X.-G. Wen, and H.~Zheng, {\it {Boundary-bulk relation in topological orders}},  {\em Nucl. Phys. B} {\bf 922} (2017) 62--76, [\href{http://arxiv.org/abs/1702.00673}{{\tt arXiv:1702.00673}}].

\bibitem{Kaidi:2022cpf}
J.~Kaidi, K.~Ohmori, and Y.~Zheng, {\it {Symmetry TFTs for Non-Invertible Defects}},  \href{http://arxiv.org/abs/2209.11062}{{\tt arXiv:2209.11062}}.

\bibitem{Antinucci:2022vyk}
A.~Antinucci, F.~Benini, C.~Copetti, G.~Galati, and G.~Rizi, {\it {The holography of non-invertible self-duality symmetries}},  \href{http://arxiv.org/abs/2210.09146}{{\tt arXiv:2210.09146}}.

\bibitem{Bhardwaj:2023wzd}
L.~Bhardwaj and S.~Schafer-Nameki, {\it {Generalized charges, part I: Invertible symmetries and higher representations}},  {\em SciPost Phys.} {\bf 16} (2024), no.~4 093, [\href{http://arxiv.org/abs/2304.02660}{{\tt arXiv:2304.02660}}].

\bibitem{Bhardwaj:2023ayw}
L.~Bhardwaj and S.~Schafer-Nameki, {\it {Generalized Charges, Part II: Non-Invertible Symmetries and the Symmetry TFT}},  \href{http://arxiv.org/abs/2305.17159}{{\tt arXiv:2305.17159}}.

\bibitem{Bartsch:2023pzl}
T.~Bartsch, M.~Bullimore, and A.~Grigoletto, {\it {Higher representations for extended operators}},  \href{http://arxiv.org/abs/2304.03789}{{\tt arXiv:2304.03789}}.

\bibitem{Bartsch:2023wvv}
T.~Bartsch, M.~Bullimore, and A.~Grigoletto, {\it {Representation theory for categorical symmetries}},  \href{http://arxiv.org/abs/2305.17165}{{\tt arXiv:2305.17165}}.

\bibitem{evans1995ocneanu}
D.~E. Evans and Y.~Kawahigashi, {\it On ocneanu's theory of asymptotic inclusions for subfactors, topological quantum field theories and quantum doubles},  {\em International journal of mathematics} {\bf 6} (1995), no.~02 205--228.

\bibitem{Izumi:2000qa}
M.~Izumi, {\it {The structure of sectors associated with Longo-Rehren inclusions. I: General theory}},  {\em Commun. Math. Phys.} {\bf 213} (2000) 127--179.

\bibitem{MUGER2003159}
M.~Müger, {\it From subfactors to categories and topology ii: The quantum double of tensor categories and subfactors},  {\em Journal of Pure and Applied Algebra} {\bf 180} (2003), no.~1 159--219, [\href{http://arxiv.org/abs/math/0111205}{{\tt math/0111205}}].

\bibitem{Huang:2023pyk}
S.-J. Huang and M.~Cheng, {\it {Topological holography, quantum criticality, and boundary states}},  \href{http://arxiv.org/abs/2310.16878}{{\tt arXiv:2310.16878}}.

\bibitem{Heckman:2024zdo}
J.~J. Heckman and M.~H\"ubner, {\it {Celestial Topology, Symmetry Theories, and Evidence for a Non-SUSY D3-Brane CFT}},  \href{http://arxiv.org/abs/2406.08485}{{\tt arXiv:2406.08485}}.

\bibitem{Braeger:2024jcj}
N.~Braeger, V.~Chakrabhavi, J.~J. Heckman, and M.~Hubner, {\it {Generalized Symmetries of Non-Supersymmetric Orbifolds}},  \href{http://arxiv.org/abs/2404.17639}{{\tt arXiv:2404.17639}}.

\bibitem{Baume:2023kkf}
F.~Baume, J.~J. Heckman, M.~H\"ubner, E.~Torres, A.~P. Turner, and X.~Yu, {\it {SymTrees and Multi-Sector QFTs}},  {\em Phys. Rev. D} {\bf 109} (2024), no.~10 106013, [\href{http://arxiv.org/abs/2310.12980}{{\tt arXiv:2310.12980}}].

\bibitem{Zhang:2023wlu}
C.~Zhang and C.~C\'ordova, {\it {Anomalies of $(1+1)D$ categorical symmetries}},  \href{http://arxiv.org/abs/2304.01262}{{\tt arXiv:2304.01262}}.

\bibitem{etingof2010fusion}
P.~Etingof, D.~Nikshych, and V.~Ostrik, {\it Fusion categories and homotopy theory},  {\em Quantum topology} {\bf 1} (2010), no.~3 209--273.

\bibitem{Ji:2021esj}
W.~Ji and X.-G. Wen, {\it {A unified view on symmetry, anomalous symmetry and non-invertible gravitational anomaly}},  \href{http://arxiv.org/abs/2106.02069}{{\tt arXiv:2106.02069}}.

\bibitem{Cardy:1989ir}
J.~L. Cardy, {\it {Boundary Conditions, Fusion Rules and the Verlinde Formula}},  {\em Nucl. Phys. B} {\bf 324} (1989) 581--596.

\bibitem{Fuchs:2000cm}
J.~Fuchs, L.~R. Huiszoon, A.~N. Schellekens, C.~Schweigert, and J.~Walcher, {\it {Boundaries, crosscaps and simple currents}},  {\em Phys. Lett. B} {\bf 495} (2000) 427--434, [\href{http://arxiv.org/abs/hep-th/0007174}{{\tt hep-th/0007174}}].

\bibitem{Gaberdiel:2002qa}
M.~R. Gaberdiel and T.~Gannon, {\it {Boundary states for WZW models}},  {\em Nucl. Phys. B} {\bf 639} (2002) 471--501, [\href{http://arxiv.org/abs/hep-th/0202067}{{\tt hep-th/0202067}}].

\bibitem{Fuchs:2004dz}
J.~Fuchs, I.~Runkel, and C.~Schweigert, {\it {TFT construction of RCFT correlators. 3. Simple currents}},  {\em Nucl. Phys. B} {\bf 694} (2004) 277--353, [\href{http://arxiv.org/abs/hep-th/0403157}{{\tt hep-th/0403157}}].

\bibitem{rowell2009classification}
E.~Rowell, R.~Stong, and Z.~Wang, {\it On classification of modular tensor categories},  {\em Communications in Mathematical Physics} {\bf 292} (2009), no.~2 343--389.

\bibitem{Mukhi:2022bte}
S.~Mukhi and B.~C. Rayhaun, {\it {Classification of Unitary RCFTs with Two Primaries and Central Charge Less Than 25}},  {\em Communications in Mathematical Physics} (2023) 1--51, [\href{http://arxiv.org/abs/2208.05486}{{\tt arXiv:2208.05486}}].

\bibitem{Rayhaun:2023pgc}
B.~C. Rayhaun, {\it {Bosonic Rational Conformal Field Theories in Small Genera, Chiral Fermionization, and Symmetry/Subalgebra Duality}},  \href{http://arxiv.org/abs/2303.16921}{{\tt arXiv:2303.16921}}.

\bibitem{Ginsparg:1988ui}
P.~H. Ginsparg, {\it {APPLIED CONFORMAL FIELD THEORY}},  in {\em {Les Houches Summer School in Theoretical Physics: Fields, Strings, Critical Phenomena}}, 9, 1988.
\newblock \href{http://arxiv.org/abs/hep-th/9108028}{{\tt hep-th/9108028}}.

\bibitem{Ishibashi:1988kg}
N.~Ishibashi, {\it {The Boundary and Crosscap States in Conformal Field Theories}},  {\em Mod. Phys. Lett. A} {\bf 4} (1989) 251.

\bibitem{Cardy:2004hm}
J.~L. Cardy, {\it {Boundary conformal field theory}},  \href{http://arxiv.org/abs/hep-th/0411189}{{\tt hep-th/0411189}}.

\bibitem{Behrend:1999bn}
R.~E. Behrend, P.~A. Pearce, V.~B. Petkova, and J.-B. Zuber, {\it {Boundary conditions in rational conformal field theories}},  {\em Nucl. Phys. B} {\bf 570} (2000) 525--589, [\href{http://arxiv.org/abs/hep-th/9908036}{{\tt hep-th/9908036}}].

\bibitem{Cordova:2022lms}
C.~Cordova and D.~Garc\'\i{}a-Sep\'ulveda, {\it {Symmetry Enriched $c$-Theorems \& SPT Transitions}},  \href{http://arxiv.org/abs/2210.01135}{{\tt arXiv:2210.01135}}.

\bibitem{Bachas:2007td}
C.~Bachas and I.~Brunner, {\it {Fusion of conformal interfaces}},  {\em JHEP} {\bf 02} (2008) 085, [\href{http://arxiv.org/abs/0712.0076}{{\tt arXiv:0712.0076}}].

\bibitem{Diatlyk:2024zkk}
O.~Diatlyk, H.~Khanchandani, F.~K. Popov, and Y.~Wang, {\it {Defect Fusion and Casimir Energy in Higher Dimensions}},  \href{http://arxiv.org/abs/2404.05815}{{\tt arXiv:2404.05815}}.

\bibitem{Wu:2023ezm}
Y.-H. Wu, H.-H. Tu, and M.~Cheng, {\it {Impurity screening by defects in (1+1)$d$ quantum critical systems}},  \href{http://arxiv.org/abs/2307.09519}{{\tt arXiv:2307.09519}}.

\bibitem{Aharony:2023amq}
O.~Aharony, G.~Cuomo, Z.~Komargodski, M.~Mezei, and A.~Raviv-Moshe, {\it {Phases of Wilson lines: conformality and screening}},  {\em JHEP} {\bf 12} (2023) 183, [\href{http://arxiv.org/abs/2310.00045}{{\tt arXiv:2310.00045}}].

\bibitem{Aharony:2022ntz}
O.~Aharony, G.~Cuomo, Z.~Komargodski, M.~Mezei, and A.~Raviv-Moshe, {\it {Phases of Wilson Lines in Conformal Field Theories}},  {\em Phys. Rev. Lett.} {\bf 130} (2023), no.~15 151601, [\href{http://arxiv.org/abs/2211.11775}{{\tt arXiv:2211.11775}}].

\bibitem{Cuomo:2022xgw}
G.~Cuomo, Z.~Komargodski, M.~Mezei, and A.~Raviv-Moshe, {\it {Spin impurities, Wilson lines and semiclassics}},  {\em JHEP} {\bf 06} (2022) 112, [\href{http://arxiv.org/abs/2202.00040}{{\tt arXiv:2202.00040}}].

\bibitem{Cuomo:2021kfm}
G.~Cuomo, Z.~Komargodski, and M.~Mezei, {\it {Localized magnetic field in the O(N) model}},  {\em JHEP} {\bf 02} (2022) 134, [\href{http://arxiv.org/abs/2112.10634}{{\tt arXiv:2112.10634}}].

\bibitem{Cuomo:2021rkm}
G.~Cuomo, Z.~Komargodski, and A.~Raviv-Moshe, {\it {Renormalization Group Flows on Line Defects}},  {\em Phys. Rev. Lett.} {\bf 128} (2022), no.~2 021603, [\href{http://arxiv.org/abs/2108.01117}{{\tt arXiv:2108.01117}}].

\bibitem{Nagar:2024mjz}
I.~Nagar, A.~Sever, and D.-l. Zhong, {\it {Planar RG flows on line defects}},  {\em JHEP} {\bf 06} (2024) 110, [\href{http://arxiv.org/abs/2404.07290}{{\tt arXiv:2404.07290}}].

\bibitem{Giombi:2022vnz}
S.~Giombi, E.~Helfenberger, and H.~Khanchandani, {\it {Line defects in fermionic CFTs}},  {\em JHEP} {\bf 08} (2023) 224, [\href{http://arxiv.org/abs/2211.11073}{{\tt arXiv:2211.11073}}].

\bibitem{Graham:2003nc}
K.~Graham and G.~M.~T. Watts, {\it {Defect lines and boundary flows}},  {\em JHEP} {\bf 04} (2004) 019, [\href{http://arxiv.org/abs/hep-th/0306167}{{\tt hep-th/0306167}}].

\bibitem{Konechny:2012wm}
A.~Konechny, {\it {Renormalization group defects for boundary flows}},  {\em J. Phys. A} {\bf 46} (2013) 145401, [\href{http://arxiv.org/abs/1211.3665}{{\tt arXiv:1211.3665}}].

\bibitem{Konechny:2020jym}
A.~Konechny, {\it {Properties of RG interfaces for 2D boundary flows}},  {\em JHEP} {\bf 05} (2021) 178, [\href{http://arxiv.org/abs/2012.12361}{{\tt arXiv:2012.12361}}].

\bibitem{Konechny:2019wff}
A.~Konechny, {\it {Open topological defects and boundary RG flows}},  {\em J. Phys. A} {\bf 53} (2020), no.~15 155401, [\href{http://arxiv.org/abs/1911.06041}{{\tt arXiv:1911.06041}}].

\bibitem{Andrei:2018die}
N.~Andrei et~al., {\it {Boundary and Defect CFT: Open Problems and Applications}},  {\em J. Phys. A} {\bf 53} (2020), no.~45 453002, [\href{http://arxiv.org/abs/1810.05697}{{\tt arXiv:1810.05697}}].

\bibitem{Konechny:2015qla}
A.~Konechny, {\it {Fusion of conformal interfaces and bulk induced boundary RG flows}},  {\em JHEP} {\bf 12} (2015) 114, [\href{http://arxiv.org/abs/1509.07787}{{\tt arXiv:1509.07787}}].

\bibitem{Kormos:2009sk}
M.~Kormos, I.~Runkel, and G.~M.~T. Watts, {\it {Defect flows in minimal models}},  {\em JHEP} {\bf 11} (2009) 057, [\href{http://arxiv.org/abs/0907.1497}{{\tt arXiv:0907.1497}}].

\bibitem{Kitaev:2005hzj}
A.~Kitaev, {\it {Anyons in an exactly solved model and beyond}},  {\em Annals Phys.} {\bf 321} (2006), no.~1 2--111, [\href{http://arxiv.org/abs/cond-mat/0506438}{{\tt cond-mat/0506438}}].

\bibitem{Simon:2022ohj}
S.~H. Simon and J.~K. Slingerland, {\it {Straightening Out the Frobenius-Schur Indicator}},  \href{http://arxiv.org/abs/2208.14500}{{\tt arXiv:2208.14500}}.

\bibitem{Cvetic:2024dzu}
M.~Cvetic, R.~Donagi, J.~J. Heckman, M.~H\"ubner, and E.~Torres, {\it {Cornering Relative Symmetry Theories}},  \href{http://arxiv.org/abs/2408.12600}{{\tt arXiv:2408.12600}}.

\bibitem{Bhardwaj:2024igy}
L.~Bhardwaj, C.~Copetti, D.~Pajer, and S.~Schafer-Nameki, {\it {Boundary SymTFT}},  \href{http://arxiv.org/abs/2409.02166}{{\tt arXiv:2409.02166}}.

\bibitem{Das:2024qdx}
A.~Das, J.~Molina-Vilaplana, and P.~Saura-Bastida, {\it {Generalized Symmetry Resolution of Entanglement in CFT for Twisted and Anyonic sectors}},  \href{http://arxiv.org/abs/2409.02162}{{\tt arXiv:2409.02162}}.

\bibitem{GarciaEtxebarria:2024jfv}
I.~n. Garc\'\i{}a~Etxebarria, J.~Huertas, and A.~M. Uranga, {\it {SymTFT Fans: The Symmetry Theory of 4d N=4 Super Yang-Mills on spaces with boundaries}},  \href{http://arxiv.org/abs/2409.02156}{{\tt arXiv:2409.02156}}.

\bibitem{Heymann:2024vvf}
J.~Heymann and T.~Quella, {\it {Revisiting the symmetry-resolved entanglement for non-invertible symmetries in $1{+}1$d conformal field theories}},  \href{http://arxiv.org/abs/2409.02315}{{\tt arXiv:2409.02315}}.

\bibitem{Kong:2022cpy}
L.~Kong and Z.-H. Zhang, {\it {An invitation to topological orders and category theory}},  \href{http://arxiv.org/abs/2205.05565}{{\tt arXiv:2205.05565}}.

\bibitem{Barkeshli:2014cna}
M.~Barkeshli, P.~Bonderson, M.~Cheng, and Z.~Wang, {\it {Symmetry Fractionalization, Defects, and Gauging of Topological Phases}},  {\em Phys. Rev. B} {\bf 100} (2019), no.~11 115147, [\href{http://arxiv.org/abs/1410.4540}{{\tt arXiv:1410.4540}}].

\bibitem{Bonderson:2007ci}
P.~Bonderson, K.~Shtengel, and J.~K. Slingerland, {\it {Interferometry of non-Abelian Anyons}},  {\em Annals Phys.} {\bf 323} (2008) 2709--2755, [\href{http://arxiv.org/abs/0707.4206}{{\tt arXiv:0707.4206}}].

\bibitem{ostrik2003module}
V.~Ostrik, {\it Module categories, weak hopf algebras and modular invariants},  {\em Transformation groups} {\bf 8} (2003) 177--206.

\bibitem{Gannon:2001ki}
T.~Gannon, {\it {Boundary conformal field theory and fusion ring representations}},  {\em Nucl. Phys. B} {\bf 627} (2002) 506--564, [\href{http://arxiv.org/abs/hep-th/0106105}{{\tt hep-th/0106105}}].

\bibitem{etingof2016tensor}
P.~Etingof, S.~Gelaki, D.~Nikshych, and V.~Ostrik, {\em Tensor categories}, vol.~205.
\newblock American Mathematical Soc., 2016.

\bibitem{PhysRevLett.67.161}
I.~Affleck and A.~W.~W. Ludwig, {\it Universal noninteger ``ground-state degeneracy'' in critical quantum systems},  {\em Phys. Rev. Lett.} {\bf 67} (Jul, 1991) 161--164.

\bibitem{Choi:2021kmx}
Y.~Choi, C.~Cordova, P.-S. Hsin, H.~T. Lam, and S.-H. Shao, {\it {Noninvertible duality defects in 3+1 dimensions}},  {\em Phys. Rev. D} {\bf 105} (2022), no.~12 125016, [\href{http://arxiv.org/abs/2111.01139}{{\tt arXiv:2111.01139}}].

\bibitem{Choi:2022zal}
Y.~Choi, C.~Cordova, P.-S. Hsin, H.~T. Lam, and S.-H. Shao, {\it {Non-invertible Condensation, Duality, and Triality Defects in 3+1 Dimensions}},  \href{http://arxiv.org/abs/2204.09025}{{\tt arXiv:2204.09025}}.

\bibitem{Turaev:1992hq}
V.~G. Turaev and O.~Y. Viro, {\it {State sum invariants of 3 manifolds and quantum 6j symbols}},  {\em Topology} {\bf 31} (1992) 865--902.

\bibitem{Barrett:1993ab}
J.~W. Barrett and B.~W. Westbury, {\it {Invariants of piecewise linear three manifolds}},  {\em Trans. Am. Math. Soc.} {\bf 348} (1996) 3997--4022, [\href{http://arxiv.org/abs/hep-th/9311155}{{\tt hep-th/9311155}}].

\bibitem{PhysRevB.103.195155}
C.-H. Lin, M.~Levin, and F.~J. Burnell, {\it Generalized string-net models: A thorough exposition},  {\em Phys. Rev. B} {\bf 103} (May, 2021) 195155.

\bibitem{Kong:2013aya}
L.~Kong, {\it {Anyon condensation and tensor categories}},  {\em Nucl. Phys. B} {\bf 886} (2014) 436--482, [\href{http://arxiv.org/abs/1307.8244}{{\tt arXiv:1307.8244}}].

\bibitem{DavydovMügerNikshychOstrik+2013+135+177}
A.~Davydov, M.~Müger, D.~Nikshych, and V.~Ostrik, {\it The witt group of non-degenerate braided fusion categories},  {\em Journal für die reine und angewandte Mathematik (Crelles Journal)} {\bf 2013} (2013), no.~677 135--177.

\bibitem{Fuchs:2012dt}
J.~Fuchs, C.~Schweigert, and A.~Valentino, {\it {Bicategories for boundary conditions and for surface defects in 3-d TFT}},  {\em Commun. Math. Phys.} {\bf 321} (2013) 543--575, [\href{http://arxiv.org/abs/1203.4568}{{\tt arXiv:1203.4568}}].

\bibitem{Bullivant:2019fmk}
A.~Bullivant and C.~Delcamp, {\it {Tube algebras, excitations statistics and compactification in gauge models of topological phases}},  {\em JHEP} {\bf 10} (2019) 216, [\href{http://arxiv.org/abs/1905.08673}{{\tt arXiv:1905.08673}}].

\bibitem{Bartsch:2022ytj}
T.~Bartsch, M.~Bullimore, A.~E.~V. Ferrari, and J.~Pearson, {\it {Non-invertible Symmetries and Higher Representation Theory II}},  \href{http://arxiv.org/abs/2212.07393}{{\tt arXiv:2212.07393}}.

\bibitem{Bartsch:2022mpm}
T.~Bartsch, M.~Bullimore, A.~E.~V. Ferrari, and J.~Pearson, {\it {Non-invertible Symmetries and Higher Representation Theory I}},  \href{http://arxiv.org/abs/2208.05993}{{\tt arXiv:2208.05993}}.

\bibitem{Cuomo:2024psk}
G.~Cuomo, Y.-C. He, and Z.~Komargodski, {\it {Impurities with a cusp: general theory and 3d Ising}},  \href{http://arxiv.org/abs/2406.10186}{{\tt arXiv:2406.10186}}.

\bibitem{Carqueville:2018sld}
N.~Carqueville, I.~Runkel, and G.~Schaumann, {\it {Orbifolds of Reshetikhin-Turaev TQFTs}},  {\em Theor. Appl. Categor.} {\bf 35} (2020) 513--561, [\href{http://arxiv.org/abs/1809.01483}{{\tt arXiv:1809.01483}}].

\bibitem{Kaidi:2021gbs}
J.~Kaidi, Z.~Komargodski, K.~Ohmori, S.~Seifnashri, and S.-H. Shao, {\it {Higher central charges and topological boundaries in 2+1-dimensional TQFTs}},  {\em SciPost Phys.} {\bf 13} (2022), no.~3 067, [\href{http://arxiv.org/abs/2107.13091}{{\tt arXiv:2107.13091}}].

\bibitem{Kawagoe:2024tgv}
K.~Kawagoe, C.~Jones, S.~Sanford, D.~Green, and D.~Penneys, {\it {Levin-Wen is a gauge theory: entanglement from topology}},  \href{http://arxiv.org/abs/2401.13838}{{\tt arXiv:2401.13838}}.

\bibitem{bischoff2022computing}
M.~Bischoff and C.~Jones, {\it Computing fusion rules for spherical g-extensions of fusion categories},  {\em Selecta Mathematica} {\bf 28} (2022), no.~2 26.

\bibitem{Kapustin:2010if}
A.~Kapustin and N.~Saulina, {\it {Surface operators in 3d Topological Field Theory and 2d Rational Conformal Field Theory}},  \href{http://arxiv.org/abs/1012.0911}{{\tt arXiv:1012.0911}}.

\bibitem{Dijkgraaf:1989pz}
R.~Dijkgraaf and E.~Witten, {\it {Topological Gauge Theories and Group Cohomology}},  {\em Commun. Math. Phys.} {\bf 129} (1990) 393.

\bibitem{Inamura:2023ldn}
K.~Inamura and X.-G. Wen, {\it {2+1D symmetry-topological-order from local symmetric operators in 1+1D}},  \href{http://arxiv.org/abs/2310.05790}{{\tt arXiv:2310.05790}}.

\bibitem{Chatterjee:2022jll}
A.~Chatterjee, W.~Ji, and X.-G. Wen, {\it {Emergent generalized symmetry and maximal symmetry-topological-order}},  \href{http://arxiv.org/abs/2212.14432}{{\tt arXiv:2212.14432}}.

\bibitem{Chatterjee:2022tyg}
A.~Chatterjee and X.-G. Wen, {\it {Holographic theory for continuous phase transitions: Emergence and symmetry protection of gaplessness}},  {\em Phys. Rev. B} {\bf 108} (2023), no.~7 075105, [\href{http://arxiv.org/abs/2205.06244}{{\tt arXiv:2205.06244}}].

\bibitem{Ji:2019ebr}
W.~Ji and X.-G. Wen, {\it {Metallic states beyond the Tomonaga-Luttinger liquid in one dimension}},  {\em Phys. Rev. B} {\bf 102} (2020), no.~19 195107, [\href{http://arxiv.org/abs/1912.09391}{{\tt arXiv:1912.09391}}].

\bibitem{Ji:2019eqo}
W.~Ji and X.-G. Wen, {\it {Non-invertible anomalies and mapping-class-group transformation of anomalous partition functions}},  {\em Phys. Rev. Research.} {\bf 1} (2019) 033054, [\href{http://arxiv.org/abs/1905.13279}{{\tt arXiv:1905.13279}}].

\bibitem{Collier:2021ngi}
S.~Collier, D.~Mazac, and Y.~Wang, {\it {Bootstrapping boundaries and branes}},  {\em JHEP} {\bf 02} (2023) 019, [\href{http://arxiv.org/abs/2112.00750}{{\tt arXiv:2112.00750}}].

\bibitem{Wang:2013yta}
J.~Wang and X.-G. Wen, {\it {Nonperturbative regularization of (1+1)-dimensional anomaly-free chiral fermions and bosons: On the equivalence of anomaly matching conditions and boundary gapping rules}},  {\em Phys. Rev. B} {\bf 107} (2023), no.~1 014311, [\href{http://arxiv.org/abs/1307.7480}{{\tt arXiv:1307.7480}}].

\bibitem{Han:2017hdv}
B.~Han, A.~Tiwari, C.-T. Hsieh, and S.~Ryu, {\it {Boundary conformal field theory and symmetry protected topological phases in $2+1$ dimensions}},  {\em Phys. Rev. B} {\bf 96} (2017), no.~12 125105, [\href{http://arxiv.org/abs/1704.01193}{{\tt arXiv:1704.01193}}].

\bibitem{Jensen:2017eof}
K.~Jensen, E.~Shaverin, and A.~Yarom, {\it {\textquoteright{}t Hooft anomalies and boundaries}},  {\em JHEP} {\bf 01} (2018) 085, [\href{http://arxiv.org/abs/1710.07299}{{\tt arXiv:1710.07299}}].

\bibitem{Numasawa:2017crf}
T.~Numasawa and S.~Yamaguchi, {\it {Mixed global anomalies and boundary conformal field theories}},  {\em JHEP} {\bf 11} (2018) 202, [\href{http://arxiv.org/abs/1712.09361}{{\tt arXiv:1712.09361}}].

\bibitem{Smith:2020rru}
P.~B. Smith and D.~Tong, {\it {Boundary RG flows for fermions and the mod 2 anomaly}},  {\em SciPost Phys.} {\bf 10} (2021), no.~1 010, [\href{http://arxiv.org/abs/2005.11314}{{\tt arXiv:2005.11314}}].

\bibitem{Smith:2020nuf}
P.~B. Smith and D.~Tong, {\it {What Symmetries are Preserved by a Fermion Boundary State?}},  \href{http://arxiv.org/abs/2006.07369}{{\tt arXiv:2006.07369}}.

\bibitem{Thorngren:2020yht}
R.~Thorngren and Y.~Wang, {\it {Anomalous symmetries end at the boundary}},  {\em JHEP} {\bf 09} (2021) 017, [\href{http://arxiv.org/abs/2012.15861}{{\tt arXiv:2012.15861}}].

\bibitem{Tong:2021phe}
D.~Tong, {\it {Comments on symmetric mass generation in 2d and 4d}},  {\em JHEP} {\bf 07} (2022) 001, [\href{http://arxiv.org/abs/2104.03997}{{\tt arXiv:2104.03997}}].

\bibitem{Li:2022drc}
L.~Li, C.-T. Hsieh, Y.~Yao, and M.~Oshikawa, {\it {Boundary conditions and anomalies of conformal field theories in 1+1 dimensions}},  \href{http://arxiv.org/abs/2205.11190}{{\tt arXiv:2205.11190}}.

\bibitem{Zeng:2022grc}
M.~Zeng, Z.~Zhu, J.~Wang, and Y.-Z. You, {\it {Symmetric Mass Generation in the 1+1 Dimensional Chiral Fermion 3-4-5-0 Model}},  {\em Phys. Rev. Lett.} {\bf 128} (2022), no.~18 185301, [\href{http://arxiv.org/abs/2202.12355}{{\tt arXiv:2202.12355}}].

\bibitem{Wang:2022ucy}
J.~Wang and Y.-Z. You, {\it {Symmetric Mass Generation}},  {\em Symmetry} {\bf 14} (2022), no.~7 1475, [\href{http://arxiv.org/abs/2204.14271}{{\tt arXiv:2204.14271}}].

\bibitem{Putrov:2024uor}
P.~Putrov and R.~Radhakrishnan, {\it {Non-anomalous non-invertible symmetries in 1+1D from gapped boundaries of SymTFTs}},  \href{http://arxiv.org/abs/2405.04619}{{\tt arXiv:2405.04619}}.

\bibitem{Ostrik:2002ohv}
V.~Ostrik, {\it {Module categories over the Drinfeld double of a finite group}},  \href{http://arxiv.org/abs/math/0202130}{{\tt math/0202130}}.

\bibitem{Shen:2019wop}
C.~Shen and L.-Y. Hung, {\it {Defect Verlinde Formula for Edge Excitations in Topological Order}},  {\em Phys. Rev. Lett.} {\bf 123} (2019), no.~5 051602, [\href{http://arxiv.org/abs/1901.08285}{{\tt arXiv:1901.08285}}].

\bibitem{Petkova:2000ip}
V.~B. Petkova and J.~B. Zuber, {\it {Generalized twisted partition functions}},  {\em Phys. Lett. B} {\bf 504} (2001) 157--164, [\href{http://arxiv.org/abs/hep-th/0011021}{{\tt hep-th/0011021}}].

\bibitem{Perez-Lona:2023djo}
A.~Perez-Lona, D.~Robbins, E.~Sharpe, T.~Vandermeulen, and X.~Yu, {\it {Notes on gauging noninvertible symmetries. Part I. Multiplicity-free cases}},  {\em JHEP} {\bf 02} (2024) 154, [\href{http://arxiv.org/abs/2311.16230}{{\tt arXiv:2311.16230}}].

\bibitem{Perez-Lona:2024sds}
A.~Perez-Lona, D.~Robbins, E.~Sharpe, T.~Vandermeulen, and X.~Yu, {\it {Notes on gauging noninvertible symmetries, part 2: higher multiplicity cases}},  \href{http://arxiv.org/abs/2408.16811}{{\tt arXiv:2408.16811}}.

\bibitem{Robbins:2024tqf}
D.~Robbins and T.~Vandermeulen, {\it {The Fusion Categorical Diagonal}},  \href{http://arxiv.org/abs/2405.08058}{{\tt arXiv:2405.08058}}.

\bibitem{Choi:2023vgk}
Y.~Choi, D.-C. Lu, and Z.~Sun, {\it {Self-duality under gauging a non-invertible symmetry}},  {\em JHEP} {\bf 01} (2024) 142, [\href{http://arxiv.org/abs/2310.19867}{{\tt arXiv:2310.19867}}].

\bibitem{Fuchs:1997kt}
J.~Fuchs and C.~Schweigert, {\it {A Classifying algebra for boundary conditions}},  {\em Phys. Lett. B} {\bf 414} (1997) 251--259, [\href{http://arxiv.org/abs/hep-th/9708141}{{\tt hep-th/9708141}}].

\bibitem{Fukusumi:2021zme}
Y.~Fukusumi, Y.~Tachikawa, and Y.~Zheng, {\it {Fermionization and boundary states in 1+1 dimensions}},  {\em SciPost Phys.} {\bf 11} (2021), no.~4 082, [\href{http://arxiv.org/abs/2103.00746}{{\tt arXiv:2103.00746}}].

\bibitem{Ebisu:2021acm}
H.~Ebisu and M.~Watanabe, {\it {Fermionization of conformal boundary states}},  {\em Phys. Rev. B} {\bf 104} (2021), no.~19 195124, [\href{http://arxiv.org/abs/2103.01101}{{\tt arXiv:2103.01101}}].

\bibitem{Felder:1999cv}
G.~Felder, J.~Frohlich, J.~Fuchs, and C.~Schweigert, {\it {Conformal boundary conditions and three-dimensional topological field theory}},  {\em Phys. Rev. Lett.} {\bf 84} (2000) 1659--1662, [\href{http://arxiv.org/abs/hep-th/9909140}{{\tt hep-th/9909140}}].

\bibitem{Lin:2021udi}
Y.-H. Lin and S.-H. Shao, {\it {$\mathbb{Z}_N$ symmetries, anomalies, and the modular bootstrap}},  {\em Phys. Rev. D} {\bf 103} (2021), no.~12 125001, [\href{http://arxiv.org/abs/2101.08343}{{\tt arXiv:2101.08343}}].

\bibitem{KRW}
J.~Kulp, B.~C. Rayhaun, and Y.~Wang In preparation.

\bibitem{Kusuki:2023bsp}
Y.~Kusuki, S.~Murciano, H.~Ooguri, and S.~Pal, {\it {Symmetry-resolved entanglement entropy, spectra \& boundary conformal field theory}},  {\em JHEP} {\bf 11} (2023) 216, [\href{http://arxiv.org/abs/2309.03287}{{\tt arXiv:2309.03287}}].

\bibitem{Fosbinder-Elkins:2024hff}
H.~Fosbinder-Elkins and J.~A. Harvey, {\it {Modular invariance groups and defect McKay-Thompson series}},  \href{http://arxiv.org/abs/2408.16263}{{\tt arXiv:2408.16263}}.

\bibitem{MRb}
S.~Möller and B.~C. Rayhaun, ``{Equivalence Relations on Vertex Operator Algebras, II: Witt Equivalence and Orbifolds}.'' {In preparation}, 2024.

\bibitem{Bae:2020pvv}
J.-B. Bae, J.~A. Harvey, K.~Lee, S.~Lee, and B.~C. Rayhaun, {\it {Conformal Field Theories with Sporadic Group Symmetry}},  {\em Commun. Math. Phys.} {\bf 388} (2021), no.~1 1--105, [\href{http://arxiv.org/abs/2002.02970}{{\tt arXiv:2002.02970}}].

\bibitem{Evans:2018qgz}
D.~E. Evans and T.~Gannon, {\it {Reconstruction and local extensions for twisted group doubles, and permutation orbifolds}},  {\em Trans. Am. Math. Soc.} {\bf 375} (2022), no.~04 2789--2826, [\href{http://arxiv.org/abs/1804.11145}{{\tt arXiv:1804.11145}}].

\bibitem{Seifnashri:2024dsd}
S.~Seifnashri and S.-H. Shao, {\it {Cluster State as a Noninvertible Symmetry-Protected Topological Phase}},  {\em Phys. Rev. Lett.} {\bf 133} (2024), no.~11 116601, [\href{http://arxiv.org/abs/2404.01369}{{\tt arXiv:2404.01369}}].

\bibitem{Cordova:2023bja}
C.~Cordova, P.-S. Hsin, and C.~Zhang, {\it {Anomalies of Non-Invertible Symmetries in (3+1)d}},  \href{http://arxiv.org/abs/2308.11706}{{\tt arXiv:2308.11706}}.

\bibitem{Inamura:2021wuo}
K.~Inamura, {\it {Topological field theories and symmetry protected topological phases with fusion category symmetries}},  {\em JHEP} {\bf 05} (2021) 204, [\href{http://arxiv.org/abs/2103.15588}{{\tt arXiv:2103.15588}}].

\bibitem{Bhardwaj:2024qrf}
L.~Bhardwaj, D.~Pajer, S.~Schafer-Nameki, and A.~Warman, {\it {Hasse Diagrams for Gapless SPT and SSB Phases with Non-Invertible Symmetries}},  \href{http://arxiv.org/abs/2403.00905}{{\tt arXiv:2403.00905}}.

\bibitem{tambara2000representations}
D.~Tambara, {\it Representations of tensor categories with fusion rules of self-duality for abelian groups},  {\em Israel Journal of Mathematics} {\bf 118} (2000) 29--60.

\end{thebibliography}\endgroup

\end{document}